\documentclass[12pt]{article}

\usepackage{graphicx}
\graphicspath{{figures/}}

\usepackage[export]{adjustbox}
\usepackage{float}

\usepackage{enumitem}
\usepackage{epsfig,amssymb,euscript}
\usepackage{amsmath,amsfonts,bbm,mathtools}

\usepackage{cite}

\usepackage{subcaption}

\allowdisplaybreaks[4]        

\usepackage[dvipsnames]{xcolor}

\usepackage{hyperref}
\usepackage{fancyhdr}
\usepackage[header,title,page,titletoc]{appendix}  
\DeclareMathOperator\arctanh{arctanh}

\newcommand{\ket}[1]{\left|#1\right>}

\newcommand{\cev}[1]{\reflectbox{\ensuremath{\vec{\reflectbox{\ensuremath{#1}}}}}}

\renewcommand{\(}{\left(}
\renewcommand{\)}{\right)}
\renewcommand{\[}{\left[}
\renewcommand{\]}{\right]}

\newcommand{\pr}[1]{\left(#1\right)}
\newcommand{\pq}[1]{\left[#1\right]}
\newcommand{\pg}[1]{\left\{#1\right\}}

\newcommand{\eg}{{\it e.g.,}\ }
\newcommand{\ie}{{\it i.e.,}\ }

\def\be{\begin{equation}}
\def\ee{\end{equation}}
\def\bea{\begin{eqnarray}}
\def\eea{\end{eqnarray}}

\def\ba{\begin{eqnarray}}
\def\ea{\end{eqnarray}}

\def\a{\alpha}
\def\b{\beta}

\def\d{\delta}
\def\D{\Delta}

\def\m{\mu}

\def\o{\omega}
\def\O{\Omega}
\def\r{\rho}
\def\s{\sigma}
\def\S{\Sigma}

\def\th{\theta}

\def\pa{\partial}

\def\del{\partial}

\numberwithin{equation}{section}

\begin{document}
\baselineskip=15.5pt
\pagestyle{plain}
\setcounter{page}{1}
\newfont{\namefont}{cmr10}
\newfont{\addfont}{cmti7 scaled 1440}
\newfont{\boldmathfont}{cmbx10}
\newfont{\headfontb}{cmbx10 scaled 1728}
\renewcommand{\theequation}{{\rm\thesection.\arabic{equation}}}
\renewcommand{\thefootnote}{\arabic{footnote}}

\vspace{1cm}
\begin{titlepage}
\vskip 2cm
\begin{center}
{\Huge{\bf Holographic and QFT Complexity \\ with angular momentum}}
\end{center}

\vskip 10pt
\begin{center}
{Alice Bernamonti$^{1,2}$, Francesco Bigazzi$^{2}$, Davide Billo$^{1,2}$, Lapo Faggi$^{1,3}$, Federico Galli$^{2,4}$}
\end{center}
\vskip 10pt
\begin{center}
\vspace{0.2cm}
\textit{$^1$ Dipartimento di Fisica e Astronomia, Universit\'a di Firenze; Via G. Sansone 1; \\I-50019 Sesto Fiorentino, Italy}\\
\textit {$^2$ INFN, Sezione di Firenze; Via G. Sansone 1; I-50019 Sesto Fiorentino, Italy}\\
\textit{$^3$ Dipartimento di Ingegneria dell'Informazione, Universit\'a di Firenze; V.le Morgagni 46; I-50134 Firenze, Italy}\\
\textit{$^4$ ISC-CNR; Via Madonna del Piano 10, I-50019 Sesto Fiorentino, Italy}
\vskip 20pt
{\small{alice.bernamonti@unifi.it, bigazzi@fi.infn.it, davide.billo@unifi.it, \\lapo.faggi@unifi.it, federico.galli@fi.infn.it}}

\end{center}

\vspace{22pt}

\begin{center}
\textbf{Abstract}
\end{center}

\noindent 
We study the influence of angular momentum on quantum complexity for CFT states holographically dual to rotating black holes. Using the holographic \textit{complexity=action} (CA) and \textit{complexity=volume} (CV) proposals, we study the full time dependence of complexity and the complexity of formation for two dimensional states dual to rotating BTZ.  
The obtained results and their dependence on angular momentum turn out to be analogous to those of charged states dual to Reissner-Nordstr\"om AdS black holes.  
For CA, our computation carefully accounts for the counterterm in the gravity action, which was not included in previous analysis in the literature. This affects the complexity early time dependence and its effect becomes negligible close to extremality. In the grand canonical ensemble, the CA and CV complexity of formation are linear in the temperature, and diverge with the same structure in the speed of light angular velocity limit. For CA the inclusion of the counterterm is crucial for both effects. We also address the problem of studying holographic complexity for higher dimensional rotating black holes, focusing on the four dimensional Kerr-AdS case. Carefully taking into account all ingredients, we show that the late time limit of the CA growth rate saturates the expected bound, and find the CV complexity of formation of large black holes diverges in the critical angular velocity limit. Our holographic analysis is complemented by the study of circuit complexity in a two dimensional free scalar model for a thermofield double (TFD) state with angular momentum. We show how this can be given a description in terms of non-rotating TFD states introducing mode-by-mode effective temperatures and times. We comment on the similarities and differences of the holographic and QFT complexity results.
\end{titlepage}
\newpage

\tableofcontents

\section{Introduction}\label{sec:intro}

In the growing connection between holography and quantum information theory, quantum computational complexity was proposed as a new measure to capture more information about the bulk spacetime than  holographic entanglement entropy alone \cite{Susskind:2014moa}. Quantum circuit complexity in particular is a measure of how difficult it is to construct a given target state from a (simple) reference state by applying a set of elementary gates, see \eg \cite{watrous2018theory,Aaronson:2016vto}. 
A variety of proposals for the bulk description of the complexity of boundary states have been advanced. The most studied holographic complexity notions are the so called \textit{complexity=volume} (CV) \cite{Susskind:2014rva,Stanford:2014jda} and \textit{complexity=action} (CA) \cite{Brown:2015bva,Brown:2015lvg}. The CV conjecture states that complexity is dual to the volume of the extremal codimension-1 bulk surface  $\mathcal{B}$ anchored to the boundary  time slice $\Sigma$  on which the state is defined,
\be \label{eq:CV}
C_{\rm V} = \max_{\partial \mathcal{ B}= \Sigma} \left[ \frac{V ({\cal B})}{G_N \ell_{\rm bulk}}\right]\,.
\ee
Here $\ell_{\rm bulk}$ is an additional length scale associated with the bulk geometry, \eg see \cite{Brown:2015bva,Couch:2018phr}. For simplicity in the main text we will set $\ell_{\rm bulk} = \ell$, the curvature radius of the asymptotically AdS geometry. The CA conjecture proposes instead that complexity is given by the gravitational action evaluated on a region of spacetime, the Wheeler-DeWitt (WDW) patch, that is the causal development of a spacelike bulk surface anchored on the boundary time slice $\Sigma$. Explicitly: 
\be \label{eq:CA}
C_{\rm A} = \frac{I_{\rm WDW}}{\pi}\,. 
\ee
The precise form of the gravitational action $I_{\rm WDW}$ in such proposal was carefully worked out in \cite{Lehner:2016vdi}. In particular this work introduced a counterterm contribution to ensure the full action is invariant under reparametrizations of the WDW null boundaries.\footnote{Recently, an alternative proposal to fix the normalization of the null boundaries by requiring the complexity of the vacuum state to vanish was put forward in \cite{Mounim:2021bba,Mounim:2021ykr}. }

The CV and CA conjectures stimulated an extensive effort aimed at investigating properties of these new gravitational observables and  at testing the validity of the proposals  \cite{Susskind:2014rva,Stanford:2014jda,Brown:2015bva,Brown:2015lvg,Couch:2018phr,Couch:2016exn,Susskind:2014jwa,Susskind:2015toa,Roberts:2014isa,Lehner:2016vdi,Mounim:2021bba,Mounim:2021ykr,Cai:2016xho,Reynolds:2016rvl,Chapman:2016hwi,Carmi:2016wjl,Moosa:2017yvt,Couch:2017yil,Cai:2017sjv,Brown:2017jil,Carmi:2017jqz,Swingle:2017zcd,Flory:2017ftd,Zhao:2017isy,Abt:2017pmf,Czech:2017ryf,Abt:2018ywl,An:2018xhv,Fu:2018kcp,Chapman:2018dem,Chapman:2018lsv,Mahapatra:2018gig,Barbon:2018mxk,Susskind:2018fmx,Susskind:2018tei,Cooper:2018cmb,Numasawa:2018grg,Brown:2018kvn,Goto:2018iay,Agon:2018zso,Chapman:2018bqj,Flory:2018akz,Flory:2019kah,Braccia:2019xxi,Sato:2019kik,Barbon:2015ria,Barbon:2015soa,Auzzi:2018zdu,Auzzi:2018pbc,Bhattacharya:2019zkb,Ghosh:2019jgd,Bernamonti:2019zyy,Bernamonti:2020bcf,Bhattacharya:2020uun,Hernandez:2020nem,Bhattacharya:2021jrn,Sato:2021ftf,Susskind:2019ddc,Barbon:2019tuq,Barbon:2020uux,Belin:2018bpg,Belin:2020oib,Hashimoto:2021umd,Iliesiu:2021ari,Chen:2020nlj,Chandra:2021kdv,Pedraza:2021mkh,Pedraza:2021fgp,Auzzi:2019fnp,Auzzi:2019mah,Auzzi:2019vyh,Auzzi:2021nrj,Barbon:2019xwc,Bolognesi:2018ion,Akhavan:2019zax,Omidi:2020oit,Yekta:2020wup}. In parallel, various approaches have been explored to define and understand the complexity of states in quantum field theory, \eg  following Nielsen's geometric approach \cite{nielsen2006quantum,nielsen2008,Nielsen:2006,Jefferson:2017sdb,Chapman:2018hou} which we review in sec.~\ref{sec:QFT}, the Fubini-Study metric approach for the space of states \cite{Chapman:2017rqy}, path integral optimization \cite{Miyaji:2016mxg,Caputa:2017urj,Caputa:2017yrh,Bhattacharyya:2018wym,Caputa:2020mgb,Boruch:2020wax,Boruch:2021hqs,Camargo:2019isp,Bhattacharyya:2019kvj}, or CFT notions of complexity \cite{Caputa:2018kdj,Magan:2018nmu,Erdmenger:2020sup,Flory:2020dja,Flory:2020eot,Bueno:2019ajd,Chagnet:2021uvi,Akhavan:2019zax}. 

Most of this research considered highly symmetric setups, mainly planar or spherically symmetric, and only few results are available in the literature for less symmetric settings. Examples include local quenches \cite{Ageev:2018nye,Ageev:2019fxn,DiGiulio:2021noo} and setups with defects \cite{Chapman:2018bqj,Braccia:2019xxi,Sato:2019kik}. 
In this work we focus on systems with rotation, which are so far less understood, and explore the influence of angular momentum on quantum complexity in holography and in QFT.  

The first estimates of the late time holographic complexity growth rate for rotating AdS black holes appeared in \cite{Brown:2015lvg,Cai:2016xho}, before a complete understanding of how to treat the action contributions of null boundaries was developed in \cite{Lehner:2016vdi}. 
For lower dimensional black holes, this computation was revisited in \cite{Auzzi:2018zdu,Auzzi:2018pbc}. They  studied  the CV and CA growth rate in three dimensional warped AdS black holes, which include rotating BTZ as a subcase. Properties of holographic complexity for exotic BTZ black holes were instead studied in \cite{Frassino:2019fgr}. 
For higher dimensional rotating black holes the technical task of evaluating CV and CA is much harder and has hindered progress until much recently. One  main obstruction resides in obtaining the explicit form of the WDW patch, necessary to evaluate CA. In fact, the first analysis of the null hypersurfaces foliation of Kerr-AdS appeared only recently in \cite{Balushi:2019pvr,Imseis:2020vsw}. A notable exception is given by odd-dimensional Myers-Perry AdS black holes with equal angular momenta in each orthogonal plane. These exhibit a symmetry enhancement that greatly simplifies the computations as compared to the general rotating solutions.  Their holographic complexity was studied in \cite{Balushi:2020wjt,Balushi:2020wkt}.  For large black holes,  \cite{Balushi:2020wjt,Balushi:2020wkt} highlighted a direct connection between CA, CV and thermodynamic volume.\footnote{See also \cite{Andrews:2019hvq} for  a related observation.}  In particular, it was argued that the complexity of formation is controlled by the thermodynamic volume rather than the entropy, with a scaling that depends on the spacetime dimension.  For the growth rate at late times, taking the large black hole limit while keeping fixed the ratio of the radial locations of the inner and outer horizons, it was shown in \cite{Balushi:2020wjt,Balushi:2020wkt} that at leading (divergent) order 
 \be
 \lim_{t \to \infty}    \frac{ d  C }{ dt } \propto   P\D V  \label{eq:manngrowth}
\ee
where $\D V  =  V_+ - V_-$ is the difference between the inner and outer horizon thermodynamic volume. 
 
In this work we expand the existing studies of quantum complexity for CFT states dual to rotating black holes, focusing on the two and three dimensional cases. On the holographic side,  we refine and complement previous analysis for the rotating BTZ black holes and address the higher dimensional case of Kerr-AdS$_4$. Differently  from the Myers-Perry AdS black hole in odd dimension considered in \cite{Balushi:2020wjt,Balushi:2020wkt}, the latter has no symmetry enhancement. Next to this we study  circuit complexity in a two dimensional free scalar model for a thermofield double (TFD) state with angular momentum. 

For the rotating BTZ black hole solution we analyze the full time dependence of complexity and the complexity of formation using the CA, CV and CV2.0 proposals, extending previous results, which mostly focused on the growth rate.
In particular for CA we also carefully take into account the role of the counterterm action, which was not included in previous analysis and has been shown to play an essential role in order to reproduce some expected features of complexity, see \eg \cite{Chapman:2018lsv}. 
We find that the counterterm affects the complexity of formation and the early time evolution of complexity.  The effects of the inclusion of the counterterm are more evident for smaller values of the angular momentum $J$, compatibly with results obtained for neutral BTZ \cite{Carmi:2017jqz}, while they become negligible as extremality is approached. The counterterm also turns out to be essential to have a matching behavior of CA and CV viewed as functions of the temperature $T$ and angular velocity $\Omega$. At the qualitative level our analysis shows that the inclusion of a rotation parameter in the BTZ solution gives for the corresponding holographic complexity a behavior similar to the one of higher dimensional charged Reissner-Nordstr\"om AdS black holes \cite{Carmi:2017jqz}.
 
In the four dimensional Kerr-AdS case we are able to make partial progress. The recent analysis \cite{Balushi:2019pvr,Imseis:2020vsw} of null hypersurfaces allows to give a description of the WDW patch only in an implicit form. As we show, this however suffices to provide a precise treatment of the late time limit of the CA growth rate. Carefully taking into account all terms  necessary to give a precise definition of the action on the WDW patch  \cite{Lehner:2016vdi}, we explicitly demonstrate that the CA growth rate saturates the bound advanced in \cite{Cai:2016xho}. We comment on the relation of our results with the observations about complexity and thermodynamic volume highlighted in  \cite{Balushi:2020wjt,Balushi:2020wkt}. 

An interesting limit to consider when analyzing spinning black holes is the one of critical angular velocity, $\Omega =1$ \cite{Hawking:1998kw}. We analyze the behavior of the different holographic complexity measures in this limit. While for BTZ this limit always gives a divergent behavior, for Kerr-AdS$_4$ the physical parameter space has a richer structure and yields to divergences in the critical velocity limit only for large black holes. 

The holographic setups we analyze are related to thermofield double states of holographic CFTs. 
Focusing on the lower dimensional BTZ case, the irrotational double sided BTZ black hole is dual to the familiar TFD state of boundary CFTs associated to the two sides of the geometry \cite{Maldacena:2001kr,Hartman:2013qma}. Analogously, the rotating BTZ black hole has a dual in the rotating TFD (rTFD) state prepared with the deformed Hamiltonian $\beta\left(H+\Omega J \right)$  \cite{Maldacena:2001kr}.  

We complement our holographic analysis with the study of circuit complexity in the simple model provided by a Gaussian rTFD constructed from two copies of a two dimensional free scalar QFT.
To perform our analysis we follow the QFT approach to complexity first  put forward in \cite{Jefferson:2017sdb}, based on Nielsen's geometric approach \cite{nielsen2006quantum,nielsen2008,Nielsen:2006}. We establish a correspondence between  our rTFD setup and the TFD state that allows us to evaluate quantum circuit complexity starting from and generalizing the TFD complexity analysis performed in \cite{Chapman:2018hou}. 
Despite dealing with a Gaussian model rather than a strongly coupled holographic CFT, we find similarities with the results obtained for holographic complexity in the complexity of formation in particular divergent limits. 

This manuscript is organized as follows. In sec.~\ref{sec:HCBTZ} we study holographic complexity for rotating BTZ solutions. We perform a thorough analysis of the effect of angular momentum on both complexity of formation and complexity full time evolution, mainly focusing on the CA and CV proposals. We extend the analysis of the CA growth rate and CV complexity of formation to Kerr-AdS black holes in sec.~\ref{sec:HCKerr}. In \ref{sec:QFT}, we parallel the holographic discussion by studying circuit complexity of a rotating TFD state within a simple free scalar field model. We summarize and discuss the main findings relating holographic and QFT complexity measures in section \ref{sec:Discussion}. Some technical details are presented in appendices \ref{app:BTZBL} and \ref{app:div}, and additional plots in appendix \ref{app:plots}. 

\textbf{Note}: Part of the analysis for the holographic complexity of rotating BTZ solutions is contained in Lapo Faggi's M. Sc. Thesis ``Holographic complexity of rotating black holes'' discussed in July 2019 at the University of Florence. 

\section{Holographic complexity: BTZ}\label{sec:HCBTZ}
 
The main goal of this section is to provide a thorough analysis for holographic complexity for the rotating BTZ solution reviewed in sec.~\ref{sec:BTZBH}. Holographic complexity for charged BTZ black holes was studied in \cite{Carmi:2017jqz}. The angular momentum provides an extra parameter against which the holographic proposals CA (sec.~\ref{sec:CABTZ}) and CV (sec.~\ref{sec:CVBTZ}) can be checked. Early results for the late time complexity growth rate in rotating BTZ appeared in \cite{Brown:2015lvg,Cai:2016xho}, before a consistent prescription for defining the gravitational action in presence of null boundaries was developed in \cite{Lehner:2016vdi}. The works \cite{Auzzi:2018zdu,Auzzi:2018pbc} and \cite{Frassino:2019fgr} --focusing respectively on warped and exotic AdS$_3$  black holes and thus including rotating BTZ as a subcase-- partially overlap with ours for what concerns the complexity growth rate. In the CA case, we carefully take into account the effect of the counterterm \cite{Lehner:2016vdi}. We also analyze other aspects, as the total complexity and its time evolution, the complexity of formation and its dependence on the angular momentum, as well as the CV 2.0 proposal \cite{Couch:2016exn}.
In view of making contact with the QFT analysis of sec.~\ref{sec:QFT}, we discuss the holographic findings also in the grand canonical ensemble. 

\subsection{BTZ black hole}\label{sec:BTZBH}
 
The BTZ black hole metric  is \cite{Banados:1992wn} (see, \eg \cite{Compere:2018aar} for a review)
\begin{equation}\label{metrica}
ds^2= -f(r)dt^2+\frac{dr^2}{f(r)}+r^2\left(d\varphi- \omega(r)\, dt\right)^2\,,
\end{equation}
where
\bea
f(r) &=&-8 M G_N + \frac{r^2}{\ell^2}+\frac{16 \, G_N^2 J^2}{r^2} \equiv \frac{(r^2-r_+^2)(r^2-r_-^2)}{\ell^2 \, r^2}  \label{blackening factor} \\
 \omega(r) &= & \frac{4 \, G_N J}{r^2} \equiv  \frac{r_+ r_-}{\ell \, r^2} \,, \label{eq:BTZangularvelocity}
\eea
and $\varphi\sim \varphi+2\pi$. The metric is stationary, axially symmetric and asymptotically AdS$_3$ with radius $\ell$. $M$ and $J$ denote the ADM mass and angular momentum of the solution, which we express in terms of the radii $r_{\pm}$ as\footnote{Given  the symmetry of the solution, we here take for simplicity to $J \ge 0$. }
\begin{equation}\label{MandJ}
M=\frac{r_+^2 + r_-^2}{8 \, G_N \, \ell^2},\qquad \qquad J=\frac{r_+ r_-}{4 \, G_N \, \ell}\,.
\end{equation} 
For $J<M \ell$ and $M>0$ the metric describes a black hole  with horizons
\begin{equation}\label{Acaso6}
r_{\pm}=2 \ell \sqrt{ G_N M }\sqrt{1\pm\sqrt{1-\left(\frac{J}{M \ell}\right)^2}}\,. 
\end{equation}
The surface $r=r_+$ is an event horizon shielding a causal singularity at $r=0$, and $r= r_-$ a Cauchy horizon. 
In \eqref{metrica}, $\omega(r)$ denotes the angular velocity of the solution, and the coordinates system is asymptotically non rotating, \textit{i.e.}, $\omega \sim 0 $ for $r\to \infty$. The $g_{tt}$ component of the metric vanishes at the critical radius $r_{\rm erg} \equiv \ell\sqrt{8G_N M} = \sqrt{r_+^2 + r_-^2}$, with $r_-<r_+<r_{\rm erg}$. This defines the ergo-region and in its interior $\partial_t$ is spacelike: in this region observers are necessarily dragged along by the black hole rotation.

The BTZ Hawking temperature, entropy and angular velocity of the event horizon are
\bea
T &=&\frac{r_+^2-r_-^2}{2\pi \, \ell^2 \, r_+} \label{temperatura} \\
S &=&\frac{ \pi \, r_+}{2 \, G_N} \\
\Omega_H &\equiv & \omega(r_+) = \frac{r_-}{\ell r_+} \label{OmegaH} \, .
\eea
Since $\omega(r) \to 0$ as $r \to \infty$, the latter is also exactly the angular velocity of the rotating Einstein universe conformal to the AdS boundary. 
The limiting case $J=M \ell$, for which $T = 0$ and $\ell \, \Omega_H=1$, describes an extremal black hole, whose Einstein universe at infinity effectively rotates at the speed of light. The $M= J=0$ solution is instead known as the ``zero mass black hole''.  Empty global AdS$_3$ is recovered setting $J=0, M = -1/(8G_N)$, and in the following we will study variations of holographic complexity with respect to this vacuum solution. The metric with $J\neq 0, M = -1/(8G_N)$ also parametrizes global AdS$_3$, but in oblate coordinates.

A useful parametrization of the BTZ geometry is given by the ingoing/outgoing Eddington-Finkelstein coordinates
\begin{equation}\label{eq:EF}
\left\{
\begin{aligned}
& v=t+r^*(r)\\
 &\Phi =\varphi +\tilde{r}(r)\\
\end{aligned}
\right.
\qquad \text{ and }\qquad
\left\{
\begin{aligned}
&u=t-r^*(r)\\
&\Psi =\varphi -\tilde{r}(r)\\
\end{aligned}
\right. ,
\end{equation}
where the tortoise coordinates are defined by
\begin{equation}\label{tortoise}
\begin{aligned}
&\frac{d r^*(r)}{dr}= \frac{1}{f(r)} \qquad \qquad\frac{d \tilde{r}(r)}{d r}=\frac{\omega(r)}{f(r)} 
\end{aligned}\, .
\end{equation}
They can be worked out explicitly
\bea
\label{rtartaruga}
r^*(r) &=&\frac{\ell^2}{2(r_-^2-r_+^2)}\left(r_+\log \frac{r+r_+}{|r-r_+|} -r_-\log \frac{r+r_-}{|r-r_-|} \right)  \\
\tilde{r}(r)&=&\frac{\ell^2}{2(r_-^2-r_+^2)}\left(r_-\log \frac{r+r_+}{|r-r_+|} -r_+\log \frac{r+r_-}{|r-r_-|} \right)\, 
\eea
and satisfy $r^*(\infty) = \tilde{r}(\infty)=0$.  The ingoing coordinates metric reads
\begin{equation}\label{metricEF}
ds^2=-f(r)dv^2+2 \, dr dv+r^2\left(d\Phi- \omega(r) dv\right)^2 
\end{equation}
and is regular across the outer event  horizon. Given $\varphi\sim\varphi+2 \pi$, we also have $\Phi\sim\Phi+2\pi$.

\begin{figure}[]
\centering
\includegraphics[width=.3\textwidth]{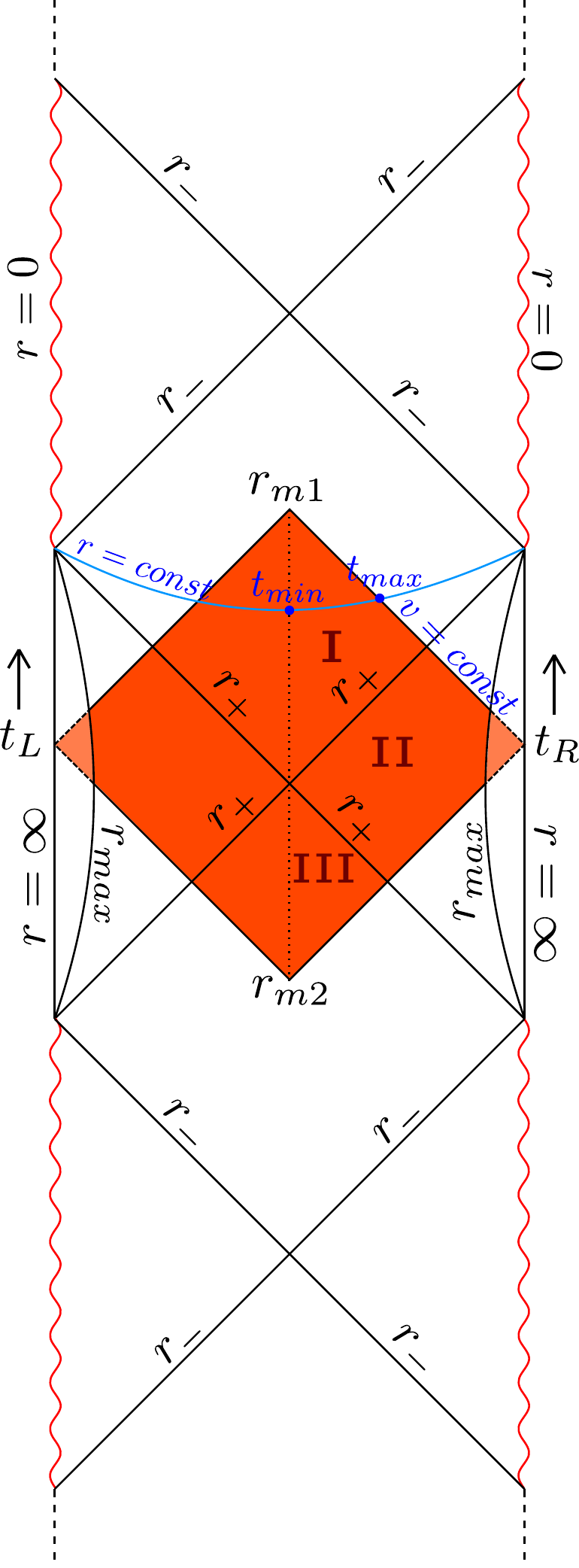} \qquad  \qquad\qquad\qquad  \includegraphics[width=0.3\textwidth]{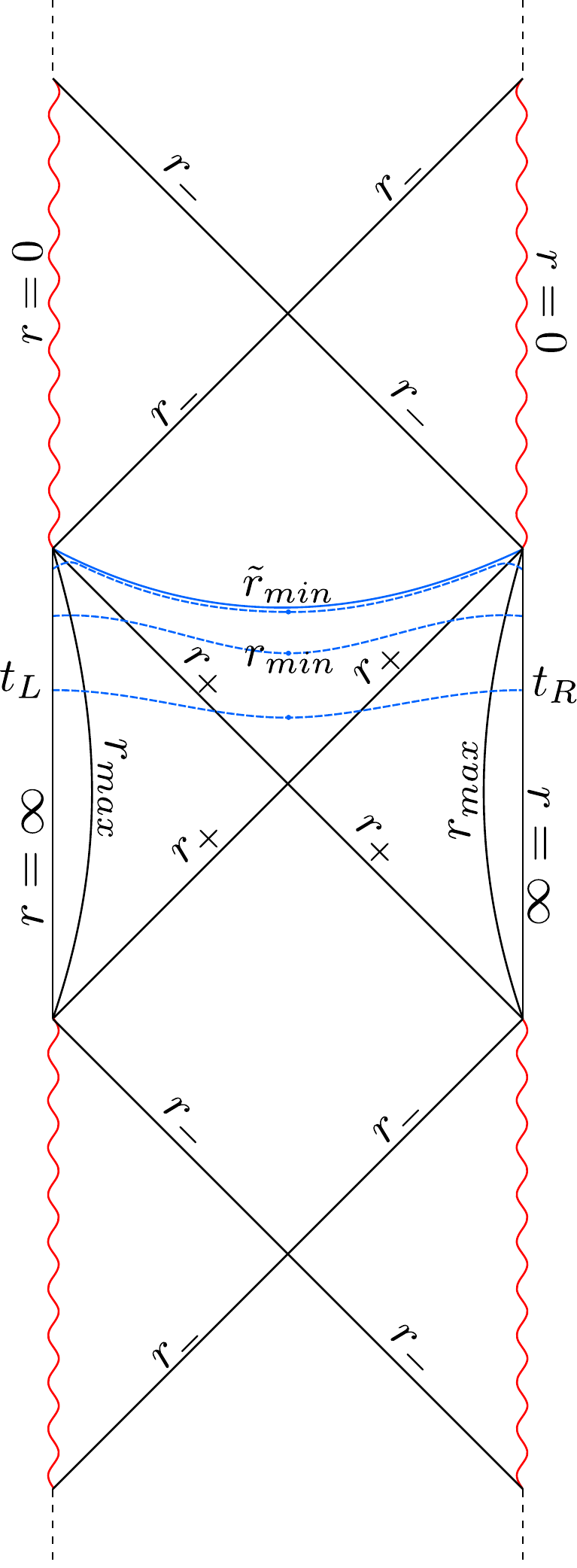}
\caption{Penrose diagram of rotating BTZ.  Left: The WDW patch of a boundary time slice with $t_L = t_R$ (see sec.~\ref{sec:WDW} ). $r_{\rm max}$ denotes the radial bulk cutoff.  By symmetry $t_{min}=0$.  Right: The blue slices are the maximal ones of the CV conjecture. The minimal radius $r_{min}$ is defined through $\left.\dot{r}\right|_{r_{min}}=0$. For $t_b\to \infty$, the maximal bulk slice is at constant $r =\tilde{r}_{min}$. }
 \label{figuraPenrose1}
\end{figure}

The BTZ black hole causal structure is depicted in the Penrose diagram \ref{figuraPenrose1}. Since we are here interested in a time-dependent gravitational system dual to pure states in CFT undergoing thermalization, we choose to time-evolve forward in both exterior regions \cite{Hartman:2013qma}.
%

\subsubsection{Wheeler-DeWitt patch}\label{sec:WDW}
The Wheeler-DeWitt patch is defined as the domain of dependence of any bulk codimension-1 spacelike slice anchored on a given boundary time slice, and is bounded by null surfaces. For the rotating BTZ these are described by the congruences of null geodesics $ v=\textrm{const.}\,, \Phi=\textrm{const.} $ and $  u=\textrm{const.}\,,\Psi=\textrm{const.}$  defined in (\ref{eq:EF}), which are twist-free and  ``surface forming'', \textit{i.e.},  are generators of a family of null hypersurfaces \cite{Townsend:1997ku}.   
The corresponding  WDW patch is drawn in fig.~\ref{figuraPenrose1} and is analogous to that of charged black holes considered in \cite{Carmi:2017jqz}.  We choose a symmetric time evolution $t_L=t_R=\frac{t_b}{2}$ and focus our attention on $t_b>0$.\footnote{Notice we define the boundary time $t_b$ at $r=\infty$, while \cite{Auzzi:2018zdu,Auzzi:2018pbc} define it at the cutoff surface $r=r_{\rm max}$.}  We denote the future (past) tip of the WDW patch $r_{m1}$ ($r_{m2}$). Using the right-left symmetry one can focus on the right boundary and observe that the tips lie on the same constant $v$ (respectively $u$) slices as $(\frac{t_b}{2}, r = \infty)$ and thus  
\begin{equation}\label{eq:rm1rm2}
\frac{t_b}{2} = r^*(r_{m1}) \,, \qquad\qquad \frac{t_b}{2} = - r^*(r_{m2}) \, . 
\end{equation}
Given \eqref{rtartaruga}, these can be solved numerically as a function of boundary time $t_b$, and also imply
\begin{equation}\label{eq:derrm1rm2}
\frac{d r_{m1} }{d t_b}=
\frac {f(r_{m1})}{2}\,, \qquad\qquad \frac{d r_{m2} }{d t_b}=
- \frac {f(r_{m2})}{2}\,.
\end{equation}
Focusing for example on the right future boundary,  the normal one-form to the null surface in  Schwarzschild-like coordinates is
\be
k_\mu = \alpha \partial_\mu v= \left(\alpha\,,\,\frac{\alpha}{f}\,,\,0\right) \, . 
\ee
with $\alpha$  an arbitrary normalization constant.  Via the relation $\del_\lambda  \equiv k^{\mu} \del_\mu$  this defines  a parametrization  of the null generators  of the WDW boundary, which is affine $k^\mu\nabla_\mu k^\nu = 0$.
The unit-normalized spacelike vector parametrizing the transverse direction is instead 
\begin{equation}
e^\mu=(0,0,1) \, 
\end{equation}
and such that $e^\mu k_{\mu} =0$.  Analogous expressions hold for the other null surfaces bounding the WDW patch.

\subsection{Complexity=Action}\label{sec:CABTZ}

We start our analysis studying in more details the  holographic complexity$=$action proposal \eqref{eq:CA} of   \cite{Brown:2015bva,Brown:2015lvg}, which  entails evaluating the action %
\bea  
I_{\rm WDW} &= & I_{\rm bulk} + I_{\rm GHY} + I_{\rm joints} + I_{\kappa} + I_{\rm ct} \nonumber  \\
&= & \frac{1}{16 \pi G_N} \int d^{d+1} x \,\sqrt{|g|}\, \left(R + \frac{d(d-1)}{\ell^2}\right) + \frac{1}{8 \pi G_N}\int_{\rm regulator} d^d y \,\sqrt{|h|}\, K \nonumber \\
&& + \frac{1}{8 \pi G_N} \int_{\rm joints} d^{ d-1} y \, \sqrt{\sigma}\, a_{\rm joint} + \frac{1}{8 \pi G_N }\int_{\partial {\rm WDW}} d\lambda\, d^{d-1} y\, \sqrt{\gamma}\, \kappa \label{eq:graction} \\
&&+ \frac{1}{8 \pi G_N} \int_{\partial {\rm WDW}} d \lambda\, d^{d-1} y\, \sqrt{\gamma}\, \Theta\, \log\left( L_{\rm ct} \Theta \right) \,. \nonumber
\eea
This includes: $I_{\rm bulk}$, the Einstein-Hilbert action with negative cosmological constant and $I_{\rm GHY}$, the Gibbons-Hawking-York term defined on the AdS boundary regulator surface. In the second line: $I_{\rm joints}$, the contribution of the intersection of the null boundaries of the WDW patch with other hypersurfaces (which we specify better below), and $I_{\kappa}$, which has support on the null boundaries of the WDW patch and vanishes when these are affinely parameterized, as in our case. 
The term in the last line $I_{\rm ct}$ is known as the \textit{counterterm}  \cite{Lehner:2016vdi}. It is also localized on the boundary of the WDW patch and is expressed in terms of $\Theta$, its expansion. This was first proposed in \cite{Lehner:2016vdi} and removes the ambiguity intrinsic to the parametrization of the WDW null boundaries, but it introduces an arbitrary length scale $L_{\rm ct}$.
In static background geometries, the role of this counterterm does not influence significantly the holographic CA, see \cite{Carmi:2017jqz}. Nevertheless, for dynamical spacetimes as the ones analyzed in \cite{Chapman:2018dem,Chapman:2018lsv}, the situation is different: there the inclusion of the counterterm in the total gravitational action is a key ingredient in order to obtain results consistent with general properties of circuit complexity. For example, in the one-sided geometry of \cite{Chapman:2018dem}, the counterterm is essential to obtain the expected late time growth rate in $d>3$ and a positive rate in $d=3$. In the two-sided case, the counterterm is needed to replicate the switchback effect \cite{Chapman:2018lsv}.
The inclusion of the counterterm also modifies the structure of divergences of holographic complexity, as first pointed out in \cite{Reynolds:2016rvl}, and was observed to play a crucial role in the cancellations occurring for CA in the study of the first law of complexity \cite{Bernamonti:2019zyy,Bernamonti:2020bcf}.

\subsubsection{Action evaluation}\label{paolo}

Let us now evaluate the various contributions to the gravitational action (\ref{eq:graction}). We follow the conventions of \cite{Carmi:2017jqz,Chapman:2018dem}.

\paragraph{Bulk term.} 

We first write explicitly
\be \label{eq:Ibulk}
I_{\rm bulk}=\frac{1}{16 \pi \, G_N}\int_{\rm WDW} d^3x\,\sqrt{-g}\left(R+ \frac{2}{\ell^2}\right)= - \frac{1}{2 \, G_N \ell^2}\int    dt  \, dr  \,r\,. 
\ee
where we used the on-shell relations $R=-\frac{6}{\ell^2}$ and $R=6 \Lambda$ and performed the angular integration.
Exploiting the left-right symmetry of the WDW patch, we divide its right half in three zones ${\rm I} - {\rm III}$, as labeled in fig.~\ref{figuraPenrose1}, each with its own integration extrema. For instance in region ${\rm I}$, for fixed $r_{m1} \leq r \leq r_+$, we have $t_{\rm min} \leq t \leq t_{\rm max}$. By symmetry $t_{\rm min}=0$, while $t_{\rm max}$ can be determined observing that the locations $(t_{\rm max}, r)$ and $ (t_b /2,r=\infty)$ share the same $v$ coordinate. This fixes $t_{\rm max} = t_b/2- r^*(r)$ in region ${\rm I}$. 
All together, we obtain 
\be 
I_{\rm bulk}= 2 \left(I_{\rm bulk}^{\rm I}+I_{\rm bulk}^{\rm II}+I_{\rm bulk}^{\rm III}\right)
\ee
with 
\bea
I_{\rm bulk}^{\rm I} &=&-\frac{1}{2 \, G_N \ell^2}\int_{r_{m1}}^{r_+}\,dr\,r\left(\frac{t_b}{2} -r^*(r)\right)\\
I_{\rm bulk}^{\rm II}&=&\frac{1}{G_N \ell^2}\int_{r_+}^{r_{\rm max}}\,dr\, r \, r^*(r) \\
I_{\rm bulk}^{\rm III}&=&\frac{1}{2 \, G_N \ell^2}\int_{r_{m2}}^{r_+}\,dr\,r\left(\frac{t_b}{2} + r^*(r)\right)\,,
\eea
where $r_{m1}, r_{m2}$ are given implicitly by eq.~\eqref{eq:rm1rm2} and $r_{\rm max}$ denotes a radial cutoff introduced to regularize these expressions. Thus
\be \label{azionebulk}
I_{\rm bulk}= \frac{1}{G_N \ell^2} \left\{ \frac{t_b}{4} \(r_{m1}^2 - r_{m2}^2\) + \int_{r_{m1}}^{r_{\rm max}}dr\,  r \, r^*(r) + \int_{r_{m2}}^{r_{\rm max}}dr\,  r \, r^*(r) \right\}
\ee
%
The UV divergent terms of the bulk action do not contribute to the complexity growth rate. In fact $r_{m1}$ and $r_{m2}$ evolve according to equation \eqref{eq:derrm1rm2}, but $r_{\rm max}$ is constant in time. As we shall see the same remains true also for the other contributions to the WDW action \eqref{eq:graction}.

Performing explicitly the  integrals  in \eqref{azionebulk} we obtain the expression
\begin{align}\label{Acaso2}
I_{\rm bulk}=\frac{1}{4\,G_N} \left\{ 2 \left(r_{m1}+r_{m2} -4 r_{\rm max} \right) - r_+ \log \frac{(r_+ + r_{m1}) (r_+ + r_{m2})}{(r_+ - r_{m1}) (r_+ - r_{m2})} \right\} + O\(\frac{1}{r_{\rm max}}\) \,,
\end{align}
where we used \eqref{eq:rm1rm2} and expanded in  $r_{\rm max} \to \infty$.  This correctly reduces to the non rotating BTZ result of  \cite{Carmi:2017jqz} for $r_- , r_{m1}\to 0$.
 %
 
\paragraph{GHY terms.}

Next we evaluate the GHY term in \eqref{eq:graction} for the timelike cutoff surface at $r=r_{\rm max}$
\begin{equation}
I_{\rm GHY}=\frac{1}{8 \pi \, G_N}\int_{r=r_{\rm max}} d^2y \,\sqrt{-h} \, K \, .
\end{equation}
Here $K = h^{ab}K_{ab}$ is the trace of the extrinsic curvature $K_{ab}=\frac{\partial x^\mu}{\partial y^a} \frac{\partial x^\nu}{\partial y^b} \nabla_\mu n_\nu$,  and $n_\nu$ the outward directed normal to the cutoff surface. These read 
\bea \label{eq:rnormal}
n_\mu dx^\mu = \frac{dr}{\sqrt{f(r_{\rm max})}} \, ,\qquad \qquad K = \frac{2 \, r_{\rm max}^2-r_+^2-r_-^2}{\ell^2 \, r_{\rm max}\sqrt{f(r_{\rm max})}}\,.
\eea
Taking into account the right-left symmetry of the problem, and the fact that the time integration along the cutoff surface  $r=r_{\rm max}$ is  restricted by the null boundaries of the WDW, we have
\bea
I_{\rm GHY} &=& \frac{\(2 r_{\rm max}^2 -r_+^2-r_-^2 \)}{2 \, G_N \ell^2}  \int_{\frac{t_b}{2}+r^*(r_{\rm max})}^{\frac{t_b}{2}-r^*(r_{\rm max})} dt =  - \frac{r^*(r_{\rm max})\(2r_{\rm max}^2-r_+^2-r_-^2\) }{G_N \ell^2} \nonumber  \\    
&=& \frac{2 \, r_{\rm max}}{G_N} + O\( \frac{1}{r_{\rm max}}\)\,.\label{GHY term} 
\eea
The GHY term only yields a divergent contribution to the total action, and thus does not contribute to the complexity growth rate.  

\paragraph{Joints terms.}\label{joints sezione}

There are different  joints with null surfaces contributing to the action  \eqref{eq:graction}. Null-null joints at the future and past tip of the WDW patch, and time-null joints formed at the intersection of the WDW patch with the cutoff surface at $r_{\rm max}$. Adopting the  conventions of  \cite{Carmi:2017jqz,Chapman:2018dem}, we have the following rules
\be
\begin{array}{lll}\label{eq:regoleJoints}
  \text{Time-Null joint:} & a_{\rm joint}=\epsilon\log\left|n_1\cdot k_2\right| & \text{with }\epsilon=-\text{sign}\left(n_1\cdot k_2\right)\text{sign}\left(\hat{t}_1\cdot k_2\right) \\
  \text{Null-Null joint:} & a_{\rm joint}=\epsilon\log\left|\frac{k_1\cdot k_2}{2}\right| & \text{with }\epsilon=-\text{sign}\left(k_1\cdot k_2\right)\text{sign}\left(\hat{k}_1\cdot k_2\right)\, .
\end{array}
\ee
Here $k_i$, $n_i$ are respectively null and spacelike normal one-forms outward-directed from the relevant boundary of the WDW patch. The auxiliary null and timelike vectors $\hat{k}_i$, $\hat{t}_i$ are defined in the tangent space of the appropriate boundary region, pointing outward from it and orthogonal to the joint. 
  
Let us start from the future null-null joint at the tip of the WDW patch, where $t=0$  and $r=r_{m1}$. This contributes to the total gravitational action \eqref{eq:graction} with 
\bea
I_{\rm joints}^{\rm Null-Null} &=&\frac{1}{8 \pi G_N}\int_{r=r_{m1}}\,dy\,\sqrt{\sigma}\log\left|\frac{k_L \cdot k_R}{2}\right|  \nonumber \\
&=& \frac{1}{4\,G_N} \, r_{m1}\log\left(-\frac{\alpha^2}{f(r_{m1})}\right) \, . 
\eea
To obtain this result we used  $\sigma=r^2$  and the following right and left null normals at the future joint of the WDW patch\footnote{Remember that, as  shown in fig.~\ref{figuraPenrose1},  in region ${\rm I}$ $t$ increases from the left to the right, $r$ decreases going up,  and $f(r)<0$.} 
\begin{equation}
 k_{R\, \mu }  = \left(\alpha\,,\,  \frac{\alpha}{f}\,,\,0\right)\, , \qquad  k_{L\, \mu} =  \left(- \alpha\,,\,\frac{\alpha}{f}\,,\,0\right)   \, . \label{eq:tipnormals}
\end{equation}
Adding the analogous contribution coming from the bottom joint, we have for null-null joints
\be
I_{\rm joints}^{\rm Null-Null} = \frac{1}{4\,G_N} \left\{  r_{m1}\log\left(-\frac{\alpha^2}{f(r_{m1})}\right) + r_{m2}\log\left(-\frac{\alpha^2}{f(r_{m2})}\right)\right\} \,. 
\ee

Next, we evaluate the time-null joints term at the cutoff surface. Consider the right cutoff surface $r=r_{\rm max}$ and the joint term in its future at $t=\frac{t_b}{2} - r^*(r_{\rm max})$.  Using the normal $n_\mu$ from \eqref{eq:rnormal} and $k_{\mu R}$ from \eqref{eq:tipnormals}, gives
\bea
I_{\rm joints}^{\rm Time-Null} &=& -\frac{1}{8 \pi G_N}\int_{r=r_{\rm max}} \,dy\,\sqrt{\sigma}\,\log\left| n \cdot k\right|  \nonumber \\
&=& -\frac{1}{4G_N } \, r_{\rm max} \log\left(\frac{ \alpha \, \ell }{r_{\rm max}}\right) + O\(\frac{1}{r_{\rm max}}\) \, . 
\eea
This divergent term is independent from the boundary time. 

The other three time-null joints at the cutoff surface yield identical contributions.    
All together, including the null-null terms, we therefore have 
\be \label{eq:BTZjoints}
I_{\rm joints} = - \frac{1}{4\,G_N} \left\{4 r_{\rm max} \log\left(\frac{ \alpha \, \ell }{r_{\rm max}}\right) - r_{m1}\log\left(-\frac{\alpha^2}{f(r_{m1})}\right) - r_{m2}\log\left(-\frac{\alpha^2}{f(r_{m2})}\right)\right\} +\, O\(\frac{1}{r_{\rm max}}\) \, .
\ee
%

\paragraph{Counterterms.}\label{counteterm sezione}

To evaluate the last contribution to the gravitational action \eqref{eq:graction}, let us consider first the right future null boundary of the WDW patch.   
The counterterm action $I_{\rm ct}$ for this contribution  evaluates to
\bea
I_{\rm ct}^{R F} &=& \frac{1}{8 \pi \, G_N} \int \,d \lambda\, d y\, \sqrt{\gamma}\, \Theta\, \log\left( \lvert L_{\rm ct} \, \Theta \rvert  \right) \nonumber \\
&=&\frac{1}{4 \, G_N}\int_{r_{m1}}^{r_{\rm max}}\, dr \, \log\left( \frac{L_{\rm ct} \, \alpha}{r}\right)   \nonumber \\
&=&\frac{1}{4 \, G_N} \left\{ r_{\rm max} \left[1 + \log \left( \frac{L_{\rm ct} \, \alpha}{r_{\rm max}}\right) \right] -r_{m1} \left[1 + \log \left( \frac{L_{\rm ct} \, \alpha}{r_{m1}}\right) \right]  \right\} \,,
\eea
In deriving this expression we used that the normal vector to the surface implicitly defines a parametrization through  $\del_\lambda = k^\mu \del_\mu $, together with the explicit form of the one-dimensional induced metric  $\gamma = e^\mu e^\nu g_{\mu \nu} = r^2$, which defines $\Theta   = \partial_{\lambda}\log\sqrt{\gamma}$. In particular, this yields  $dr=\alpha d\lambda$ and    $\Theta=   \alpha\partial_{r}\log\sqrt{\gamma}=\frac{\alpha}{r}$. 

Given the left-right symmetry, the left future null boundary gives an identical contribution. It is also straightforward to check that the past boundaries lead to an analogous result with $r_{m1} \to r_{m2}$. Putting everything together:
\be\label{eq:BTZct}
I_{\rm ct} = \frac{1}{2 \, G_N} \left\{ 2 \, r_{\rm max} \left[1 + \log \left( \frac{L_{\rm ct} \, \alpha}{r_{\rm max}}\right) \right] -r_{m1} \left[1 + \log \left( \frac{L_{\rm ct} \, \alpha}{r_{m1}}\right) \right] -r_{m2} \left[1 + \log \left( \frac{L_{\rm ct} \, \alpha}{r_{m2}}\right) \right] \right\}\, .
\ee
The counterterm will thus give a non-vanishing contribution both to CA itself and to its growth rate. We will analyze in what follows how this counterterm contribution modifies the results of  \cite{Auzzi:2018pbc}, obtained without the counterterm action later introduced in \cite{Lehner:2016vdi}.

\subsubsection{CA results}\label{sezione risultati}

Combining the results of the previous subsection, \eqref{Acaso2}, \eqref{GHY term}, \eqref{eq:BTZjoints} and \eqref{eq:BTZct}, the total holographic complexity for the rotating BTZ black hole reads
\begin{align}\label{eq:azionetot}
C_{\rm A}(t_b) = &\frac{1}{4 \pi G_N} \Bigg\{ 4 r_{\rm max} \left( 1+ \log \frac{L_{\rm ct}}{\ell} \right)  - r_+ \log \frac{(r_+ + r_{m1}) (r_+ + r_{m2})}{(r_+ - r_{m1}) (r_+ - r_{m2})}  \\
& - r_{m1} \log \frac{L_{\rm ct}^2 (r_+^2 - r_{m1}^2) (r_{m1}^2 - r_-^2)}{r_{m1}^4 \, \ell^2}- r_{m2} \log \frac{L_{\rm ct}^2 (r_+^2 - r_{m2}^2) (r_{m2}^2 - r_-^2)}{r_{m2}^4  \, \ell^2} 
\Bigg\} + O\(\frac{1}{r_{\rm max}}\) \,.  \nonumber
\end{align}
 As expected, in presence of the counterterm action any dependence from the normalization $\alpha$ of the null normals to the boundaries of the WDW patch drops. On the other hand, the result depends on the arbitrary constant $L_{\rm ct}$.
 
 The time dependence of CA as written in \eqref{eq:azionetot} is implicitly given by the time dependence of the tip locations, $r_{m1}$ and $r_{m2}$, through  \eqref{eq:rm1rm2} and \eqref{eq:derrm1rm2}.  The  CA growth rate  can be  more directly obtained from the expressions in the previous subsection. 
 As anticipated, the purely divergent GHY term does not contribute to the complexity growth rate. The contribution from the bulk term is most easily evaluated using the intermediate expression \eqref{azionebulk}, while the ones of the  joints and counterterm action follows from \eqref{eq:BTZjoints} and \eqref{eq:BTZct}. All together this yields the growth rate
\begin{equation}\label{eq:ratetotale}
\begin{aligned}
\frac{d C_{\rm A}}{d t_b}=&\frac{1}{4 \pi G_N} \Bigg\{ \frac{r_{m1}^2-r_{m2}^2}{\ell^2}  \\
&+ \frac{f(r_{m1})}{2} \log\left(-\frac{\alpha^2}{f(r_{m1})}\right)-\frac{r_{m1}}{2}  f'(r_{m1}) 
 - \frac{f(r_{m2})}{2} \log\left(-\frac{\alpha^2}{f(r_{m2})}\right) + \frac{r_{m2}}{2} f'(r_{m2})  \\ 
& - f(r_{m1}) \log\left( \frac{L_{\rm ct} \, \alpha}{r_{m1}}\right) + f(r_{m2}) \log\left( \frac{L_{\rm ct} \, \alpha}{r_{m2}}\right) \Bigg\}\, .
\end{aligned}
\end{equation}
The first term is the bulk contribution, while the last line is the contribution of the counterterm. The latter combines  with contributions  from the joints in the second line to give a result that does not depend on $\alpha$. Using the explicit expression for $f'(r)$
\begin{align}\label{eq:ratetotalev2}
\frac{d C_{\rm A}}{d t_b} = \frac{1}{4 \pi G_N}  \Bigg\{ \frac{r_+^2 r_-^2}{ \ell^2} \, \frac{r_{m2}^2 - r_{m1}^2}{r_{m1}^2 r_{m2}^2}  -  \frac{f(r_{m1})}{2} \log\left( - \frac{L_{\rm ct}^2 \, f(r_{m1})}{r_{m1}^2}\right) +  \frac{f(r_{m2})}{2} \log\left( -\frac{L_{\rm ct}^2 \, f(r_{m2})}{r_{m2}^2}\right) \Bigg\}\, .
\end{align}

In the limit $J/(M \ell) \to 0$, from these expressions one smoothly recovers  the non rotating results \cite{Carmi:2017jqz,Auzzi:2018pbc}.\footnote{\cite{Auzzi:2018pbc} used a different regularization of the WDW patch. However, the structure of divergences does not play a role in the complexity growing rate, and the two results coincide.}  It is also easy to verify that in a small $J/(M \ell)$ expansion, these expressions do not have linear order terms. This is consistent with general results obtained in the study of the first law of complexity \cite{Bernamonti:2019zyy,Bernamonti:2020bcf,Hashemi:2019aop}. 
\paragraph{Growth rate.}

Let us start analysing the late time limit $t_L=t_R=\frac{t_b}{2}\to \infty$ of the complexity growth rate. In this limit  $r_{m1}\to r_-$ and $r_{m2}\to r_+$ (see fig.~\ref{figuraPenrose1})
and only the bulk contribution in the first line of \eqref{eq:ratetotale} survives
\begin{equation}\label{lateTimeRate}
\lim_{t_b \to \infty} \frac{dC_{\rm A}}{dt_b} = \frac{r_{+}^2-r_{-}^2}{4 \pi G_N \ell^2}  \,. 
\end{equation}
Notice that this clearly vanishes in the extremal limit $r_+ \to r_-$.
In terms of the mass $M$ and angular momentum $J$ (\ref{MandJ}), the above formula reads
\begin{equation}\label{eq:BTZCAratelatetime}
\lim_{t_b \to \infty} \frac{dC_{\rm A}}{dt_b} =  \frac{2}{\pi}\left(M-\Omega_H J\right) \,, 
\end{equation} 
where $\Omega_H$ is the angular velocity of the horizon in \eqref{OmegaH}. 

 This limiting value corresponds to saturation of the Lloyd's computational (upper) bound conjectured in \cite{Brown:2015lvg} as well as the one proposed in \cite{Cai:2016xho}, which for the rotating BTZ are actually equivalent.\footnote{For higher dimensional rotating black holes the two bounds are in general not equivalent. Indeed,  for the 4d Kerr-AdS solution we consider in sec.~\ref{sec:KerrAdS},  we find compatibility with the bound of \cite{Cai:2016xho} but not with that of \cite{Brown:2015lvg}.} 
Nonetheless,  both bounds are violated at intermediate times here. In fact  the late time value is approached from above, as for all two-sided black holes studied in \cite{Carmi:2017jqz} and in contrast with the one-sided black holes of \cite{Chapman:2018dem}. 
 To show this we can follow the same strategy as in \cite{Carmi:2017jqz} and decompose the inverse blackening factor as
\begin{equation}
\frac{1}{f}=\frac{1}{r_+-r_-}\left(\frac{r_+}{r F(r_+)(r-r_+)}-\frac{r_-}{r F(r_-)(r-r_-)}+H(r)\right),
\end{equation}
in terms of the strictly positive functions
\be\label{f grande}
F(r) \equiv \frac{f(r)}{(r-r_+)(r-r_-)} = \frac{(r+r_+)(r+r_-)}{\ell^2 \, r^2}
\ee
and
\bea
H(r) &\equiv & \frac{F(r_+) r-F(r) r_+}{r F(r) F(r_+)(r-r_+)}-\frac{F(r_-) r-F(r) r_-}{r F(r) F(r_-)(r-r_-)} \nonumber  \\ 
&=&\ell^2 \frac{(r_+-r_-)(r_+r_-+r(r_++r_-))}{2r(r+r_+)(r+r_-)(r_++r_-)} \,,
\eea
which is regular in $r_+$ and $r_-$, and decays as $1/r^2$ for $r\to\infty$.
We then solve up to first subleading order in the late time limit the equations (\ref{eq:rm1rm2}) for $r_{m1}$ and $r_{m2}$:
\begin{equation}\label{eq:rm1rm2 a late time}
r_{m1}\simeq r_{-}\left(1+c_-e^{-\frac 1 2 F(r_-)(r_+-r_-) t_b}\right),\qquad\qquad r_{m2}\simeq r_{+}\left(1-c_+e^{-\frac 1 2 F(r_+)(r_+-r_-) t_b}\right)\,,
\end{equation}
where $c_+$ and $c_-$ are  positive constants
\begin{equation}
c_-=\left(\frac{r_+-r_-}{r_-}\right)^{\frac{F(r_-)}{F(r_+)}}e^{-F(r_-)\int_{r_-}^{\infty}H\left(\tilde{r}\right)d\tilde{r}} \,, \qquad c_+=\left(\frac{r_+-r_-}{r_+}\right)^{\frac{F(r_+)}{F(r_-)}}e^{F(r_+)\int_{r_+}^{\infty}H\left(\tilde{r}\right)d\tilde{r}} \,.
\end{equation}
Substituting the above expressions for $r_{m1}$ and $r_{m_2}$  in (\ref{eq:ratetotale}) we find the late time behavior
\begin{equation}\label{Acaso33}\begin{aligned}
\frac{d C_{\rm A}}{d t_b} \simeq   \frac{2}{\pi}\left(M-\Omega J\right) + \frac{(r_+ - r_-)^2}{16 \pi G_N \ell^2} \, t_b  \pq{c_+ r_+ F^2(r_+)e^{-\frac 1 2 F(r_+)(r_+-r_-) t_b}-c_- r_- F^2(r_-) e^{-\frac 1 2 F(r_-)(r_+-r_-) t_b}}.
\end{aligned}
\end{equation}
At late times the exponent with smaller coefficient dominates, and thus the asymptotic value \ref{eq:BTZCAratelatetime} is reached from above if $F({r_+})< F({r_-})$. This is indeed the case here, as from \eqref{f grande} we get $\frac{F({r_+})}{F({r_-})}=\frac{r_-}{r_+}$. 
Notice that both the late time limit and this result  do not depend on the presence of the counterterm action. The counterterm only enters in this expansion at orders that are subleading with respect to our analysis. Indeed,  \cite{Auzzi:2018pbc} evaluated $\frac{d C_{\rm A}}{d t_b}$ without the inclusion of the countertem action and also similarly  found that the late time limit is approached from above. 

In the opposite limit, at $t_b=0$, the complexification rate is zero, independently from the presence of the counterterm. This can be easily tracked to the fact that $r_{m1}=r_{m2}$, making the counterterm contribution vanish.  

\begin{figure}[ht]
\begin{center}
\includegraphics[width=.32\linewidth]{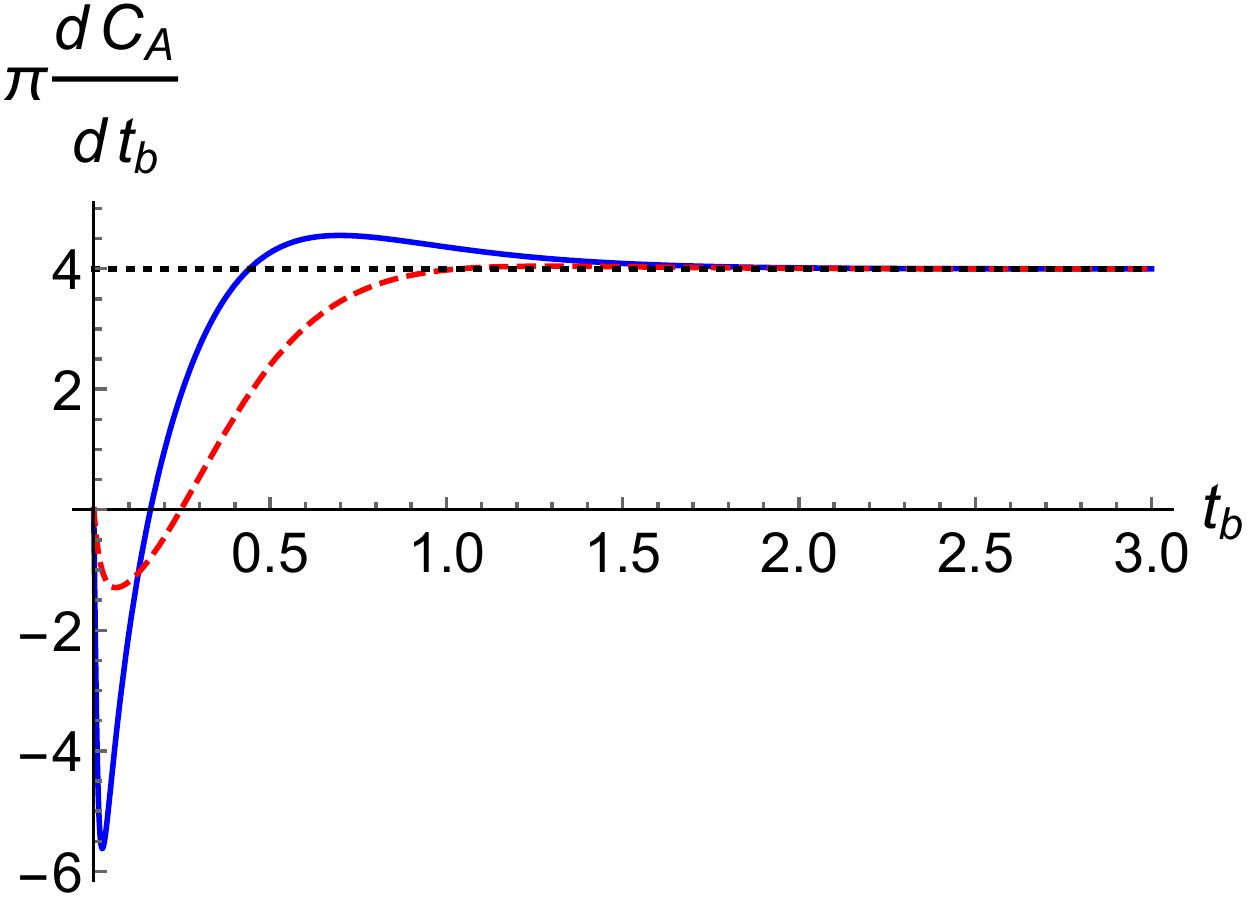}  \hfill 
 \includegraphics[width=.32\linewidth]{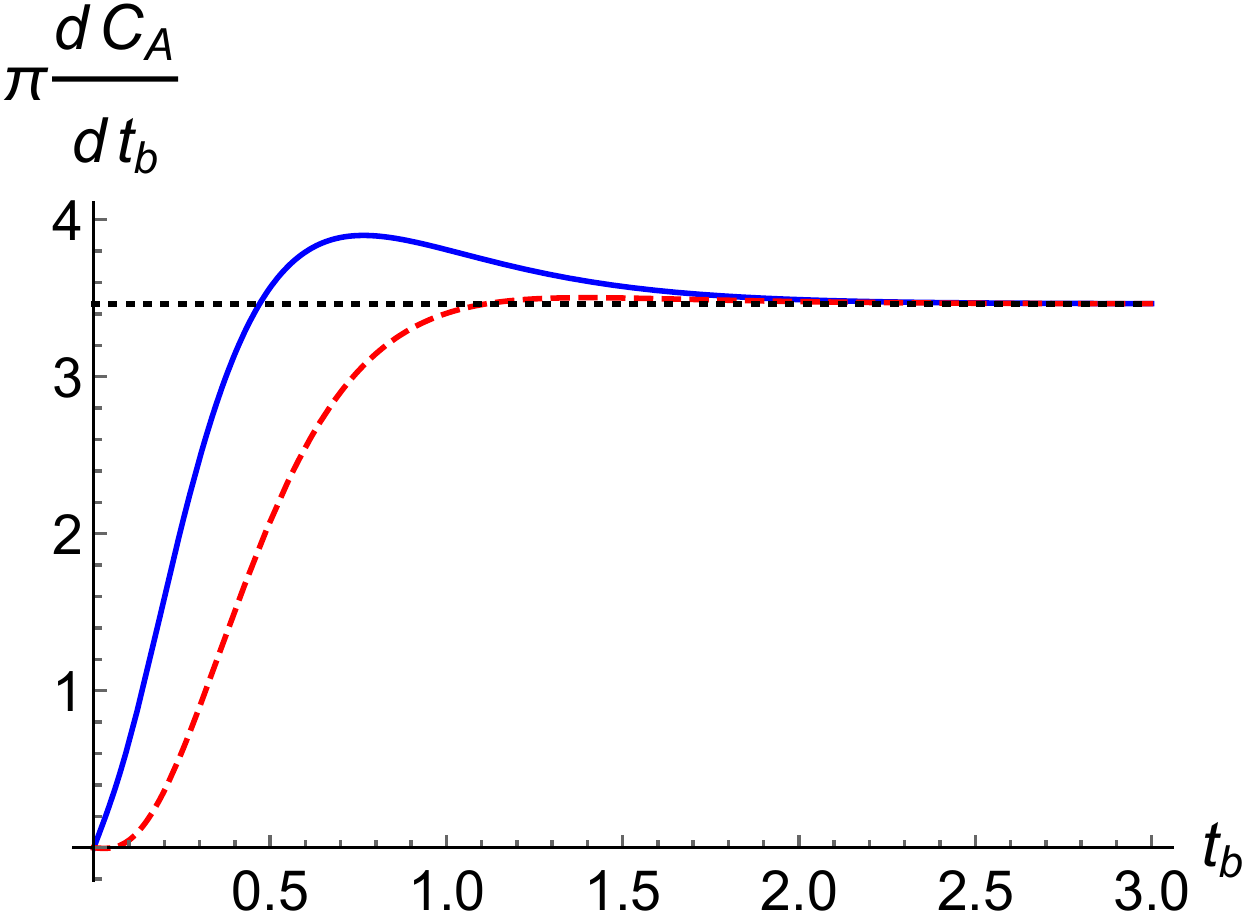} \hfill 
 \includegraphics[width=0.32\linewidth]{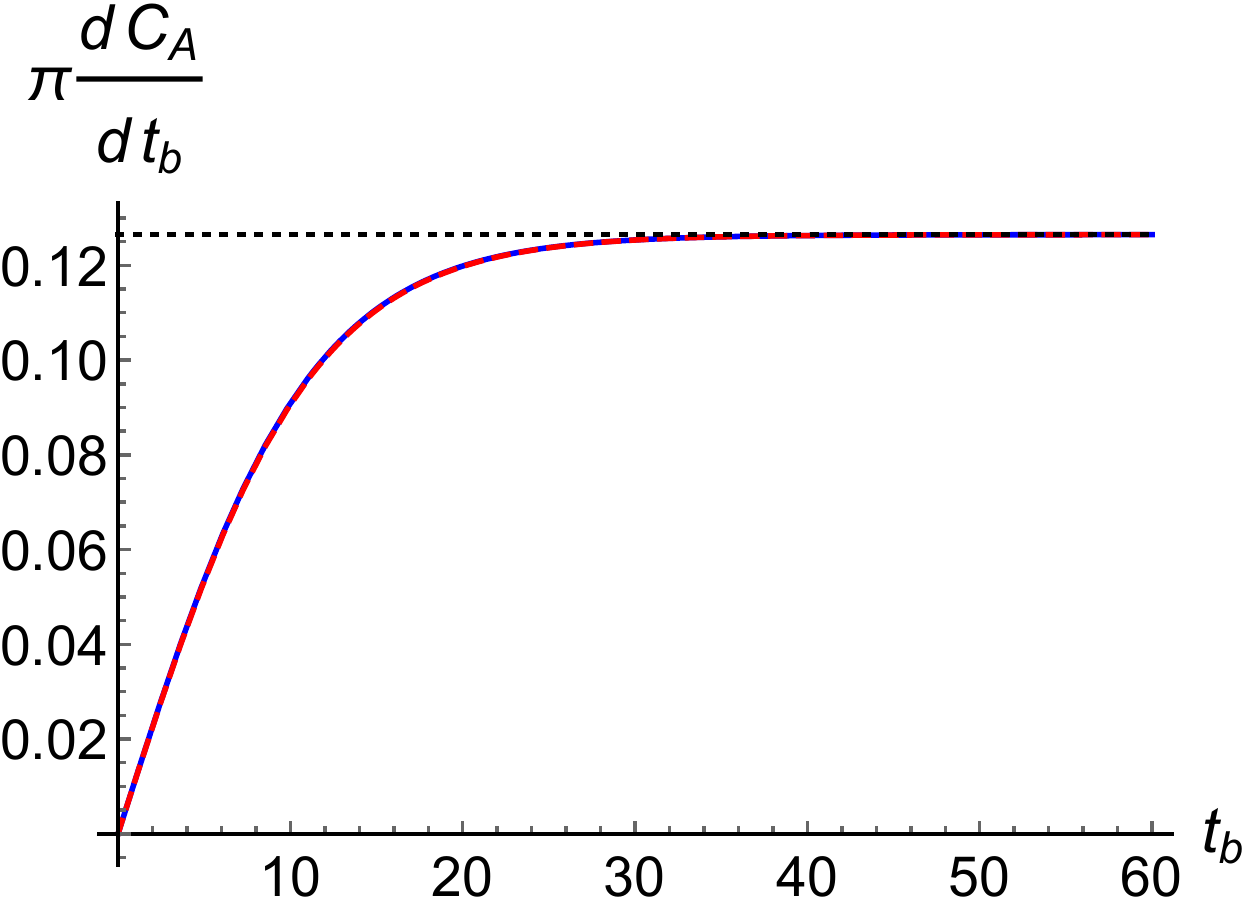}
\end{center}
\caption{ $\pi \frac{d C_{\rm A}}{d t_b}$ with $M=2$, $ \ell = G_N = \alpha = L_{\rm ct}=1$ and (left) $J=0.1$, (center) $J=1$, (right) $J=1.999$. The CA rate with the counterterm is depicted in blue solid, in red dashed without. In black dotted the late time value \eqref{eq:BTZCAratelatetime}. 
}\label{gr1}
\end{figure}

For intermediate times,  we can analyze semi-analitically the effects of the counterterm  on the complexity growth rate, by solving numerically \eqref{eq:rm1rm2} for $r_{m1}$ and $r_{m2}$.  Including the counterterm, the result depends on  $L_{\rm ct}$ but is insensitive to the choice of normalization $\alpha$, while the opposite holds if we drop the counterterm. We perform a qualitative comparison between the two cases in fig.~\ref{gr1} where we report some sample plots as we fix $M \ell$ and vary $J$. 

For small values of $J$, the presence of the counterterm produces a large negative peak at early times. This is followed by a rapid growth that at intermediate times generally yields a larger complexification rate as compared to the case without counterterm. As the amount of angular momentum increases, the rapid growth overcomes the negative peak and the complexity starts immediately to increase (the rate of growth becomes everywhere positive). As the extremal limit $J \to M \ell $ is approached, the effect of the counterterm becomes less and less important and the counterterm contribution becomes negligible.

The qualitative dependence on the angular momentum  closely parallels the dependence observed for (higher dimensional) charged, non-rotating, AdS black holes  \cite{Carmi:2017jqz}. Also notice that approaching the irrotational limit the early time negative peak turns into the negative divergence characteristic of neutral  AdS black holes  \cite{Carmi:2017jqz}. The inclusion of the counterterm is essential to have this divergence  (see fig.~18 and  fig.~2 in \cite{Carmi:2017jqz}). 

\begin{figure}[ht]
\begin{center}
\includegraphics[width=.32\linewidth]{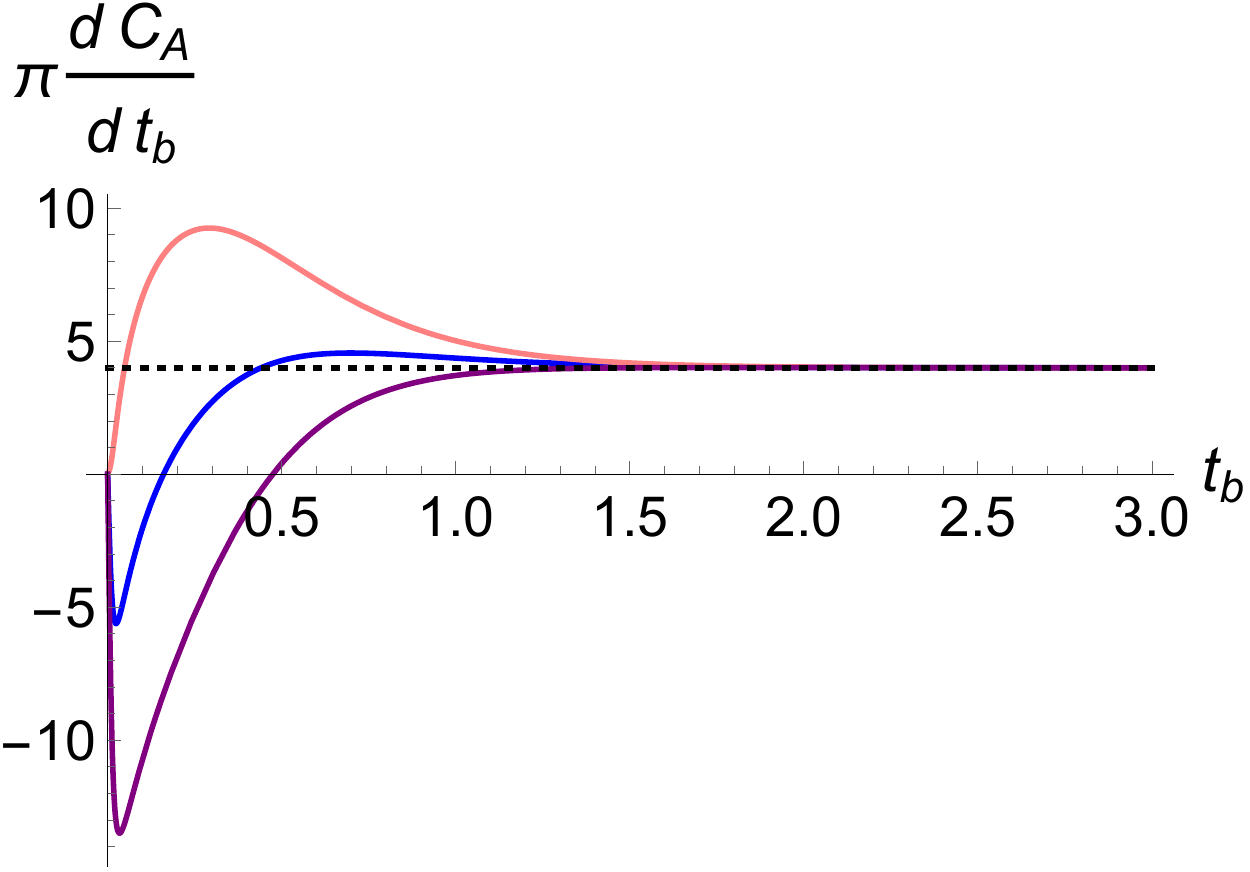}  \hfill 
 \includegraphics[width=.32\linewidth]{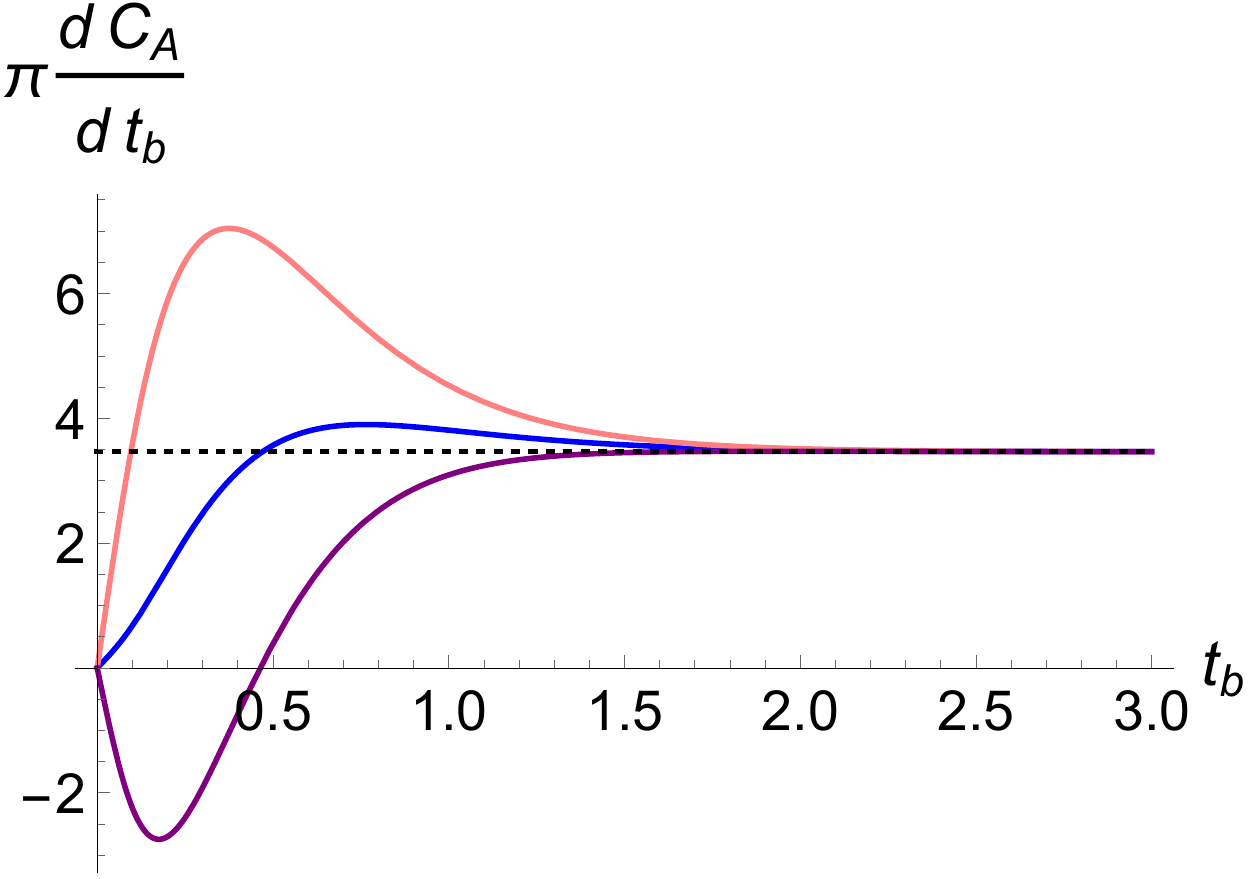} \hfill 
 \includegraphics[width=0.32\linewidth]{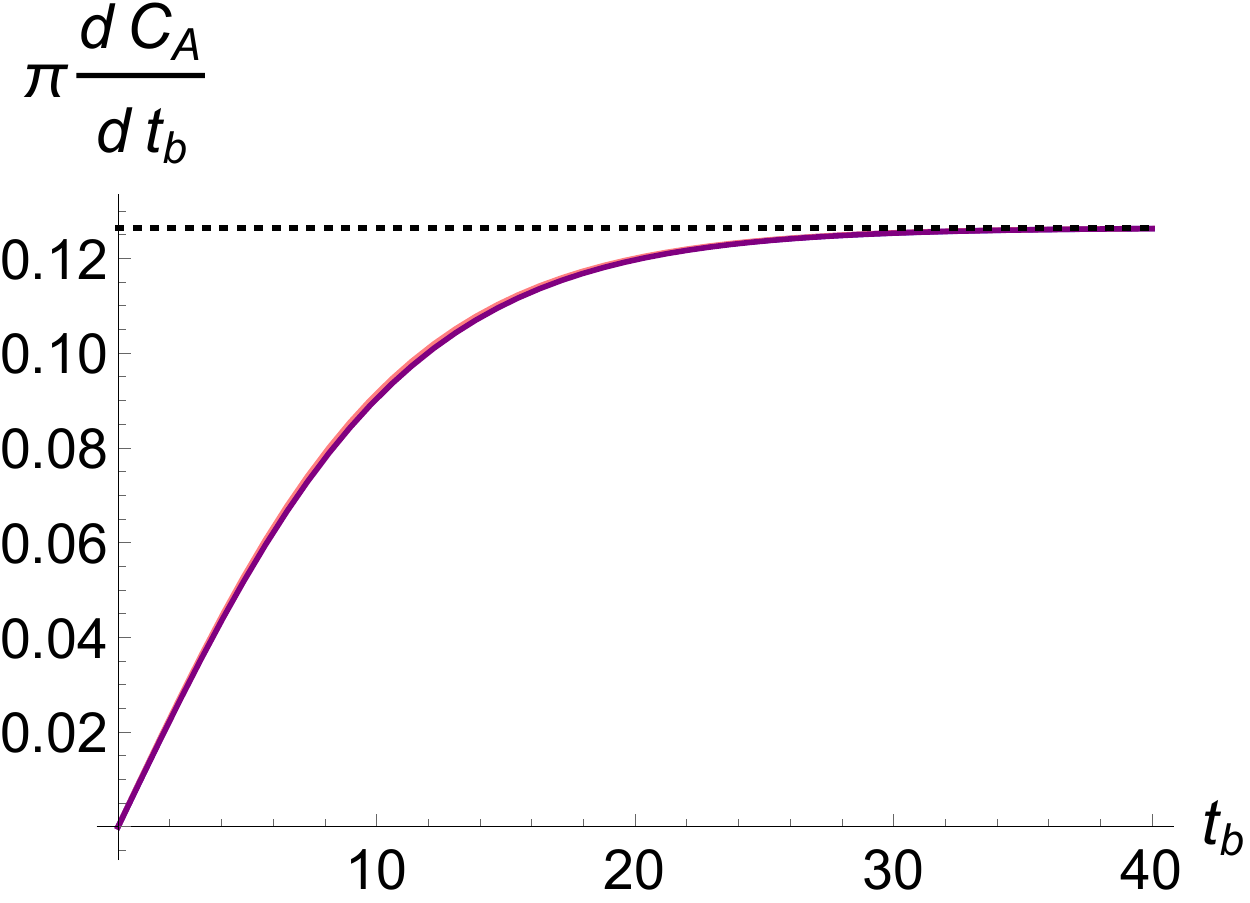}
\end{center}
\caption{ $\pi \frac{d C_{\rm A}}{d t_b}$ in eq.~\eqref{eq:ratetotalev2} with $M=2$, $ \ell = G_N =1$ and (left) $J=0.1$, (center) $J=1$, (right) $J=1.999$. In the three panels we explore the dependence on the counterterm scale: $L_{\rm ct} = 0.1$ (pink), $L_{\rm ct} = 1$ (blue), $L_{\rm ct} = 10$ (purple), increasing from the top down. In black dotted the late time value \eqref{eq:BTZCAratelatetime}. 
}\label{gr1ct}
\end{figure}

In fig.~\ref{gr1ct} we illustrate the dependence on the counterterm scale $L_{\rm ct}$ in the result of eq.~\eqref{eq:ratetotalev2}. The three panels correspond to those in fig.~\ref{gr1} and $L_{\rm ct}$ increases from the top down. At the qualitative level, a larger counterterm scale effectively acts as a reduction of the angular momentum $J$, and viceversa. For instance in the left panel it is manifest that increasing $L_{\rm ct}$ makes the negative peak at early times deeper. In the right panel, we see explicitly that the counterterm contribution becomes negligible as the three curves are essentially superposed. 
%

\paragraph{Complexity variation.} 

Next we study holographic complexity variations with respect to the Neveu-Schwarz vacuum of the boundary theory, dual to global AdS$_3$.
Since we are considering a double sided BTZ geometry we subtract twice the gravitational action computed on the WDW patch in vacuum AdS$_3$
\be
\Delta C_{\rm A}(t_b) \equiv C_{\rm A}(t_b) - 2 C_{\rm A}^{\rm AdS} \, .
\ee 
 Physically this quantity describes how difficult it is to prepare the thermofield double state at the boundary time $t_b/2$, describing two entangled copies of the boundary CFT, with respect to preparing the vacuums of the same unentangled copies. At $t_b= 0$ this defines the complexity of formation.\footnote{Another option is to consider variations with respect to the Ramond vacuum in the boundary theory, corresponding to the zero mass BTZ geometry. For the neutral non-rotating case, \cite{Chapman:2016hwi}  found a vanishing complexity of formation. Here subtracting the zero mass BTZ  result would only shift the complexity of formation by $\frac{\ell}{2 G_N}$, \ie at $t_b=0$ we would have $C_{\rm A} - 2 C_{\rm A}|_{{J=M=0}} =  \Delta C_{\rm A} + \frac{\ell}{2 G_N}$.}
 
The AdS$_3$ result reads \cite{Chapman:2016hwi,Carmi:2017jqz}
\be
C_{\rm A}^{\rm AdS}=\frac{1}{4 \pi G_N}  \Bigg \{ 2\, r^{\rm AdS}_{\rm max} \left(1+\log \frac{ L_{\rm ct}}{\ell} \right) + \pi \, \ell \Bigg\}  \, . 
\ee
The regulator surface $r^{\rm AdS}_{\rm max}$ is in principle different from the BTZ one, $r_{\rm max}$. To relate the two, one uses the standard holographic procedure, writing the two metrics in a Fefferman-Graham (FG) expansion \cite{fefferman1985elie,Fefferman:2007rka} and imposing the same UV cutoff. This  exercise shows that the two cutoffs, $r_{\rm max}$ in BTZ and $r_{\rm max}^{\rm AdS}$ in AdS, differ by a linear term in the FG cutoff. Such a linear correction does not yield any finite term to CA in the limit where the cutoff is removed. Thus for our purpose we can simply identify $ r_{\rm max}^{\rm AdS} = r_{\rm max} $. 
 
Subtracting (twice) the AdS$_3$ result from \eqref{eq:azionetot} then renders a finite variation
\begin{align} \label{eq:BTZDeltaCA}
\Delta C_{\rm A}(t_b ) = &- \frac{1}{4 \pi G_N} \Bigg\{2\pi \ell + 2\(r_{m1} + r_{m2}\) \log \frac{L_{\rm ct}}{\ell}+ r_+ \log \frac{(r_+ + r_{m1}) (r_+ + r_{m2})}{(r_+ - r_{m1}) (r_+ - r_{m2})} \nonumber  \\
& + r_{m1} \log \frac{ (r_+^2 - r_{m1}^2) (r_{m1}^2 - r_-^2)}{r_{m1}^4 } + r_{m2} \log \frac{ (r_+^2 - r_{m2}^2) (r_{m2}^2 - r_-^2)}{r_{m2}^4  } 
\Bigg\} \,,
\end{align}
where we dropped all terms that vanish as we take the UV cutoff to zero. 

\begin{figure}[H]
\centering
\includegraphics[width=.32\linewidth]{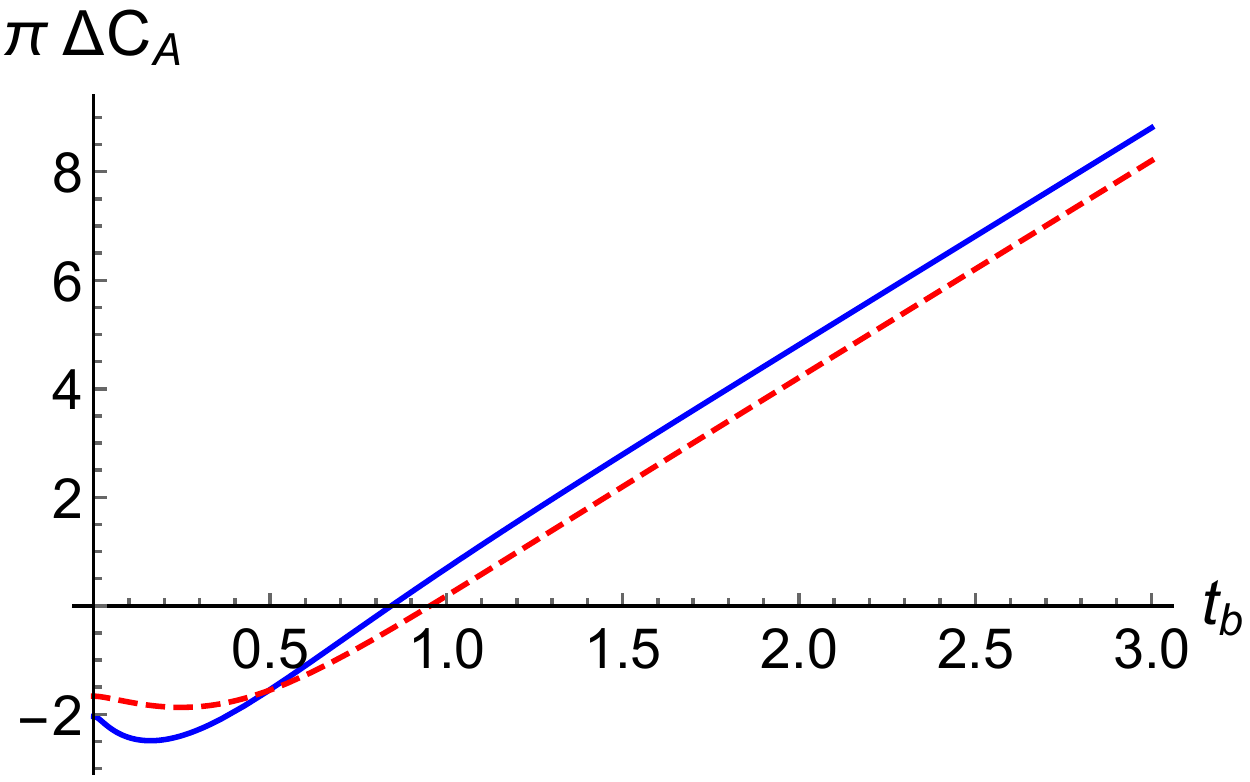} \hfill \includegraphics[width=.32\linewidth]{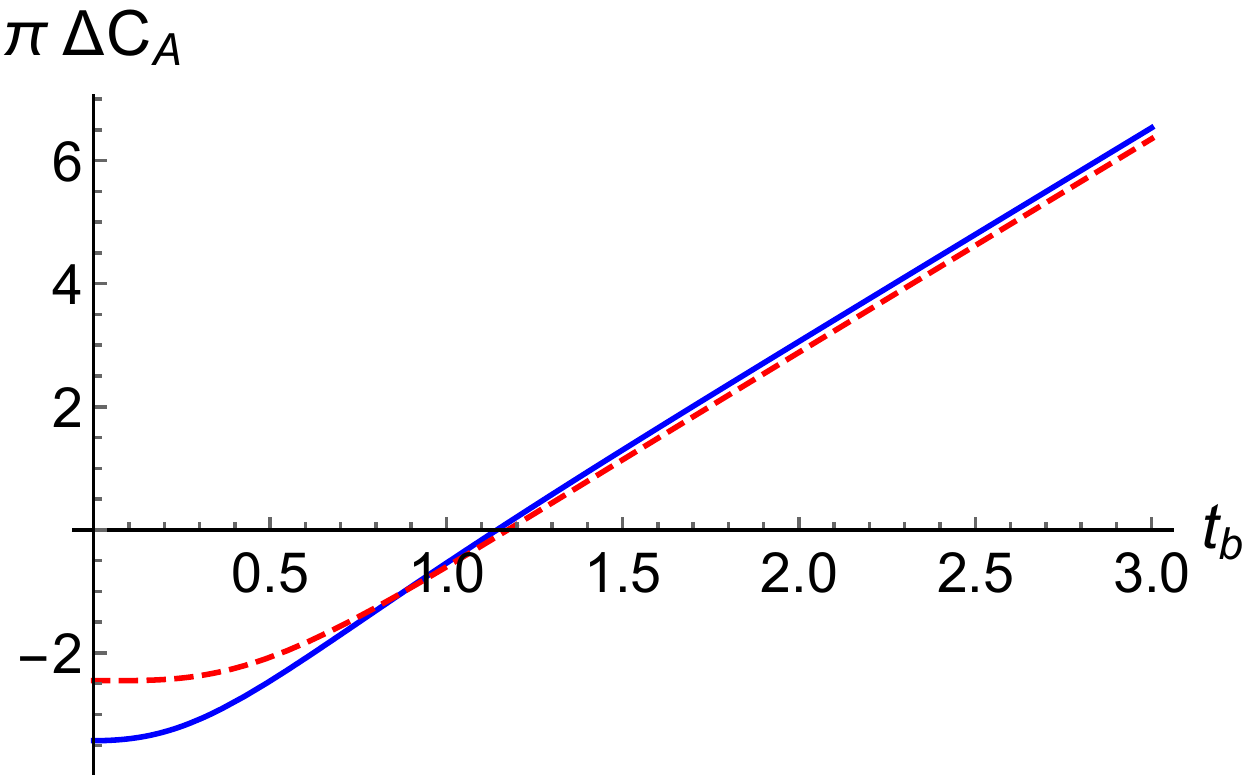} \hfill \includegraphics[width=0.32\linewidth]{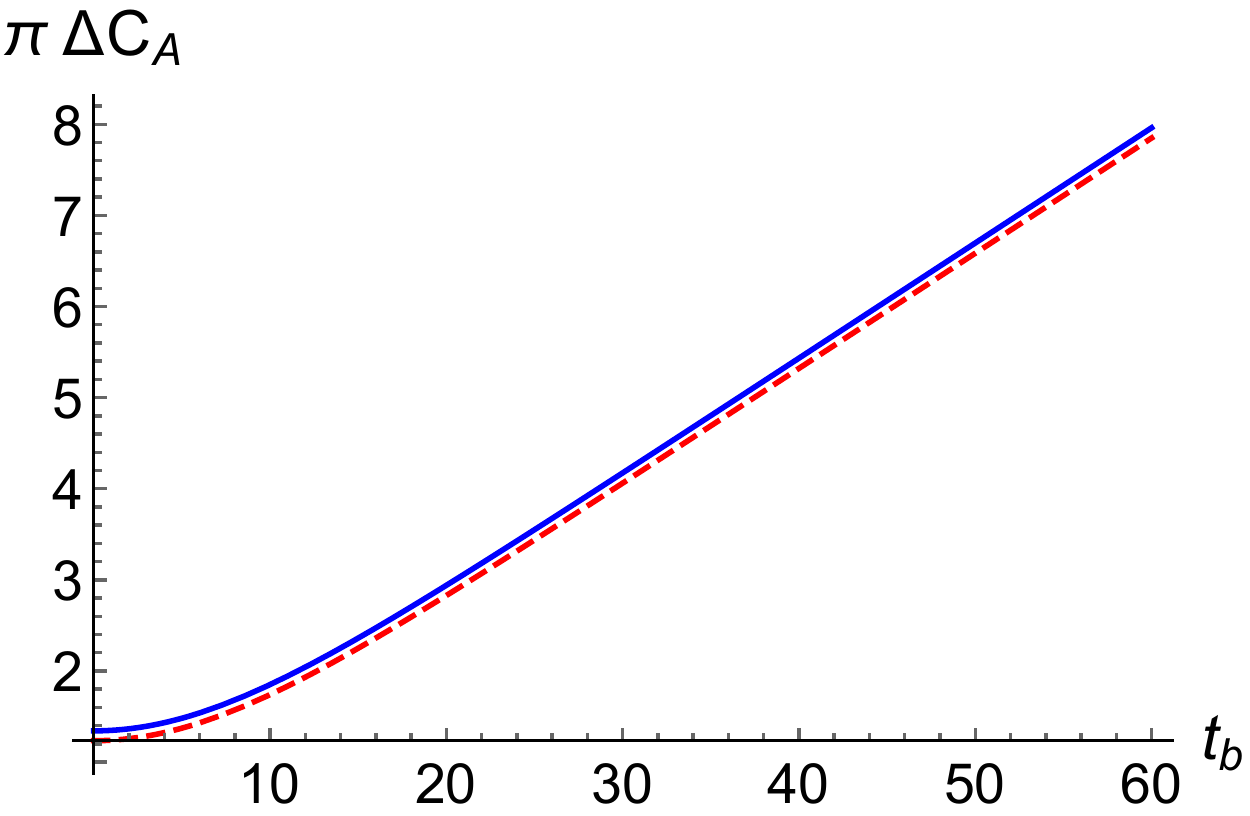}
\centering
\caption{ $\pi \Delta C_{\rm A} (t_b)$ with $M=2$, $ \ell = G_N = \alpha = L_{\rm ct}=1$ and (left) $J=0.1$, (center) $J=1$, (right) $J=1.999$. CA with the counterterm is depicted in blue solid, in red dashed without.
} \label{gr3}
\end{figure}

As for the rate of complexification, we can solve numerically the equations defining the joint terms (\ref{eq:rm1rm2}) obtaining the following results.
In fig.~\ref{gr3}, we plot the  complexity variation \eqref{eq:BTZDeltaCA}, as we vary the ratio $J/(M \ell)$.  
These reflect what observed in analyzing the growth rate. At late times, for any value of $J$,  $ \Delta C_{\rm A} $ grows linearly, with a slope that does not depend on the counterterm. The early time behavior depends instead on the value of $J$. For small enough angular momentum $\Delta C_{\rm A}$ initially decreases before increasing monotonically in time, while for larger values of the angular momentum immediately increases. The exact evolution depends also on the presence of the counterterm, or lack thereof. This also produces a finite difference in the complexity variation at $t_b=0$, \ie  the complexity of formation. This contrasts with the non-rotating case where the counterterm contribution vanishes at $t_b=0$ \cite{Carmi:2017jqz}.

\paragraph{Complexity of formation.}  

\begin{figure}[th]
\centering
\includegraphics[width=.5\linewidth]{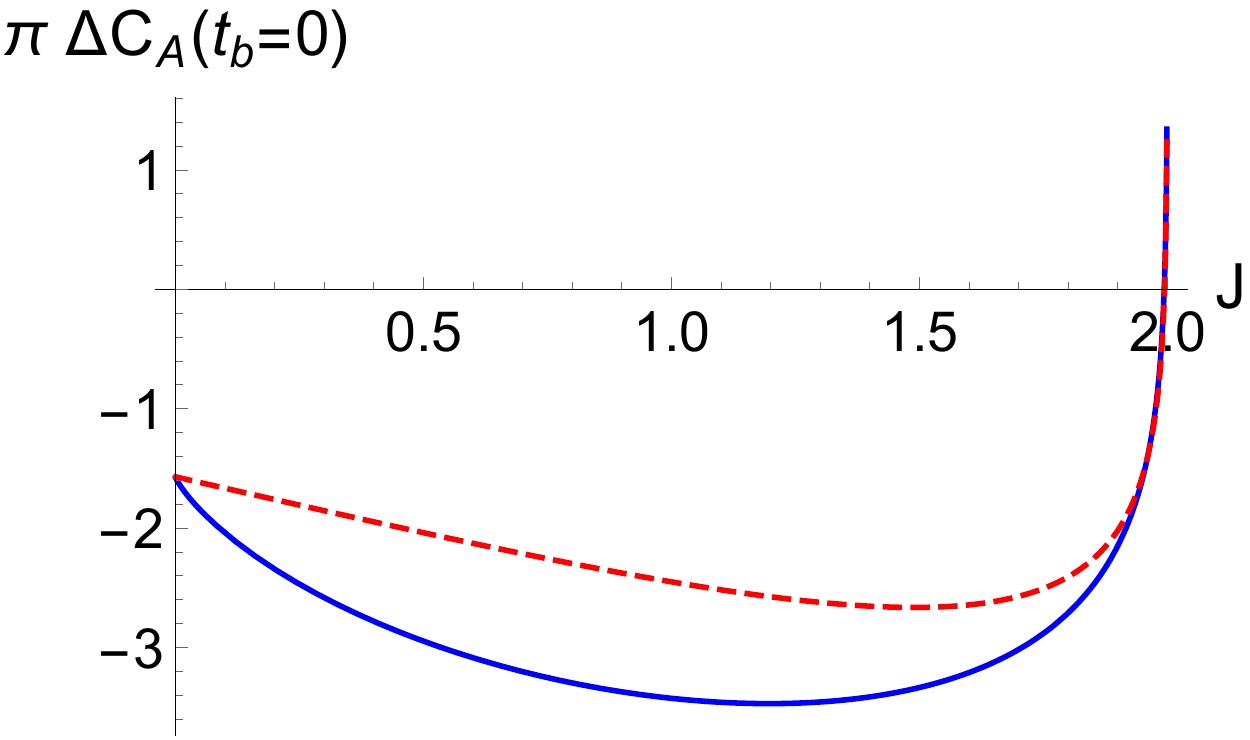}
\centering
\caption{\label{fig:CAformation}\small $\pi \Delta C_{\rm A} (t_b =0)$ with $M=2$, $ \ell = G_N = \alpha = L_{\rm ct}=1$ as a function of $J$. In blue solid with the counterterm, in red dashed without. In the extremal limit $J \to M \ell $, the complexity of formation diverges. 
}\label{figuraDivergenza}
\end{figure}

Focusing on the complexity of formation, we can study semi-analytically its dependence on the angular momentum $J$, see fig.~\ref{fig:CAformation}.
The $J =0$ complexity of formation is just the non-rotating BTZ value \cite{Chapman:2016hwi} 
\begin{equation}
\Delta C_{\rm A}^{J=0}(t_b=0)=-\frac{\ell}{2 G_N} \,, 
\end{equation} 
independently from the inclusion of the counterterm action. As a function of $J$, the complexity of formation initially decreases, with the counterterm giving a larger negative $\Delta C_{\rm A} (0)$.  As $J$ increases the complexity of formation value increases, turns positive and diverges in the extremal limit $J \rightarrow M \ell$ (or $r_+ \to r_-$) as 
\be \label{eq:CAformationDivergence}
\Delta C_{\rm A}(t_b=0) \sim - \frac{1}{2 \pi G_N} \left( r_- \log \frac{8(r_+ -r_-)}{r_-} + \pi \ell + 2 \,  r_- \log \frac{L_{\rm ct}}{\ell} \right)\,. 
\ee
The divergence comes from both bulk and joint terms, while the counterterm only provides a finite contribution.  Again, the extremal limit behavior is analogous to the one of charged black holes \cite{Carmi:2017jqz}  telling us that according to this measure the boundary CFT state with $T=0, \ell \O_H = 1$ is infinitely more complex than at finite temperature and angular velocity. 

Finally, let  us remark that despite the complexity of formation is quickly increasing as one approaches the extremal limit, for fixed $J$ near extremality the growth in time of $\Delta C_{\rm A}$ remains finite and in fact slower than for smaller values of $J$ (see again fig.~\ref{gr1}).

\paragraph{Grand canonical ensemble.}

For later comparison with the TFD model of sec.~\ref{sec:QFT}, we here also consider the grand canonical ensemble and study the results in terms of the thermodynamic variables $(T, \O_H)$. These are related to the horizons radii by
\be  \label{TOtorpm}
r_+=\frac{2\pi \ell^2 T}{1-(\ell \, \O_H  )^2}\,, \qquad\qquad  r_-=\frac{2\pi \ell^3 \,  T \,  \O_H}{1-(\ell \, \O_H )^2}\,.
\ee

Notice in the grand canonical ensemble the difference in free energies $\Delta G$ between rotating BTZ and AdS is 
\be
\Delta G \equiv G_{\rm AdS} - G_{\rm BTZ} =  - \frac 1 8 + \frac{(\pi \ell T)^2}{2\(1-\ell^2 \Omega_H^2\)}\,,
\ee
with BTZ (AdS) being the dominant phase when $\Delta G > 0$ ($<0$) \cite{Detournay:2015ysa}. This specifies a region in the $(T, \O_H)$ parameter space where rotating BTZ is the dominant gravitational solution. In the following figures we  plot $(T, \O_H)$ within the entire parameters range, including the region where AdS would dominate the grand canonical ensemble. 

\begin{figure}[h]
\centering
\includegraphics[width=.45\linewidth]{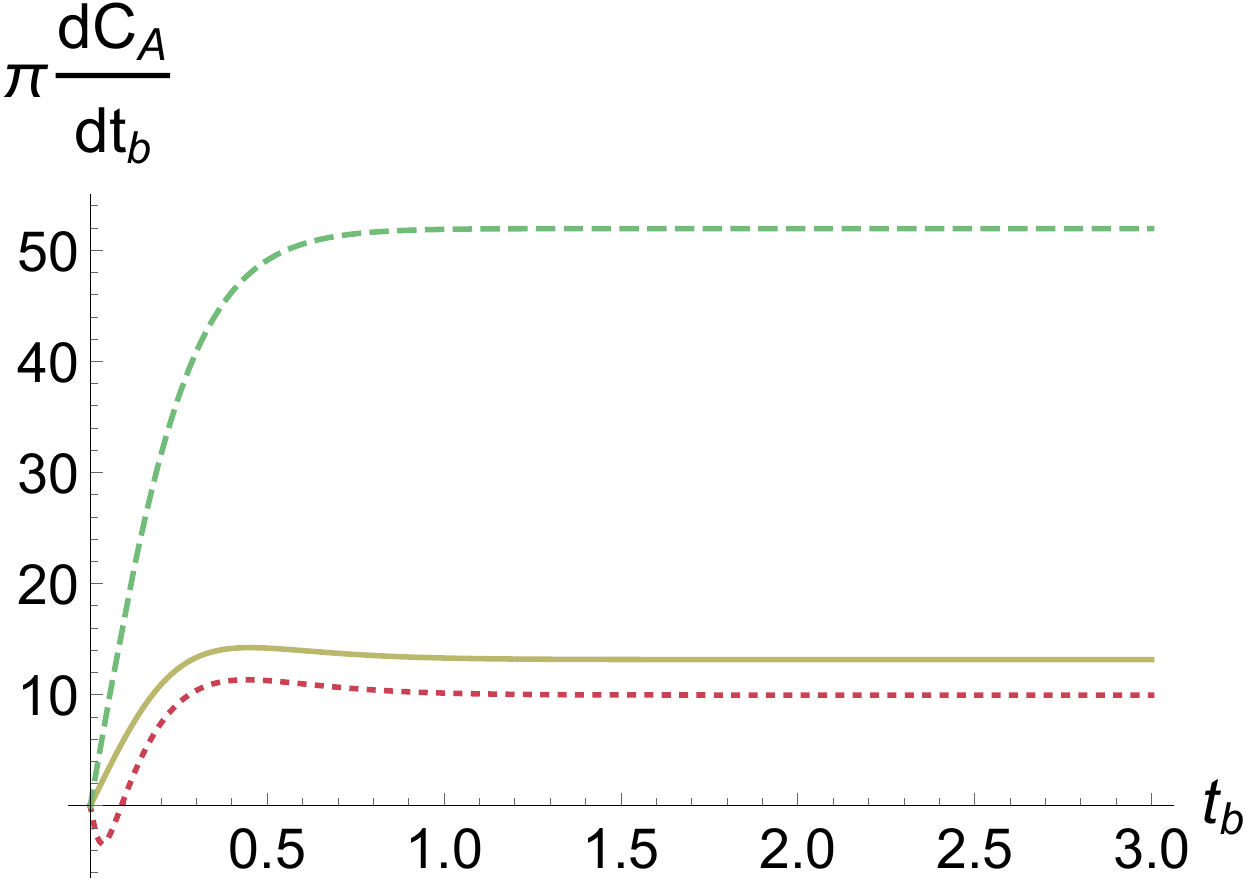} \hfill
\includegraphics[width=.45\linewidth]{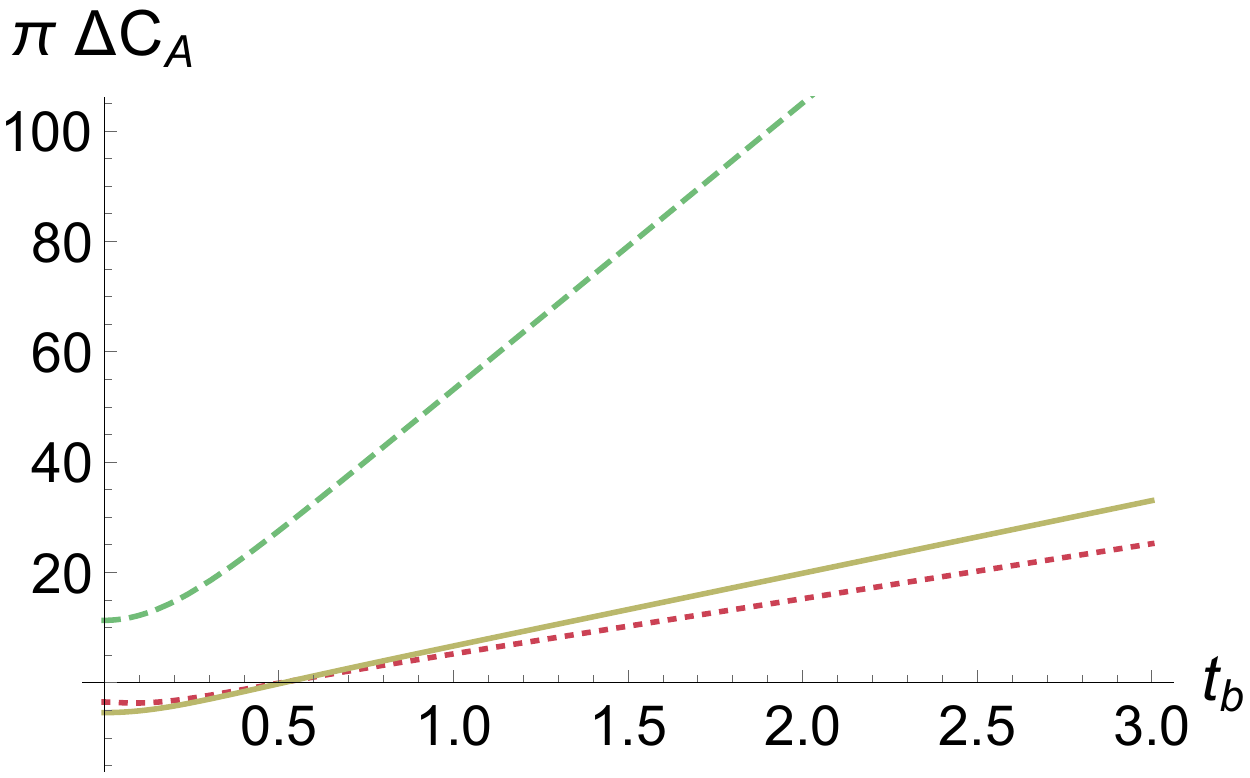}  \\
\includegraphics[width=.45\linewidth]{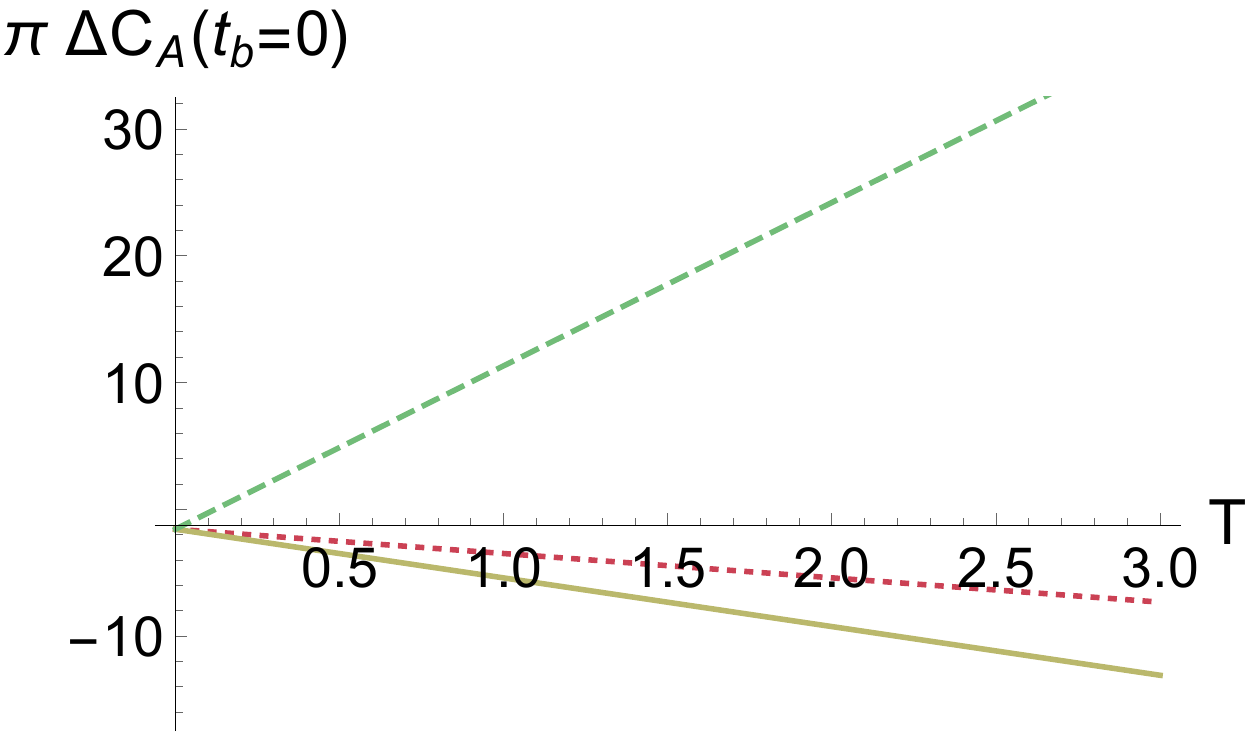} \hfill
\includegraphics[width=.45\linewidth]{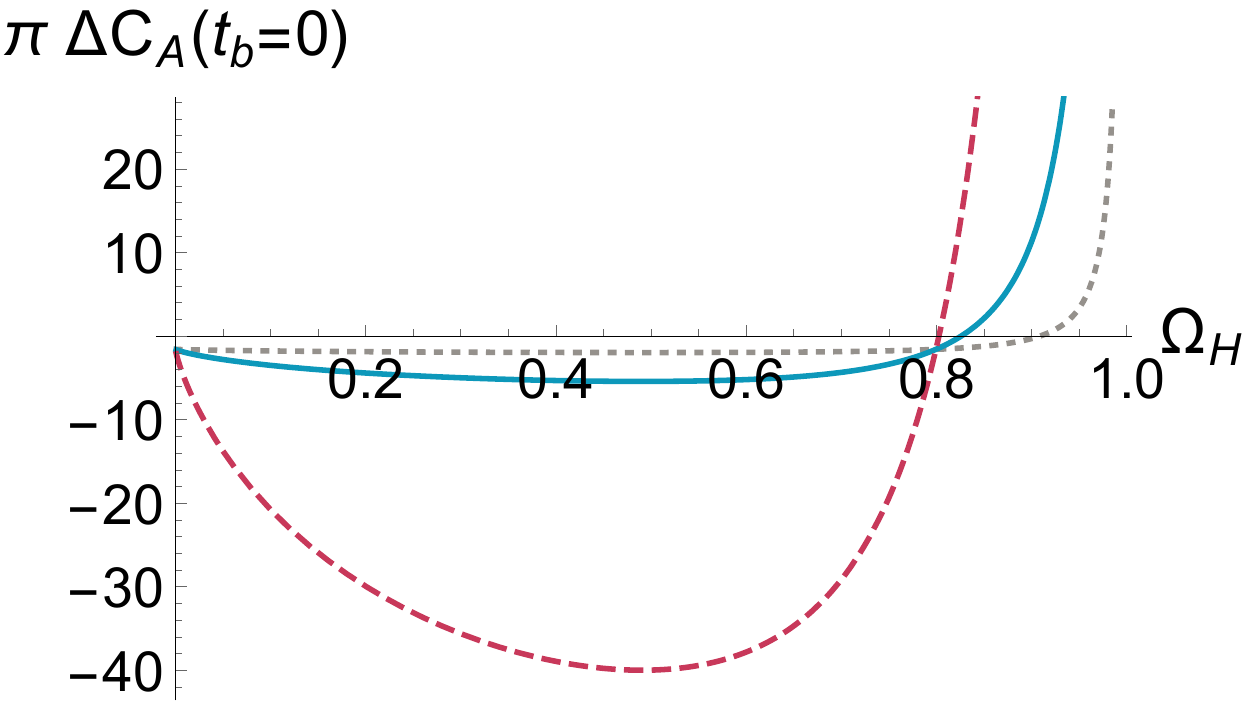}  
\centering
\caption{Above: time dependence of (left) $\pi \frac{d C_{\rm A}}{dt_b}$ and (right) $\pi \Delta C_{\rm A}$ for  $ \ell = G_N = T = 1$ and $\O_H = 0.1$ (dotted magenta), $\O_H =0.5$ (solid olive), $\O_H = 0.9$ (dashed green). \\ Below: complexity of formation $\pi \Delta C_{\rm A}(t_b=0)$ for $ \ell = G_N =1$ and (left) $\O_H = 0.1$ (dotted magenta), $\O_H =0.5$ (solid olive), $\O_H = 0.9$ (dashed green), (right) $T=0.1$ (dotted gray), $T=1$ (solid blue), $T =10$ (dashed red).}\label{fig:GCcfCA}
\end{figure}

In fig.~\ref{fig:GCcfCA}, we plot the CA  growth rate, complexity variation and complexity of formation. The complexity growth rate increases with $\Omega_H$. In particular, the limiting value \eqref{lateTimeRate} diverges when $\ell \, \Omega_H   \to 1$, as it is apparent  substituting \eqref{TOtorpm} into \eqref{lateTimeRate}. This is the critical angular velocity limit, in which the Einstein universe conformal to the AdS boundary rotates at the speed of light  \cite{Hawking:1998kw}.  For fixed angular velocity $\ell \,\O_H$, the complexity of formation is linear in the temperature. However, the slope changes sign in CA and is only positive for large enough angular velocity $\ell \,  \O_H$. For $\ell \, T \to 0$ all curves tend to:
\be
\Delta C_{\rm A}(t_b =0) \sim - \frac{\ell}{2 G_N}\,,
\ee
independently of $\Omega_H$. This correctly coincides with the complexity of formation of BTZ first evaluated in \cite{Chapman:2016hwi}. 

At fixed $\ell \, T$,  $\Delta C_{\rm A}(t_b=0)$ diverges for  $\ell \, \O_H \to 1$. In this limit $r_-, r_+$ and $r_{m1} = r_{m2}$ go to infinity, but as apparent from eq.~\eqref{TOtorpm} their differences $r_+ - r_-$, etc. are finite. For the tips location we have explicitly: 
\be
r_{m1} = r_{m2} \approx  \frac{\pi \ell^2 T}{1- \ell \, \Omega_H }\,.
\ee
Substituting into eq.~\eqref{eq:BTZDeltaCA}, we obtain the leading divergence in the critical velocity limit $\ell \, \O_H \to 1$
\be \label{eq:OmegaoneCA}
\Delta C_{\rm A}(t_b =0) \sim  \frac{\ell^2 T}{2 G_N (1-\ell \, \Omega_H )} \log \frac{\ell^2}{4 L_{\rm ct}^2 (1 -\ell \,\Omega_H)} \,. 
\ee

\begin{figure}[h]
\centering
\includegraphics[width=.45\linewidth]{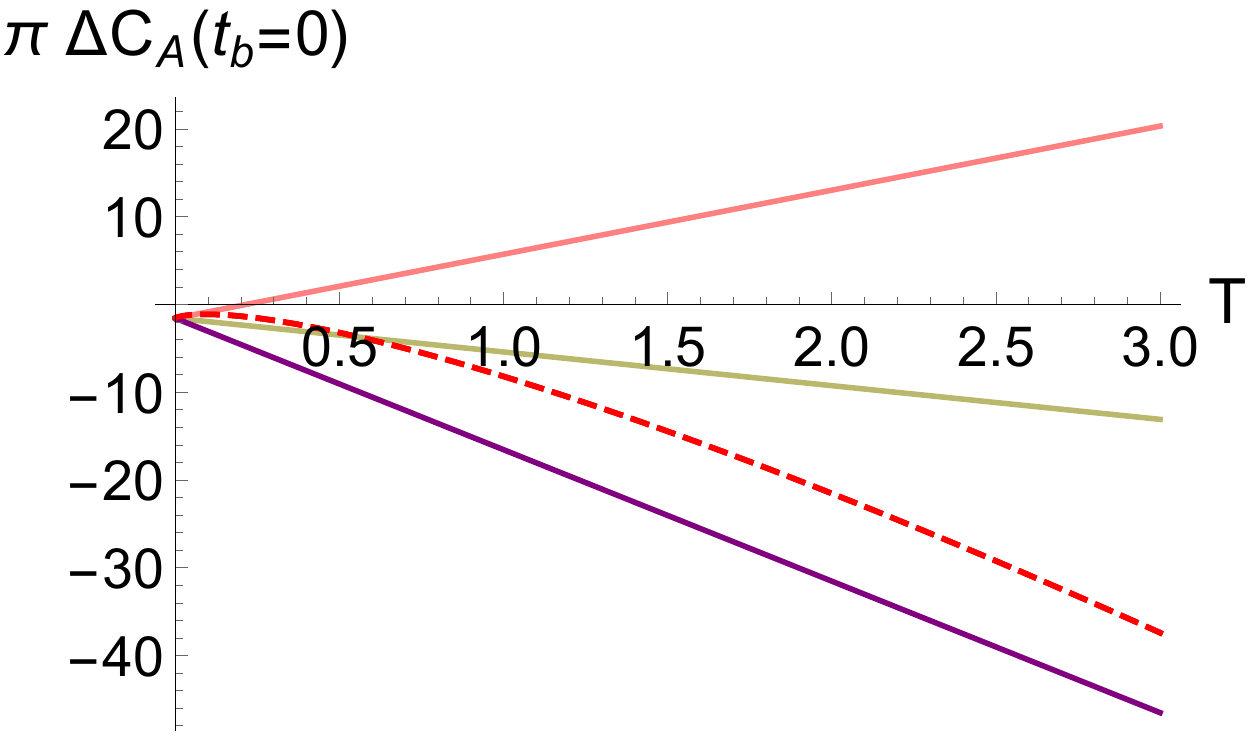} \hfill
\includegraphics[width=.45\linewidth]{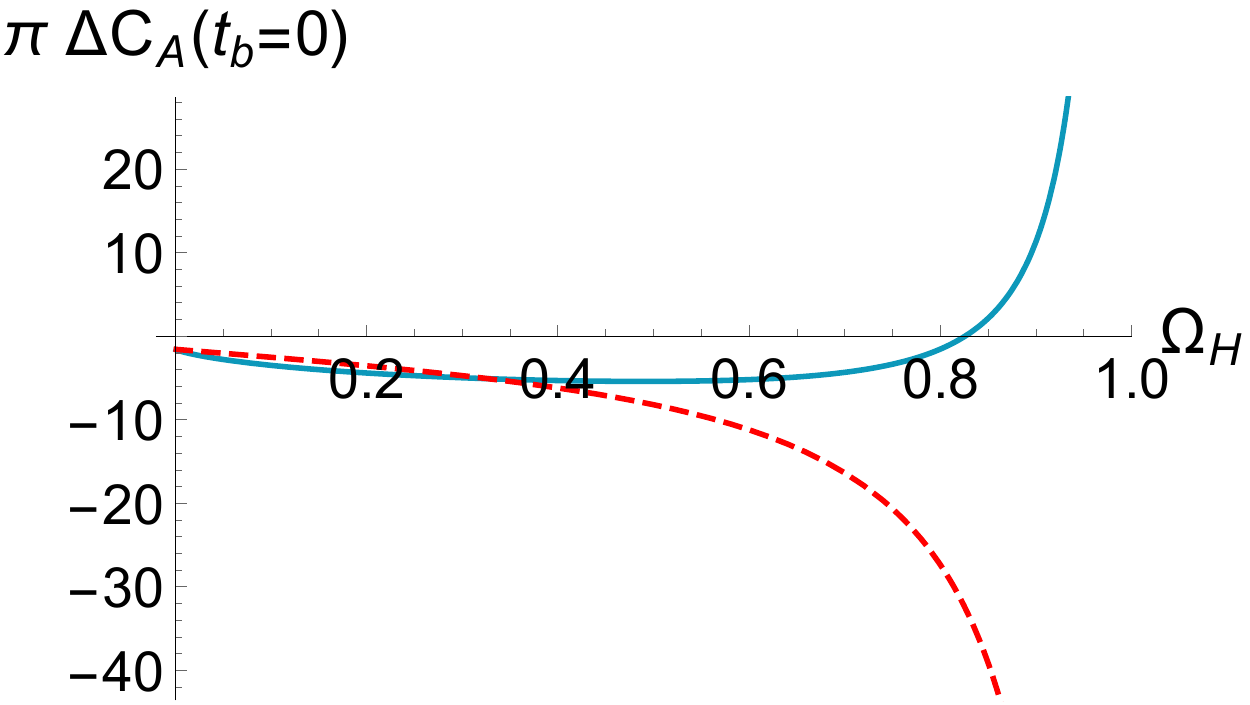}  
\centering
\caption{The plots illustrate the dependence on the counteterterm and its scale of the complexity of formation $\pi \Delta C_{\rm A}(t_b=0)$ for $ \ell = G_N =1$. (Left) $\O_H =0.5$ and in solid lines, $L_{\rm ct} = 0.1,1,10$ increasing from the top down, (right) $T=1$. In red dashed in both plots, the complexity of formation without the counterterm.}\label{fig:GCcfCALct}
\end{figure}

The linearity of the complexity of formation in the temperature is stable against variations of the counteterterm scale $L_{\rm ct}$, as illustrated in the left panel of fig.~\ref{fig:GCcfCALct}. There we also see that without the counterterm contribution the dependence would not be linear, see dashed red curve. The right panel instead illustrates that interestingly in the critical limit $\ell \, \O_H \to 1$ the complexity of formation is positively divergent with the counterterm action included, while it diverges negatively otherwise. Indeed, for the latter we can derive in this limit
\be \label{eq:OmegaoneNOCT}
\Delta C_{\rm A}(t_b =0)  - \Delta C_{\rm A}^{\rm CT}(t_b =0) \sim   - \frac{\ell^2 T}{2 G_N (1-\ell \, \Omega_H )} \log \frac{4 \pi^2 \ell^2 T^2}{\alpha^2 (1 -\ell \,\Omega_H)} \,. 
\ee
%

\subsection{Complexity=Volume}\label{sec:CVBTZ}

The growth rate  of CV for the rotating BTZ black hole was analyzed in \cite{Auzzi:2018zdu} as a subcase of warped AdS$_3$ black holes. In this section we extend the existing results considering the total complexity and its time evolution, as well as the complexity of formation and its  $J$ dependence.

\subsubsection{Volume evaluation}

In order to compute complexity according to the CV proposal \eqref{eq:CV} we look for the maximal codimension-1 spacelike slice in the bulk that ends on a time-slice on the boundary where the CFT lives, see figure \ref{figuraPenrose1}.  As before, we choose a symmetric time evolution $t_L=t_R=\frac{t_b}{2}$ and exploit the left-right symmetry of the extremal slice. 
The maximal hypersurface in the bulk shares the same axial symmetry of the Eddington-Finkelstein metric \eqref{metricEF}.  We can therefore use a parametrization of the form:
\begin{equation}
x^\mu\left(\lambda\,,\,\Phi\right)=\left(v(\lambda)\,,\,r(\lambda)\,,\,\Phi\right) \, ,
\end{equation} 
and the volume functional, expressed in terms of the induced metric $h$, thus reads
\be \label{eq:volBTZEF}
V=\int  d\lambda \int d \Phi\, \sqrt{h} =2\pi \int d\lambda \,r\sqrt{2 \dot{v}\dot{r}-f(r) \dot{v}^2} \, . 
\ee
Except for the explicit expression of the blackening factor $f(r)$, \eqref{eq:volBTZEF} has the same form as in AdS-Schwarzschild black holes analyzed in \cite{Carmi:2017jqz}. In the rest of this section we will therefore just review the main steps entering the CV analysis and refer to \cite{Carmi:2017jqz} for the details.

Noticing that  the volume functional does not depend explicitly on $v$ and  gives rise to a conserved quantity $E$, and using the freedom in fixing the parametrization $\lambda$,  one arrives to the following equations for the extremal hypersurface
\begin{equation}\label{Acaso16} 
\begin{aligned}
&\,E=r^2 \left(f \dot{v}-\dot{r} \right)\, , \\
&\,r^2\dot{r}^2=f(r)+\frac{E^2}{r^2} \, . 
\end{aligned}
\end{equation}  
From these  and  given the left-right symmetry of the problem, we get  the volume 
\begin{equation}
V=2\pi \int d\lambda =4\pi \int_{r_{\rm min}}^{r_{\rm max}}\frac{dr}{\dot{r}} = 4\pi \int_{r_{\rm min}}^{r_{\rm max}}\frac{r^2}{\sqrt{f(r)r^2+E^2}} \, dr \, .
\end{equation}
The range of integration goes from the regulator surface at the asymptotic boundary to the minimal radius $r_{min}$,  identified by $\left.\dot{r}\right|_{r_{min}}=0$, see fig.~\ref{figuraPenrose1}. From (\ref{Acaso16}) one can check that this is related to the  constant $E$ as\footnote{Notice that $r_-<r_{\rm{min}}< r_+$ and thus $f(r_{min}) <0$.}
\begin{equation}\label{Acaso18}
E=-\sqrt{-f(r_{\rm min})r_{\rm min}^2}. 
\end{equation}
$r_{\rm{min}}$ and $E$  are in turn related to the boundary time $t_b$ as  
\begin{equation}\label{Acaso23}
\frac{t_b}{2}=\int_{r_{\rm min}}^{\infty} \frac{E}{f(r)\sqrt{f(r)r^2+E^2}} \, dr\, .
\end{equation}
This follows from the definition of the ingoing null coordinate $v$ and equations \eqref{Acaso16} 
\begin{equation}\label{Acaso19}
\frac{t_b}{2}+ r^*(\infty)-r^*(r_{\rm min})=\int_{v_{\rm min}}^{v_{\infty}}dv=\int_{r_{\rm min}}^{\infty}\left[\frac{E}{f(r)\sqrt{f(r)r^2+E^2}}+\frac{1}{f(r)}\right]dr
\end{equation}
noticing that $r^*(\infty)-r^*(r_{\rm min})=\int_{r_{\rm min}}^{\infty}dr/f(r)$. We do not report  the explicit expression here, but eq.~\eqref{Acaso23} can be integrated  in terms of elliptic integrals of the third kind. 

For our proposes here  we can replace the upper limit of integration  with $r_{\rm{max}}$,  as this only gives corrections to $V$ that vanish in the limit where the regulator is removed, and use the resulting expression to rewrite the volume as
\begin{equation}
V=4\pi \int_{r_{\rm min}}^{r_{\rm max}}\left[\frac{\sqrt{f(r)r^2+E^2}}{f(r)}+\frac{E}{f(r)}\right] dr - E\left(\frac{t_b}{2}+r^*(\infty)-r^*(r_{\rm min})\right) \, .
\end{equation}
Evaluating the time derivative while keeping in mind that both $E$ and  $r_{\rm min}$ depend on the boundary time $t_b$ then yields
\begin{equation}\label{Acaso20}
\frac{dV}{dt_b}=-2\pi E= 2\pi \sqrt{- f(r_{\rm min})r_{\rm min}^2}\,.
\end{equation} 
 
\subsubsection{CV results}

Having computed the volume, the expression for CV only requires to fix the arbitrary scale $\ell_{\rm bulk}$ appearing in equation \eqref{eq:CV}. We  set it to be equal to the AdS radius $\ell$, as common in the literature \cite{Chapman:2016hwi}. We then have 
\begin{equation}\label{Acaso21}
C_{\rm V}=\frac{4\pi}{G_N \ell}\int_{r_{\rm min}}^{r_{\rm max}}\frac{r^2}{\sqrt{f(r)r^2+E^2}} \, dr 
\end{equation}
and 
\be
 \frac{dC_{\rm V}}{dt_b}=-\frac{2\pi E}{G_N \ell}=\frac{2\pi}{G_N \ell}  \sqrt{- f(r_{\rm min})r_{\rm min}^2} \, .
\ee
Eq.~\eqref{Acaso21} can also be evaluated explicitly in terms of elliptic integrals.

\paragraph{Growth rate.} 

\begin{figure}[ht]
\begin{center}
\includegraphics[width=.5\linewidth]{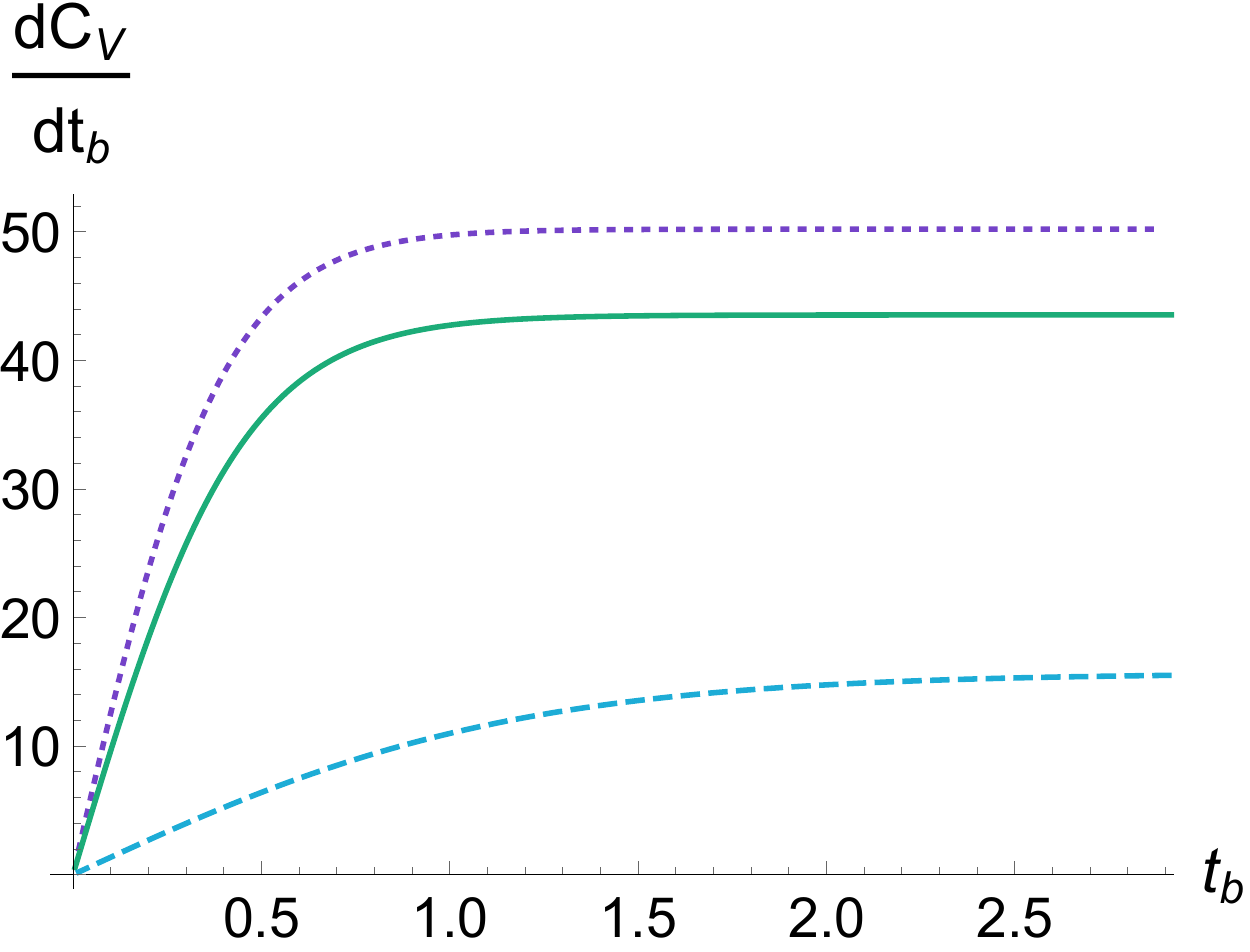}
\end{center}
\caption{$\frac{dC_{\rm V}}{dt_b} $ with $M=2$, $\ell= G_N=1$ for $J =0.1$ (purple dotted), $J=1$ (green solid) and $J=1.9$ (blue dashed).} \label{figc1}
\end{figure}

Let us start from the time dependence of the rate of complexification. At generic values of $t_b$,  this can be studied semi-analytically, inverting numerically equation \eqref{Acaso23}  to extract $r_{min}(t_b)$. From fig.~\ref{figc1} we observe first of all a dependence on the value of the angular momentum that is qualitatively similar to the one of CA. The growth rate decreases  as we increase $J$.

On the other hand there are also substantial differences with CA. First of all the rate of growth is always positive, that is, the maximal slice keeps increasing in volume as time evolves. The other main difference concerns the late time growth rate. As  we are now going to discuss in some details, in the case of CV the asymptotic value of the growth rate is given by the same function of $M, \Omega_H$ and $J$ as for CA, but this value is now approached from below. 

At late times the maximal surface will be almost tangent to a special bulk slice of constant radius $r=\tilde{r}_{min}$ \cite{Stanford:2014jda}. The value of $\tilde{r}_{min}$ can be obtained following the same strategy as in \cite{Carmi:2017jqz}. Starting from \eqref{Acaso18} we can define 
\be
W(r)\equiv \sqrt{-f(r)r^2} \, ,
\ee
 so that $r_{min}$ will be the larger positive root of the equation\footnote{Explicitly, the two roots   are  $r_{1,2}=\sqrt{\frac{\left(r_+^2+r_-^2\right)\pm\sqrt{(r_+^2 - r_-^2)^2-4E^2 \ell^2}}{2}}.\,$}
\begin{equation}
E^2-W^2(r)=0\,,
\end{equation}
For $t_b=0$, by symmetry, the extremal hypersurface coincides with the slice $t=0$ and in this case $r_{min}=r_+$, with $E=0$. As time increases, the negative $E$ will decrease, and so will do $r_{min}$ until the two roots coincide for $r_{min}=\tilde{r}_{min}$.\footnote{A way to understand this is to look at \eqref{Acaso23}. Despite $r_{\rm{min}}$ being a root of the factor $\sqrt{E^2-W^2(r)}$ appearing at the denominator,  for finite $t_b$ the integral converges, since the integrand goes like $1/\sqrt{r-r_{\rm{min}}}$ for $r \to r_{min}$. However, at late time,  $t_b\to \infty$, the integral has to diverge.  Since $r_+ > r_{\rm{min }}> r_-$, this is only possible if the two roots coincide in this limit, so that the integrand goes like $1/(r-r_{\rm{min}})$.}  The corresponding late time values are 
\begin{equation}
\tilde{r}_{\rm min}^2=\frac{r_+^2+r_-^2}{2}\,,\qquad\qquad\tilde{E}^2=\frac{\left(r_+^2-r_-^2\right)^2}{4 \ell^2}\,.
\end{equation}
Notice that this also implies that  $\tilde{r}_{\rm min}$  represents an extremal point for $W(r)$. Expanding \eqref{Acaso21} around the extremum $W(\tilde r_{\rm{\rm min}})$, we thus obtain the following late time rate of complexification 
\begin{equation}
\frac{dC_{\rm V}}{dt_b}=\frac{\pi}{G_N \ell^2}\left(r_+^2-r_-^2\right)  - \frac{ 4\pi \left( r_+^2+r_-^2 \right)}{G_N \ell^2\left(r_+^2-r_-^2\right)}\left({r}_{\rm min}^2-\tilde{r}_{\rm min}^2\right)+ \dots \, . 
\end{equation}
The limiting value, which expressed in terms of $M$, $\Omega_H$ and $J$ takes the form
\be
\frac{dC_{\rm V}}{dt_b} \sim 8\pi\left(M-\Omega_H J\right) \, .
\ee
This is approached from below, as opposed to CA, and smoothly reduces to the non-rotating result  \cite{Carmi:2017jqz} for $J\to 0$.

\paragraph{Complexity variation.}

\begin{figure}[h]
\begin{center}
\includegraphics[width=.45\linewidth]{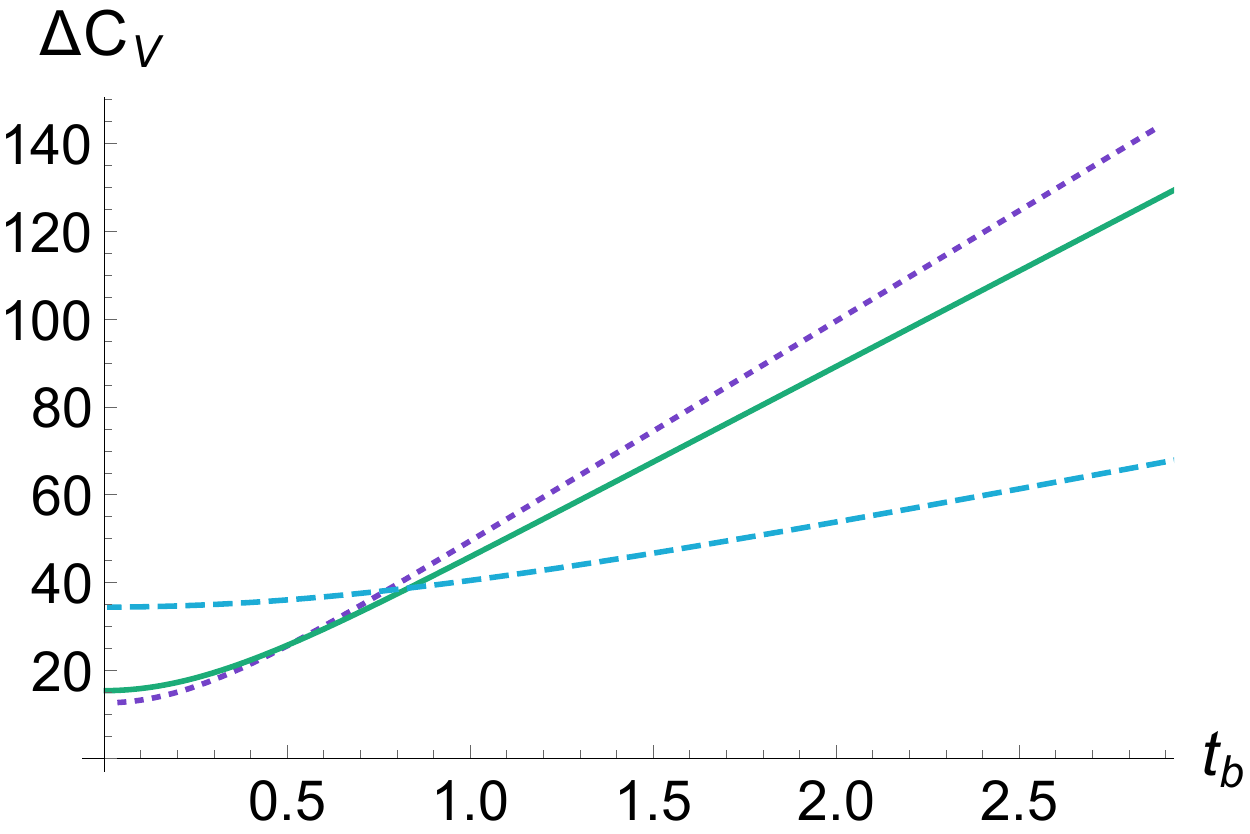} \hfill \includegraphics[width=.45\linewidth]{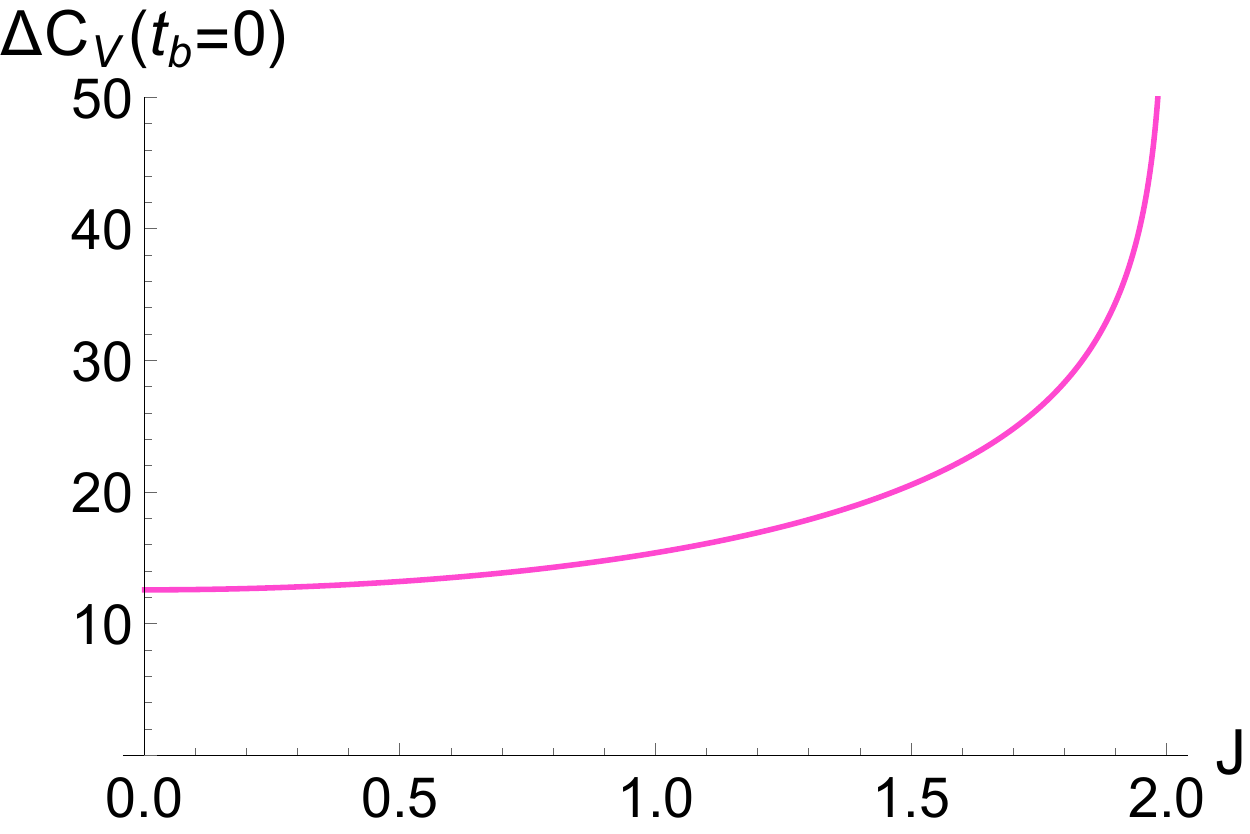}
\end{center}
\caption{(Left) $\Delta C_{\rm V}(t_b)$ for  $J =0.1$ (purple dotted), $J=1$ (green solid) and $J=1.9$ (blue dashed). (Right) $\Delta C_{\rm V}(t_b =0) $ as a function of $J$.  In both plots $M=2$, $\ell=G_N=1$.}\label{figuraDivergenza2}
\end{figure}

The rate of growth of CV is  UV-finite, but the total complexity  diverges as
\begin{equation}
C_{\rm V}^{\rm{BTZ}}\sim\frac{4\pi}{G_N} \, r_{\rm{max}} + \dots \, . 
\end{equation}
This is the same divergence one finds for (twice) global AdS$_3$\footnote{Again, in principle one should be careful and appropriately match the cutoffs for these two different spacetimes. In the same way as for CA,  also here this procedure turns out to only give negligible corrections, \ie corrections  vanishing in the limit where the cutoff is removed. Therefore we set $r_{\rm{max}}^{\rm{BTZ}} = r_{\rm{max}}^{\rm{AdS}}= r_{\rm{max}}.$} 
\begin{equation}
C_{\rm V}^{\rm{AdS}} = \frac{2\pi}{G_N }\int_{0}^{r_{\rm max}} \frac{r}{\sqrt{\ell^2 + r^2}} =  \frac{2\pi}{G_N} \( r_{\rm max}  - \ell\) \, . 
\end{equation}
Subtracting twice the constant AdS value 
\begin{equation}
\Delta C_{\rm V}=C_{V}^{\rm BTZ}-2 C_{V}^{\rm AdS}
\end{equation}
gives a finite result,  which we plot as a function of boundary time in figure \ref{figuraDivergenza2} (left).  
%

\paragraph{Complexity of formation.}

For $t_b=0$, this defines the CV complexity of formation and since the radial integration runs all the way to the bifurcation surface $r_{\rm min} = r_+$, it can also be evaluated explicitly: 
\be \label{eq:CVformationSt}
\Delta C_{\rm V} (t_b =0)= \frac{4\pi r_+}{G_N} \left\{ \frac{\ell}{r_+} +  K\(\frac{r_-^2}{r_+^2}\) - E\(\frac{r_-^2}{r_+^2}\) \right\}
\ee
in terms of complete elliptic integrals.

In the extremal limit $J\to M \ell$ (or $r_+ \to r_-$), this yields  a logarithmic divergence
\be \label{eq:CVformationDivergence}
\Delta C_{\rm V}  (t_b =0) \sim \frac{4\pi r_-}{G_N} \( - \frac 1 2 \log \frac{r_+ -r _-}{8 r_-} + \frac{\ell}{r_-} -1\)\,. 
\ee
The leading term is $(2\pi)^2$ the one of CA discussed above, and analogous to the divergence observed in the extremal limit for charged black holes \cite{Carmi:2017jqz}. 
A plot showing the dependence on the angular momentum $J$ of the complexity of formation is reported in fig.~\ref{figuraDivergenza2} (right).

As for CA, $\frac{d C_{\rm V}}{dt_b}\sim 0$ near extremality. Thus, even though the complexity of formation increases rapidly as we approach the extremal limit, at fixed values of $J$ near extremality $\Delta C_{\rm V}$ increases only slightly in time (see fig.~\ref{figuraDivergenza2}).
Finally, we notice that CV, and in particular the complexity of formation, is always positive, which was not the case for  CA.

In appendix~\ref{app:BTZBL}, for later reference with the Kerr-AdS discussion of sec.~\ref{sec:CVKerr}, we present a slightly different computation in which we evaluate the CV complexity of formation in Boyer-Lindquist-like coordinates \cite{Hawking:1998kw}. 
 
\paragraph{Grand canonical ensemble.}
 
\begin{figure}[h]
\centering
\includegraphics[width=.45\linewidth]{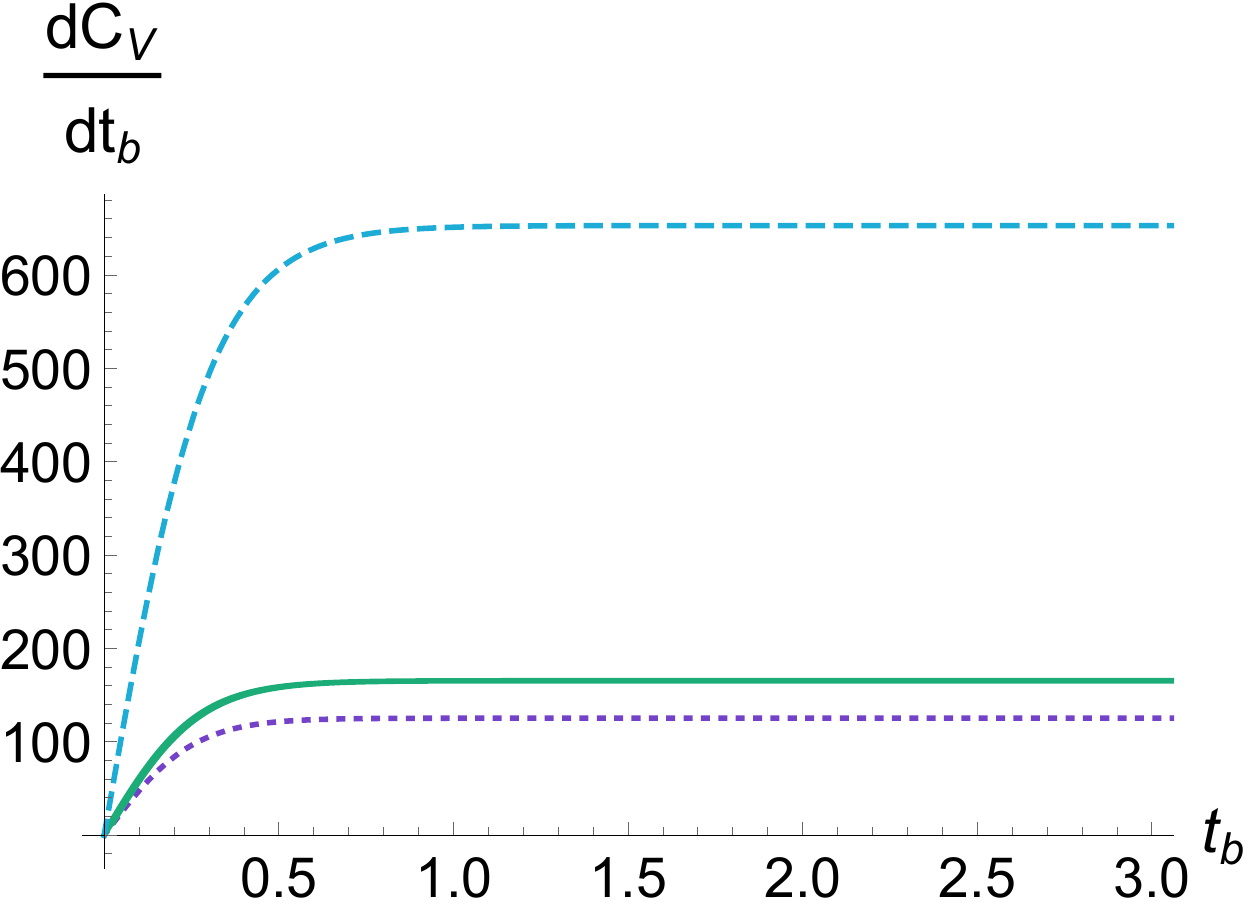} \hfill
\includegraphics[width=.45\linewidth]{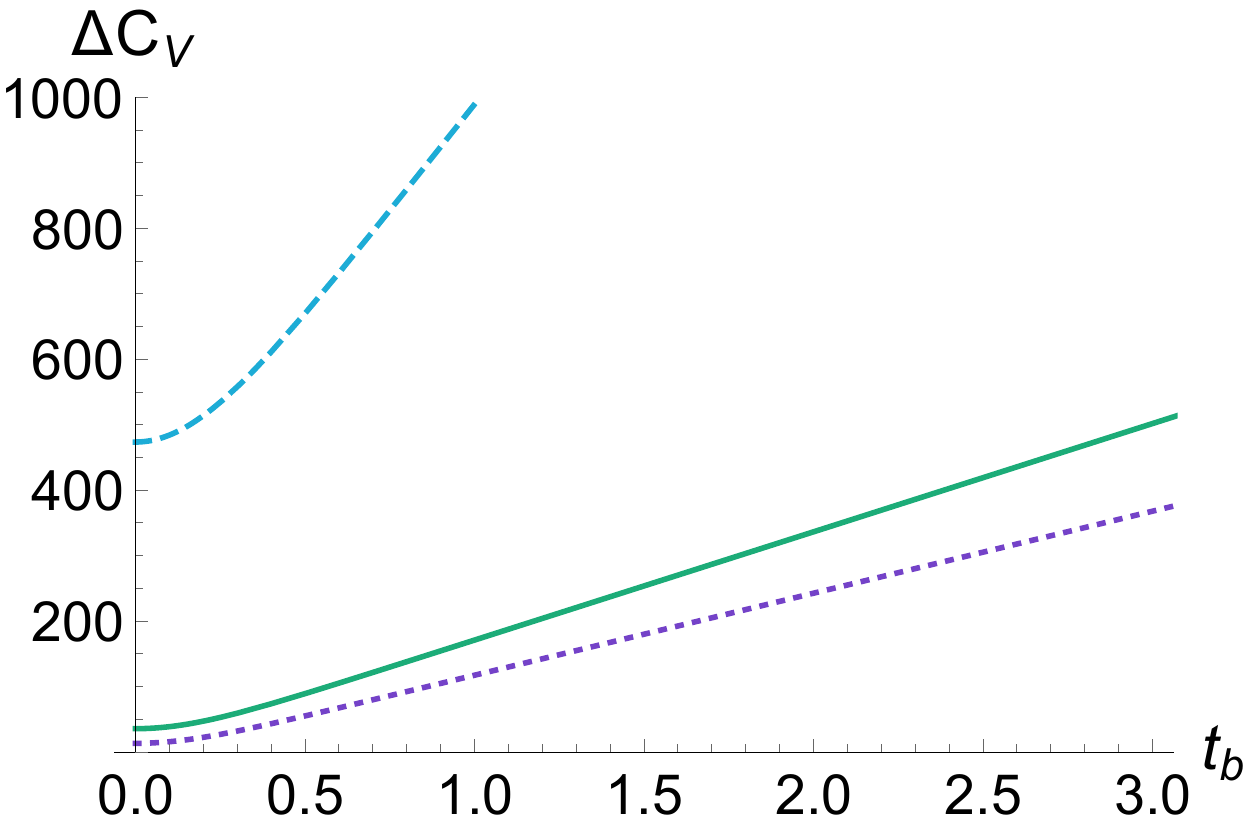}  \\
\includegraphics[width=.45\linewidth]{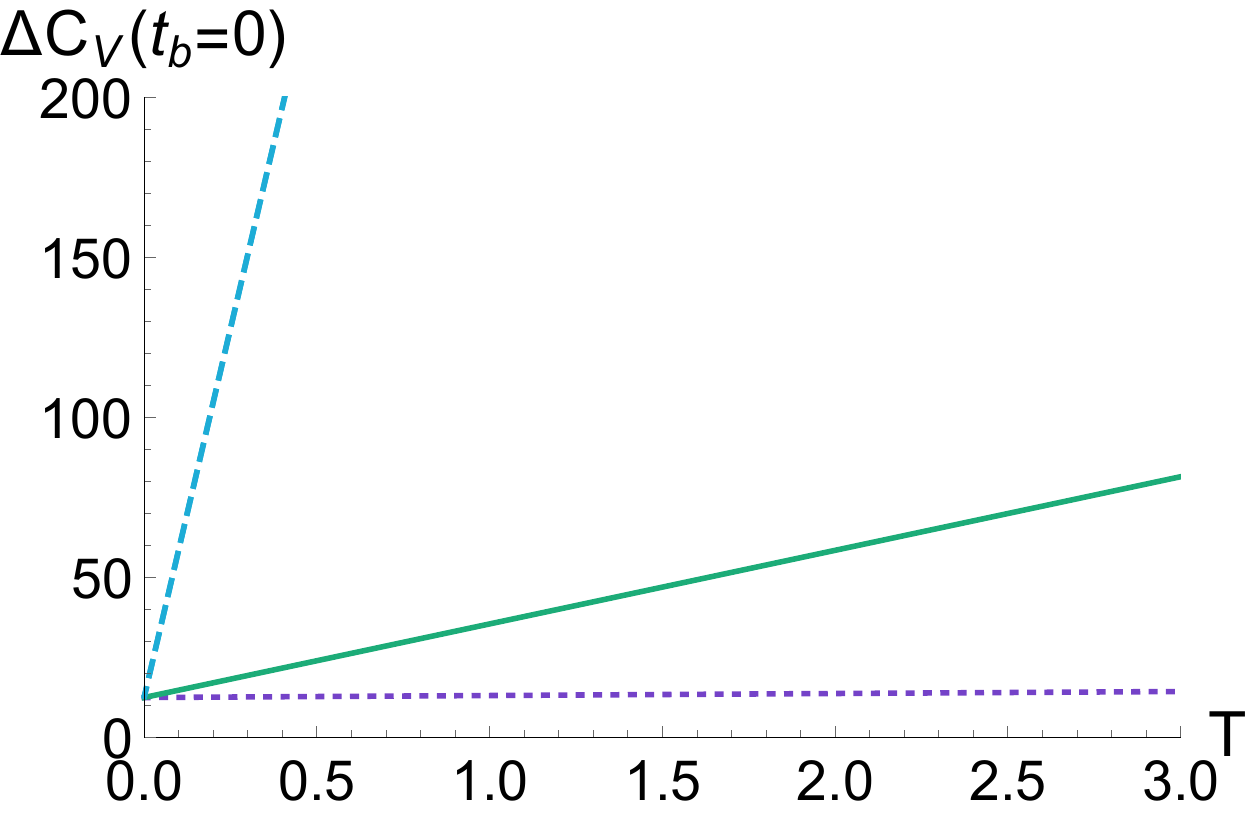} \hfill
\includegraphics[width=.45\linewidth]{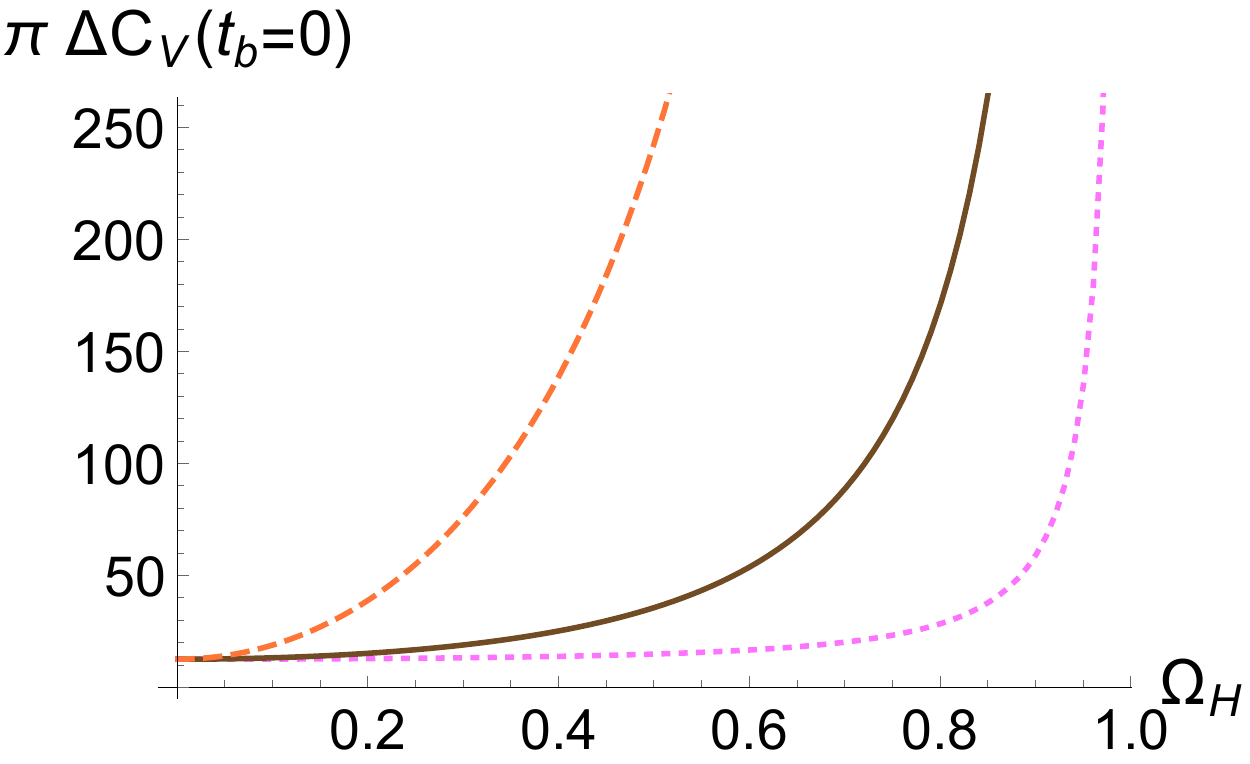}  
\centering
\caption{Above: Time dependence of (left) $\frac{d C_{\rm V}}{dt_b}$ and (right)  $\Delta C_{\rm V}$ for $ \ell = G_N = T = 1$  and $\O_H = 0.1$ (dotted purple), $\O_H =0.5$ (solid green), $\O_H = 0.9$ (dashed blue).  \\ 
Below: $\Delta C_{\rm V}(t_b=0)$ for $ \ell = G_N =1$ and (left) $\O_H = 0.1$ (dotted purple), $\O_H =0.5$ (solid green), $\O_H = 0.9$ (dashed blue), (right) $T=0.1$ (dotted pink), $T=1$ (solid brown), $T=10$ (dashed orange). }\label{fig:GCcfCV}
\end{figure}

We plot in fig.~\ref{fig:GCcfCV} the CV growth rate, complexity variation and complexity of formation in terms of the thermodynamic variables $(T, \O_H)$. 

As for CA, for fixed angular velocity $\ell \, \O_H$, the complexity of formation is linear in the temperature. This is apparent in the CV result \eqref{eq:CVformationSt} over which we have analytic control and which yields: 
\be \label{eq:CVBTZTO}
\Delta C_{\rm V} (t_b = 0)= \frac{4\pi \ell}{G_N} \left\{ 1+  \frac{2 \pi \ell \, T }{1- \ell^2 \, \O_H^2} \left[ K\( \ell^2 \, \O_H^2 \) - E\( \ell^2 \, \O_H^2\) \right] \right\} \, .
\ee
Contrary to the CA proposal,  $\Delta C_{\rm V} (t_b = 0)$ is however everywhere positive and increasing with $T$. At fixed $ \ell \, T$, the CV complexity of formation is also divergent in the limit of critical angular velocity $\ell \, \O_H \to 1$,  as can be obtained expanding \eqref{eq:CVBTZTO} 
\be \label{eq:OmegaoneCV}
\Delta C_{\rm V} (t_b = 0) \sim  \frac{2 \pi^2 \ell  \, }{G_N}   \frac{\ell \, T }{1- \ell \, \O_H}  \( \log \frac{8}{1- \ell \, \O_H} -2 \)\,. 
\ee
This is precisely the same divergence structure as for CA \eqref{eq:OmegaoneCA}. 
%

\paragraph{CV 2.0 proposal.}

Here we will briefly consider an alternative proposal for complexity advanced in \cite{Couch:2016exn}, the so called CV 2.0 conjecture. This conjecture proposes that complexity of a state should be dual to the spacetime volume of the WDW patch multiplied by the pressure\footnote{Another CA 2.0 proposal was advanced in \cite{Fan:2018wnv}. This simply coincides with the one considered here in the case of pure Einstein gravity.}
\begin{equation}
C_{\rm V\,2.0} = P\, V_{\rm WDW},
\end{equation}
where the pressure is identified with the cosmological constant according to  $P=-\frac{\Lambda}{8\pi G_N}$, as proposed by  \cite{Dolan:2012jh,Kubiznak:2014zwa,Kubiznak:2016qmn}. 

We can immediately evaluate CV2.0 and its growth rate using our results for CA. In fact for vacuum solutions the volume of the WDW is proportional to $I_{\rm bulk}$ in \eqref{eq:Ibulk}
\begin{equation}
C_{\rm V\,2.0}=-\frac{I_{\rm bulk}}{2} =-\frac{1}{8\,G_N} \left\{ 2 \left(r_{m1}+r_{m2} -4 r_{\rm max} \right) - r_+ \log \frac{(r_+ + r_{m1}) (r_+ + r_{m2})}{(r_+ - r_{m1}) (r_+ - r_{m2})} \right\}  \, ,
\end{equation}
from which follows  the growth rate
\begin{equation}\label{Acaso32}
\frac{d C_{\rm V\,2.0}}{dt_b}=\frac{r_{m2}^2-r_{m1}^2}{8 \, G_N \ell^2} \,.
\end{equation}
The late time limit
\be
\lim_{t_b \to \infty} \frac{d C_{\rm V\,2.0}}{dt_b} = \frac{r_{+}^2-r_{-}^2}{8 \, G_N \ell^2}
\ee
 agrees with the one found in  \cite{Couch:2016exn}, and in particular equals
\begin{equation}\label{Acaso31}
\frac{d C_{\rm V\,2.0}}{dt_b} = P\, \left(V_+ - V_-\right),
\end{equation}
where
 \be
 V_\pm = \pi r_{\pm}^2 \, 
 \ee
  are the thermodynamic volumes associated with the outer and inner horizons. 
The late time limit is reached from below. In fact, as time increases $r_{m1}$  tends to $r_-$ from above while $r_{m2}$ tends to $r_+$ from below, see fig.~\ref{figuraPenrose1}. One can check this explicitly using the  late time expansion for  $r_{m1}$ and $r_{m2}$ given in  \eqref{eq:rm1rm2 a late time}.

The divergence of $C_{\rm V\,2.0}$ exactly cancels when subtracting two copies of AdS$_3$ CV\,2.0 complexity
\begin{equation}\begin{aligned}
\Delta C_{\rm V\,2.0}\left( t_b \right)&= C_{\rm V\,2.0}^{\rm BTZ}\left( t_b\right)-2C_{\rm V\,2.0}^{\rm AdS} =-\frac{1}{2}\left[I_{\rm bulk}^{\rm BTZ}\left( t_b\right)-2I_{\rm bulk}^{\rm AdS}\right]\\
&=\frac{1}{8\,G_N} \left\{2 \pi \ell - 2 \( r_{m1} + r_{m2}  \) + r_+ \log \frac{(r_+ + r_{m1}) (r_+ + r_{m2})}{(r_+ - r_{m1}) (r_+ - r_{m2})} \right\} \,.
\end{aligned}
\end{equation}

\begin{figure}[ht]
\centering
\includegraphics[width=.32\linewidth]{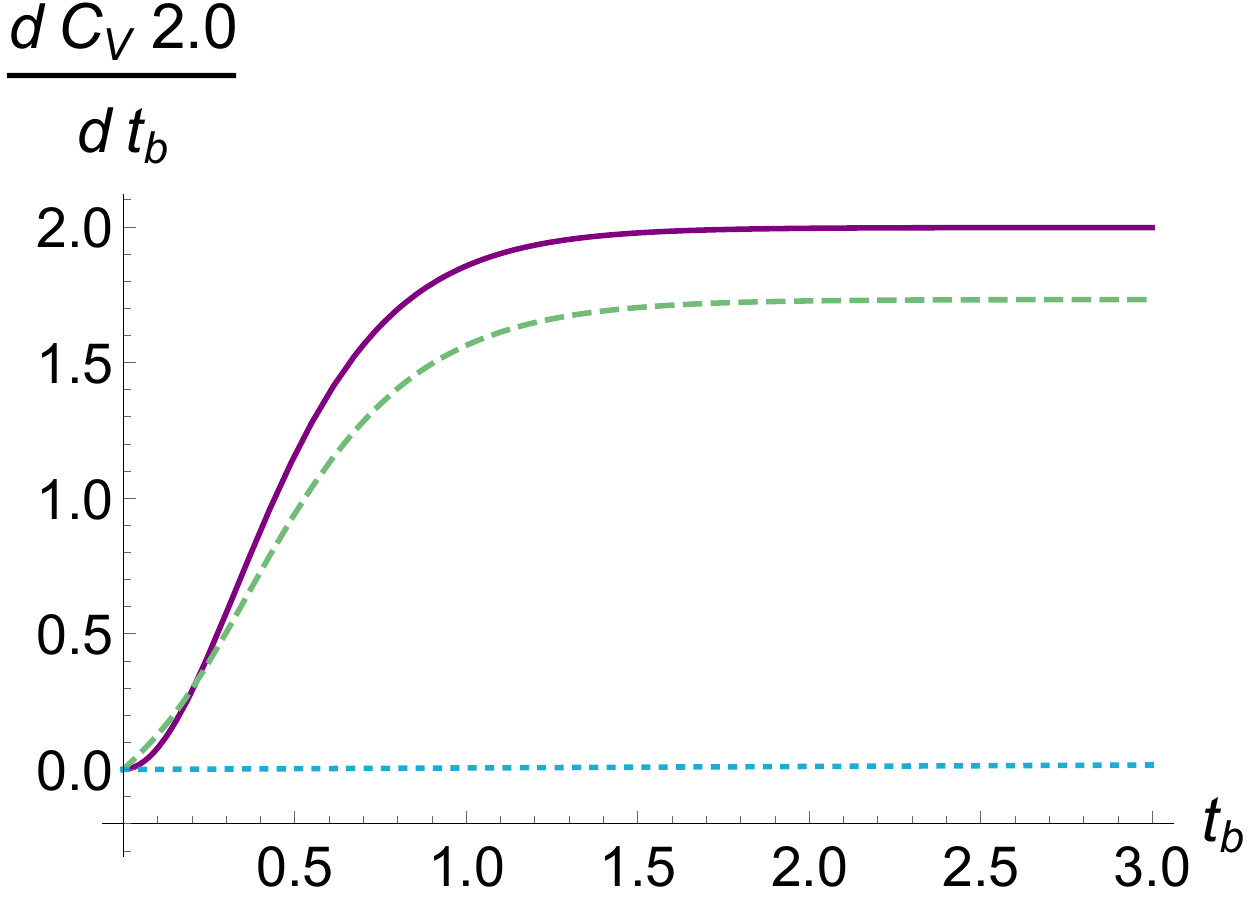} \hfill
\includegraphics[width=.32\linewidth]{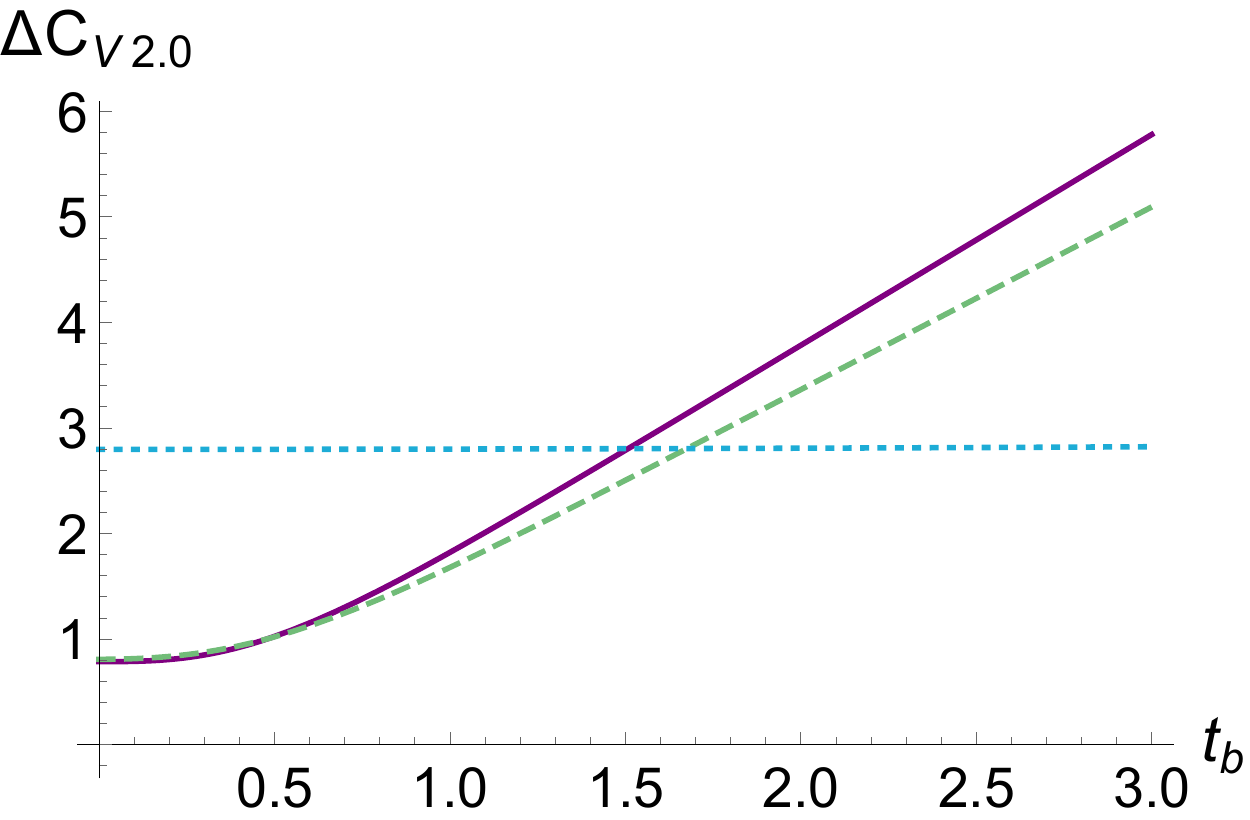}  \hfill 
\includegraphics[width=.32\linewidth]{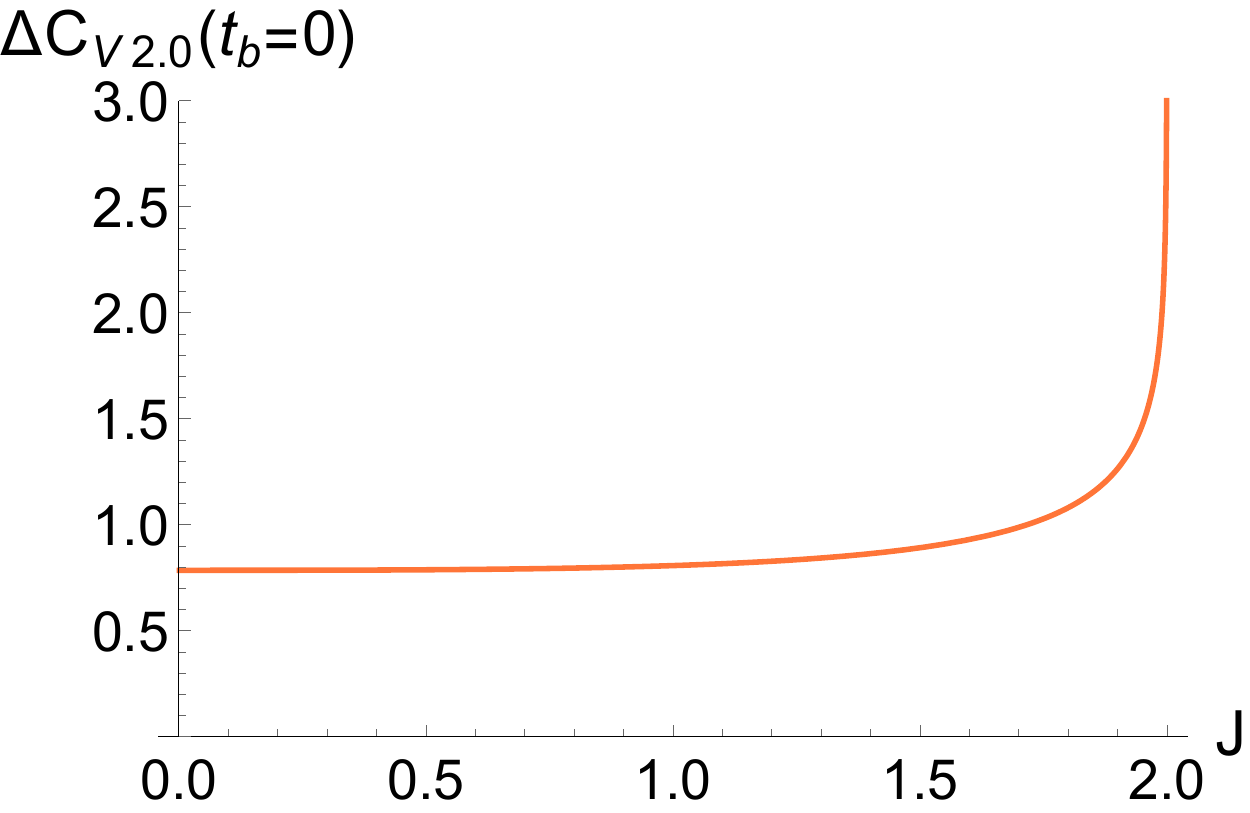}
\centering
\caption{(Left) $\frac{d C_{\rm V\,2.0}}{dt_b}$, (center)  $\Delta C_{\rm V\,2.0}(t_b)$ and (right) $\Delta C_{\rm V \, 2.0}(t_b=0)$ as a function of $J$ for $M=2$, $ \ell = G_N =1$. In the left and center panels: J=0.1 (purple solid), J = 1 (green dashed) and J =1.999 (blue dotted). 
}\label{figc4}
\end{figure}

We study this difference and the growth rate semi-analytically. Sample plots are shown in fig~\ref{figc4}.
Comparing with figure~\ref{figc1},  we see that the results for CV complexity and CV2.0 are qualitatively similar. The late time limit, although different in the two cases because of the overall relative factor in the definitions, is reached from below. Moreover, both quantities grow monotonically with the boundary time. In the near extremal limit the $C_{\rm V\,2.0}$ complexity tends to a constant in time, as in the CA and CV conjectures.  Also the complexity of formation $\Delta C_{\rm V\, 2.0}(t_b=0)$ as function of the angular momentum $J$ (see fig.~\ref{figc4}) resembles the results for CV (see fig.~\ref{figuraDivergenza2}), is always positive and diverges in the extremal limit.
%

\section{Holographic Complexity: Kerr-AdS}\label{sec:HCKerr}

In this section we extend part of the above holographic analysis to four-dimensional Kerr-AdS black holes. The axial, rather than spherical, symmetry of the solution complicates the explicit evaluation of holographic complexity. Indeed, as we comment below, even the null hypersurfaces foliation of Kerr-AdS spacetimes needed to construct the WDW patch was only worked out recently and is only known in implicit form \cite{Balushi:2019pvr}. In sec.~\ref{sec:CAKerr} and sec.~\ref{sec:CVKerr} we study respectively the CA growth rate and CV complexity of formation in this higher dimensional setup. 

\subsection{Kerr-AdS black hole}\label{sec:KerrAdS}
 
The 3+1-dimensional Kerr-AdS metric in Boyer-Lindquist coordinates reads (see \textit{e.g.} \cite{Caldarelli:1999xj})
\be\label{KerrAdSmetric}
ds^2=-\frac{\D}{\r^2}\pr{dt-\frac{a}{\Xi}\sin^2 \theta d\varphi}^2 + \frac{\r^2}{\D}dr^2 +\frac{\r^2}{\D_{\theta}}d\theta^2 + \frac{\D_{\theta}}{\r^2} \sin^2 \theta\pr{a dt-\frac{r^2 +a^2}{\Xi}d\varphi}^2
\ee
with
\be
\begin{array}{lll}
\Delta=(r^2+a^2)\left(1+\frac{r^2}{\ell^2}\right)-2m r\,, & \quad & \rho^2=r^2+a^2\cos^2\theta\,, \\
\Delta_{\theta}=1-\frac{a^2}{\ell^2}\cos^2\theta\,, & \quad & \Xi= 1- \frac{a^2}{\ell^2}\, ,
\end{array}
\ee
where $m, a$ denote respectively the mass and rotational parameters.

\begin{figure}[t]
\center
 \includegraphics[width=.3\linewidth]{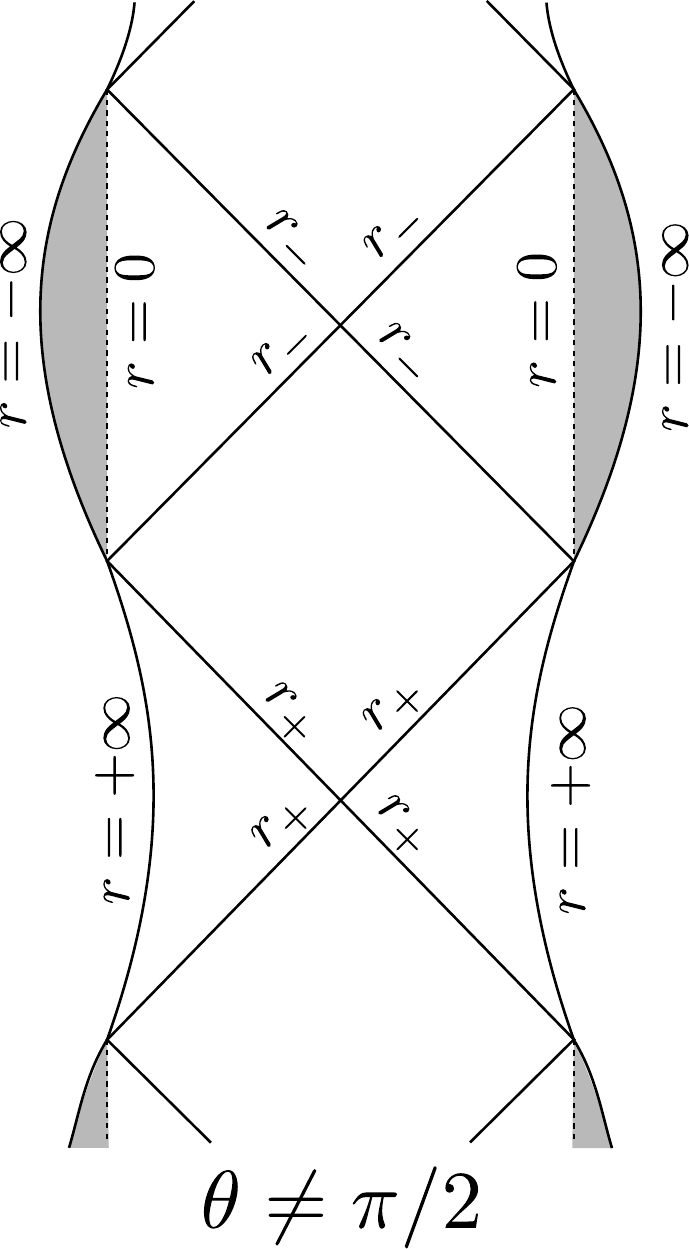} \qquad \qquad\qquad\qquad \includegraphics[width=.25\linewidth]{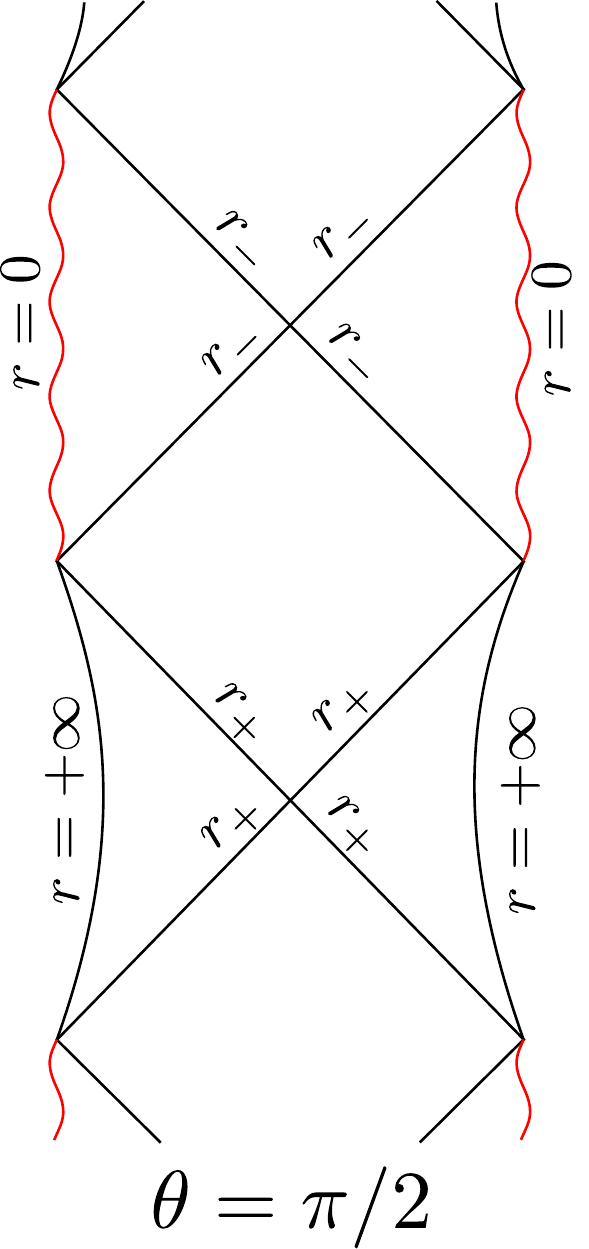}
 \caption{Fixed $\theta$ projection diagrams of Kerr-AdS$_4$. }
\label{fig:KAdSPenrosediagram}
\end{figure}

There is a singularity with the topology of a ring at $r=0,\theta=\frac{\pi}{2}$, where $\rho^2 =0$,  and the spacetime structure is fixed by the positive zeros of $\D$.  
Defining 
\be\label{mextr}
m_{\rm extr} (a) \equiv \frac{\ell}{3\sqrt 6} \(\sqrt{\( 1+ \frac{a^2}{\ell^2}\)^2 + \frac{12}{\ell^2} a^2} + \frac{2 a^2}{\ell^2} +2 \)\sqrt{\sqrt{\( 1+ \frac{a^2}{\ell^2}\)^2 + \frac{12}{\ell^2} a^2} - \frac{a^2}{\ell^2} -1}\,,
\ee
the value of $m$ for which $\D = 0$ has a double positive root, the geometry describes a naked singularity for $m < m_{\rm extr}$, an extremal black hole for $m = m_{\rm extr}$ and a black hole with outer event horizon $r = r_+ $ and inner Cauchy horizon $r= r_-$ for $m > m_{\rm extr}$. 
Notice there is also a parameter singularity at $|a|=\ell$, where $\Xi$ vanishes. 
In the following, we will consider the black hole solution with $m > m_{\rm extr}$ and $|a| < \ell$, and also study the complexity behavior  in the critical limit $|a| \to \ell$. In this limit, the three dimensional Einstein universe at infinity where the dual CFT is defined rotates at the speed of light \cite{Hawking:1998kw,Berman:1999mh,Caldarelli:1999xj} (see also below).
In fig.~\ref{fig:KAdSPenrosediagram}, we depict two fixed $\theta$ diagrams of the Kerr-AdS$_4$ black hole. 

In terms of the metric parameters $(m,a)$, the mass $M$ and angular momentum $J$ read\footnote{As for BTZ we restrict for clarity, and without loss of generality, to $J, a \ge 0$\,. }
\be \label{eq:MJKerr}
M = \frac{m}{G_N \, \Xi^2}\,, \quad\quad J = a M\,,
\ee
while the entropy $S$, temperature $T$ and angular velocity of the event horizon $\O_{H}$ are 
\bea 
S &=& S_+ = \pi \, \frac{r_+^2 +a^2}{G_N \, \Xi} \label{eq:Skerr}\\
T&=& T_+ = \frac{r_+}{4\pi\pr{r_+^2 +a^2}}\pr{1+\frac{a^2}{\ell^2}+3\frac{r_+^2}{\ell^2}-\frac{a^2}{r_+^2}}  \label{eq:Tkerr} \\
\O_{H} &=& \frac{a \, \Xi}{r_+^2 + a^2}\,.
\eea

It is also often useful to re-express the metric in the ADM form
\be\label{ADM}
ds^2=-N^2 dt^2+\frac{\r^2}{\D}dr^2 +\frac{\r^2}{\D_{\theta}}d\theta^2 +\frac{\S^2 \sin^2\theta}{\r^2 \, \Xi^2}(d\varphi-\o dt)^2,
\ee
where
\bea
N^2&=&\frac{\r^2\Delta \, \Delta_{\theta}}{\Sigma^2} \label{eq:NKerr} \\ 
\Sigma^2&=&(r^2 +a^2)^2\Delta_{\theta}- a^2 \Delta \sin^2 \theta \label{eq:SigmaKerr} \\ 
\o &=& \frac{a \, \Xi}{\Sigma^2}\pq{\Delta_{\theta}(r^2 + a^2)-\Delta} \label{eq:omegaKerr}\,,
\eea
$N$ being the lapse function and $\o$ the angular velocity. At the outer horizon
\be
\o(r_+)= \O_{H}\,, 
\ee
while asymptotically
\be
  \O_{\infty} \equiv  \omega(r\to \infty)\sim -\frac{a}{\ell^2}  \,. 
\ee
Contrarily to the asymptotically flat case (\ie $\ell \to \infty$) and to  rotating BTZ of sec.~\ref{sec:BTZBH}, the boundary at infinity is rotating. 
This feature leads to define the angular velocity as the difference
\be \label{eq:O+Kerr} 
\O_+ \equiv \O_{H} - \O_{\infty} = \frac{a}{\ell^2} \, \frac{r_+^2 + \ell^2}{r_+^2 + a^2}\, ,
\ee
which is the angular velocity of the rotating Einstein universe at infinity \cite{Hawking:1998kw,Caldarelli:1999xj}. Notice the latter rotates at the speed of light $\ell \, \O_+ = 1$ either at the critical value $a = \ell$ or if $ r_+^2 = a \, \ell $.  This critical limit was throughly studied in  \cite{Hawking:1998kw}.

\begin{figure}[h]
\begin{center}
\includegraphics[width=.30\linewidth, valign=t]{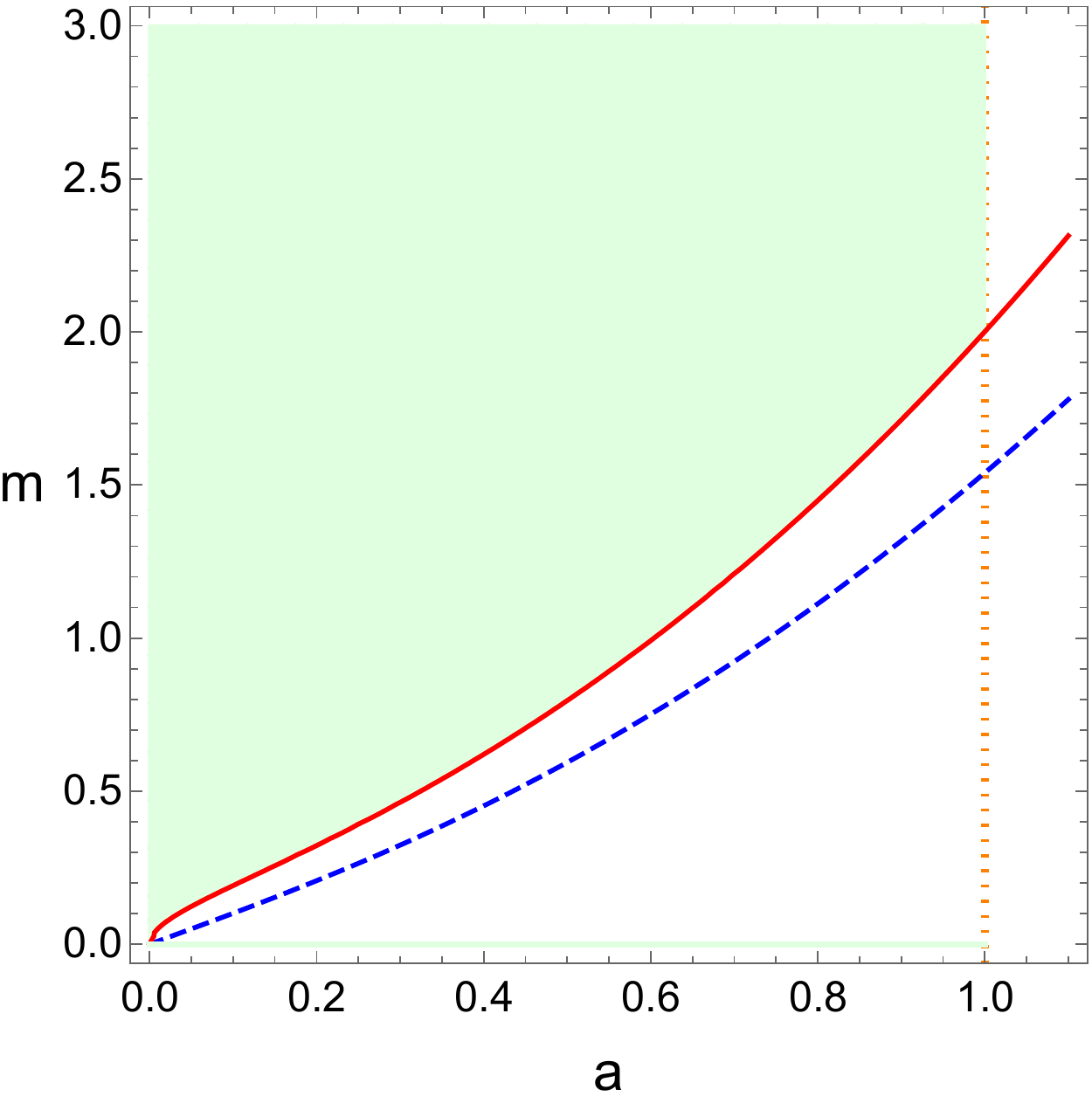}  \hfill  \includegraphics[width=.30\linewidth, valign=t]{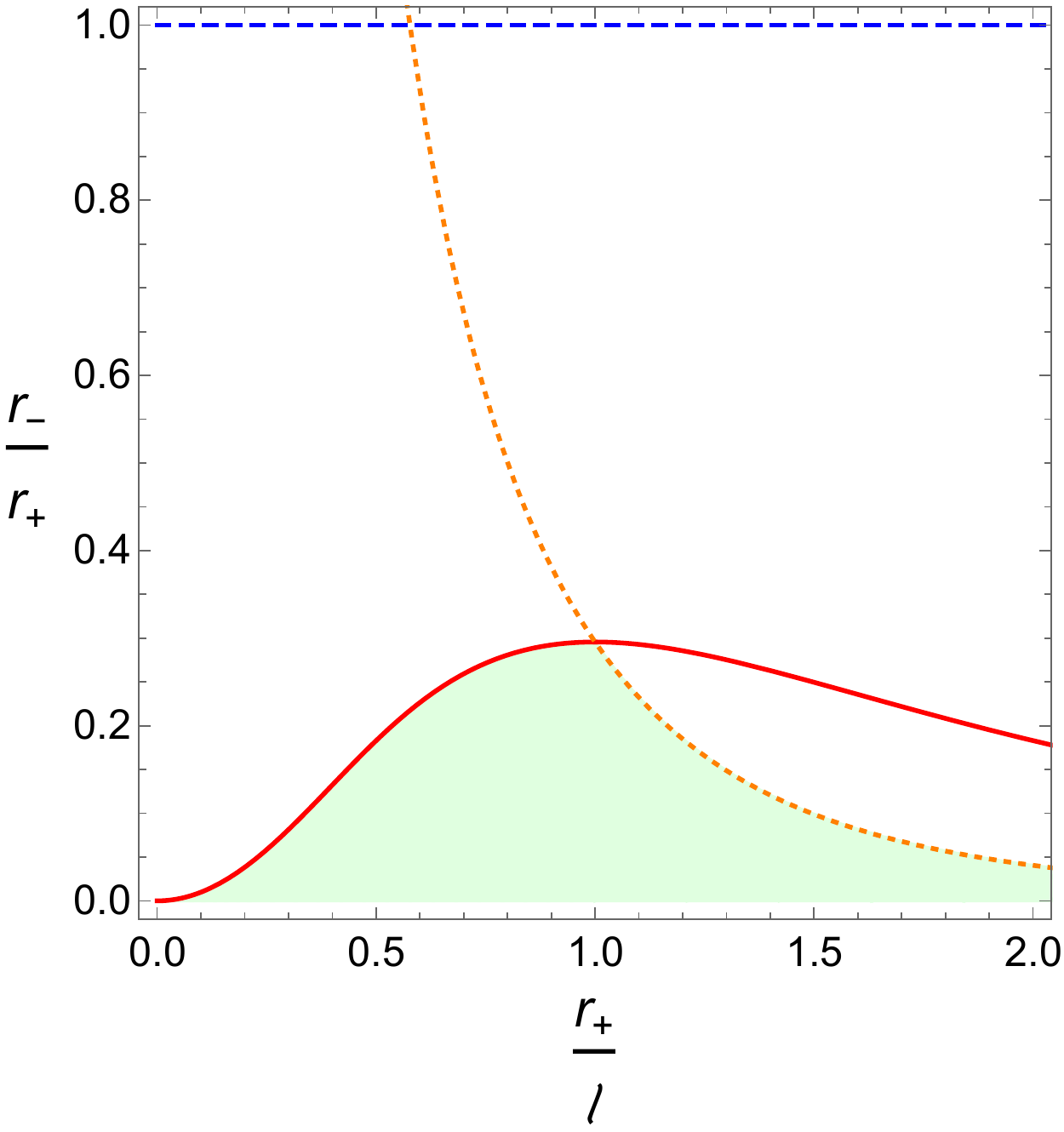}  \hfill  \includegraphics[width=.30\linewidth, valign=t]{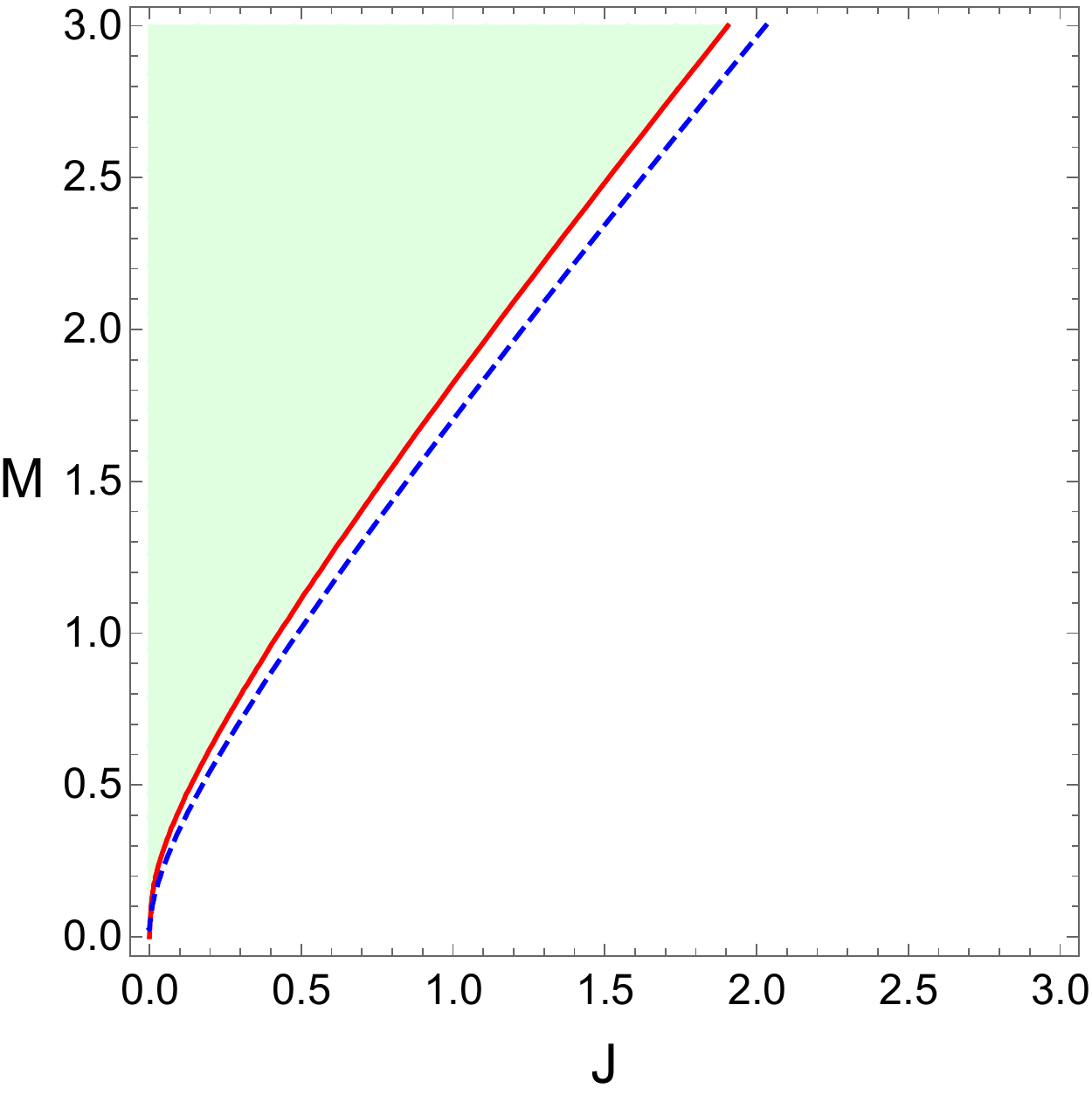}
\end{center}
\caption{In filled green the allowed region satisfying conditions 1) -- 3) in the space of parameters, both in the $(a,m)$-plane (left), $\(\frac{r_+}{\ell}, \frac{r_-}{r_+}\)$-plane (center), and $(J,M)$-plane (right) for $\ell =1$. In the left panel, we also include the empty AdS$_4$ line of solutions with $m=0$ in oblate coordinates. In dashed blue the extremality curve, in dotted orange the parameter singularity $a = \ell$, in solid red the other instability bound $ r_+^2 = a \, \ell$. 
} \label{fig:KerrParamters}
\end{figure}

In addition to the above mentioned constraints: 1) $m > m_{\rm extr}$ and 2) $a < \ell$, we here further restrict the analysis to satisfying everywhere 3) $\ell \, \O_+ < 1$, which in fact also automatically implies 1) - 2). It is only in this case in fact that a timelike Killing vector can be globally defined outside the outer horizon, and the black hole is in thermodynamic equilibrium with rotating thermal radiation all the way to radial infinity \cite{Hawking:1998kw}. Relatedly, as soon as this bound is violated, the black hole exhibits superradiant  instability.  That is the black hole is unstable to losing energy  and angular momentum in gravitational and scalar modes that are reflected and amplified in the AdS potential \cite{Hawking:1999dp,Kunduri:2006qa}. 
In dual terms, 3) represents the speed of light upper bound to the rotation of the  boundary state, and is necessary to define a consistent thermodynamics. 
In fig.~\ref{fig:KerrParamters}, we depict in filled green the allowed region in the space of solutions satisfying 1) -- 3), in terms of both the $(a,m)$ (left panel) and $\(\frac{r_+}{\ell}, \frac{r_-}{r_+}\)$ (central panel) variables. 
 This region is bounded by the intersection of the speed of light critical curves: $a =\ell$ in dotted orange and $ r_+^2 = a \, \ell$ in solid red. In particular, extremal black holes always rotate faster than the speed of light and are hence unstable towards radiating away their angular momentum.  Notice also that, within the solutions satisfying  1) -- 3),  large Kerr-AdS black holes with $r_+ \gg \ell$ have $r_- \ll r_+$. 
In fig.~\ref{fig:KerrParamters} (right panel) we also depict how the allowed  $(a,m)$ region maps to  the $(J,M)$  space. It is worth noticing that the $a=\ell$ curve maps to $M,J \to \infty$.  
Finally notice that the set of solutions with $m =0, a< \ell$ parametrize empty AdS$_4$ in oblate coordinates. We thus include also this set of solutions in the allowed green region in the figure (fig.~\ref{fig:KerrParamters} left panel).  These solutions correspond  to complex  values of the $\(\frac{r_+}{\ell}, \frac{r_-}{r_+}\)$ variables and are thus not depicted in the center panel of fig.~\ref{fig:KerrParamters}. 

In order to  study holographic complexity we choose, as for the BTZ black hole, to time evolve forward in both exterior regions in a symmetric fashion, \ie we pick time to flow upward on both sides of the Kerr-AdS black hole diagram in fig.~\ref{fig:KAdSPenrosediagram}  with $t_L =t_R =\frac{t_b}{2}$.

\subsubsection{WDW patch}\label{sec:WDWKerr}
The WDW patch is a codimension-0 region bounded by null hypersurfaces defined by a constraint equation $\Phi(x)= $ constant with null normal, \textit{i.e.} $g^{\a\b}\pa_{\a}\Phi\pa_{\b}\Phi=0$. We need null hypersurfaces that penetrate the horizon, and hence look for a suitable set of Eddington-Finkelstein-like coordinates
\be \label{eq:nullcoord}
v=t+r^* (r, \theta), \qquad u=t-r^* (r, \theta) \,. 
\ee
The hypersurfaces we are seeking are then those defined by $v=$ const and $u=$ const. Taking for concreteness $v$, the null condition reads
 \be \label{eq:PDEWDWKerr}
\D(\pa_r r^*)^2+\D_{\theta}(\pa_x{\theta} r^*)^2= \frac{(r^2 +a^2)^2}{\D}-\frac{a^2 \sin^2\theta}{\D_{\theta}} \, .
\ee
To solve this PDE \cite{Balushi:2019pvr} recently adapted to AdS the approach of \cite{Pretorius:1998sf} for null hypersurfaces in asymptotically flat Kerr geometry.\footnote{Note the extra $\Xi$ factor appearing in \cite{Balushi:2019pvr}.} In particular, one introduces an auxiliary function $\zeta=\zeta(r,\theta)$ and defines
\bea
Q^2 &\equiv &(r^2 +a^2)^2 -a^2 \zeta  \D \label{eq:QKerr} \\ 
P^2 &\equiv & a^2\pr{\zeta \, \D_{\theta}-\sin^2\theta} \label{eq:PKerr} \,. 
\eea
The PDE \eqref{eq:PDEWDWKerr} is then solved by
\be \label{eq:partialrPDE}
\pa_r r^*=\frac{Q}{\D} \qquad \qquad  \pa_{\theta}r^*=\frac{P}{\D_{\theta}} 
\ee
provided $d^2 r^*= 0$ for consistency, which in turn implies  the auxiliary function $\zeta$ must have a  differential of the form
\be \label{eq:zeta}
  d \zeta  =\frac{1}{\mu}\(- \frac{dr}{Q} + \frac{d\theta}{P}\) \, .
\ee
with $\mu = \mu (r,\theta)$. 

 This in principle allows solving for the relevant hypersurfaces and induced metric, but the solution is implicit away from the special limit $m=0$ \cite{Balushi:2019pvr}. Nonetheless,  as we will see,  to obtain the CA late time growth rate we do not need to explicitly solve the equation for $r^*$.

The null normal one forms  associated  to the null hypersurfaces defined by  \eqref{eq:nullcoord}  are  readily obtained. Focusing  on the  future part of the WDW patch and  choosing them to be outward directed from the boundary of the WDW patch, we get
\be \label{eq:kkerr}
k_{R \m}=\pa_{\mu}v=\( 1,\frac{Q}{\D},\frac{P}{\D_{\theta}},0 \) , \qquad k_{L \m}=-\pa_{\mu} u = \( - 1,\frac{Q}{\D},\frac{P}{\D_{\theta}},0 \) \, .
\ee
The associated null vectors implicitly define a  parametrisation  along the null direction of the WDW  in terms of the null generator $\lambda$, \ie$ \del_\lambda \equiv k^\mu \del_\mu$.

Similarly one can define two spacelike vectors linked to the spacelike intrinsic coordinates $y^A$ on the boundary of the  WDW patch 
 \be
 e^\mu_A = \frac{\partial x^\mu}{\partial y^A} \, .
 \ee
Noticing that  \eqref{eq:zeta} implies that $\zeta$ is constant along the null generators, $k^\mu \del_\mu \zeta = 0$, one could take 
\be \label{eq:zetavector}
e_\zeta^{\mu} = \mu \( 0, - \frac{ P^2 Q \Delta  }{\Sigma^2 },\frac{ P Q^2 \Delta_\theta }{\Sigma^2 },0\) \, .
\ee
Given the symmetry of the solution, one can  take the second intrinsic coordinate to just coincide with $\phi$\footnote{There are some subtleties with the regularity of this choice at the horizon, and in principle one would need to define a shifted angular variable similarly to the one  defined  for BTZ  in \eqref{eq:EF} (see  \cite{Balushi:2019pvr}). This however does not have any bearing on the result of our analysis.}
\be
e_\phi^\mu = \( 0,0,0,1\)  
\ee
These are both orthogonal to $k_\mu$   and define the non-degenerate transverse metric associated to the WDW patch  
\be\label{eq:gamma}
\gamma_{AB} = g_{\mu\nu}  e^\mu_A e^\nu_B =  \( \begin{array}{cc}
\frac{\mu^2 \rho^2 P^2 Q^2}{\Sigma^2}& 0\\ 
  0 &\frac{\Sigma^2 \sin^2\theta}{\Xi^2 \rho^2} \\ 
\end{array} \)
\ee
with determinant 
\be
\sqrt{\gamma}  =\frac{ \mu P  Q \sin \theta}{\Xi}\, . 
\ee

The last piece of information we will need for our analysis concerns the future and past    null-null joints of the WDW patch.
In this axially symmetric solution, we expect them to be $\theta$-dependent, meaning that at each instant of boundary time $t_b$ the codimension-2 intersection is not a round sphere. We then parametrize the joints radial location as $r_{m1} = r_{m1} (\theta)$  and $ r_{m2} = r_{m2} (\theta)$, respectively for the future and past joints. Analogously to the BTZ case of eq.~\eqref{eq:rm1rm2}, these are defined by
\be  \label{eq:tipdef}
\frac{t_b}{2} +r^*(\infty,\theta) = r^* \(r_{m1}, \theta\) \,, \quad\quad \frac{t_b}{2} - r^*(\infty,\theta) = - r^* \(r_{m2}, \theta\)\,,
\ee
where $r^*(\infty,\theta) $ denotes the $r \to \infty$ limit of $r^*$. 
Differentiating with respect to the boundary time $t_b$ and using \eqref{eq:partialrPDE}, we obtain
\bea 
\frac{\pa r_{m1}}{\pa t_b}&=& 
\left. \frac{\D}{2 Q} \right|_{r_{m1}} \label{jointevo1} \\
\frac{\pa r_{m2}}{\pa t_b}&=&
\left. - \frac{\D}{2Q} \right|_{r_{m2}} \label{jointevo2}  \, .
\eea
Notice that at  late times $r_{m1}\rightarrow r_-$, $r_{m2}\rightarrow r_+$, and  since $\D(r_{\pm})=0$, these vanish.

\subsection{Late time CA growth rate}\label{sec:CAKerr}

In this section we evaluate the late time limit of the complexity=action growth rate and show that it matches the extension of Lloyd bound to settings with angular momentum advanced in \cite{Cai:2016xho}. Our result is derived using the gravitational action prescription developed in \cite{Lehner:2016vdi}, which appeared after \cite{Cai:2016xho} and which, in particular, carefully takes into account the contribution of null-null joints.  

We first evaluate the bulk term and null-null joints, which are the only non-zero contributions to the late boundary time derivative of the gravitational action. We then explain why the other terms are irrelevant.

\paragraph{Bulk term.} In complete analogy to what we did in sec.~\ref{paolo}, we write
\be 
I_{\rm bulk} = \frac{1}{16\pi \, G_N} \int_{\rm WDW} d^4 x \sqrt{-g}\(R + \frac{6}{\ell^2}\) = - \frac{3}{4 \, G_N \, \ell^2 \Xi} \int  d\theta \, dr \,  dt \, \rho^2  \sin \theta \, , 
\ee
and exploit the left-right symmetry of the WDW patch and split each half in three regions, so that
\be 
I_{\rm bulk}= 2 \left(I_{\rm bulk}^{\rm I}+I_{\rm bulk}^{\rm II}+I_{\rm bulk}^{\rm III}\right)
\ee
with 
\begin{align}
I_{\rm bulk}^{\rm I} &=- \frac{3}{4 \, G_N \, \ell^2 \Xi} \int_0^{\pi}d\theta \sin\theta \int_{r_{m1}}^{r_+}dr \(r^2+a^2\cos^2\theta\) \(\frac{t_b}{2} +r^*(\infty,\theta) - r^*(r, \theta) \)\\
I_{\rm bulk}^{\rm II}&=\frac{3}{2 \, G_N \, \ell^2 \Xi} \int_0^{\pi}d\theta\sin\theta \int_{r_+}^{r_{\rm max}}dr  \(r^2+a^2\cos^2\theta\) \(r^*(r, \theta)-r^*(\infty,\theta) \)\\
I_{\rm bulk}^{\rm III}&=\frac{3}{4 \, G_N \, \ell^2 \Xi} \int_0^{\pi}d\theta\sin\theta \int_{r_{m2}}^{r_+}dr \(r^2+a^2\cos^2\theta\)  \(\frac{t_b}{2}-r^*(\infty,\theta) + r^*(r, \theta) \) \,.
\end{align}
Here $r_{\rm max} = r_{\rm max}(\theta)$ is a time independent UV cutoff that regularizes the bulk integral (see app.~\ref{app:div}).  
The UV-divergent contribution to $I_{\rm bulk}$, \ie $I_{\rm bulk}^{\rm II}$, is then time independent and the CA growth rate is UV-finite. 
 $I_{\rm bulk}^{\rm I}$ and $I_{\rm bulk}^{\rm III}$ depend on $t_b$  both directly, through the appearance of $t_b$ in the integrand,  and via the time dependence of the WDW tips, which represent the extrema of the radial integral.   However, the contribution from the latter vanishes (see \eqref{eq:tipdef}-\eqref{jointevo2}). 
Therefore for the bulk action growth rate we obtain
\be
\frac{dI_{\rm bulk}}{dt_b}=- \frac{1}{4 \, G_N \, \ell^2 \Xi} \int_0^{\pi}d\theta\sin\theta\pg{ r^3_{m2}(\theta)-r^3_{m1}(\theta) +3a^2\cos^2\theta \left[ r_{m2}(\theta)-r_{m1}(\theta)\right] },
\ee
which at late time approaches
\be  \label{eq:bulkKerr}
\lim_{t_b \to \infty }\frac{dI_{\rm bulk}}{dt_b} =- \frac{r_+^3-r_-^3 +a^2\pr{r_+-r_-}}{2 \, G_N  \(\ell^2 -a^2\) }\,.
\ee

\paragraph{Null-null joints.} Next we evaluate the contribution of joints at the 2-dimensional top and bottom corners of the WDW patch. For the future null-null joint at  $r = r_{m1}$, we have
\be\label{jointKAdS} 
I^{\rm Null-Null}_{\rm joints}=\frac{1}{8\pi G_N} \int_{r_{m1}} d^2y\,  \sqrt{\gamma} \log \left\lvert \frac{k_L \cdot k_R}{2}\right\lvert \, . 
\ee
The two intrinsic variables are given by $\phi$ and $\zeta$, the integration in $\phi$ can thus be performed straight away. Using   \eqref{eq:kkerr}, which yields 
\be 
 \frac{k_L \cdot k_R}{2}  = \frac{1}{2}\(\frac{1}{N^2} + \frac{Q^2}{\Delta \rho^2}  +\frac{P^2}{\Delta_\th \rho^2}   \)  = \frac{1}{N^2} \, , 
 \ee
one thus have
\be 
\frac{dI^{\rm Null-Null}_{\rm joints}}{dt_b}=\frac{1}{4 G_N}  \, \[ \left.  \frac{\partial}{\partial t_b}  \int d\zeta \sqrt{\gamma}\log\pr{-\frac{1}{N^2}} \right] \right|_{r=r_{m1}}  ,
\ee
were we noticed that at  the joint, $r = r_{m1}$,   $N^2$ is negative.

Using \eqref{eq:zetavector} to change integration variable from $\zeta$ to $\theta$ and the fact that the entire time dependence comes from the dependence on $t_b$ of the joint location $r_{m1}(\theta)$, we arrive to 
\be
\frac{dI^{\rm Null-Null}_{\rm joints}}{dt_b}=\frac{1}{8 G_N}  \, \left.  \int d\theta \sin \theta \,    \frac{\D}{Q}  \frac{\partial}{\partial r} \( \frac{\Sigma^2}{ Q  \Delta_\theta \Xi}  \,  \log\pr{-\frac{1}{N^2}}  \)  \right|_{r=r_{m1}}  \, .  \label{eq:nnder}
\ee

In the late time limit $r_{m1} \to r_-$ and the expressions appearing in \eqref{eq:nnder} are thus evaluated at the innner horizon, implying in particular $\Delta(r_{m1} ) \to \Delta(r_- ) =0$.
With this in mind,  it is then easy to convince oneself that in the late time limit the only terms that survive are those where the derivative acts on the $\D$ factor contained within the lapse function $N$ (see  \eqref{eq:NKerr}, \eqref{eq:SigmaKerr}  and \eqref{eq:QKerr}).  
In particular, the derivative acting on the  $\log$ term yields $\Delta' / \Delta$. Using the explicit expressions one can then check that 
\be
\left. \frac{\Sigma^2}{ Q  \Delta_\theta \Xi} \right|_{r= r_-}=   \frac{ (r_-^2 + a^2)^2}{ \Xi}
\ee
and the numerator of this expression simplifies with  the $1/Q |_{r= r_-}$ in front of the derivative in  \eqref{eq:nnder}. The remaining $\Delta$ simplifies with the one coming from the derivative of the $\log$ factor and one is left with $\Delta' / \Xi$ and the angular integral. Performing the integration in $\theta$ and adding the contribution for  the past null-null joint, one then gets 
\be
\lim_{t_b \to \infty} \frac{dI^{\rm Null-Null}_{\rm joints}}{dt_b} = \frac{\D'(r_+)-\D'(r_-)}{4 \, G_N \,  \Xi} = \frac{2r_+^3 -2 r_-^3 + \(a^2 + \ell^2\) \(r_+ -r_-\) }{2 \, G_N \, \(\ell^2 - a^2\) } \label{eq:jointsKerr} \,.
\ee
%

\paragraph{Vanishing contributions.} We now show that all remaining terms in the WDW action  \eqref{eq:graction}, namely  the GHY terms at the regularized boundary, the  null-timelike joint contributions, and the counterterm do not contribute to the CA growth rate at late time. In particular, while the GHY and joints are time-independent, and hence give a vanishing growth rate at any time, the counterterm growth rate only vanishes as $t_b \to \infty$. 

First, we consider the GHY term associated to the cutoff surface near the right boundary. Schematically this reads
\be
I_{\rm GHY}= \frac{1}{8 \pi G_N} \int_0^{2\pi}d\varphi\int_0^{\pi}d\theta\int_{t_{\rm min}}^{t_{\rm max}}dt  \,\sqrt{|h|}\, K \Bigg|_{r=r_{\rm max}} \,. 
\ee
The integrand and the cutoff surface are  time-independent. The integral in $t$ similarly gives a factor that does not depend on $t_b$, as the range of integration only depends on the choice of $r_{\rm max}$ 
\bea
t_{\rm max}-t_{\rm min}&=&\pq{\frac{t_b}{2}+r^*(\infty,\theta)-r^*(r_{\rm max},\theta)}-\pq{\frac{t_b}{2}-r^*(\infty,\theta)+r^*(r_{\rm max},\theta)} \nonumber \\
&=& 2 \left[ r^*(\infty,\theta)-r^*(r_{\rm max},\theta) \right]\, , \eea
so that 
\be 
\frac{dI_{\rm GHY}}{dt_b}=0\, . 
\ee

With a completely similar reasoning, given the time-null joints lie on the cutoff surface at $r_{\rm max}$, we have
\be 
\frac{dI_{\rm joints}^{\rm Time-Null}}{dt_b}=0 \, .
\ee

Finally let us consider the  counterterm action
\be
I_{\rm ct}  = \frac{1}{8 \pi G_N} \int_{\partial {\rm WDW}} d \lambda\, d^{2} y\, \sqrt{\gamma}\, \Theta\, \log\left( L_{\rm ct} \Theta \right) \, , 
\ee
with $\Theta = \del_\lambda \log \sqrt{\gamma}$, and take for concreteness the right future boundary of the WDW-patch.  
The integration  runs  over $\zeta, \phi$ and the null coordinate $\lambda$, which can be expressed in terms of the auxiliary null vector $\tilde N^\mu$ associated with $k_\mu$, such that $ \tilde N^\mu k_\mu = -1$. 
For the right future boundary of the WDW patch under consideration, in particular
\be \label{eq:lambdafed}
 d\lambda  = - \tilde N_{\mu} dx^\mu = -\frac{N^2}{2} (dt - dr^*)  \, . 
 \ee

Given the independence of the integrand from $\phi$ one can perform the corresponding integration. 
Using the condition of being on the boundary of the WDW patch, $dv= dt + dr^* =0$,  with  \eqref{eq:zeta}  and  \eqref{eq:lambdafed} we can change the remaining integration variables to $r^*$ and $\theta$ using
\bea
d\lambda  &=&   N^2  dr^* \, ,\\
d\zeta  &=& \frac{1}{\mu Q^2} \( - \Delta dr^* + \frac{\Sigma^2}{P \Delta_\th} d\theta\) \, .
\eea
This gives an integral of the form 
\bea
I_{\rm ct} 
=  \frac{1}{4 G_N}  \int  d \theta \, dr^* \frac{ \Delta \rho^2 \sin\theta}{\Xi Q }   \,\Theta \log \( L_{\rm ct}\Theta  \)  \, ,
\eea
where the   time dependence   is implicitly contained in the extremum of integration corresponding to the tip of the WDW,   \ie in $r^*(r_{m1},\theta)$. Taking the time derivative with \eqref{eq:tipdef}, we obtain
\be
\frac{dI_{\rm ct}}{dt_b} = \frac{1}{8 G_N}  \int  d \theta  \,\frac{ \Delta \rho^2 \sin\theta}{\Xi Q }   \,\Theta \log \( L_{\rm ct}\Theta  \)\Big|_{r=r_{m1}} \, .
\ee
For us here it is then enough to notice that in the late time limit  $r_{m1} \to r_-$ and the integrand vanishes because of the $\Delta$ factor. Therefore
\be
\lim_{t_b \to \infty} \frac{dI_{\rm ct}}{dt_b} =  0 \, .
\ee 

\paragraph{Complexification rate.}

Combining the non-vanishing bulk contribution \eqref{eq:bulkKerr} and the null-null joint one   \eqref{eq:jointsKerr}, we obtain the late time CA growth rate
\be
\lim_{t_b \to \infty} \pi \frac{dC_{\rm A}}{dt_b} = \frac{r_+^3-r_-^3 +\ell^2\pr{r_+-r_-}}{2 \, G_N  \(\ell^2 -a^2\) } = (M-\O_+ J)-(M-\O_- J) \, . \label{eq:lateCAK}
\ee
In writing the second equation we used the relations $\Delta(r_{\pm})=0$, the definitions \eqref{eq:MJKerr} and \eqref{eq:O+Kerr}, and the analogue quantities defined at the inner horizon. 
Similarly to the lower dimensional BTZ case studied in the previous section, the limit \eqref{eq:lateCAK} saturates  the bound of \cite{Cai:2016xho} (but not the one of  \cite{Brown:2015lvg} here). While in the BTZ case we were able to show that the limiting value was approached from above and the bound violated at intermediate times, we cannot draw a conclusion with our analysis of Kerr-AdS. Nonetheless, we expect the bound to be generically violated, as this is what happens in the irrotational limit \cite{Carmi:2017jqz}. 

In the critical limit $a \to \ell$, in which the conformal boundary rotates at the speed of light, the late time limit \eqref{eq:lateCAK} diverges. This is apparent from the intermediate expression in eq.~\eqref{eq:lateCAK} since $r_- < r_+$ (see central panel of fig.~\ref{fig:KerrParamters}), and mimics the behavior we observed in BTZ. Notice though that for small black holes with $r_+ < \ell$ the growth rate late time limit does not diverge in the speed of light limit, which corresponds in this regime to $ r_+^2 \to a \, \ell$.  

The works \cite{Balushi:2020wjt,Balushi:2020wkt} studied holographic complexity for odd-dimensional Myers-Perry black holes with equal angular momenta in each orthogonal plane. For large black holes, they highlighted a direct connection between CA, CV and thermodynamic volume.  For the growth rate at late times, taking the large black hole limit with $r_- / r_+$ held fixed, it was shown in \cite{Balushi:2020wjt,Balushi:2020wkt} that at leading (divergent) order $T_+S_+-T_-S_- \propto  P\(  V_+ - V_-\)$ and
 \be
 \lim_{t_b\to \infty} \frac{ d  C_{\rm A/V} }{ dt_b } \propto   P\D V  \label{eq:manngrowth}
\ee
where $\D V  =  V_+ - V_-$ is the difference between the inner and outer horizon thermodynamic volume.
 
Using   $T_+ =T$ in  \eqref{eq:Tkerr}, $S_+ =S$ in    \eqref{eq:Skerr} and analogous ones to define $T_-$  and $S_-$ at $r_-$,  the limiting value for the Kerr-AdS complexity growth rate \eqref{eq:lateCAK} can  be rewritten in the following form
\be
(M-\O_+J)-(M-\O_-J)=T_+S_+-T_-S_- -  P\(  V_+ - V_-\) \, ,
\ee
in terms of pressure 
\be 
P=   \frac{3}{8\pi G_N \ell^2}\,  ,  
\ee
and thermodynamic volumes
\be
V_{\pm}=\frac{4\pi}{3} \( \frac{r_{\pm}(r_{\pm}^2+a^2)}{\Xi}  + a \, G_N \, J \)  \,. 
\ee
Explicitly, and factorizing common factors we have: 
\bea
T_+ S_+  - T_-S_- &= & \frac{r_+ \ell^2}{ 2 G_N \Xi \(\ell^2 -  r_+ r_- \) } \(     2 \frac{r_+^2}{\ell^2}  + 1  -    \frac{r_+^3 r_-}{\ell^4} +   \frac{r_-}{r_+}   - 2\frac{r_-^3}{r_+\ell^2} -  \frac{ r_-^4}{\ell^4} \)   \\
2 P \Delta V&= & \frac{r_+ \ell^2}{2 G_N \Xi  \(\ell^2 -  r_+ r_- \) } \(      2 \frac{r_+^2}{\ell^2}   +  2\frac{r_+  r_- }{\ell^2}   - 2 \frac{   r_-^2}{\ell^2}  -2\frac{r_-^3}{r_+\ell^2}   \) \,. 
\eea
As apparent in fig.~\ref{fig:KerrParamters}, the large black hole limit $r_+/\ell \gg 1$ can here only be taken consistently in the regime $r_- / r_+ \ll 1$. In particular, to remain within the allowed parameters region we need to take $r_+/\ell \gg 1$ while taking  $r_-/ r_+$ to zero as  $r_-/ r_+ \sim \ell^4/r^4_+$ or faster. Taking this limit then implies that the only divergent term inside the parenthesis of both expressions is $2 \frac{r_+^2}{\ell^2}$, giving  at leading order 
\be
T_+ S_+  - T_-S_-  = 2 P \Delta V
\ee
compatibly with the claim \eqref{eq:manngrowth} in \cite{Balushi:2020wjt,Balushi:2020wkt}. 

\subsection{CV complexity of formation} \label{sec:CVKerr}

In this section we evaluate the complexity of formation according to the CV proposal, that by symmetry is the volume of the $t=0$ slice of Kerr-AdS \ref{KerrAdSmetric}. 
Such maximal volume slice anchored on the $t_b=0$ surface on the boundary intersects the bifurcation surface, and is straightforward to evaluate.


In Boyer-Lindquist coordinates, the regularized volume of the maximal $t=0$ slice of \ref{KerrAdSmetric} gives
\be
C_{\rm V} (t_b =0) = 
 \frac{4\pi}{G_N \, \ell} \int_0^{\pi}d\theta\sin\theta \int_{r_+}^{r_{\rm max}}dr \,  \frac{\r}{\Xi} \, \sqrt{\frac{(r^2+a^2)^2}{\D}-\frac{a^2 \sin^2\theta}{\D_{\theta}}}\,.
\ee
We follow the standard holographic procedure to fix the UV cutoff $\delta$ in Fefferman-Graham coordinates, which corresponds to a $\theta$-dependent IR bulk cutoff $r_{\rm max}$ in Boyer-Lindquist coordinates  (see app.~\ref{app:div}, eq.~\eqref{eq:UVcut}):
\be
r_{\rm max} = \frac{\ell^2}{\delta} - \frac{\delta}{4} \(1 + \frac{a^2}{\ell^2} \sin^2 \theta\)  + \frac{m}{3 \, \ell^2} \, \delta^2 +\dots \,. \label{eq:rmaxBL}
\ee
From this we evaluate the complexity of formation, \textit{i.e.}  the additional complexity arising in preparing the rotating entangled thermofield double state with two copies of the boundary CFT, as compared to preparing the individual vacuum states of the two copies. In these coordinates in which the boundary asymptotic metric has rotation, the natural vacuum to consider is the solution   \ref{KerrAdSmetric} with  $m=0$, that is AdS$_4$ in oblate coordinates. We thus consider the variation 
\bea\label{CVformationKAdS}
\D C_{\rm V} (t_b =0)= \frac{4\pi}{G_N \,\ell}\int_0^{\pi}d\theta\sin\theta \Bigg\{ \int_{r_+}^{r_{\rm max}} dr \, \frac{\r}{\Xi} \, \sqrt{\frac{(r^2+a^2)^2}{\D}-\frac{a^2 \sin^2\theta}{\D_{\theta}}} \nonumber\\ 
 -\int_{0}^{r_{\rm max}^{m=0}}dr \, \frac{\r}{\Xi}\, \sqrt{\frac{(r^2+a^2)^2}{\D |_{m=0}} -\frac{a^2 \sin^2\theta}{\D_{\theta}}} \Bigg\} \,. 
\eea
Once we take $\delta \to 0$, the difference between $r_{\rm max}$ and $r_{\rm max}^{m=0}$ and the $\theta$-dependence in \eqref{eq:rmaxBL} have no influence on $\D C_{\rm V}$ (see app.~\ref{app:div}). We can thus use a unique $\theta$-independent cutoff $r_{\rm max}$ in both expressions and perform the integration numerically. 

To make further contact with the BTZ case for which we mainly focused on the coordinates system \eqref{metrica}, we need to consider Schwarzschild-like coordinates. This is the situation in which the boundary CFT background metric is not rotating and all rotation is in the states. We saw for BTZ in app.~\ref{app:BTZBL} this accounts for an additional finite contribution in $\D C_{\rm V}(t_b =0)$, as compared to Boyer-Lindquist-like coordinates. For Kerr-AdS we verify explicitly in app.~\ref{app:div} that, differently from BTZ, $\D C_{\rm V}(t_b =0)$ evaluated in Schwarzschild-like coordinates subtracting twice the complexity of a fixed time slice in global AdS coincides precisely with the result obtained in eq.~\eqref{CVformationKAdS}.

\begin{figure}[t]
\begin{center} 
 \includegraphics[width=.45\textwidth]{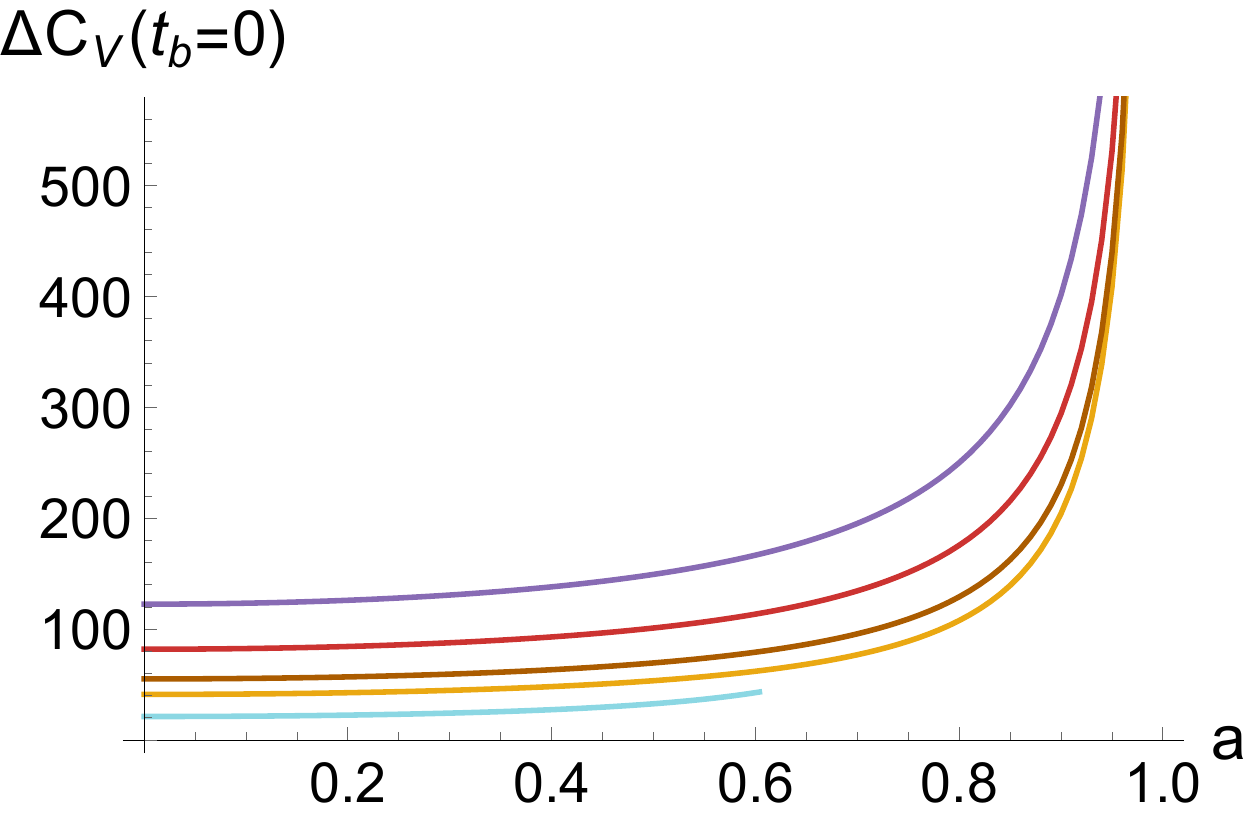} 
\end{center}
\caption{ $\Delta C_{\rm V}(t_b =0) $ as a function of   $m$  and  $a$, for $G_N=\ell=1$.  From the bottom up  $m=1,3,5,10,20$.} \label{fig:DeltaCVKerrma}
\end{figure}

\begin{figure}[ht]
\begin{center} 
\includegraphics[width=.45\textwidth]{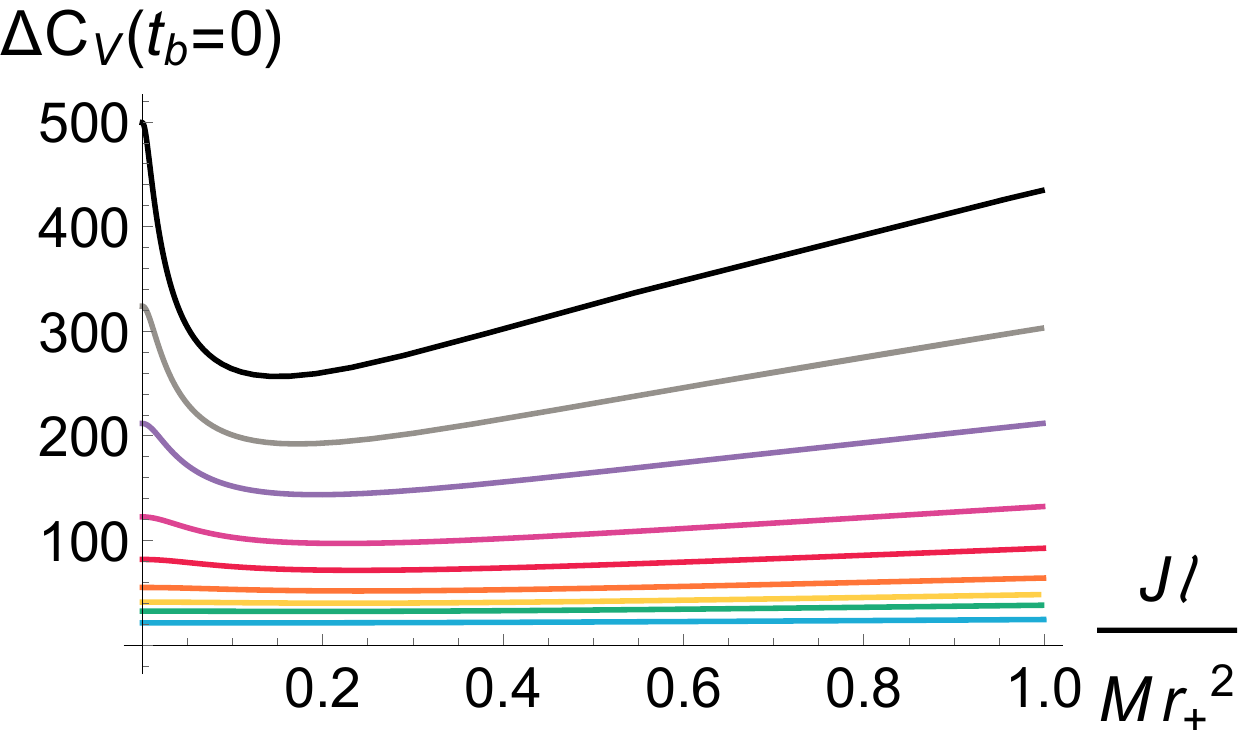}  \hfill   \includegraphics[width=.45\textwidth]{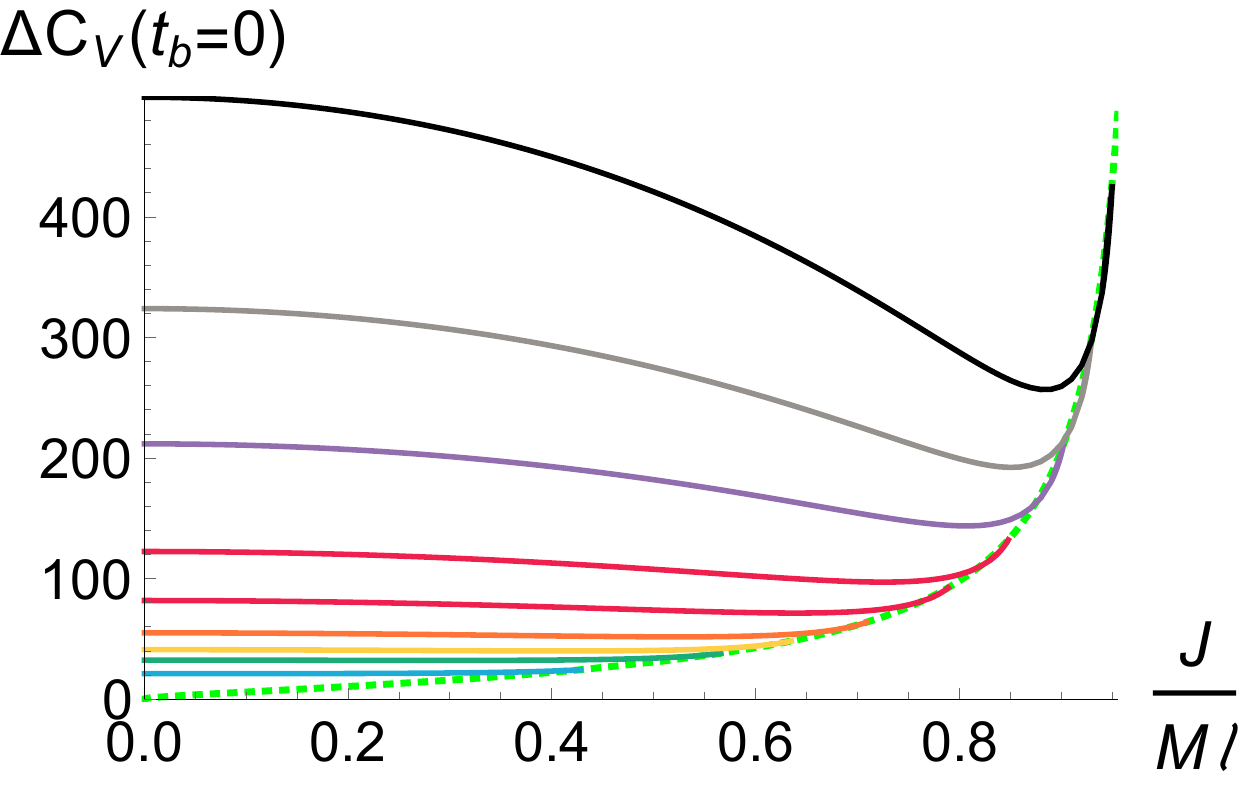}  
\end{center}
\caption{ $\Delta C_{\rm V}(t_b =0) $ as a function of   $M$  and  $J$, for $G_N=\ell=1$. From the bottom up $M = 1,2,3,5,10,20, 50,100, 200$.  As made explicit in the left panel, $J$ spans the full range  allowed by the  condition $\ell \Omega_+ <1$, or equivalently $r_+ < \sqrt{ a \ell }$, which provides the most stringent bound for any finite $M$. In the right panel $\Delta C_{\rm V}(t_b =0) $  is plotted as a function of the dimensionless combination $J/M\ell = a/\ell$. Each curve stops at the limiting point  $r_+ = \sqrt{ J \ell  /M}$, corresponding to the superimposed green dotted curve.  $\Delta C_{\rm V}(t_b =0)$ remains finite,  but  as $J/ M \ell $ approaches 1    it  rapidly increases as a function of $M$. }\label{fig:DeltaCVKerr}
\end{figure}

We plot the results in fig.~\ref{fig:DeltaCVKerrma} as a function of the parameters $a$ and $m$.  
The behavior of $\Delta C_{\rm V}(t_b =0)$ has analogies with  the one observed for the BTZ black hole (see  fig.~\ref{figuraDivergenza2}), but the comparison  requires some care. In particular from  fig.~\ref{fig:DeltaCVKerrma} one can observe a divergent behavior in the limit $a \to \ell$, which is qualitatively similar to  the one observed in the BTZ case in the limit $J /M \to \ell$. Notice however that  in the allowed region of parameters for Kerr-AdS,  the limit  $a \to \ell$ can only be taken  for large enough values of $m$, as for smaller  $m$ the  bound $r_+^2 <a \, \ell$ is stronger  than the bound  $a < \ell$ (see fig.~\ref{fig:KerrParamters}). This is reflected in fig.~\ref{fig:DeltaCVKerrma} by the lower mass curve ending at some finite value of $a$, as well as in the  in the plots of the complexity of formation as a function of  $M$  and $J$ in fig.~\ref{fig:DeltaCVKerr}.  In terms of these variables  for any finite $M$  the only relevant bound is $r_+^2 <a \, \ell$, as shown in  fig.~\ref{fig:KerrParamters}.
These differences between Kerr-AdS and BTZ, reflect the fact that the multiple conditions one has to impose on the Kerr-AdS parameters to avoid super-luminal rotation translate into the single bound $J/M \leq \ell$  for the rotating BTZ black hole.

In the limit of large odd-dimensional Myers-Perry black holes $r_+/\ell \gg 1$ with equal angular momenta in each orthogonal plane and at fixed ratio $r_-/r_+$, \cite{Balushi:2020wjt,Balushi:2020wkt} found the complexity of formation  is controlled by the thermodynamic volume rather than by the entropy, with a scaling that depends on the spacetime dimensionality $D$. In particular they verified that for $\frac{r_+}{\ell} \gg 1$ (see eq.~(4.20) in \cite{Balushi:2020wkt})
\be \label{eq:CVMann}
\Delta C_{\rm V}(t_b =0) \sim S \log \frac{\Omega_H}{T}  + \tilde f\(\frac{r_-}{r_+}\) V^{\frac{D-2}{D-1}} \,. 
\ee
Here $\tilde f$ is a function of the fixed ratio $\frac{r_-}{r_+}$ and $V$ the thermodynamic volume. \cite{Balushi:2020wjt,Balushi:2020wkt}  were able to determine the $V$ dependence of $\Delta C_{\rm V}$ of large odd-dimensional Myers-Perry black holes by studying this quantity both in the non-rotating limit $\frac{r_-}{r_+} \ll 1$, where $V^{\frac{D-2}{D-1}} \sim S \sim \(\frac{r_+}{\ell}\)^{D-2}$, and in the extremal limit $\frac{r_-}{r_+} \sim 1$, where $V \sim S^{\frac{D+1}{D-1}} \sim \(\frac{r_+}{\ell}\)^{D+1}$. 
Our findings are compatible with \eqref{eq:CVMann}, but we are not able to verify independently this scaling for the $D=4$ Kerr-AdS solution. This is because within the region of parameters space covered by the physical solutions (see center panel of fig.~\ref{fig:KerrParamters}), taking $\frac{r_+}{\ell} \gg 1$ consistently forces also $\frac{r_-}{r_+} \ll 1$, \textit{i.e.} it automatically implies the irrotational limit. In this region, the $\frac{r_+}{\ell}$ scaling is everywhere fixed: $V^{2/3} \sim S \sim \(\frac{r_+}{\ell}\)^2$, and one cannot distinguish between the two thermodynamic variables.  

\begin{figure}[ht]
\begin{center} 
\includegraphics[width=.45\textwidth]{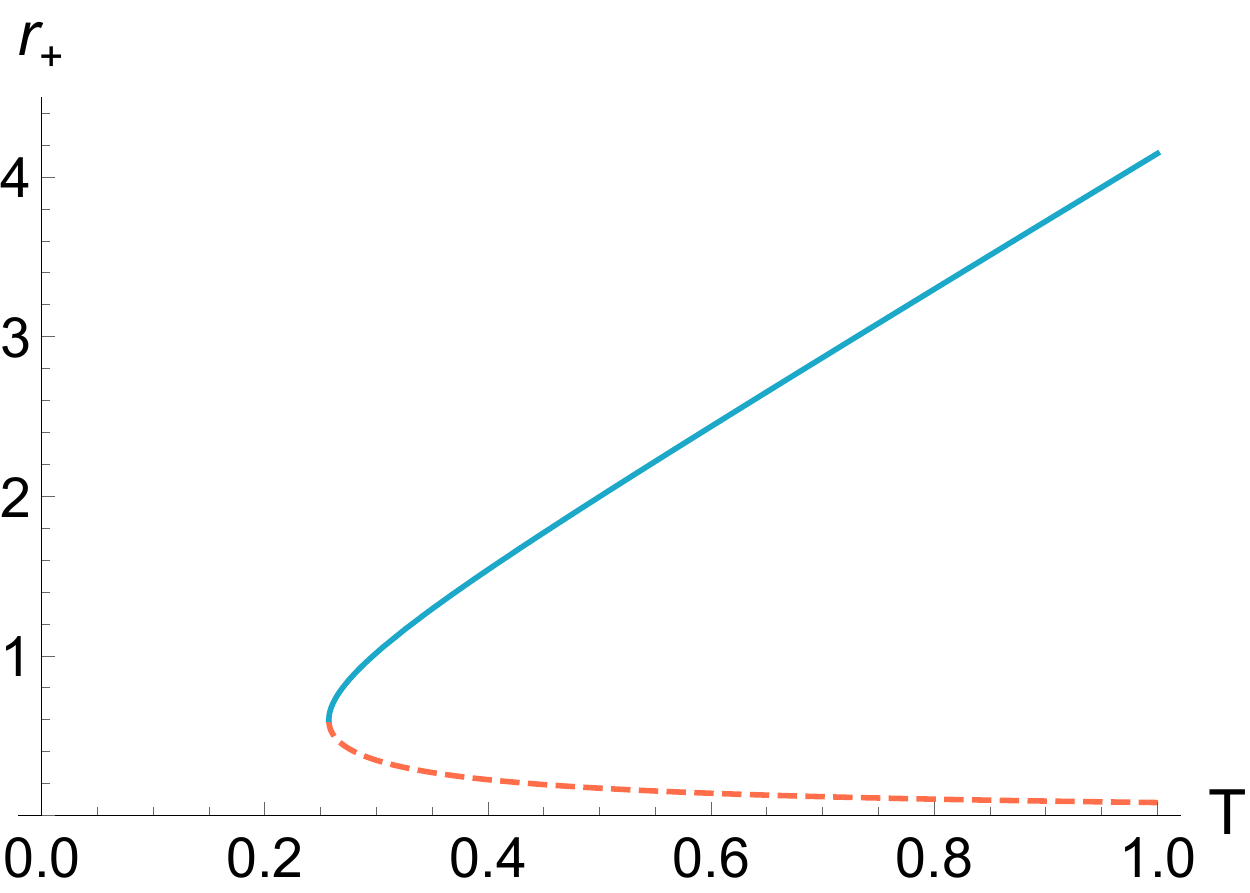}  
\end{center}
\caption{At fixed angular potential, here $\Omega_+ = 0.5$ and $\ell =1$, for each value of $T$ there are a small and large black hole branch. Within the physical parameter space of fig.~\ref{fig:KerrParamters}, there exists a minimum value of the temperature for such   black hole solutions.}\label{fig.KerrBranches}
\end{figure}

\paragraph{Grand canonical ensemble.} 
To express expliclty the complexity of formation in terms of the thermodynamical variables $T$ and $\O_+$, we use equations \eqref{eq:Tkerr} and \eqref{eq:O+Kerr}. From these we observe first of all that  there exist two branches of small and large black holes, as well as a minimal value of the temperature that can be attained within the physical space of solutions (see fig.\ref{fig.KerrBranches}). This minimal value ranges between $\sqrt 3 /(2\pi) \geq \ell \, T_{\rm min} >1/(2\pi)$ for $0 \leq \ell\, \Omega_+ < 1$. 

\begin{figure}[h]
\begin{center} 
\includegraphics[width=.45\textwidth]{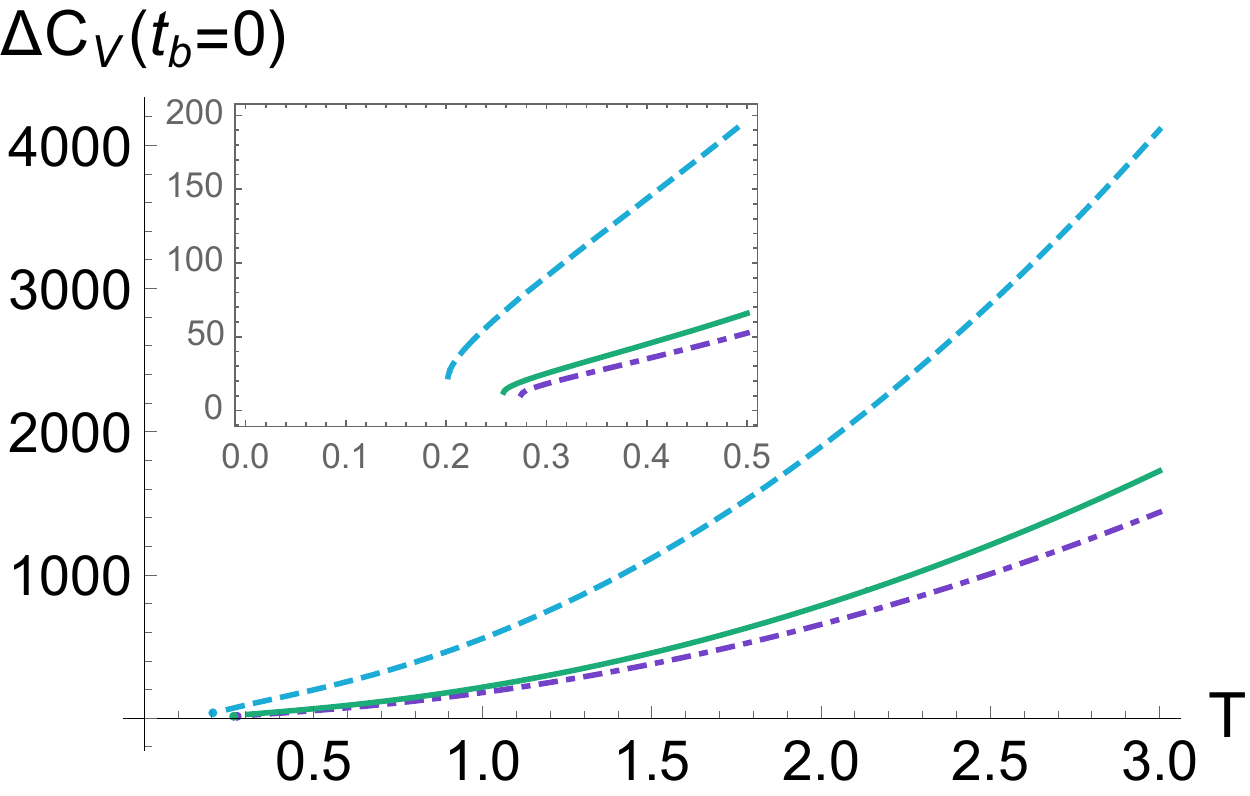}  \hfill   \includegraphics[width=.45\textwidth]{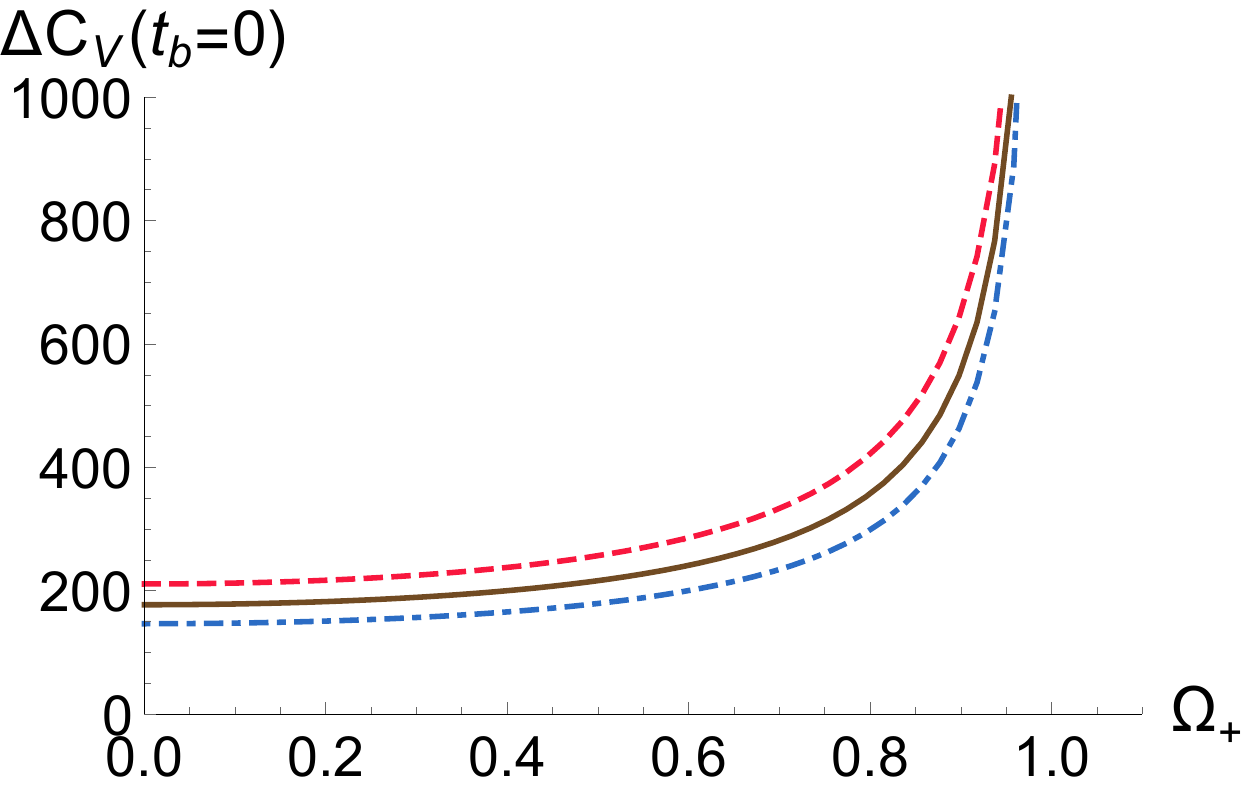}  \\
\includegraphics[width=.45\textwidth]{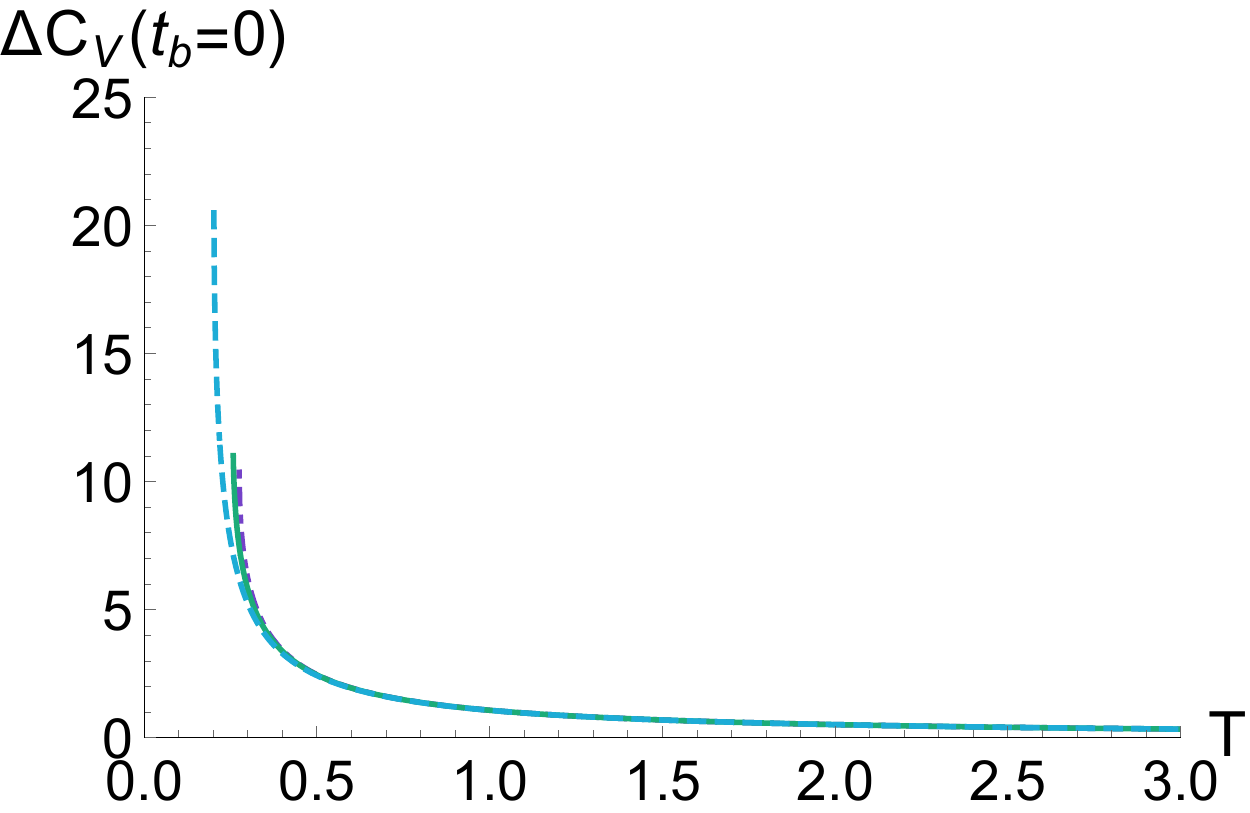}  \hfill   \includegraphics[width=.45\textwidth]{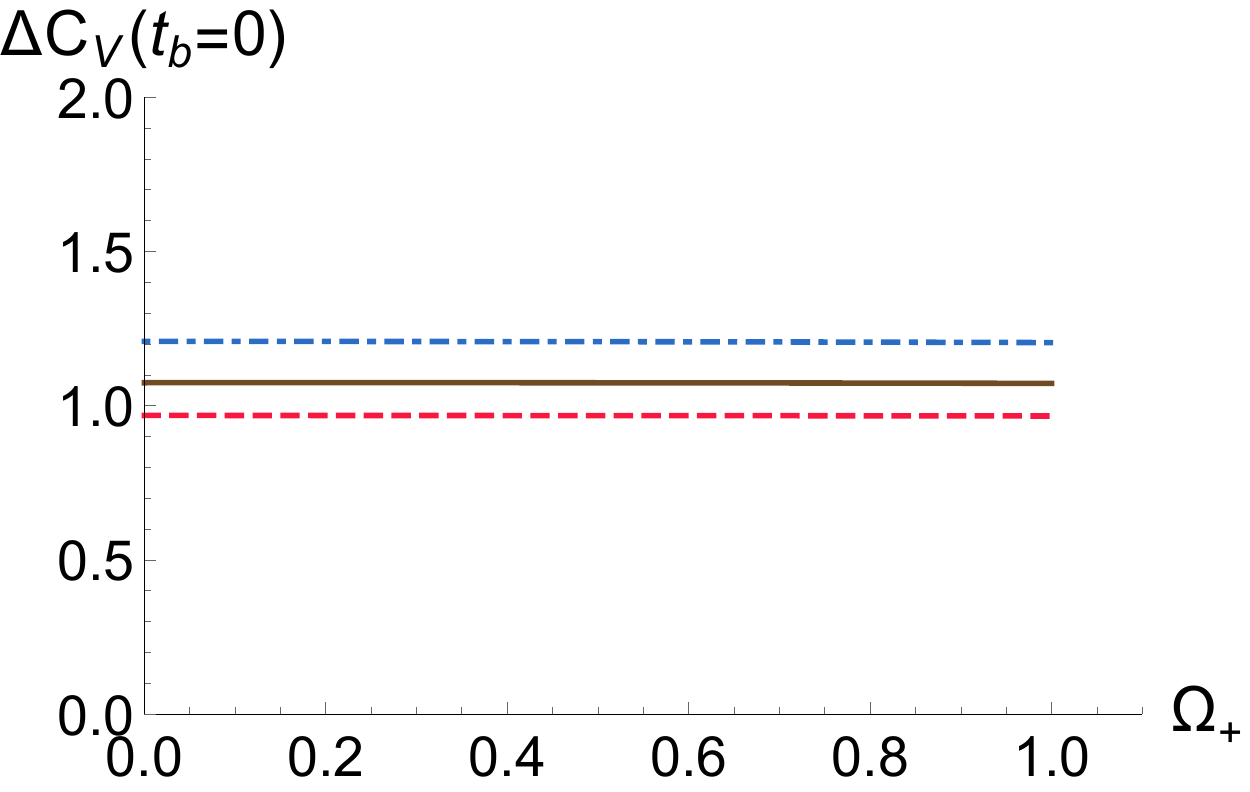} 
\end{center}
\caption{ $\Delta C_{\rm V}(t_b =0) $ for $G_N=\ell=1$ in the grand canonical ensemble for large black holes (upper panels) and small black holes (lower panels). (Left) Temperature dependence for $\O_+ = 0.1, 0.5, 0.9$ from the bottom up. (Right) Angular velocity dependence for $T = 0.9, 1, 1.1$ from the bottom up. }\label{fig:DeltaCVKerrGC}
\end{figure}

For both branches, we plot the CV complexity of formation at fixed $\ell \, \Omega_+$ (left panels) and fixed $\ell \, T$ (right panels) in fig.~\ref{fig:DeltaCVKerrGC}. As for BTZ (see fig.~\ref{fig:GCcfCV}), $\Delta C_{\rm V}(t_b =0)$ of large black holes is always positive, increases with the temperature and diverges in the critical angular velocity limit $\ell\, \Omega_+ \to 1$. However, as opposed to BTZ,  the dependence on the temperature is not linear, and fixed $\ell \, T$  curves approach different values  as $\ell\, \Omega_+ \to 0$. The CV complexity of formation for small black holes instead behaves very differently: it decreases and goes to zero as $\ell \, T \to \infty$ and it only decreases slightly as $\ell \, \Omega_+$  varies from 0 to 1 at fixed $\ell  \, T$. 
%

\section{Circuit complexity: rotating TFD state}\label{sec:QFT}

After working out different holographic measures of complexity in rotating black hole settings, we would like to study the corresponding complexity in the boundary theory.  For concreteness we focus on the holographic dual of rotating BTZ, \textit{i.e.} the rotating TFD state \cite{Israel:1976ur,Maldacena:2001kr,Hartman:2013qma}
\begin{equation}\label{eq:RTFD}
\ket{rTFD}=\frac{1}{\sqrt{ Z\left(\beta,\Omega\right) }}\sum_n e^{-\beta\left(E_n+\Omega J_n\right)/2}e^{-i (E_n+\Omega J_n) t }\ket{E_n,J_n}_L \ket{E_n,J_n}_R \, ,
\end{equation}
describing an entangled state of the two identical CFT$_2$ on the right and left asymptotic boundaries of the black hole geometry. 
Here $E_n$ and $J_n$ label energy and momentum eigenstates,  $\beta$ matches the inverse Hawking temperature of the dual black hole and $\Omega$  is the angular velocity. In writing the dynamics in \eqref{eq:RTFD}, we have taken a symmetric time $t_R= t_L =t/2$, as to match the holographic model, and evolved with the deformed Hamiltonian on both sides. Another possibility would be to evolve with the undeformed Hamiltonian only, that is 
\begin{equation}\label{eq:RTFDund}
\ket{rTFD}=\frac{1}{\sqrt{ Z\left(\beta,\Omega\right) }}\sum_n e^{-\beta\left(E_n+\Omega J_n\right)/2}e^{-i E_n t }\ket{E_n,J_n}_L \ket{E_n,J_n}_R \, .
\end{equation}
We will consider the two options in what follows. 

In both cases, turning-off the potential $\Omega$, one obtains 
\begin{equation} \label{eq:TFD}
\ket{TFD}=\frac{1}{ \sqrt{ Z\left(\beta\right)}}\sum_n e^{-\beta  E_n  /2} e^{- i E_n t }\ket{E_n}_L \ket{E_n}_R ,
\end{equation}
representing the TFD state dual to the (non-spinning) BTZ black hole  \cite{Israel:1976ur,Maldacena:2001kr}.

Ideally, one would like to evaluate complexity for this state in a holographic CFT$_2$, but a general definition of complexity in QFT (and CFT) is still lacking and the majority of results available so far concerns Gaussian states in free theories (see \eg \cite{Jefferson:2017sdb,Khan:2018rzm,Chapman:2017rqy,Molina-Vilaplana:2018sfn,Hackl:2018ptj,Alves:2018qfv,Camargo:2018eof,Chapman:2018hou,Guo:2018kzl,Ali:2018fcz,Bhattacharyya:2018bbv, Jiang:2018nzg,Caceres:2019pgf,Chapman:2019clq,Doroudiani:2019llj,Bueno:2019ajd,Ali:2019zcj,Ge:2019mjt,DiGiulio:2020hlz,Ruan:2020vze,DiGiulio:2021oal,DiGiulio:2021noo}).\footnote{
An interesting approach based on the Euler-Arnold formalism to study complexity in chaotic quantum systems was developed in \cite{Balasubramanian:2019wgd,Balasubramanian:2021mxo}. 
}
In order to make a qualitative comparison with the holographic results, we will follow the approach of  \cite{Chapman:2018hou} and consider as a toy model that of a free scalar field.
 As we will show explicitly, it is then easy to give an effective description of the rotating TFD state \eqref{eq:RTFD} in terms of the non-rotating one \eqref{eq:TFD}, and make use of the available Gaussian state results \cite{Chapman:2018hou}. This is analogous to what happens for the charged TFD studied in \cite{Chapman:2019clq}, which can also be given an effective description in terms of \eqref{eq:TFD}.
 
\paragraph{Rotating TFD.} 

We consider a simple model where right and left degrees of freedom are described by two identical copies of a  (1+1)-dimensional free scalar QFT on a circle of length $L$, each with  Hamiltonian  
\be\label{HQFT}
H =\int_{-\frac{L}{2}}^{\frac{L}{2}}dx \pq{\frac{\pi^2}{2}+\frac{m^2}{2}\phi^2 +\frac{1}{2}(\pa_x \phi)^2} = \sum_k \omega_k \, \( a^\dagger_k a_k +\frac{1}{2} \) \, ,
\ee
and  angular momentum operator
\be \label{JQFT}
J= -\int_{-\frac{L}{2}}^{\frac{L}{2}}dx ~\pi \pa_x \phi = \sum_k p_k  \( a^\dagger_k a_k + \frac{1}{2}\)\, \,. 
\ee
In writing the r.h.s. of these expressions we have used the mode decompositions at $t=0$
\bea
\phi = \sum_k \frac{1}{\sqrt{2 L \, \omega_k   }}  \(  e^{i p_k  x } a_k + e^{-i p_k x } a^\dagger_k \) \qquad 
\pi = - i  \sum_k  \sqrt{ \frac{\omega_k}{2 L}}   \( e^{i p_k  x } a_k - e^{-i p_k  x } a^\dagger_k \) 
\eea
with   $p_k =  \frac{2\pi}{L} k$  and  $\omega_k = \sqrt{  p^2_k + m^2 }$. 
For each  mode, modulo the shift in the zero-point energy,  both $H$ and $J$ are proportional to the particle number operator $N_k =  a^\dagger_k a_k$, 
\be
N_k \ket{n}_k = n_k \ket{n}_k ,  \qquad   \quad  \ket{n}_k =\frac{(a^{\dagger}_k)^n}{\sqrt{n!}}\ket{0} \, .
\ee
Mode-by-mode we can therefore simultaneously label Hamiltonian and momentum eigenstates in terms of the particle number eigenstates $\ket{n}_k$
\be 
H \ket{n}_k =E_{k,n}\ket{n}_k =\o_k \( n+\frac{1}{2} \)\ket{n}_k, \qquad J \ket{n}_k = J_{k,n} \ket{n}_k = p_k \, \pr{n  +\frac{1}{2} } \ket{n}_k . 
\ee

Given the free QFT structure, which yields modes factorization,  the TFD state can be written as the product of TFD states of  single right-left couples of harmonic oscillators, each  labeled by the mode number $k$
\be\label{eq:TFDfactor} 
\ket{rTFD}=\bigotimes_k \ket{rTFD}_k \, . 
\ee
Making  the eigenvalues structure explicit, the single mode states then  take the form 
\bea
\ket{rTFD}_k &=& \frac{1}{\sqrt{Z_k(\b,\O)}}\sum_n e^{- \( \frac{\b}{2} + it \) \( E_n +\O J_n\)}\ket{n}_{k,L} {\ket{n}}_{k,R}\\
&=& \frac{1 }{\sqrt{Z_k(\b,\O)}}  \sum_n e^{- \(\frac{\b}{2}+it \) \( \o_k  + \O \,   p_k   \) \( n  + \frac{1}{2} \) } \ket{n}_{k,L} {\ket{n}}_{k,R} \, ,
\eea
with normalization factor
\be 
Z_k(\b,\O) = \frac{e^{- \frac{   \beta  }{2}  \(  \o_k  + \O \,   p_k \) } }{1-e^{-\beta\( \o_k  + \O \,   p_k   \)} } \, .
\ee
Defining for every single mode an effective inverse temperature and time as
\be\label{identifications}
\b_k =\b\( 1+\Omega \,  \frac{p_k}{\o_k} \) , \qquad t_k =t \( 1+\O \, \frac{p_k}{\o_k} \)  \, ,
\ee
it is then immediate to see that the rotating TFD state can be effectively written as a TFD state with no rotation 
\be\label{eq:TFDeff} 
\ket{rTFD}_k= \frac{ 1 }{\sqrt{Z_k(\b_k, \Omega=0 )  }} \sum_{n} e^{-\pr{\frac{\b_k }{2}+it_k }\o_k \( n  + \frac{1}{2} \)}\ket{n}_{k,L} \ket{n}_{k,R}\, .
\ee
We shall notice that as long as  $|\O| < 1$ the effective inverse temperature \eqref{identifications} is non-negative, and only vanishes  in the limiting case where $|\O| \to 1$ with $m\to 0$. Also, $t=0$ maps to $t_k=0$,  and this will be important when computing the complexity of formation.
A completely similar reasoning goes through if we choose to time-evolve with the undeformed Hamiltonian as in \eqref{eq:RTFDund}. The only difference being that the effective representation \eqref{eq:TFDeff} would only involve an effective inverse temperature, but not an effective time. 
This simple identification, valid for each mode $k$, allows to borrow and adapt the results of \cite{Chapman:2018hou} for non-rotating TFD states.

Before reviewing the results of \cite{Chapman:2018hou}, let us mention that a similar identification can be performed in the charged, non-rotating, case  \cite{Chapman:2019clq}.  There however the absolute value of the chemical potential, through the identification of the effective temperature, sets a lower bound for the mass parameter $m$. This in particular prevents from taking the $m\to0$ limit in the charged case.

\paragraph{TFD complexity.}
 We have shown that single mode rotating TFD states admit an effective description in terms of non-rotating TFD states. Here we briefly review the complexity analysis of  \cite{Chapman:2018hou} for the TFD state \eqref{eq:TFD}.  

The  analysis of \cite{Chapman:2018hou}  follows and extends the work of \cite{Jefferson:2017sdb}, which adapted  Nielsen's approach  to complexity \cite{nielsen2006quantum,nielsen2008,Nielsen:2006}  to free scalar fields.  The latter starts with a continuum representation of the unitary transformation 
\be
U(\s) = \cev{\mathcal{P}} \exp \[ -i \int^\s_0\!\!\! d s\, \sum_I Y^I(s)\,K_I \] \quad \text{with} \quad  	U(0)= \mathbbm{1}, \quad \text{and} \quad U(1)= U_{\rm T}\, , 
\ee
acting on states and connecting the reference and target states 
\be
\ket{\psi_{\rm T}} = U(1) \ket{\psi_{\rm R}}\, .
\ee
The unitary is constructed in terms of  a basis of Hermitian operators $K_I$, the gate's  generators, applied along the circuit parametrized by $s$ as specified by the control functions $Y^{I}$. For practical reasons, the  set of generators is normally taken to be finite and to realize a closed algebra.  Nielsen's approach then assigns a cost to each circuit through a functional 
\be \label{cost_D}
\mathcal{D}[U]= \int^{1}_0 ds ~ F \( U(s), Y^I(s)  \)
\ee
specified in terms of  a local cost function  $F$, and defines the complexity of going from a reference to a target state as the cost associated to the circuit that minimizes the functional, namely
\be 
\mathcal{C}(U_{\rm T})=\min_U \mathcal{D}[U].
\ee 
In this approach  $U(\sigma)$ defines a trajectory in the space of unitaries, with $Y^I(\s)$ the components of its tangent vector. The problem of computing complexity is then analogous to solving for the motion of a particle in the geometry emerging from the group structure provided by the gate set, with Lagrangian specified by $F$.

In \cite{Chapman:2018hou},  the target state was the  non-rotating  TFD state, which is the product of single modes TFD states, each corresponding to a TFD state of a pair of  harmonic oscillators at fixed $k$
\be
\ket{TFD}=  \bigotimes_k \ket{TFD}_k=\bigotimes_k \frac{1}{\sqrt{Z(\b )}} \sum_{n} e^{-\( \frac{\b}{2}+it\)  \o_k \( n + \frac{1}{2} \) } \ket{n}_{k,L} \ket{n}_{k,R}\, .
\ee
Following \cite{Jefferson:2017sdb}, the reference state was chosen to  be a completely unentangled state  obtained as the ground state of (two copies of) a ultralocal Hamiltonian where the spatial derivative term is absent.  That is, the ground state of an Hamiltonian with a fixed frequency $\mu$ for all modes 
\be 
H =\sum _{k}\mu \( a^\dagger_k a_k + \frac{1}{2}\)\, .
\ee 
To connect the TFD state to the reference state,  \cite{Chapman:2018hou} considered circuits built with gates $K_I$  quadratic in the canonical variables associated to each of the entangled  pairs of harmonic oscillators making the TFD state. Introducing a UV regulator in the field theory yields a finite number of such gates.  A simple way  to regularize the theory in the setup at hand is to consider a finite number of modes $\tilde N$.\footnote{Notice that our regularization procedure is slightly different from the one adopted in \cite{Chapman:2018hou}, where the UV regularization comes from putting the theory on a spatial lattice. 
}     In such a case, in the analysis of   \cite{Chapman:2018hou} the relevant group structure turns out to be $Sp(2\tilde N,\mathbb{R})$.
The construction of the generators also introduces an arbitrary   gate scale $\m_g$, which together with the reference state scale $\m$ and the mode frequency $\o_k$ characterize the complexity model. 

The cost function on which  \cite{Chapman:2018hou} focused their analysis is the so called $\kappa=2$
\be 
 F_{\kappa=2}=\sum_I \left|Y^I\right|^2 
\ee
which is independent of the specific basis  for the gates generators.
Importantly, for this cost function, when the reference and gate scales are set equal, $\mu_g =\mu$,  the optimal circuit does not mix modes with different $k$, and  the minimal length circuit for each mode is generated by repeatedly applying a single generator \cite{Chapman:2018hou}.  In geometrical terms, in this case the optimal circuit computing complexity for each $k$ corresponds to a straight-line geodesic  on $Sp(2,\mathbb{R})$. The resulting complexity evaluated in \cite{Chapman:2018hou} is 
\be \label{eq:kappa2}
\mathcal{C}_{\kappa=2} = \frac{1}{4}\sum_{k}\log^2\( f^{(+)}_k + \sqrt{\(f^{(+)}_k\)^2 -1} \)+  \log^2\(f^{(-)}_k + \sqrt{\(f^{(-)}_k\)^2 -1} \)
\ee
with
\bea \label{eq:fk}
f_{k}^{(\pm)} &=& \frac{1}{2}\( \frac{\mu}{\omega_k} + \frac{\omega_k}{\mu} \) \cosh 2 \alpha_k  \pm   \frac{1}{2}\( \frac{\mu}{\omega_k} - \frac{\omega_k}{\mu} \) \sinh 2 \alpha_k \cos \omega_k t \,  \label{eq:fk} ,\\
\alpha_k  &=& \frac{1}{2} \log \(  \frac{1+ e^{-\beta \omega_k /2} }{ 1- e^{-\beta \omega_k /2}}\) \label{eq:alphak}\, . 
\eea
Let us reiterate that the mode factorization for the circuit allows to obtain the TFD complexity as the the sum of complexities evaluated for each mode separately. This is crucial in view of using the effective description of the rotating TFD \eqref{identifications}-\eqref{eq:TFDeff} to evaluate complexity in terms of the non-rotating TFD results.  In the rest of our work we will  thus only consider the  situation where the gate scale is set equal to the reference scale.

In \cite{Chapman:2018hou},  the basis-dependent cost function 
\be
F_1=\sum_I \left| Y^I \right| \, 
\ee
was also considered to evaluate the length of the straight-line circuit. That is, \cite{Chapman:2018hou} did not solve explicitly for the optimal circuit for the   $F_1$ cost function, but simply evaluated the length of the straight-line circuit with this measure. Nonetheless, this still provides an upper bound on computational complexity of the TFD state. 
Interestingly, \cite{Chapman:2018hou}   found that the straight-line circuit provides a  qualitative matching with the holographic complexity results for the TFD state when working in the so called physical basis.\footnote{In this basis the $K_I$ are constructed with the canonical variables associated to the single harmonic oscillator Hamiltonian, retaining the original left and right splitting of the TFD construction, see \cite{Chapman:2018hou}.}

In what follows we will then only explore the corresponding result for the  $F_1$  cost:
\be
\begin{aligned} \label{eq:c1}
\mathcal{C}_{1}=\frac{1}{ 2 } ~\sum_{k}& ~~ \sqrt{2} \left | \log\pr{f^{(+)}_k   +\sqrt{\( f^{(+)}_k \)^2 -1}}\cos \theta^{(+)}_k   +  \log\pr{f^{(-)}_k +\sqrt{ \( f^{(-)}_k \)^2  -1}}\cos \theta^{(-)}_k \right|   \\
&+ \left | \log\pr{f^{(+)}_k +\sqrt{\( f^{(+)}_k \)^2 -1}}\sin\theta^{(+)}_k  +  \log\pr{f^{(-)}_k +\sqrt{ \( f^{(-)}_k \)^2 -1}}\sin \theta^{(-)}_k \right|  \\
&+ \left |  \log\pr{ f^{(+)}_k +\sqrt{ \( f^{(+)}_k \)^2  -1}}\cos\theta^{(+)}_k  -  \log\pr{f^{(-)}_k +\sqrt{ \( f^{(-)}_k \)^2  -1}}\cos \theta^{(-)}_k \right|  \\
&+ \left |  \log\pr{f^{(+)}_k +\sqrt{  \( f^{(+)}_k \)^2   -1}}\sin\theta^{(+)}_k  -  \log\pr{ f^{(-)}_k +\sqrt{\( f^{(-)}_k \)^2  -1}}\sin \theta^{(-)}_k   \right| \, ,
\end{aligned}
\ee
with 
\be \label{eq:thetak}
\tan \theta^{(\pm)}_k =   \frac{1}{2}\( \frac{\mu}{\omega_k} + \frac{\omega_k}{\mu} \) \cot \omega_k t  \pm \frac{1}{2}\( \frac{\mu}{\omega_k}  - \frac{\omega_k}{\mu} \) \frac{1}{\tanh  2\alpha_k  \sin \omega_k t } \, . 
\ee
\\

We will also be interested in the complexity of formation,  the difference between the rotating TFD state complexity at $t=0$ and that of  two copies of the vacuum state\footnote{The vacuum states and the corresponding complexity are simply recovered taking $\beta \to \infty$.}%
\be 
\D\mathcal{C} \equiv \mathcal{C}(\ket{rTFD(0)}  -\mathcal{C}(\ket{0}_L \ket{0}_R) \, ,
\ee
This takes a particular simple form for the two cost functions we are considering and is independent from the reference scale $\mu$, namely
\be
\Delta\mathcal{C}_{1} = 2 \sum_{k }  |\alpha_k|\,  , \qquad \qquad \Delta\mathcal{C}_{\kappa=2} = 2 \sum_{k}\alpha_k^2   \, . 
\ee

This concludes our summary of the main results of \cite{Chapman:2018hou} that we will use next to evaluate the complexity of rotating TFD states making use of the effective description \eqref{identifications}-\eqref{eq:TFDeff} of the single mode rotating TFD in terms of a non rotating TFD state. 
 
\subsection{Complexity of formation }
 
In this subsection we analyze the complexity of formation of the rotating TFD state for the two cost functions described above. 
Using the effective description \eqref{identifications}-\eqref{eq:TFDeff}, the complexity of formation reads 
\be\label{cformationLAMBDA} 
\D\mathcal{C}_1=2\sum_{k=-N/2}^{N/2}|\a_k |, \qquad \qquad \D\mathcal{C}_{\kappa=2}=2\sum_{k=-N/2}^{N/2}\a_k^2
\ee
with
\be \label{ak}
\alpha_k  = \frac{1}{2} \log \(  \frac{1+ e^{-\beta_k \omega_k /2} }{ 1- e^{-\beta_k \omega_k /2}}\) =\arctanh e^{-\beta_k \omega_k /2} \geq 0 \, . 
\ee
Notice we have written the above formulae making explicit a UV cutoff on momenta. These are actually  UV-finite quantities, and  $N$ can be taken to infinity. However, we are not able to sum the series analytically in general and thus we use a large, but finite, $N$ to evaluate them  and produce plots. 
The sum runs on positive and negative momenta, which contributions are related by $\a_{-k}(\Omega) = \a_{k}(-\Omega)$.  For simplicity and  without loss of generality, we will also henceforth assume $0 \leq \Omega < 1$. 

In view of comparing with the holographic results, we will be particularly interested in the conformal limit $m\to 0$. In this limit
the complexity of formation exhibits a IR divergence due to the zero mode
\be
\alpha_0  = \frac{1}{2} \log \(  \frac{1+ e^{-\beta \, m / 2} }{ 1- e^{-\beta \,  m /2}}\)  \sim -  \frac{1}{2} \log \beta \,   m \, , 
\ee
In the remainder, to evaluate complexity numerically and produce plots for the conformal limit, we will  introduce a IR regulator by using a small but non-vanishing mass.  Finally, in the rest of our analysis we  will consider the compact spatial dimension of the system $L$ to be fixed.  We will measure dimensional quantities with respect to the scale set by $L$, which we therefore simply set to $L=1$.  

\begin{figure}[t]
\centering
\includegraphics[width=.45\textwidth]{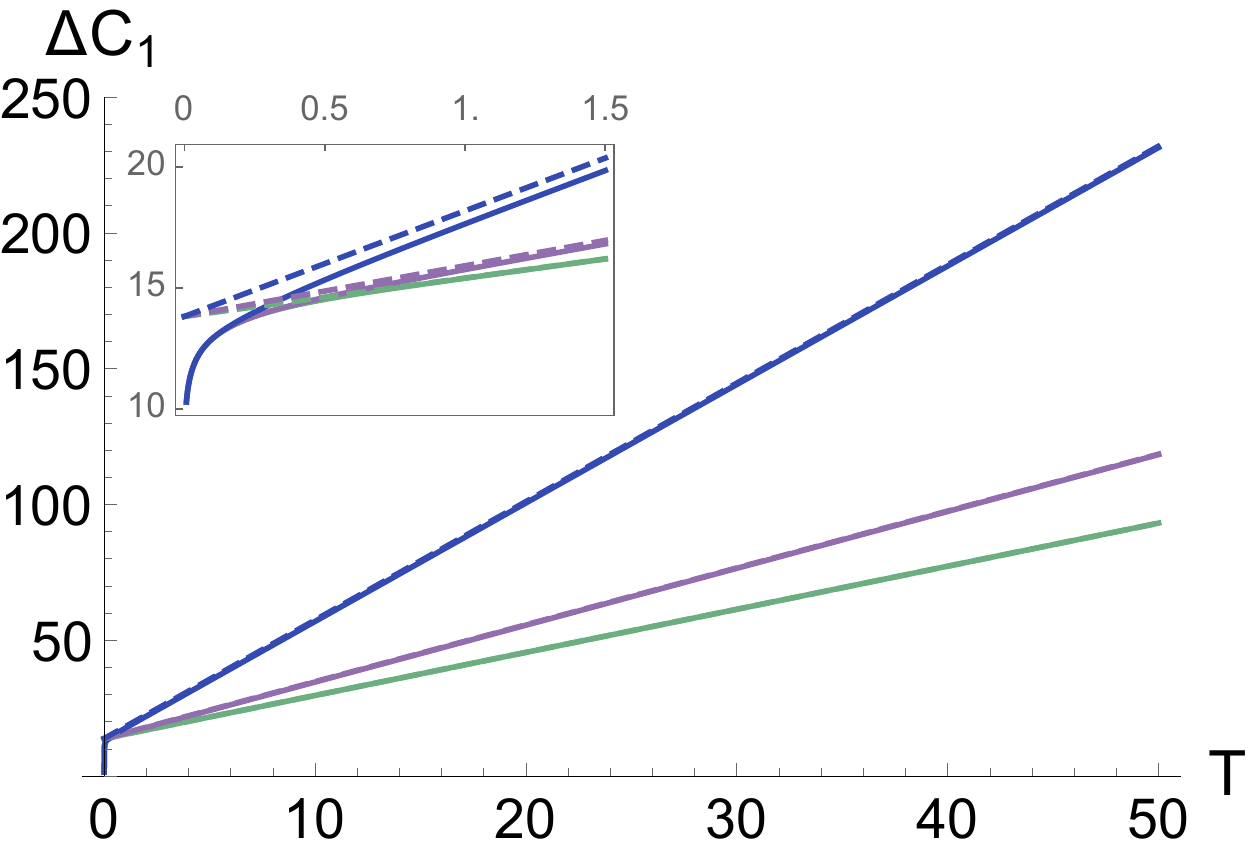} \hfill
\includegraphics[width=.45\textwidth]{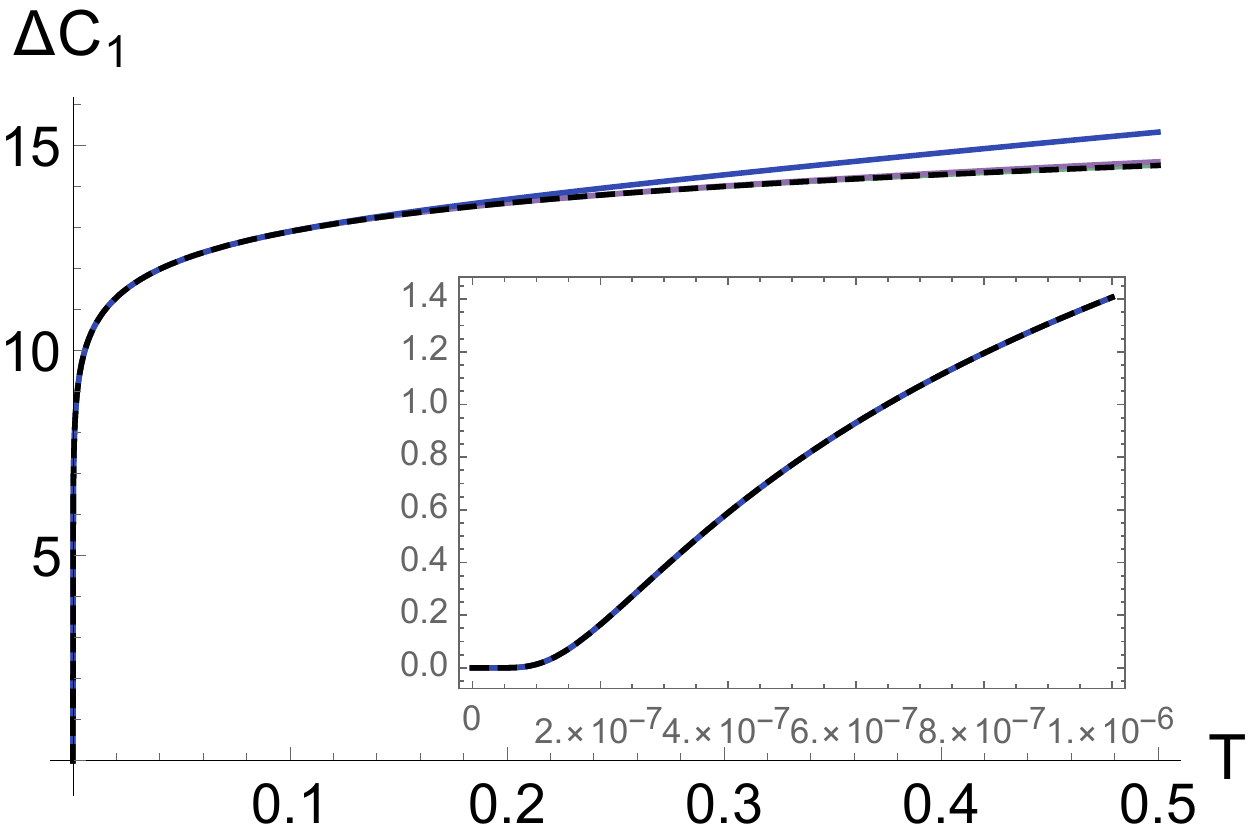}\\
\includegraphics[width=.45\textwidth]{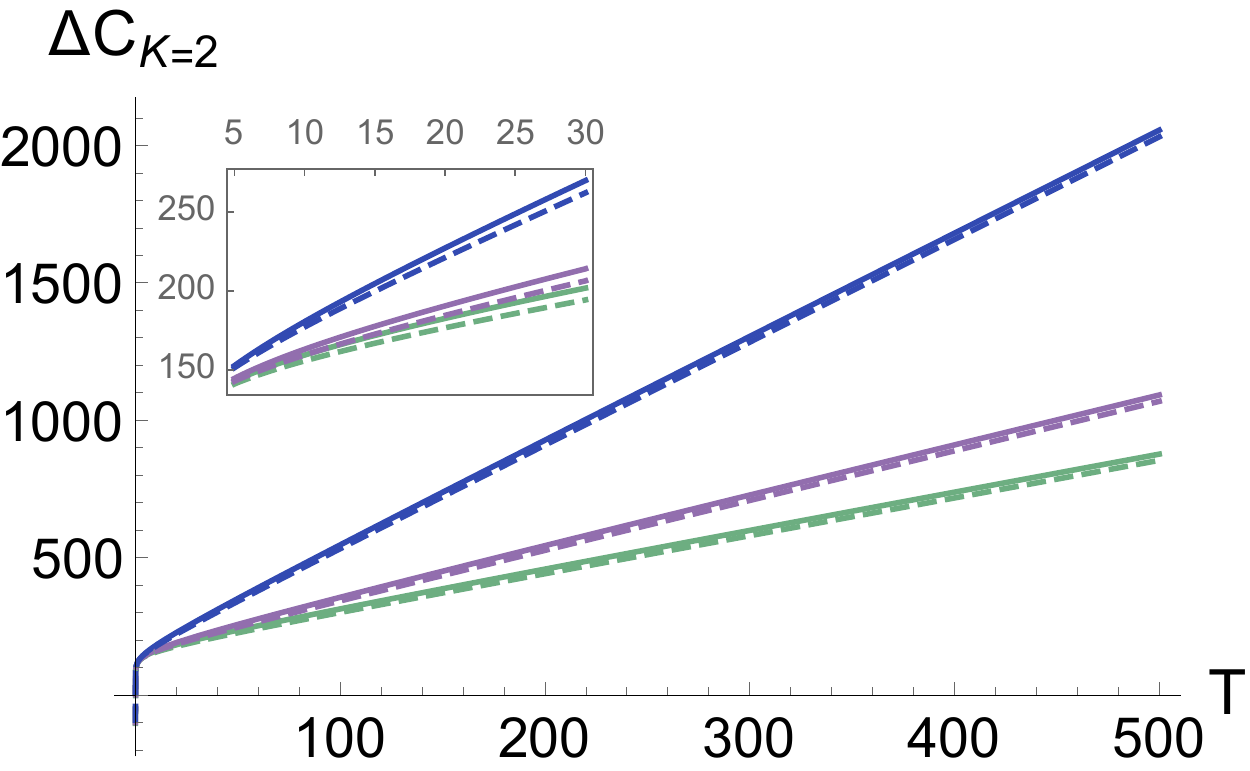} \hfill
\includegraphics[width=.45\textwidth]{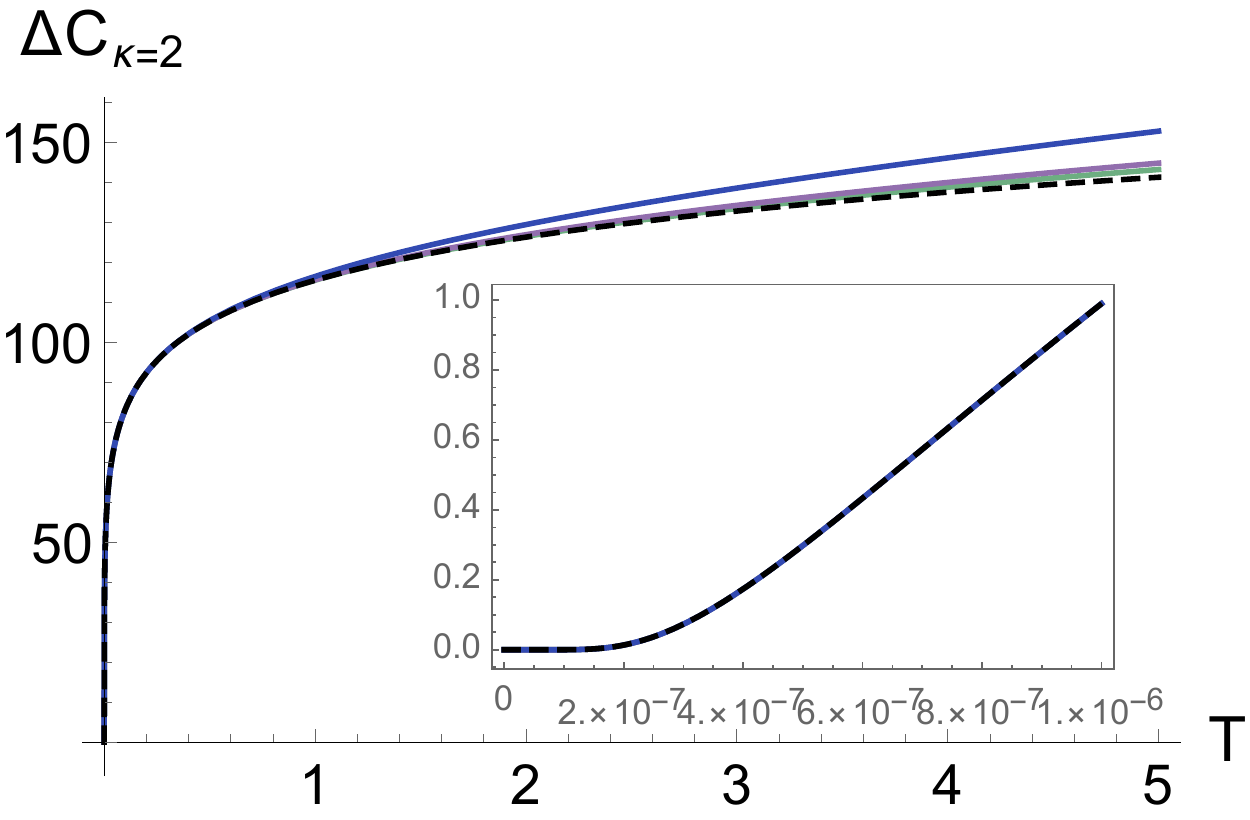}
\centering
\caption{Complexity of formation  for different values of $\Omega$ as a function of  the temperature $T$, with   $L=1$, $N=20000$ and $m=10^{-6}$. From the bottom up  $\O=0.1$ (green), $\O=0.5$ (violet) and $\O=0.8$ (blue). 
The left  panels illustrate how  for sufficiently large temperatures the complexity of formation agrees well with the analytic expression \eqref{eq:largeTF1} and \eqref{FK2formationconf} (dashed lines). 
The right panels show that for sufficiently low temperatures the $\Omega$-independent zero-mode contribution  \eqref{zeromodelowT} (dashed black) dominates the sum over modes.
}\label{fig:FORMconformalTdep}
\end{figure}

\paragraph{Dependence on $T$. }  At low temperature the complexity of formation is dominated by the zero-mode, which  is the least suppressed as $T\to 0$ 
\be\label{zeromodelowT} 
\D\mathcal{C}_1\sim 2 \arctanh e^{- \frac{m}{2T}}  , \qquad \qquad \D\mathcal{C}_{\kappa=2}\sim 2 \arctanh^2 e^{- \frac{m}{2T}} \, .
\ee
As the temperature increases, the contributions of the other modes become relevant.  As opposed to the non-rotating case,  for each positive-negative mode pair labeled by $k>0$, we have $\alpha_k < \alpha_{-k}$ at finite $T$, and the larger $\Omega$ the more important is the negative mode contribution with respect to the positive one.
This can be seen from the explicit expression \eqref{ak} and the definition of the effective inverse temperature \eqref{identifications}. In fact, the angular velocity $\Omega$ translates into a smaller effective temperature for positive modes as compared to negative ones. The effect is  apparent for $m=0$, where we can write for positive (negative) modes
\be  
\alpha_k  = \arctanh e^{-\frac{\pi |k| (1\pm\O)}{T}} \, . 
\ee
The resulting  dependence on the temperature for a small value of the mass is reported in figure \ref{fig:FORMconformalTdep}. 
There we observe that  after an intermediate temperature regime, the complexity of formation shows a transition to a linear  regime at high temperature.
In the conformal limit $m\to 0$ we  are able to extract the high temperature behavior  analytically.  We isolate the contribution  coming from the zero mode,  and remove   the momenta cutoff  from the sum, \ie we work with the full $N = \infty$ series
\be\label{F1formationconf} 
\D\mathcal{C}_1 \sim  \log\pr{\frac{4T}{m}}+2\sum_{k=1}^{\infty}\pg{\arctanh\pq{e^{-\frac{\pi k (1+\O)}{T}}}+\arctanh\pq{e^{-\frac{\pi k (1-\O)}{T}}}} \,.  
\ee
We then  re-express the series using the  Euler-MacLaurin formula
\be 
\sum_{k=a}^{b}f_k=\int_{a}^{b}dk f(k)+\sum_{j=1}^p \frac{B_j}{j!}\pg{\frac{\pa^{(j-1)}}{\pa k^{(j-1)}}f(k)\Big|_{k=b}-\frac{\pa^{(j-1)}}{\pa k^{(j-1)}}f(k)\Big|_{k=a}}+R_p,
\ee
where $B_j$ are Bernoulli numbers,  $R_p$ the remainder and $p$ a positive integer. Rather than using the explicit expression  for the reminder,  we use the fact that  $R_p$  satisfies the general bound
\be 
|R_p|\leq \frac{2\zeta(p)}{(2\pi)^p} \int_a^b\Big| \frac{\pa^p}{\pa k^p}f(k)\Big|,
\ee
where $\zeta$ is the Riemann zeta function. We then select a value of $p$ such that the r.h.s. shows no divergences as  $T \to \infty$, which ensures that $R_p$  is not divergent either.  The remaining terms in the Euler-MacLaurin formula can  then be evaluated explicitly and the structure of divergences in $T$ obtained.   
In the case at hand, with $p=2$, the finite sum gives a divergence logarithmic in $T$, which combines with an analogous one coming from the integral to cancel the $\log T$ coming from the zero mode contribution to  $\D\mathcal{C}_1$. The only remaining divergent term comes from the integral, and gives 
\be  \label{eq:largeTF1}
\D\mathcal{C}_1 \sim \frac{\pi}{2}\frac{T  }{1-\O^2}+\log \frac{1}{m } +O(1).
\ee
In writing this expression we kept  explicit the  zero-mode logarithmic divergence in $m\to 0$.
With an identical  strategy,  we can isolate the high-temperature divergence structure  for the $F_{\kappa=2}$ cost function in \eqref{cformationLAMBDA}, which in the conformal limit can be written  as %
\be\label{FK2formationconf} 
\D\mathcal{C}_{\kappa=2} \sim \frac{1}{2}\log^2\pr{\frac{4T}{m}}+2\sum_{k=1}^{N/2}\pg{\arctanh^2\pq{e^{-\frac{\pi k (1+\O)}{T}}}+\arctanh^2\pq{e^{-\frac{\pi k (1-\O)}{T}}}} \,  .
\ee
The analysis is similar,  with the  $\alpha^2_k$ contributions resulting in a richer structure of divergences.  Using the Euler-MacLaurin formula, the integral still gives both the leading divergence and subleading ones. The latter combine with analogous terms coming from the finite sum and the resulting expression is
\be
\begin{aligned}  \label{eq:largeTK2}
\D\mathcal{C}_{\kappa=2} \sim & \frac{7\zeta(3)}{2\pi}\frac{T}{1-\O^2}-\frac{1}{2}\log^2 T -\( \log^2\pq{\frac{1}{\pi^2(1-\O^2)}}+\frac{11+9\log 2}{6} \) \log T\\
&+ \log T   \log\frac{1}{m} +\frac{1}{2}\log^2\frac{1}{m}+2\log 2 \log\frac{1}{m} +O(1),
\end{aligned}
\ee
where  again we isolated the zero-mode divergences.  
The leading divergence is again linear in $T$ with the same $\Omega$ dependence  for the linear coefficient for both  costs. The  linear behavior  also matches the one observed for both the CA and CV holographic prescriptions, see eq.~\eqref{eq:CVBTZTO}  and  fig.~\ref{fig:GCcfCA},\ref{fig:GCcfCV}.   
Next to the leading linear divergence, the   $F_{\kappa=2}$ cost has  a number of  subleading divergences which are completely absent for the $F_{1}$ cost.  There is also a mixed $\log T \log m $ term  originating from  the zero-mode.

\begin{figure}[ht]
\centering
\includegraphics[width=.45\textwidth]{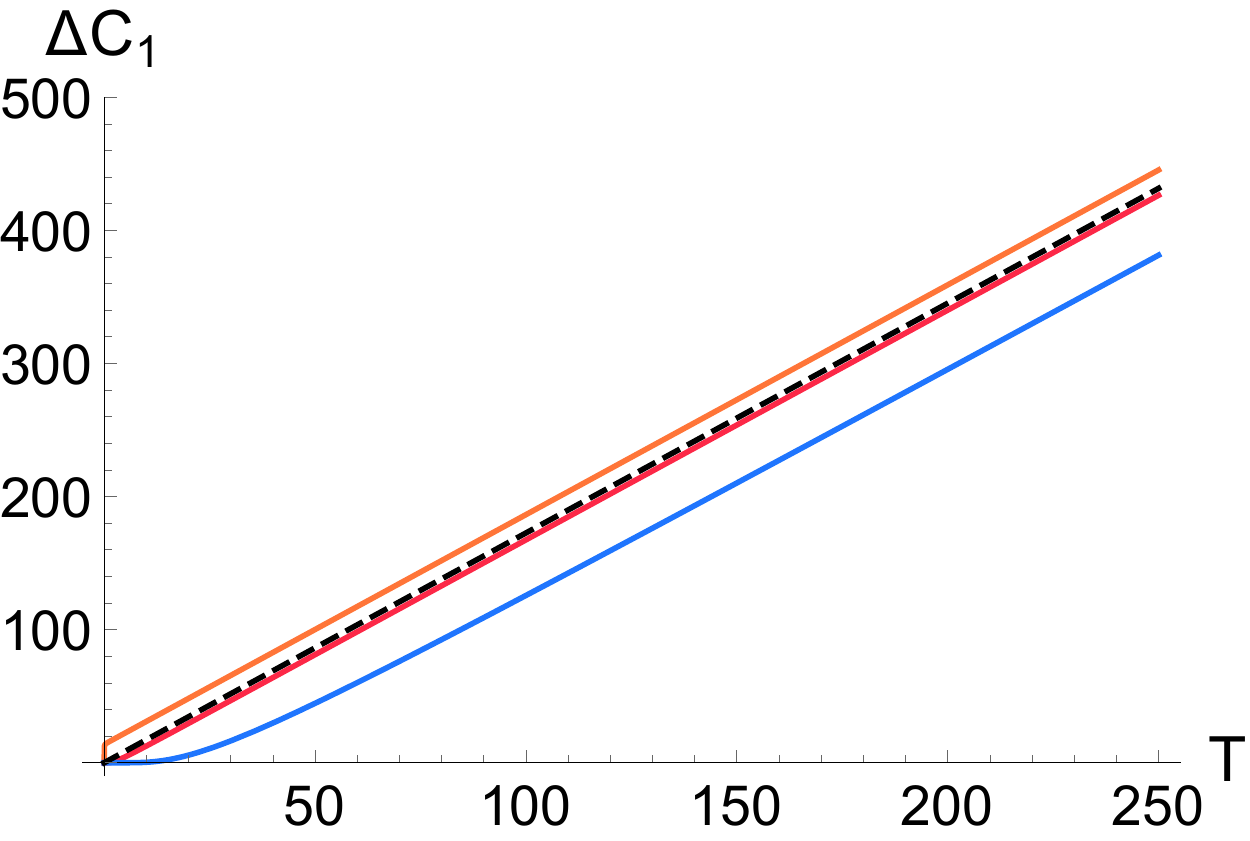} \hfill
\includegraphics[width=.45\textwidth]{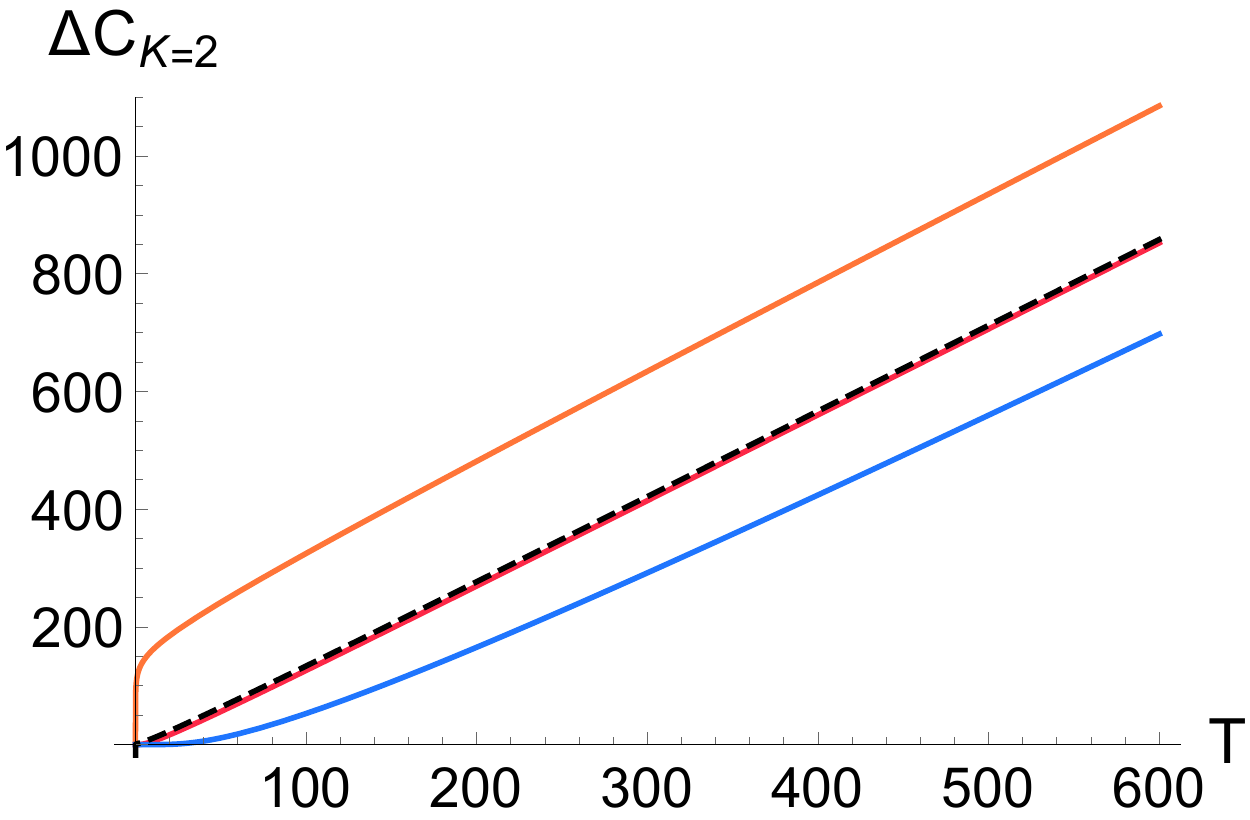}
\centering
\caption{Complexity of formation for different values of $m$ as a function of the  temperature $T$, with  $L=1$, $N=20000$ and $\O=0.3$. From the top down  $m=10^{-6}$ (orange), $m=10$ (red) and $m=100$ (blue). The large $T$ behavior is linear with a slope matching the one obtained in the massless limit and reported in  eq.~\eqref{eq:largeTF1} and \eqref{eq:largeTK2}  for   $\D\mathcal{C}_1$ and  $\D\mathcal{C}_{\kappa=2}$  respectively (dashed black). }
\label{fig:FFORMmassiveTdep}
\end{figure}

For finite values of $m$, as shown in  \ref{fig:FFORMmassiveTdep}, the  leading divergence of the  complexity of formation remains linear in $T$ and with the  same slope obtained in the  conformal case in eq.~\eqref{eq:largeTF1}  and \eqref{eq:largeTK2}.   
%

\begin{figure}[ht]
\centering
\includegraphics[width=.45\linewidth]{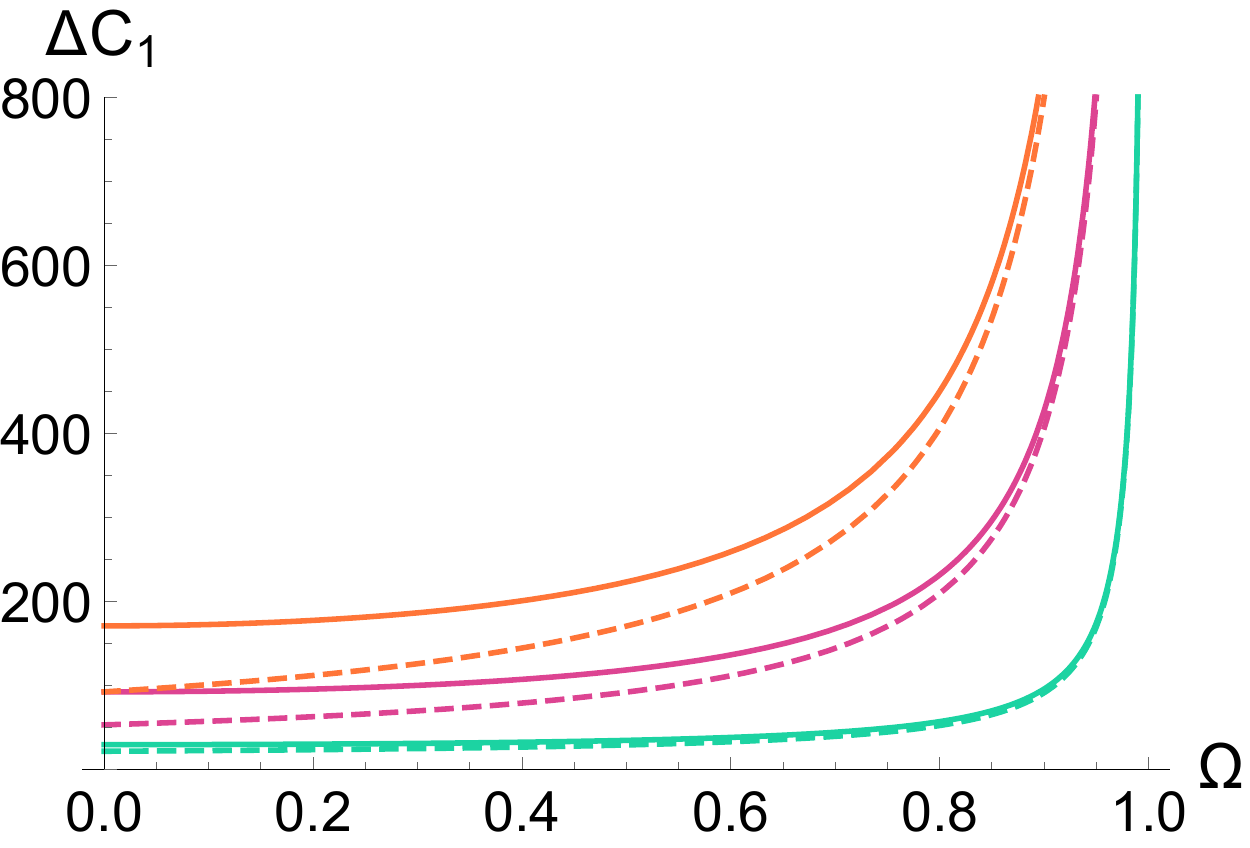} \hfill
\includegraphics[width=.45\linewidth]{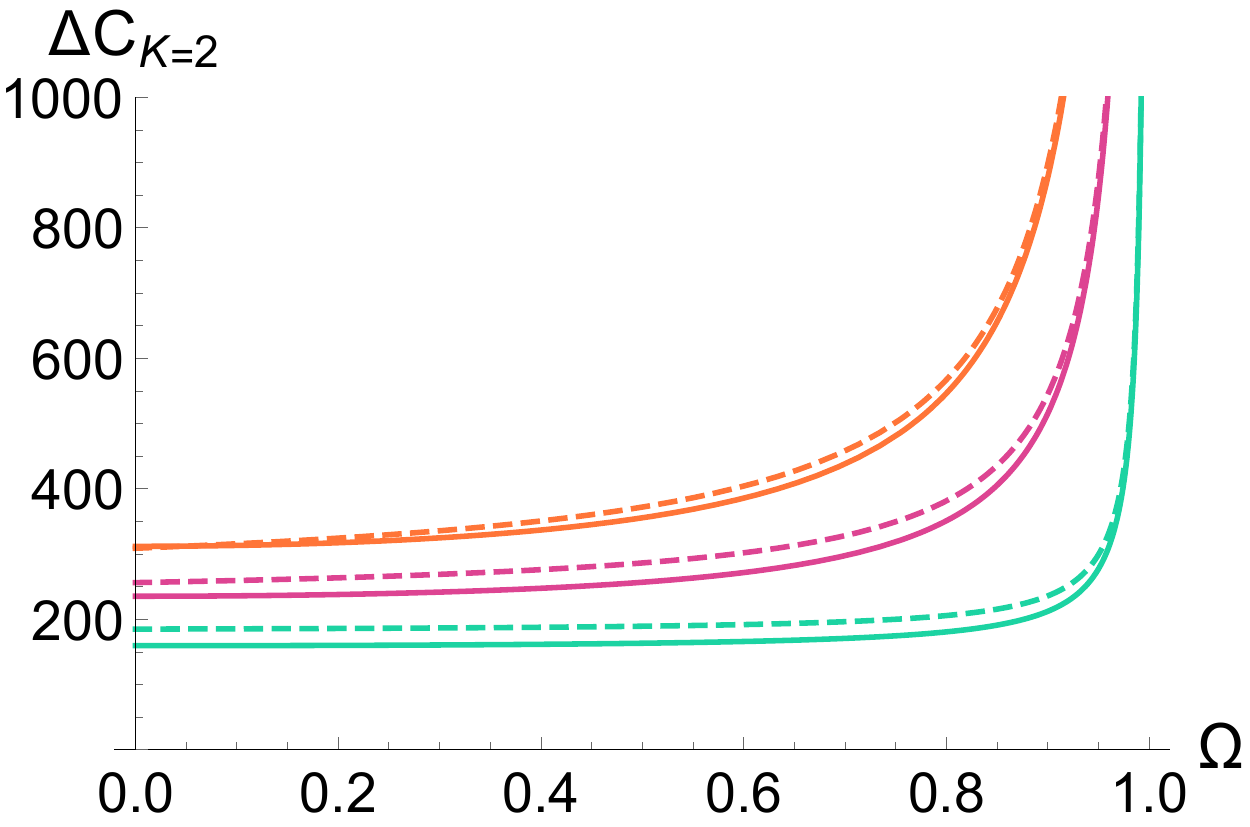}
\caption{Complexity of formation as a function of the angular velocity $\O$.  For any curve, $L=1$, $N=20000$ and $m=10^{-6}$. From the bottom up: $T=10$ (solid green), $T=50$ (solid pink) and $T=100$ (solid orange).  Dashed curves represent  the  $\Omega \to 1$ divergences plus the zero-mode divergence obtained analytically  in  \eqref{Omegaonelimitc1}  and \eqref{OmegaonelimitK2}. Notice that the complexity of formation evaluated with the two costs approach the limiting value from opposite directions. }\label{fig:FFORMtempOmegadep}
\end{figure}

\begin{figure}[ht]
\centering
\includegraphics[width=.45\linewidth]{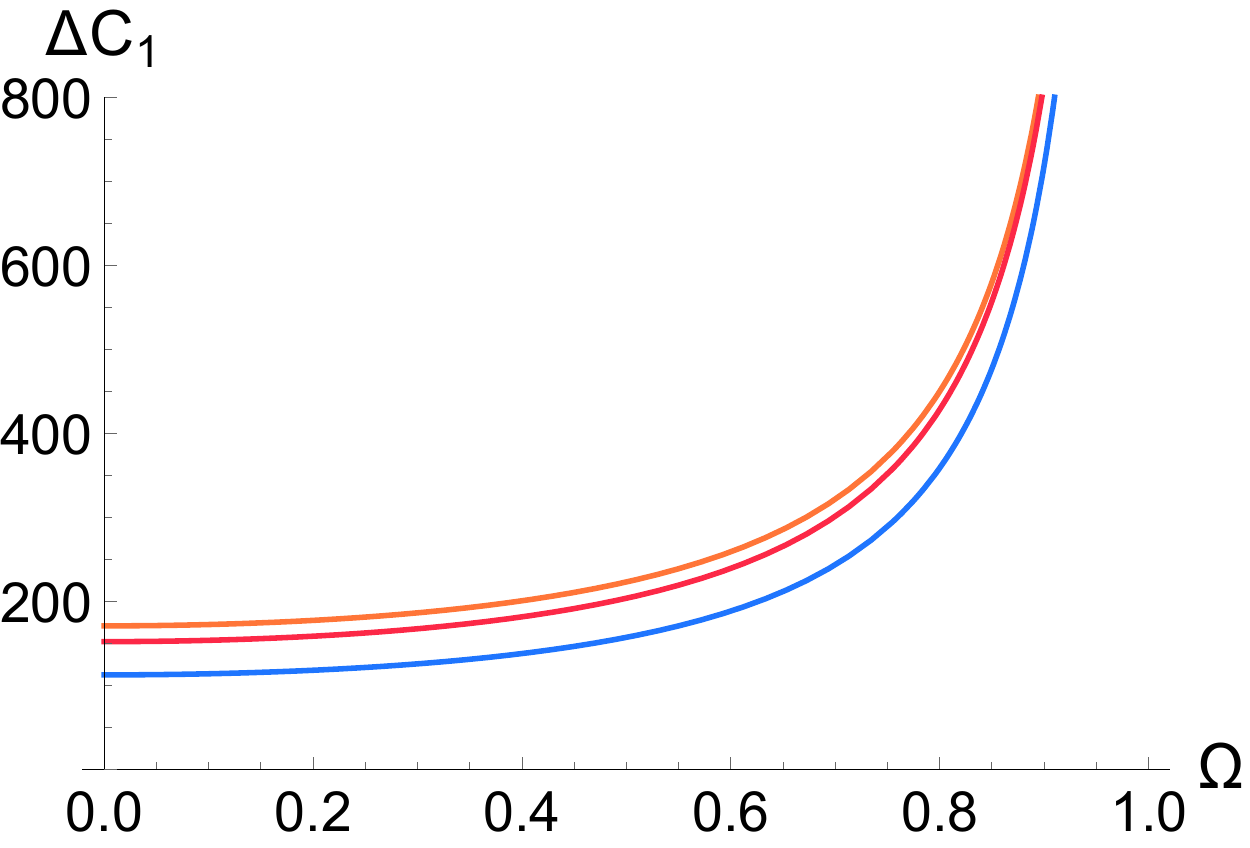} \hfill
\includegraphics[width=.45\linewidth]{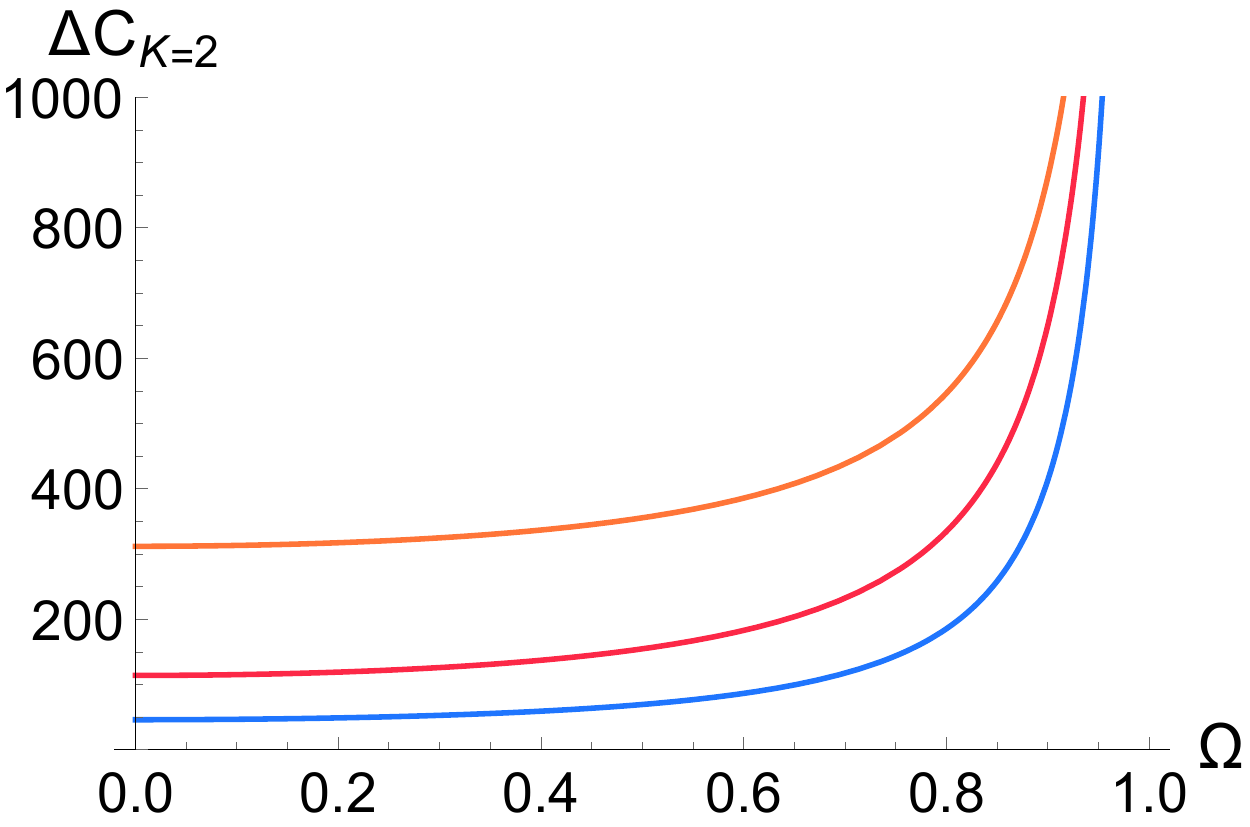}
\centering
\caption{Complexity of formation as a function of the angular velocity $\O$ for different values of $m$. For all curves $L=1$,    $N=20000$ and $T=100$. From the top down $m=10^{-6}$ (orange), $m=10$ (red) and $m=100$ (blue). The  divergent behavior as $\Omega \to 1$ is apparent for both cost functions.}\label{fig:FFORMmassiveOmegadep}
\end{figure}

\paragraph{Dependence on $\Omega$.}
The dependence on the angular velocity is reported in figure \ref{fig:FFORMtempOmegadep}  for different values of the temperature and  a fixed  mass close to the conformal limit, and  in fig.~\ref{fig:FFORMmassiveOmegadep}  for different values of the mass at fixed temperature. The plots show a clear divergent behavior in the critical limit $\O\to 1$.
Again, we can extract the divergences analytically in the $m\to 0$ limit.  As apparent from \eqref{F1formationconf} and \eqref{FK2formationconf},  the divergence structure is linked to the one  for the high-temperature limit.
The main differences are that when taking  $\O\to 1$  only the second $\arctanh$ terms in \eqref{F1formationconf} and \eqref{FK2formationconf} will be divergent, and that the zero mode contribution is independent from $\Omega$ while was diverging in $T$. Thus the  recombination of the various contributions and the final result are slightly different from the high temperature case.   In particular, for $\Omega \to   1$ one gets
\be  \label{Omegaonelimitc1}
\D\mathcal{C}_1 \sim \frac{\pi}{4}\frac{T}{1 - \O} +\frac{1}{2} \log\pr{  1 -\O }+\log\pr{\frac{1}{m}}+O(1)
\ee
and
\be \label{OmegaonelimitK2}
\begin{aligned} 
\D\mathcal{C}_{\kappa=2} \sim & \frac{7\zeta(3)}{4 \pi}\frac{T}{1- \O}- \frac{1}{2}\log^2\pr{\frac{1}{1 - \O}}-\pq{\log\frac{T}{\pi}+\frac{11+15\log 2}{12}}\log\pr{\frac{1}{1 - \O}} \\
&+\frac{1}{2}\log^2\frac{1}{m}+(\log T +2\log 2)\log \frac{1}{m}+O(1).
\end{aligned}
\ee
The leading divergence is thus the same for the  two  cost functions. 
 
Comparing to the holographic analysis, we can see that the leading divergence  in this case differs from the holographic one. In particular, for both CA and CV we have found a divergence with an additional logarithmic factor (see  \eqref{eq:OmegaoneCA} and \eqref{eq:OmegaoneCV}) of the form
\be 
\Delta C (t_b = 0) \sim \frac{T\ell }{1- \O_H \ell } \log \frac{1}{1- \O_H \ell} \,  .
\ee

\begin{figure}[h]
\centering
\includegraphics[width=.45\linewidth]{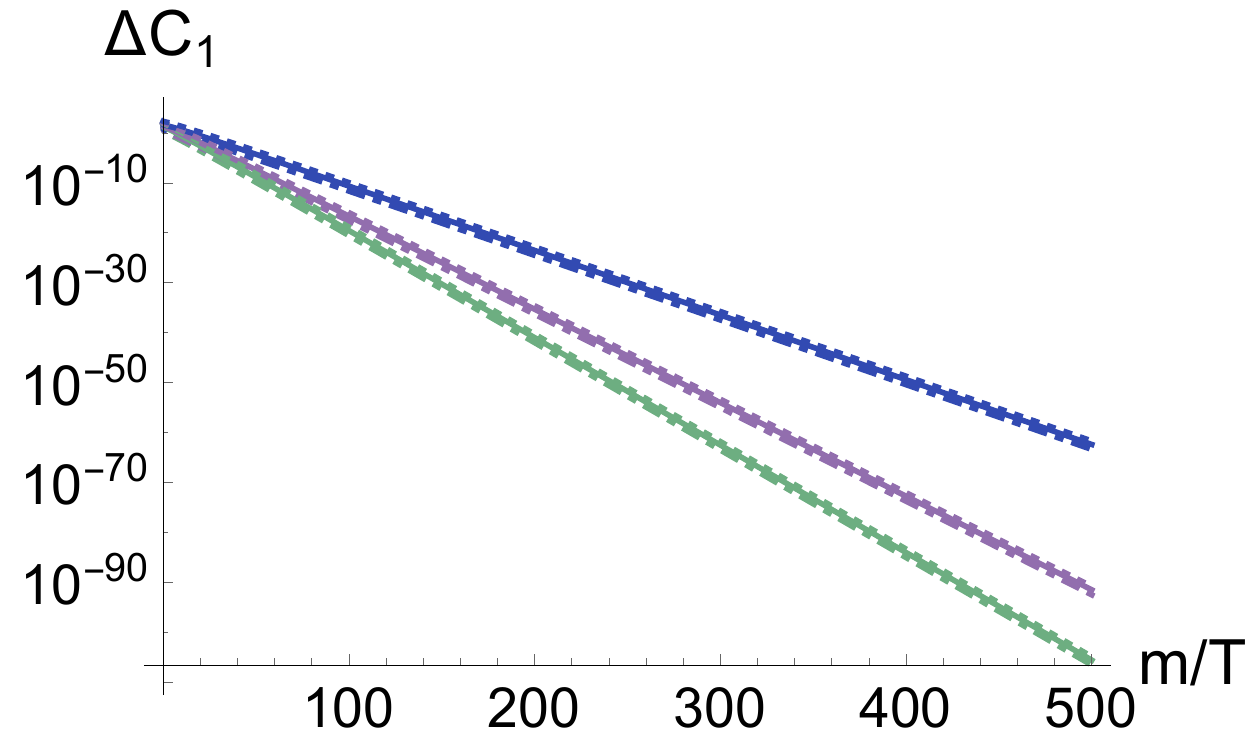} \hfill
\includegraphics[width=.45\linewidth]{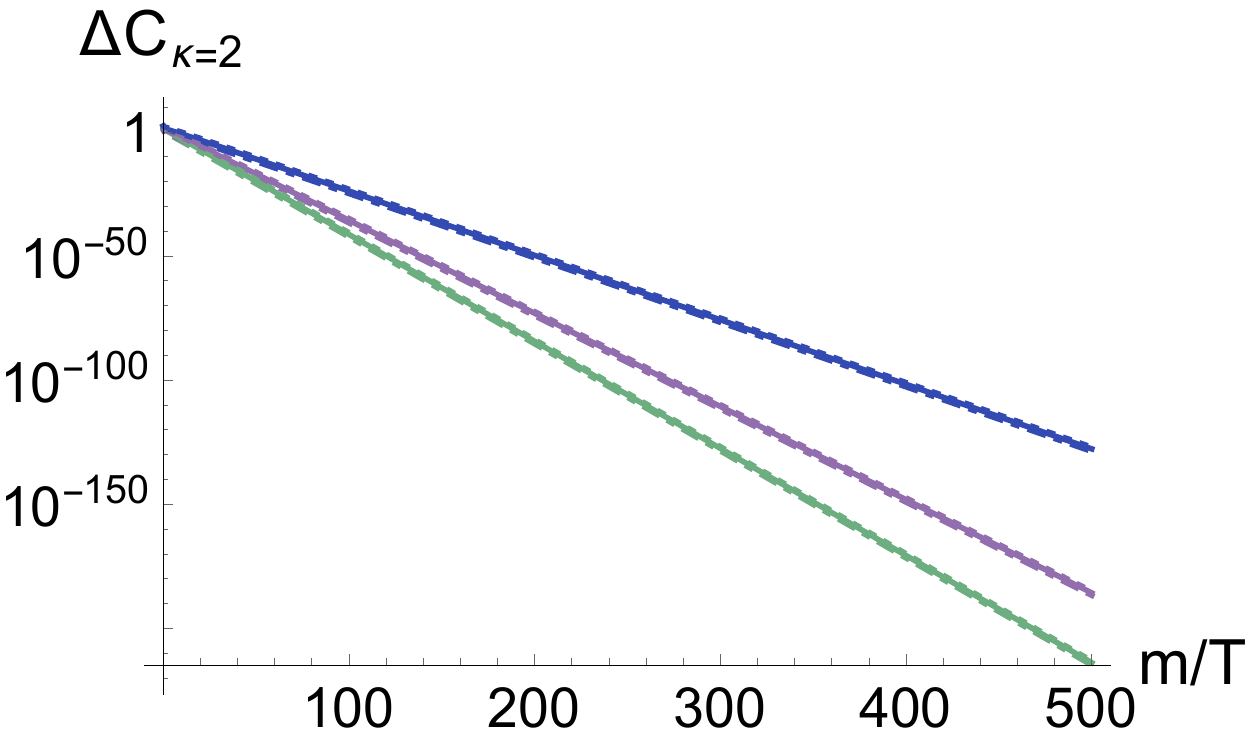} \\
\centering
\caption{Complexity of formation as a function of the mass $m$.  Plots have been produced with  $L=1$ and  $N=20000$.   Three values of the temperature are reported, $T=1$ (dotted-dashed lines), $T=10$ (solid lines) and $T=50$ (dotted lines), but are indistinguishable. For each $T$ from the bottom up $\O=0.1$ (green), $\O=0.5$ (violet) and $\O=0.8$ (blue). The  logarithmic scale shows how the complexity of formation decreases exponentially in $m/T$, with a coefficient fixed by $\Omega$. }\label{fig:FFORMOmegamassdep} 
\end{figure}

\paragraph{Dependence on $m$.}
To conclude,  we briefly comment on the dependence  on the mass parameter.  As already pointed out the complexity of formation diverges in the conformal limit $m\to0$ with a behavior set by the zero mode. At leading order 
\be 
\D\mathcal{C}_1 \sim \log\frac{4T}{m}, \qquad \qquad \D\mathcal{C}_{\kappa=2}\sim \frac{1}{2}\log^2\frac{4T}{m} \, .
\ee
As the mass is increased, the complexity of formation monotonically decreases and approaches zero. This   is illustrated in  the logarithmic plots in fig.~\ref{fig:FFORMOmegamassdep}, where we can see that at large $m$, $\D\mathcal{C}$ decreases exponentially in $m/T$ with a slope that is larger for smaller values of $\Omega$.  This can be understood  as the rotating TFD state and its complexity  getting  increasingly close  to the direct product of two copies of the vacuum state as the mass gets larger.
%

\subsection{Time dependence}
In this subsection we analyze the evolution of circuit complexity.  We will consider how complexity varies as compared to its initial, $t=0$ value  
\be 
\d\mathcal{C}(t) \equiv \mathcal{C}(\ket{rTFD(t)})-\mathcal{C}(\ket{rTFD(0)}) \, . 
\ee
As described above,  $\mathcal{C}(\ket{rTFD(t)})$ is evaluated using the results of   \cite{Chapman:2018hou}  summarized in   \eqref{eq:kappa2}-\eqref{eq:alphak} and \eqref{eq:c1},\eqref{eq:thetak} and the effective description for the rotating TFD state outlined at the beginning of this section. At the practical level this boils down to evaluate these expressions by plugging in for each mode the effective inverse temperature and time of eq.~\eqref{identifications}.  As it was the case for the complexity of formation, also $\d\mathcal{C}(t)$ is a UV-finite quantity, but we employ a cutoff $N/2$ in the sum over  momenta as in  \eqref{cformationLAMBDA}  in order to numerically evaluate $\d\mathcal{C}(t)$ and produce the plots. 
We shall notice that away from $t=0$ $\mathcal{C}(\ket{rTFD(t)})$  depends on the reference state scale $\mu$. We will set it to one for the rest of the section, and produce additional plots to illustrate the $\mu$ dependence in appendix~\ref{app:plots}.

In figures \ref{fig:Ftimeconformal}, we analyze the time dependence in the near conformal limit, at fixed angular velocity and as the temperature increases. 
For both cost functions, we observe an oscillatory behavior, with the amplitude of the oscillations increasing with the temperature. At low temperature, the zero-mode dominates the sum giving $\d\mathcal{C}(t)$. As apparent in the left panels of  fig.~\ref{fig:Ftimeconformal}  $\d\mathcal{C}(t)$ is indistinguishable from the zero-mode alone. 
As the temperature increases, the contribution of the various modes becomes relevant. The zero-mode still sets the overall shape for the time dependence, but the superposition of the other modes yields the oscillatory behavior reported in  the central and right panels in fig.~\ref{fig:Ftimeconformal}. The periodicity of the oscillation can be understood form  \eqref{eq:kappa2}-\eqref{eq:alphak} and \eqref{eq:c1},\eqref{eq:thetak}, combined with the effective mode-by-mode redefinition of time according to  \eqref{identifications}. For illustration, let us focus on the simple massless case, disregarding here the zero mode divergence.   When $\O=0$, oscillations are governed by half the circle length: the argument of the trigonometric functions governing the time evolution is of the form $2 \pi |k| t /L$, and the absolute values appearing in the cost functions effectively halve the periodicity. For non-vanishing values of $\O$, the positive and negative $k$ modes have periodicity respectively:
\be
t \sim t + \frac{L}{2 |k|}\frac{1}{1\pm\O}.  \label{eq:periodicity}
\ee
As already observed for the complexity of formation, next to having different periodicity, positive and negative modes contribute with different amplitudes. As $T$ increases, negative modes with a given $k$ have larger amplitudes as compared to the corresponding positive $k$ modes. The net effect  is that for the cases reported in fig.~\ref{fig:Ftimeconformal}, the oscillatory behavior that can be resolved by the eyes corresponds to the contribution of the lower negative $k$ modes (see also appendix  \ref{app:plots}). Indeed, one can check that the periodicity of the larger spikes showing in the central and right panels of fig.~\ref{fig:Ftimeconformal} matches (within the zero mass approximation) the one for negative modes in \eqref{eq:periodicity}. Of course also the value of the angular velocity affects the amplitude of the oscillations. This is illustrated in fig.~\ref{fig:FtimeconformalOmega}, which is to be compared with the central panels of fig.~\ref{fig:Ftimeconformal}, and which shows the amplitude increases as we raise the value of $\Omega$.

\begin{figure}[ht]
  \centering
  \includegraphics[width=0.32\textwidth]{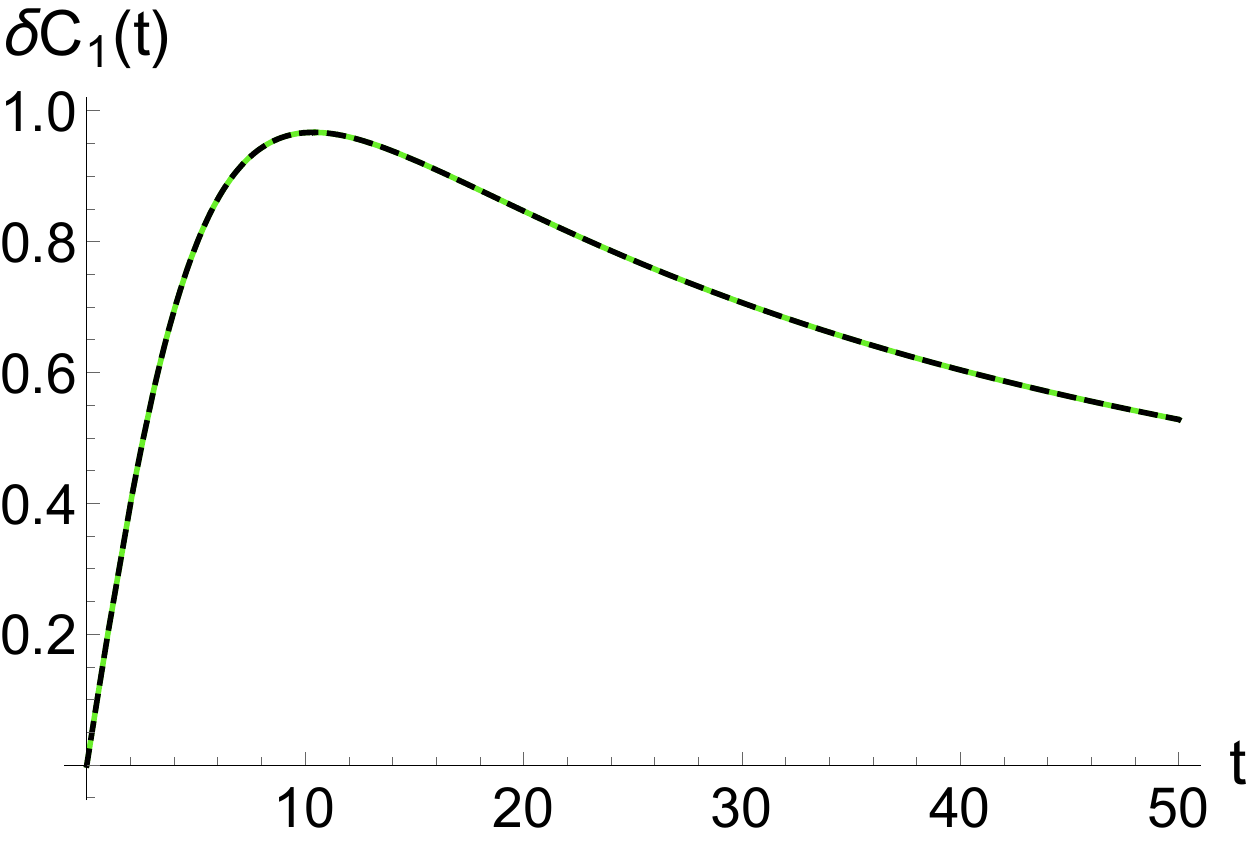}\hfill
  \includegraphics[width=0.32\textwidth]{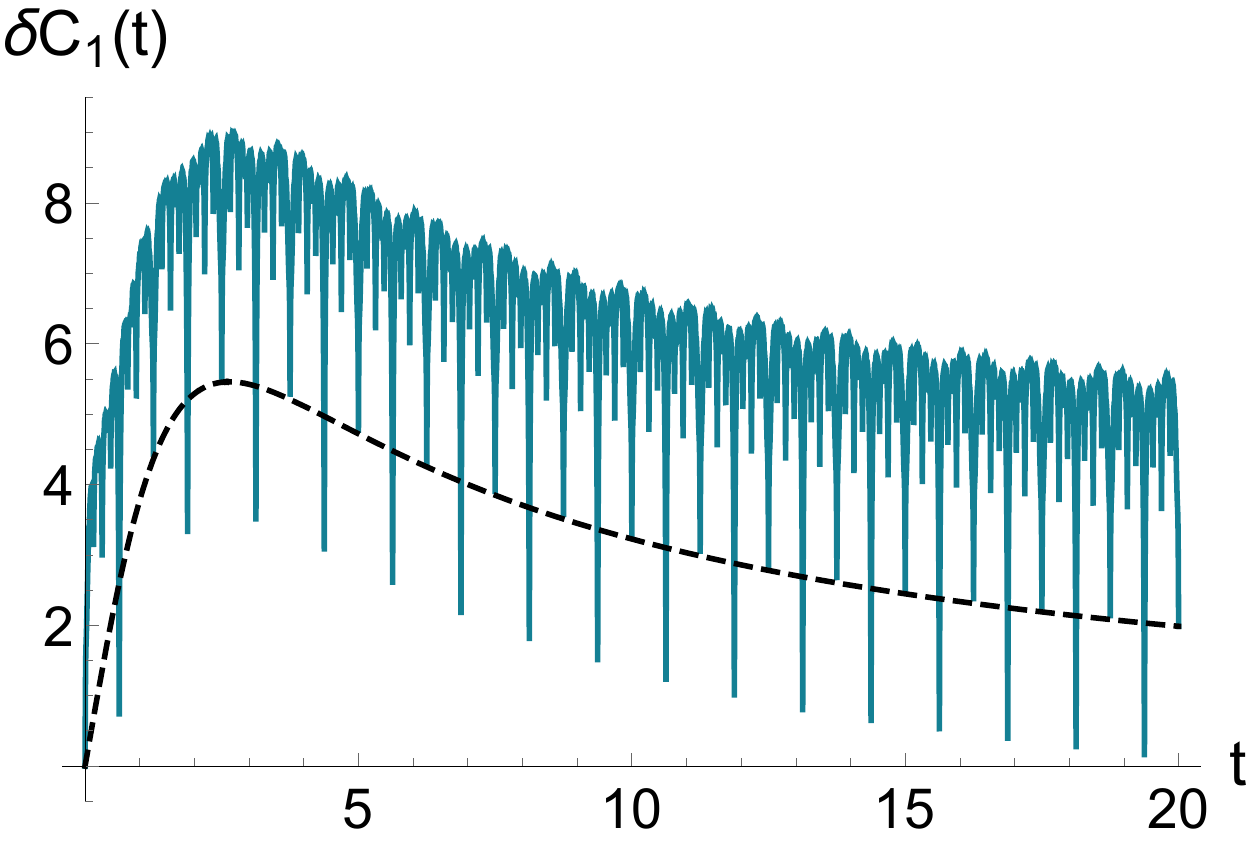}\hfill
  \includegraphics[width=0.32\textwidth]{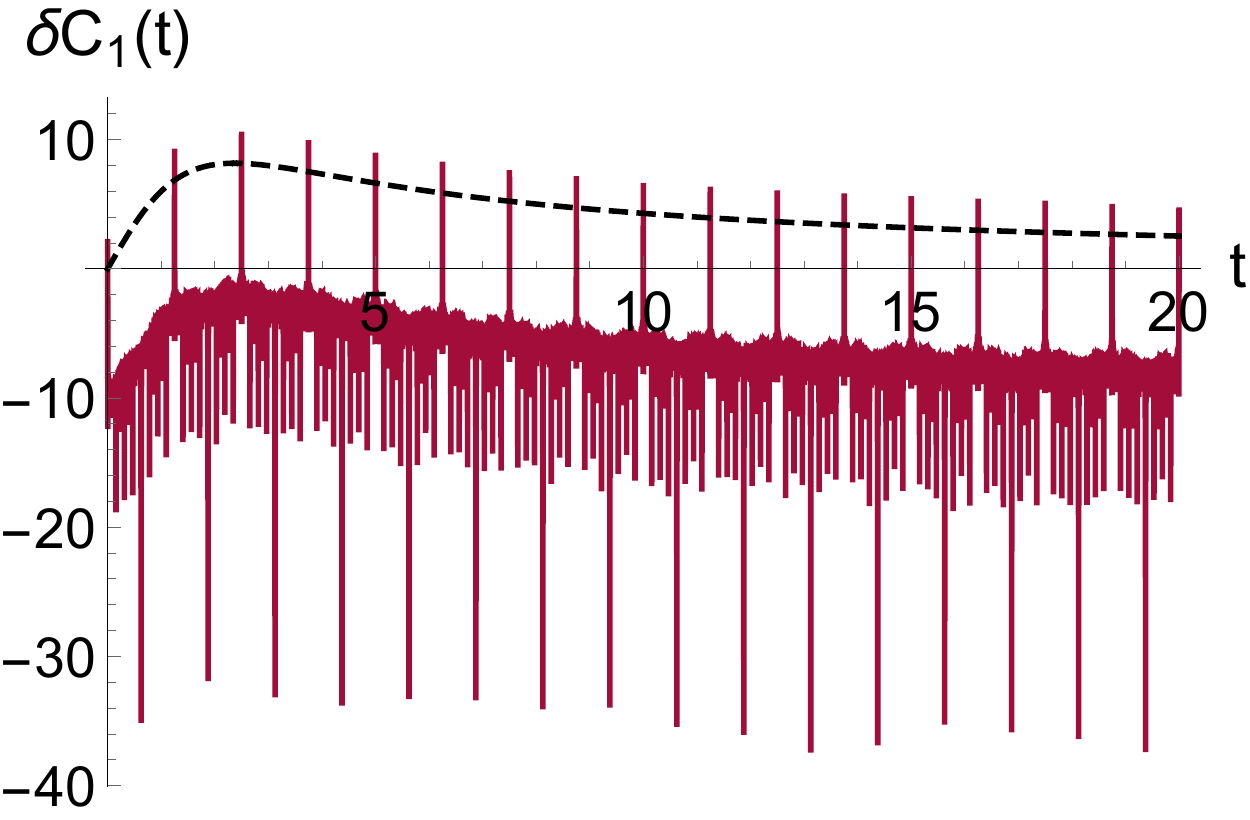} \\
     \includegraphics[width=0.32\textwidth]{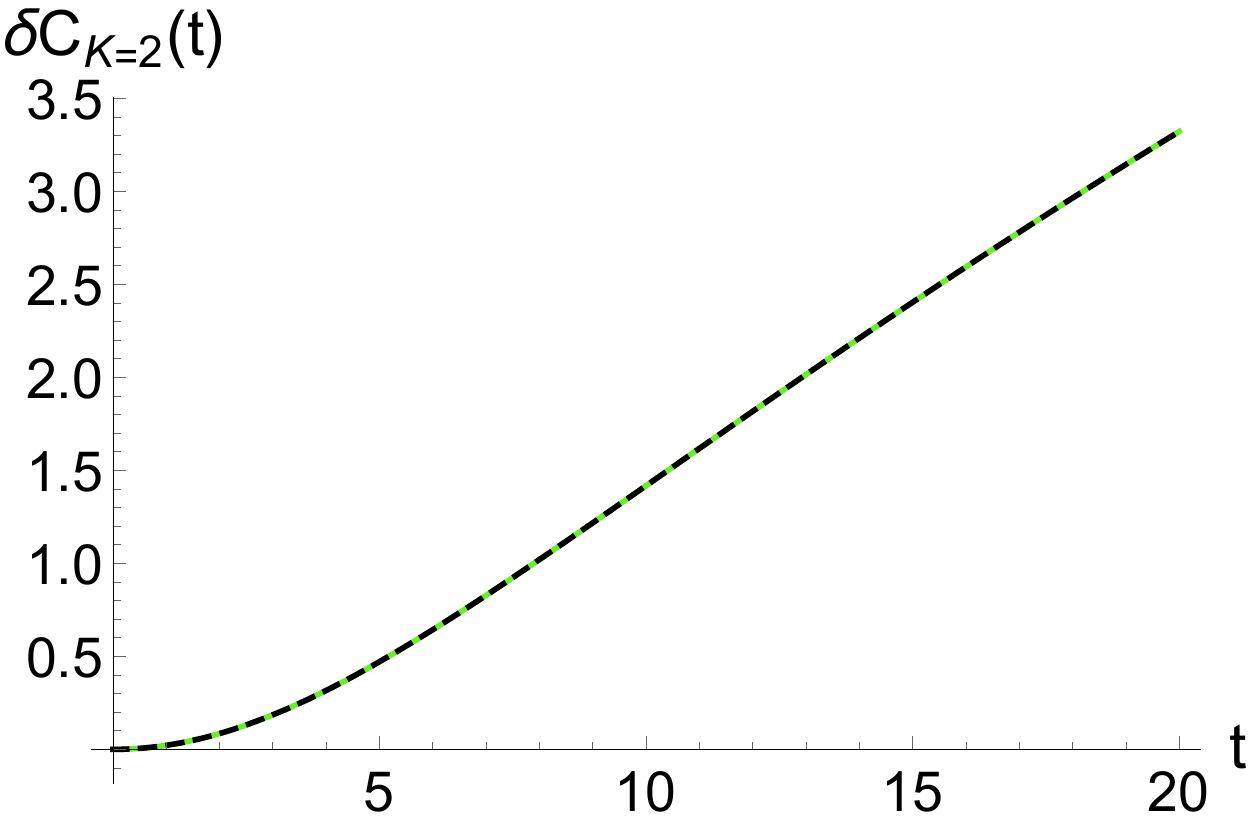}\hfill
  \includegraphics[width=0.32\textwidth]{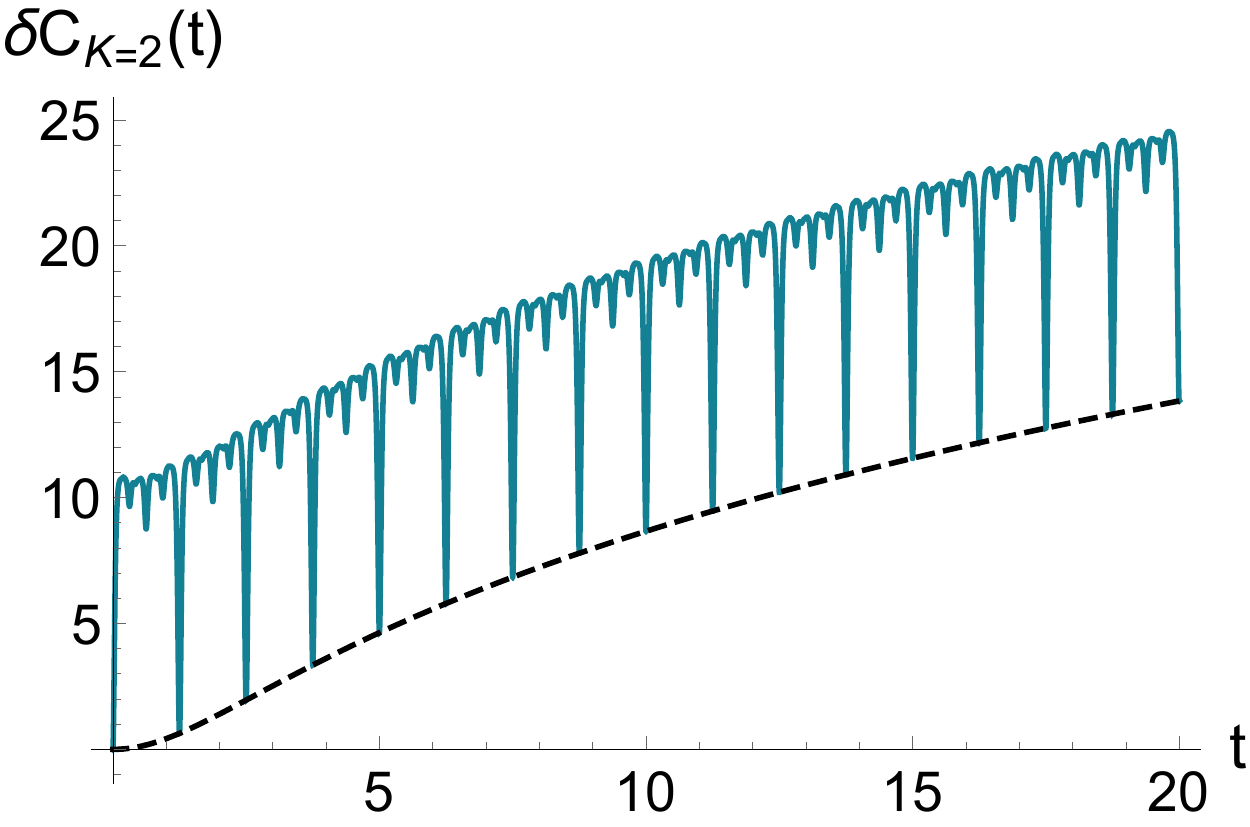}\hfill
  \includegraphics[width=0.32\textwidth]{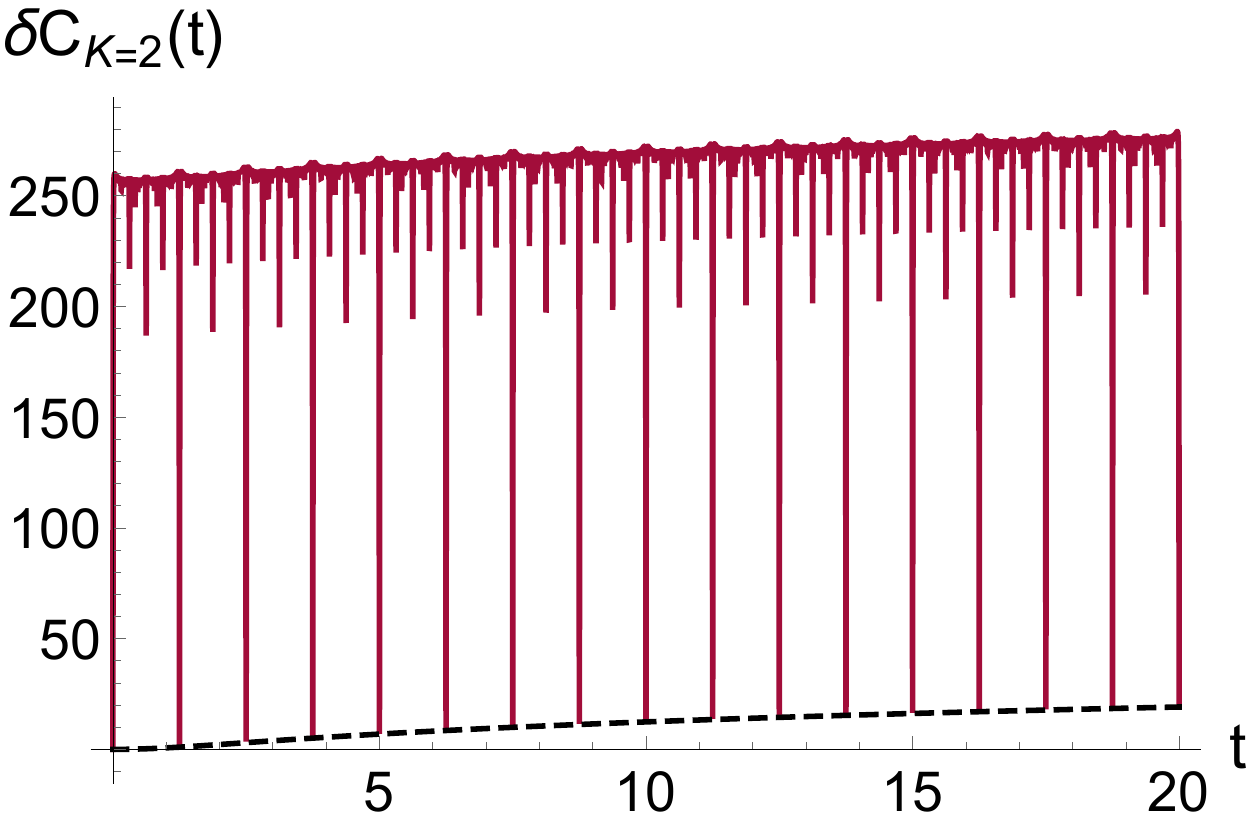}
\caption{Evolution of the complexity of the rotating TFD with the initial $t=0$ value subtracted, $\d\mathcal{C}(t) \equiv \mathcal{C}(\ket{rTFD(t)})-\mathcal{C}(\ket{rTFD(0)}) $ for different temperatures. 
Here $L=1$,   $N=1200$, $\O=0.6$ and $m=10^{-6}$. The temperature is increased from left to right:  $T=0.1, 10,100$.  The dashed black curve in each panel  is the zero-mode contribution, which dominates the low-temperature regime. }
\label{fig:Ftimeconformal}
\end{figure}
\begin{figure}[h]
  \centering
  \includegraphics[width=0.45\textwidth]{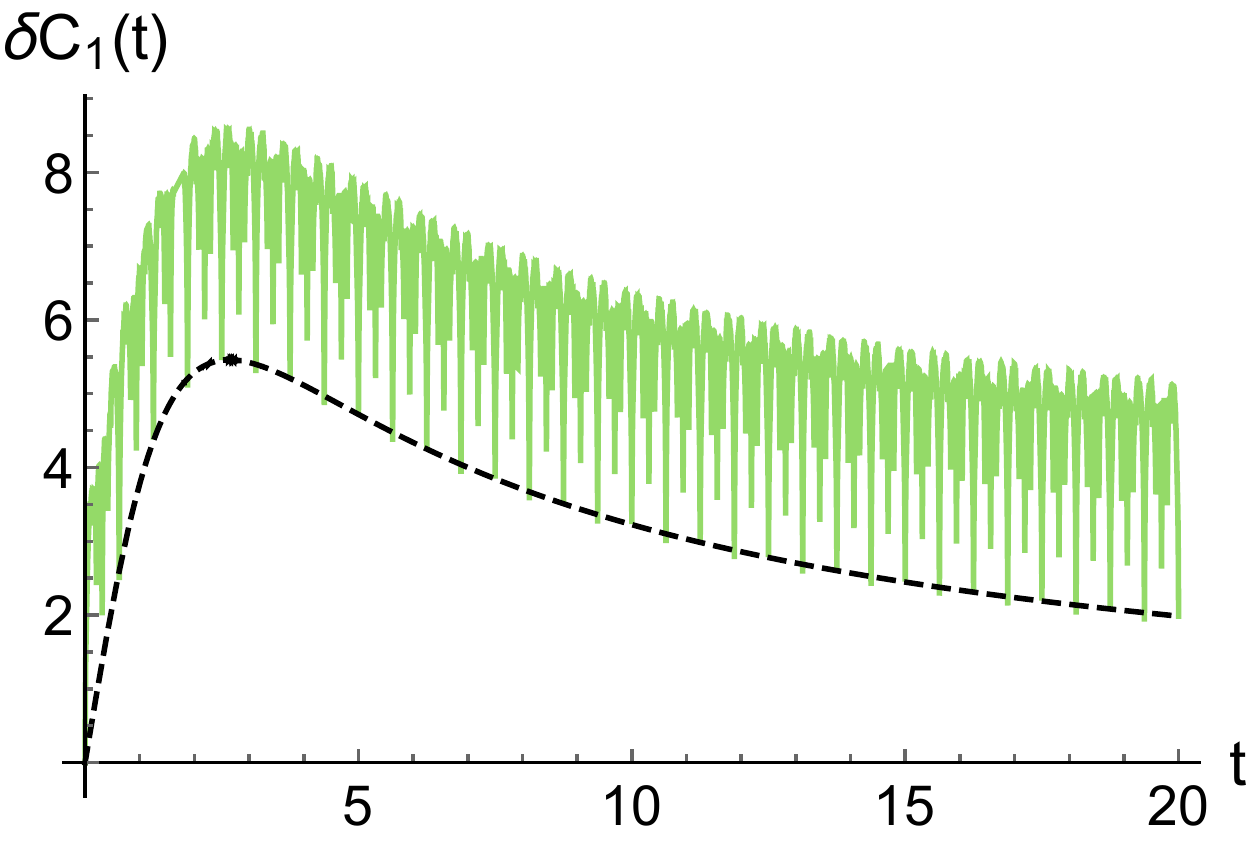}\hfill
  \includegraphics[width=0.45\textwidth]{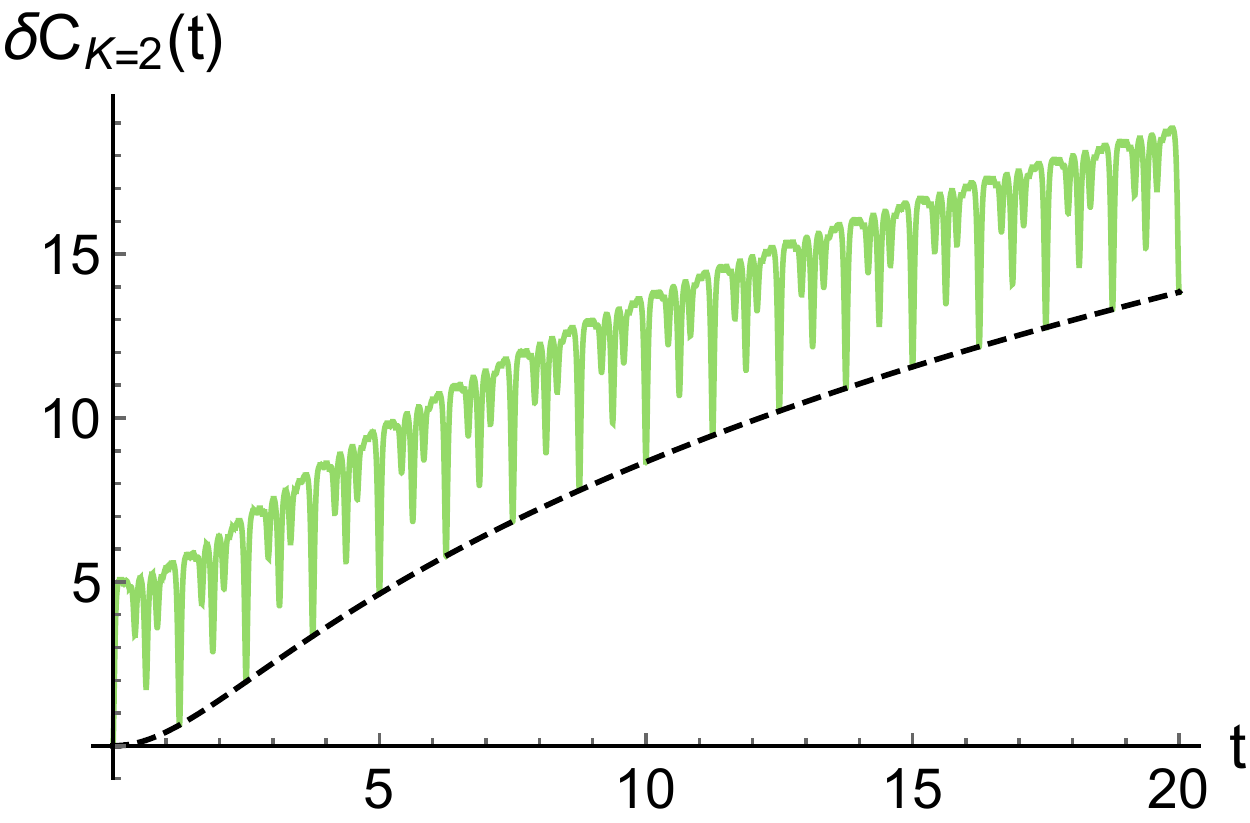}
\caption{$\d\mathcal{C}(t)$ for the same variables as in the central panels of fig.~\ref{fig:Ftimeconformal} except a for lower value of the potential $\Omega =0.2$. This reflects in a shorter periodicity and lower amplitudes.}
\label{fig:FtimeconformalOmega}
\end{figure}

There are some marked differences between the behavior of the  $F_{\kappa=2}$ and the  $F_1$ as the temperature is increased.   While the contribution of the different  modes is  always non-negative  for the $F_{\kappa=2}$, and thus the oscillatory behavior is always bounded form below by the value of the zero-mode,  for the $F_1$ these contributions can also be negative. In particular, single mode contributions with small enough $|k|$ can take negative values and the number of such modes increases with the temperature. This together with the dominance of the (positive) zero mode contribution at lower temperature yields a picture where the $F_1$ can be negative as the temperature is increased. We shall remark that this is   not in contradiction with the form of the $F_1$, as here we are looking at variations of complexity with respect to its initial value. Therefore this only indicates that  for large enough $T$ complexity can decrease  as time passes. Indeed  a similar observation was  made in the non-rotating case  \cite{Chapman:2018hou}.

\begin{figure}[h]
    \centering
  \includegraphics[width=0.32\textwidth]{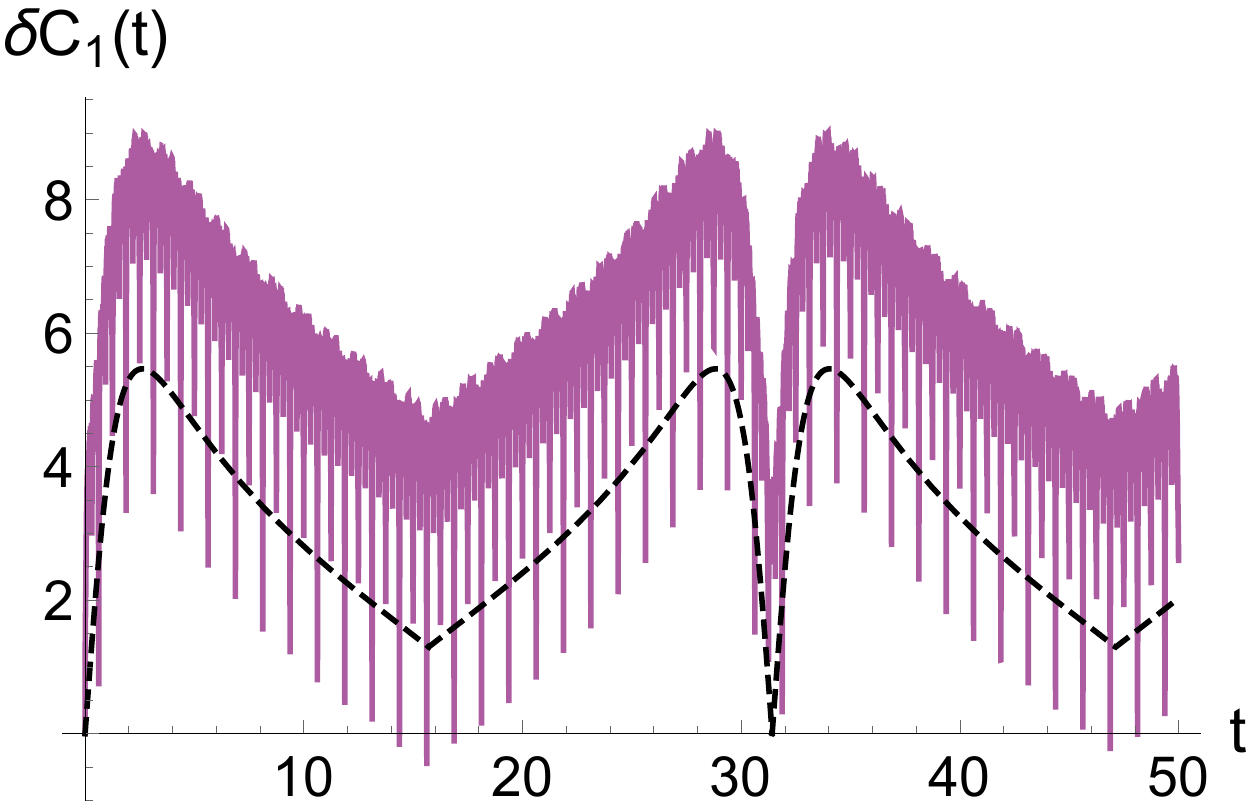} 	\hfill
  \includegraphics[width=0.32\textwidth]{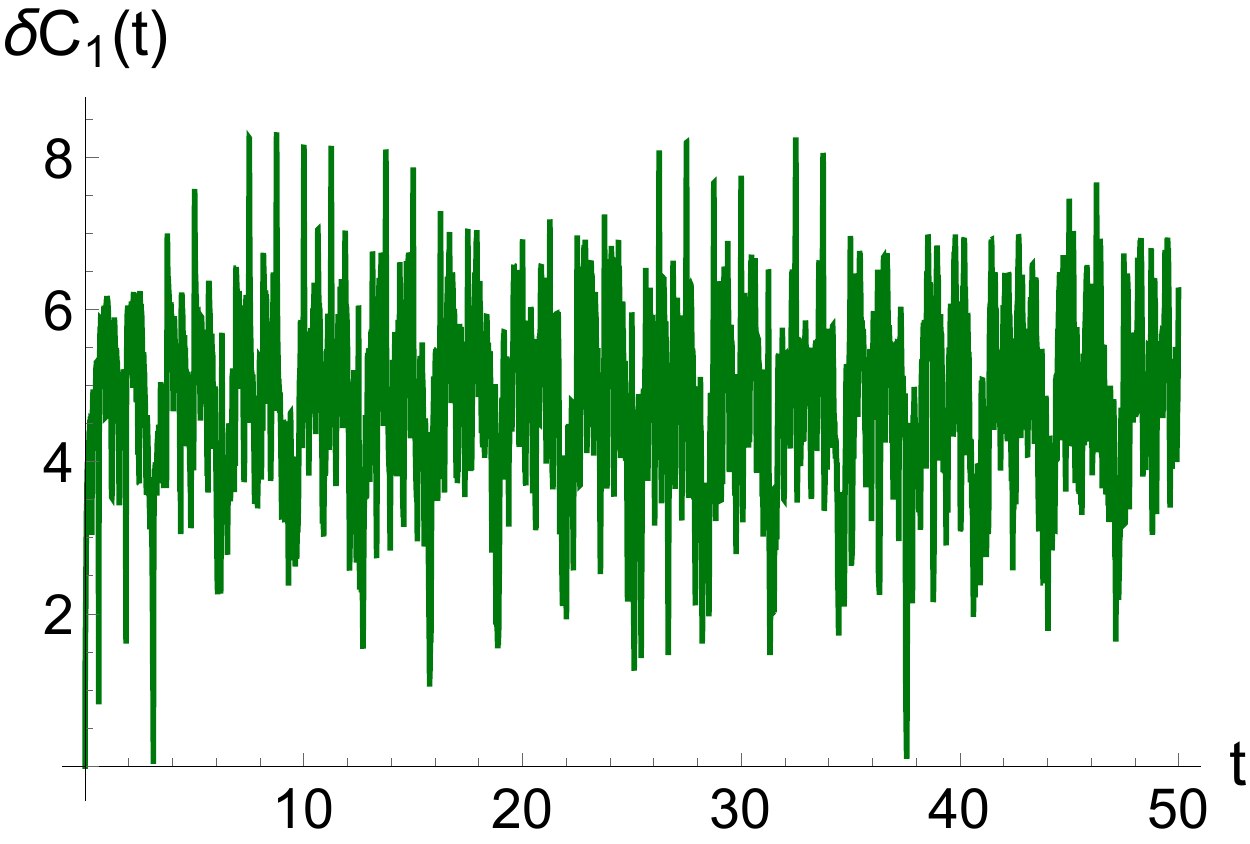}\hfill
  \includegraphics[width=0.32\textwidth]{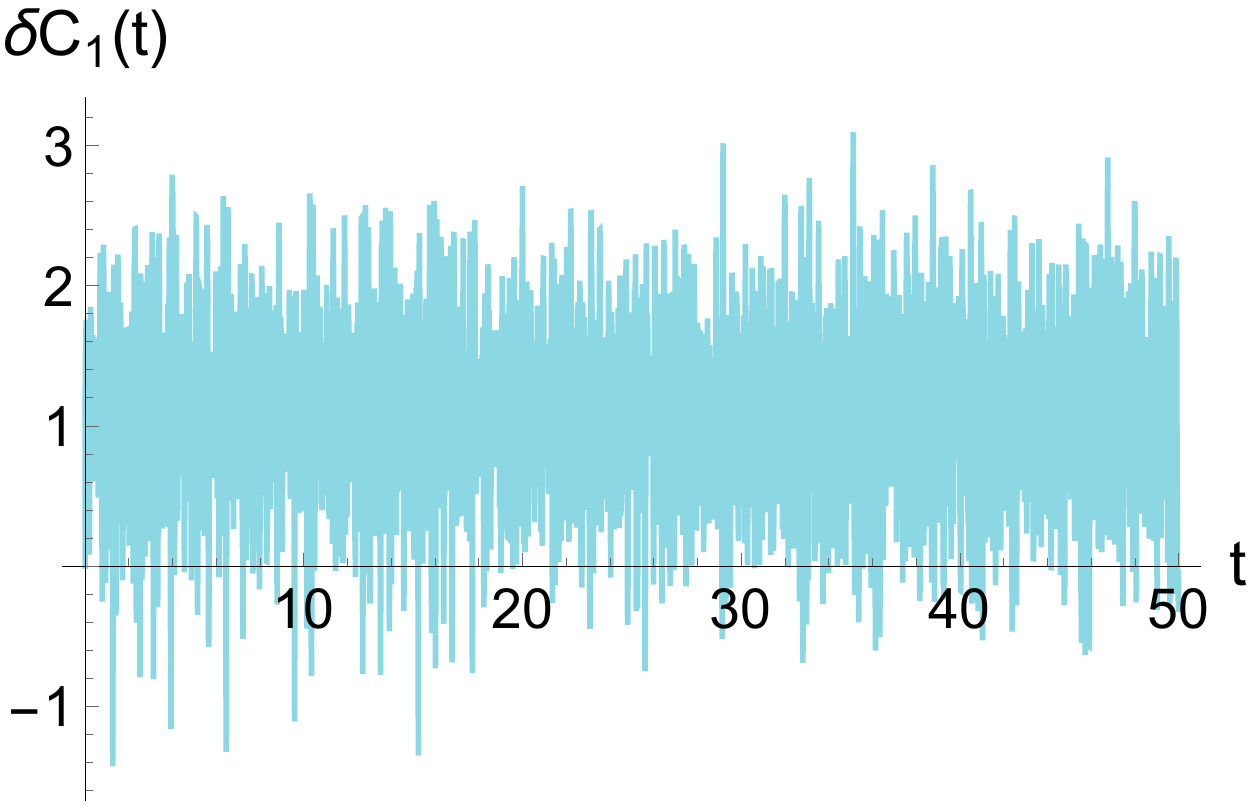}\\
  \includegraphics[width=0.32\textwidth]{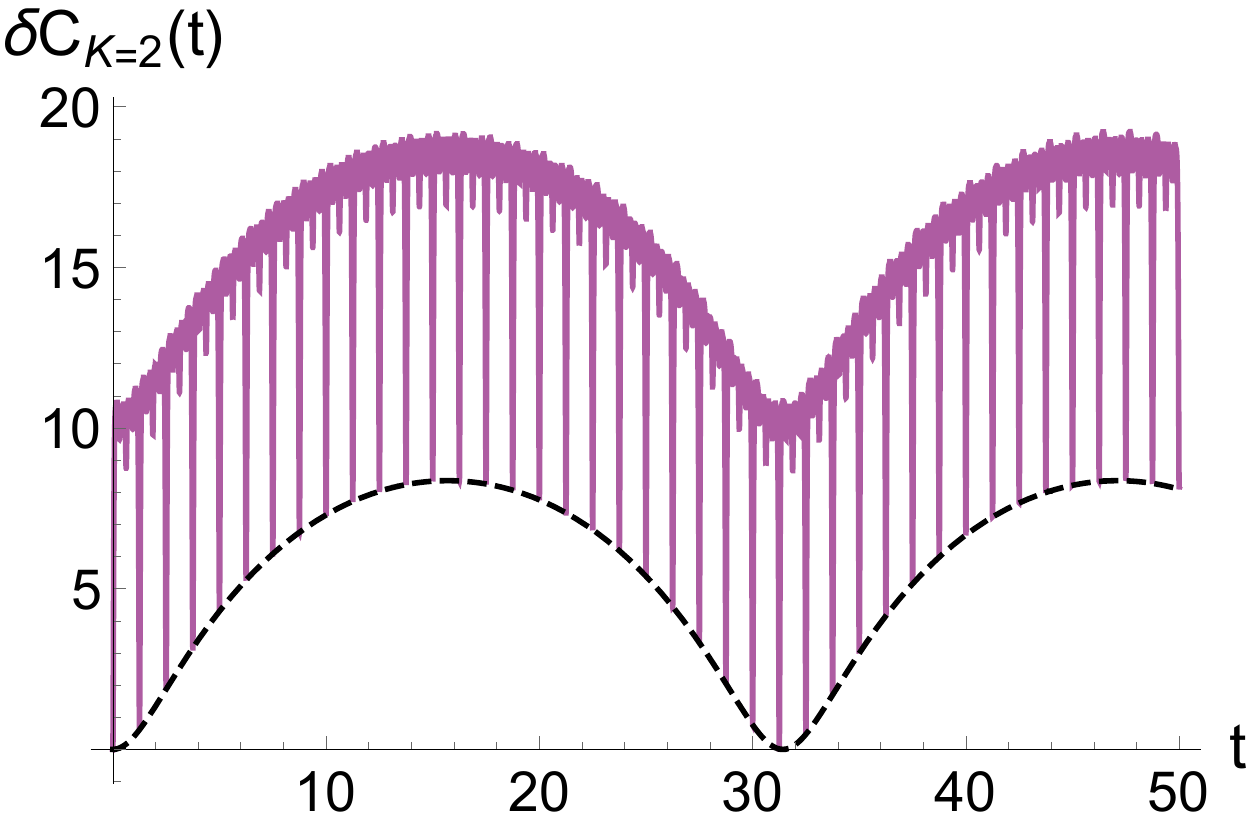} 	\hfill
  \includegraphics[width=0.32\textwidth]{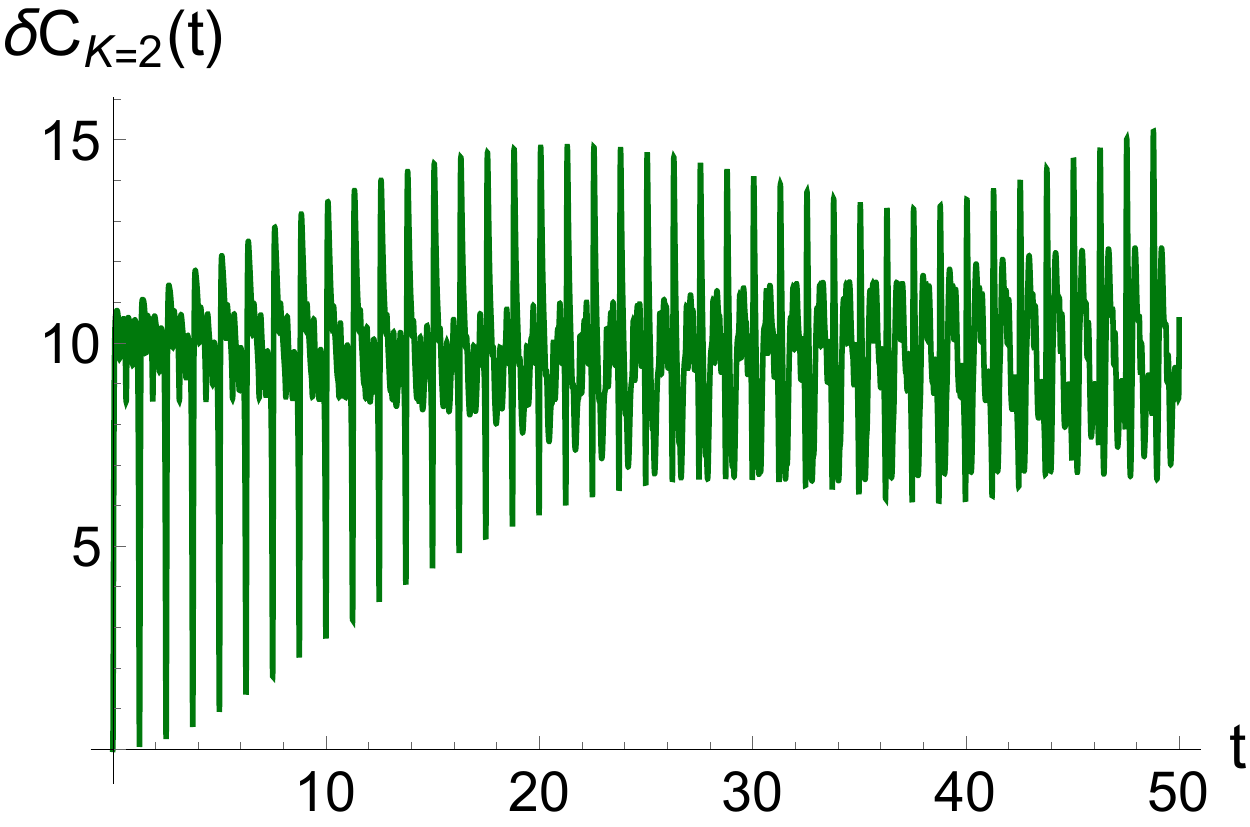}\hfill
  \includegraphics[width=0.32\textwidth]{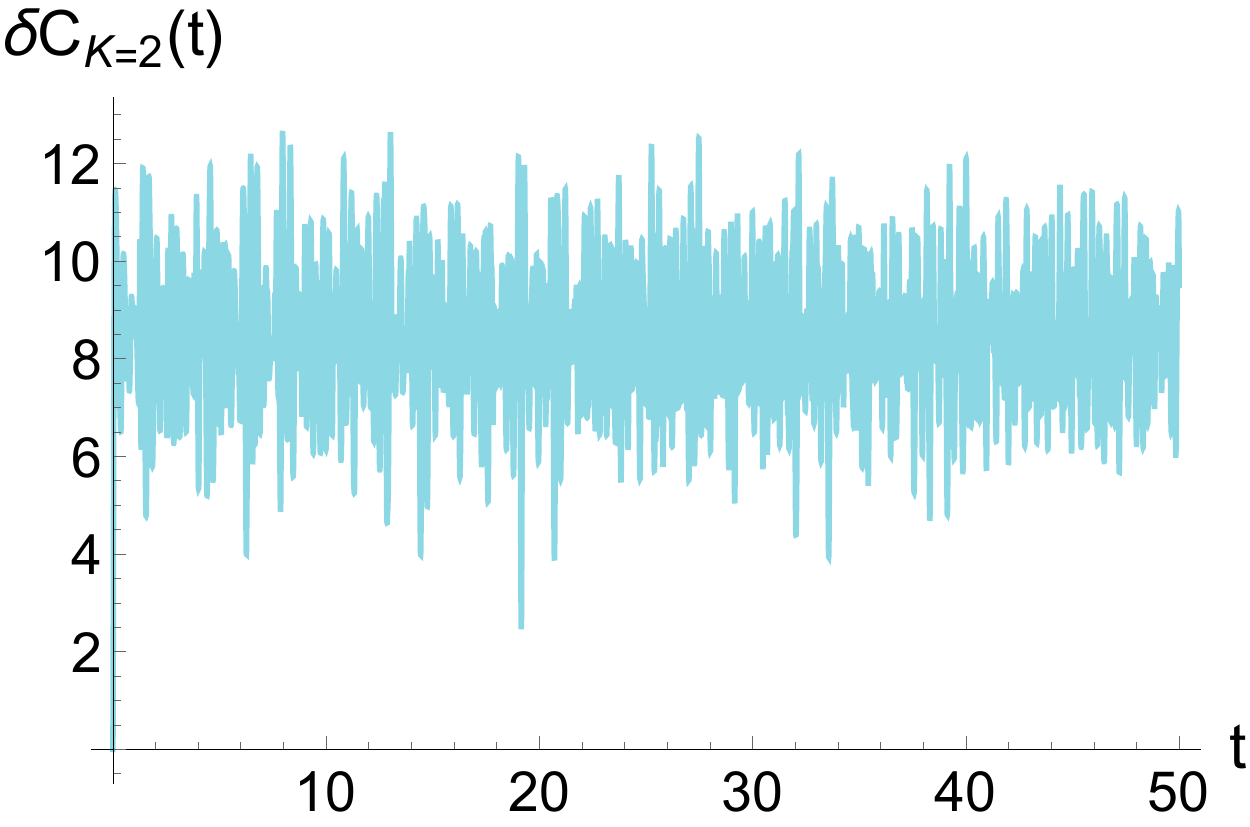}
\caption{ Evolution of  $\d\mathcal{C}(t)$ as the mass parameter is increased. From left to right $m=0.1,1,10$ both in the upper and lower panels. Here $L=1$, $N=1200$,  $\O=0.6$ and $T=10$,  as in the central panels of fig.~\ref{fig:Ftimeconformal} (where $m=10^{-6}$).  At lower values of the mass,  the zero-mode contribution (dashed  black line)  is still dominant and sets an overall oscillating behavior on top of which one can observer a pattern similar to the one in  fig.~ \ref{fig:Ftimeconformal}. As the mass increases this regular oscillatory pattern is destroyed. }
\label{fig:Ftimemass}
\end{figure}

So far we discussed the behavior of complexity in the small mass limit. A finite mass  mitigates the positive versus negative  modes amplitude suppression effect, which is practically absent for large enough values of $m$.  Similarly, the difference in periodicity which in the massless limit takes the form \eqref{eq:periodicity} gets attenuated by a finite mass and removed in the large mass limit.   This can be seen in fig.~\ref{fig:Ftimemass}. We see that for smaller values of the mass parameter the oscillatory pattern of the lower $|k|$ modes can still be resolved. Also, the zero-mode, which still gives the dominant contribution for small enough  values of $m$,  provides the enveloping oscillatory behavior with period $\pi /m$.  As the mass parameter is increased,  the zero and lower $|k|$ modes get comparable amplitudes and close-by periodicities making it hard to identify regular patterns. 

\begin{figure}[h]
\centering
  \includegraphics[width=0.45\linewidth]{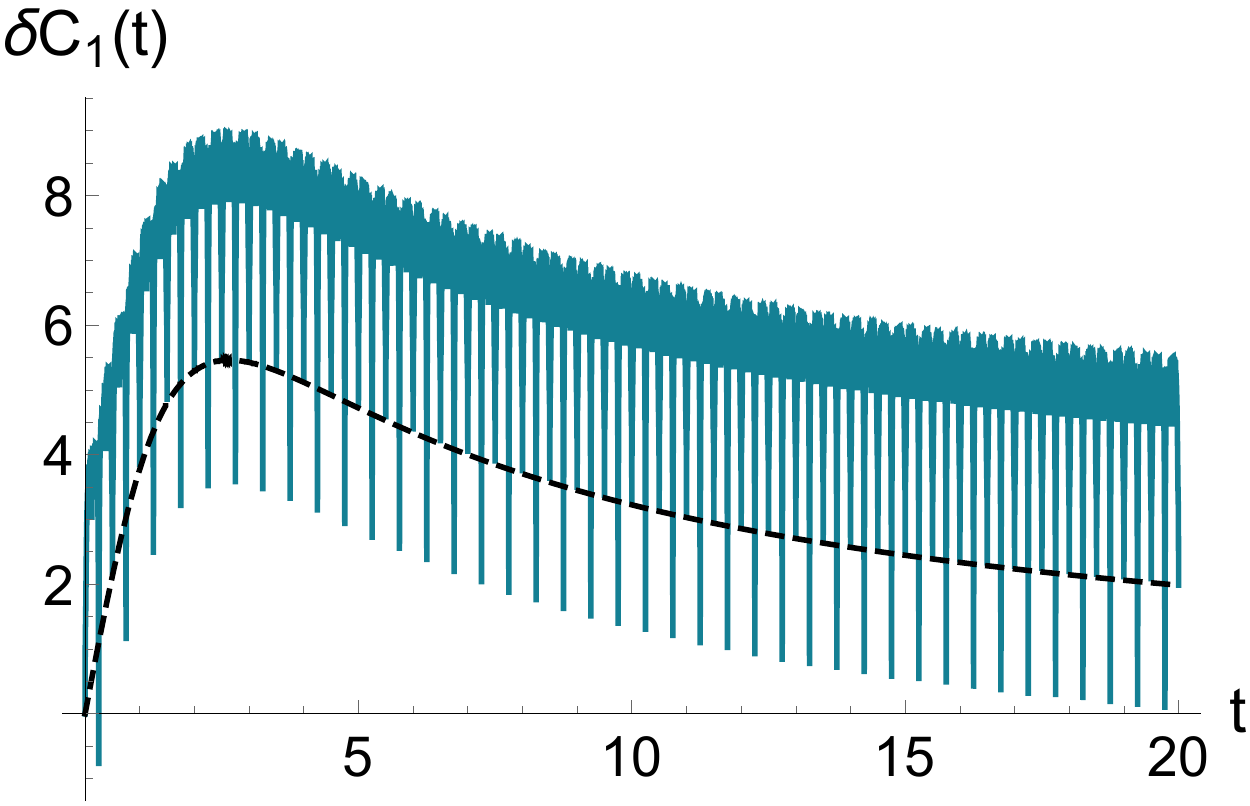} \hfill
  \includegraphics[width=0.45\linewidth]{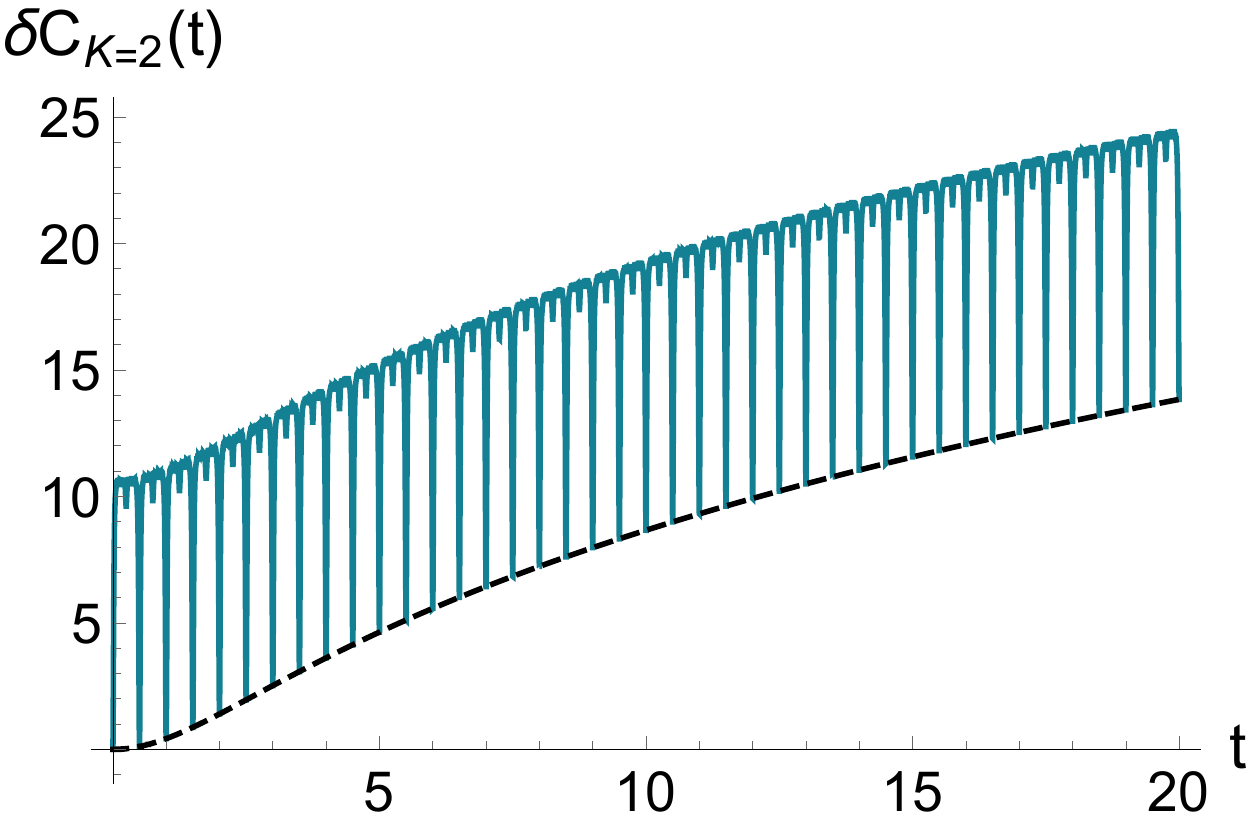} 
\caption{$\delta {\mathcal C(t)}$ evolution with undeformed Hamiltonian.  $F_1$ (left) and  $F_{\kappa=2}$  (right)  for $L=1$, $N=1200$, $\O=0.6$ and $m=10^{-6}$, $T=10$. The dashed black curve represents the zero-mode contribution. }
\label{fig:FtimeconformalHonly}
\end{figure}

\paragraph{Alternative time evolution.} Another natural way of evolving the TFD state is with the undeformed  Hamiltonian, as in \eqref{eq:RTFDund}. The result is reported in fig.~\eqref{fig:FtimeconformalHonly}. 
The main difference with respect to the previous case lies in the fact that in the mode by mode mapping of the rotating TFD  into the neutral one only involves the definition of a $k$-dependent effective temperature, but not of time.   A direct consequence is that the periodicity of oscillations is now set by $L /2$, as for $\Omega=0$. Negative and positive modes with the same $|k|$ now oscillate with the same frequency, and the potential $\Omega$ only affects the amplitudes. Also the role of the zero-mode remains completely unchanged as compared to the evolution with the deformed Hamiltonian considered above. 
%


\section{Discussion}\label{sec:Discussion}

In this work we studied various aspects of holographic complexity for states with rotation dual to AdS black holes, and extended the QFT complexity analysis to a rotating thermofield double state of a 2d free boson.\\ 
 
For the case of rotating BTZ black holes, we carried out a thorough study, refining existing results and analysing in detail the role of the counterterm action in CA \cite{Lehner:2016vdi} and the full time dependence of the holographic complexity proposals. 

The  effects of  rotation are in many respects analogous to those of charge, observed for (higher dimensional) Reissner-Nordstr\"om AdS black holes in \cite{Carmi:2017jqz}.  Both with the CA and CV prescriptions (and with CV 2.0, which substantially parallels CV) the complexification rate vanishes at $t_b =0$ and reaches asymptotically the expected late time limit \cite{Brown:2015lvg,Cai:2016xho}
\begin{equation}
\lim_{t_b \to \infty} \frac{dC }{dt_b} \sim \left(M-\Omega_H J\right) \,.
\end{equation} 
This vanishes for extremal black holes (see \eqref{lateTimeRate}).
While with CV (and CV 2.0) the growth rate is always positive and increases monotonically between these two values,  fig.~\ref{figc1}, for CA the intermediate evolution is richer, fig.~\ref{gr1}.
At fixed $\ell M$, for small values of the angular momentum $J$, ${dC_A }/{dt_b}$ develops a negative peak at early times, as observed for charged black holes \cite{Carmi:2017jqz}. The negative  peak is  followed by a phase where ${dC_A }/{dt_b}$ becomes positive and overshoots the  late time value, which is then approached from above.  Therefore the  CV and CA growth rate reach the late time value from opposite directions, in agreement with what observed for two-sided black holes, including charged ones  \cite{Carmi:2017jqz}. 

This general behavior is largely independent from the counterterm action. Indeed, for finite rotation parameter $J$ (at fixed $\ell M$), the  qualitative behavior of ${dC_A }/{dt_b}$ is the same with or without counterterm (see fig.~\ref{gr1}). The choice of the counterterm scale  $L_{\rm ct}$ gives however quantitive differences.  As illustrated in   fig.~\ref{gr1ct}, the counterterm acts as an effective reduction (enhancement) of the angular momentum for small (large) values of  $L_{\rm ct}/\ell$. In general we also observe that the role of the counterterm becomes less and less pronounced as one approaches the extremal limit, $J \to M \ell$. 
In the opposite $J \to 0$ limit, where the effect of the counterterm is most pronounced, we smoothly recover the neutral case results of \cite{Carmi:2017jqz}.  In particular the counterterm is essential to obtain the early time negative divergence in the irrotational limit.

The time dependence of the complexity variation $ \Delta C (t_b) \equiv C(t_b) - 2 C^{\rm AdS}$ directly follows from  $dC / dt_b$, except for its initial value, the complexity of formation.
 Comparing  fig.~\ref{gr3} and \ref{figuraDivergenza2}, we see that while the CA variation can assume negative values,  for CV  it is  always non-negative. In both cases, $ \Delta C (t_b)$ diverges in the extremal limit $J \to M \ell$.
The complexity of formation  $\Delta C_{\rm A}(t_b=0)$  depends  on the counterterm. This is best appreciated looking at the dependence  from the temperature $\ell \, T$, while working at fixed potential $\ell \, \Omega_H$ (fig.~\ref{fig:GCcfCA}). From fig.~\ref{fig:GCcfCALct}  it is clear that tuning the value of the counterterm scale $L_{\rm ct}$ one can change the value (and the sign, for large enough $\ell \, T$) of the complexity of formation. Another key observation one can draw from fig.~\ref{fig:GCcfCALct} is that the inclusion of the counterterm is essential to obtain for CA the same linear in $T$ behavior as for CV (see fig.~\ref{fig:GCcfCV}). Without,  the CA complexity of formation  approaches linearity in $T$ only at high enough temperature.

We shall notice that the linear in $T$ behavior of the complexity of formation  of  the rotating BTZ black hole is in tension with the hypothesis of a \emph{third law of complexity} advanced in \cite{Carmi:2017jqz}. In fact, as we take $\ell \, T\to 0$ (keeping $\ell \,\O_H$ fixed), we do not obtain a divergent complexity of formation, but a finite result. In particular, for both CA and CV, the zero temperature limit reduces to the corresponding finite complexity of formation of neutral BTZ \cite{Chapman:2016hwi}.\footnote{We do instead have a logarithmically divergent behavior in the extremal limit $J \to M\ell$, or equivalently $r_+ \to r_-$. This however implies the simultaneous limit $\ell \, T\to 0 $ and $\Omega_H \ell \to 1$.}
This contrasts with what observed  for spherical Reissner-Nordstr\"om AdS$_4$ black holes,  where $\Delta C_A (t_b =0 )$ diverges logarithmically as $\ell \, T \to 0 $ at fixed chemical potential $\mu$ \cite{Carmi:2017jqz}.  A similar logarithmic divergence was also found for higher dimensional rotating black holes. In particular for odd-dimensional Myers-Perry AdS black holes with equal angular momenta, \cite{Balushi:2020wkt} found that for fixed $\ell \, \Omega_H$ and $r_+ \gg \ell$, both the CA and CV complexity of formation exhibit a logarithmic  divergence as $\ell \, T \to 0$. Notice that this limit cannot be studied for Kerr-AdS$_4$. As we discuss in sec.~\ref{sec:CVKerr}, within the physical space of solutions satisfying $\ell \, \Omega_+ < 1$, there exists a minimum positive temperature value that can be attained, and it is therefore not possible to study  the $\ell \, T \to 0$ limit.

The limit  $\ell  \, \Omega_H \to 1$ at fixed $\ell \, T$ is an interesting one to study, as it corresponds to the critical angular velocity limit, in which the Einstein universe conformal to the AdS boundary rotates at the speed of light \cite{Hawking:1998kw}.  In this limit for BTZ, both the complexity growth rate and the complexity of formation diverge. The late time limit of the complexity growth rate diverges as
\begin{equation}
\lim_{t_b \to \infty} \frac{dC }{dt_b} \sim  \frac{\(\ell \, T\)^2}{1-\ell \, \Omega_H}
\end{equation} 
both for the CA and CV proposals, independently  from the counterterm. The complexity of formation  diverges as well. In particular the leading divergence has the  schematic form
\be  \label{eq:Omegadivholo}
\Delta C (t_b=0) \sim \frac{\ell \, T}{1- \ell \,\O_H} \log \frac{1}{1-\ell\, \O_H} \,  
\ee
for both CV and CA prescriptions (though with different prefactors, see \eqref{eq:OmegaoneCA} and \eqref{eq:OmegaoneCV}). It is also interesting to notice that the CA divergence is always positive, independently from the choice of  $L_{\rm ct}$. Without the inclusion of the counterterm, one would instead get a divergence with the same structure but opposite sign, as made explicit in eq.~\eqref{eq:OmegaoneNOCT} and fig.~\ref{fig:GCcfCALct}. \\

We were able to analyze how part of these results carry over to the Kerr-AdS$_4$ black hole and how they compare to some of the findings obtained in \cite{Balushi:2020wjt,Balushi:2020wkt} for higher odd-dimensional Myers-Perry AdS black holes with equal angular momenta.  The reduced symmetry of Kerr-AdS, as compared to these setups, makes evaluating holographic complexity for all times computationally challenging. Using the results of \cite{Balushi:2019pvr} for  null hypersurfaces foliation of Kerr-AdS, we were able to work out  the late time limit of the CA growth rate. Carefully taking into account all terms for the action defined on the WDW patch, we explicitly showed that this is given by the difference in internal energies between the inner and outer horizons, as first advanced in \cite{Cai:2016xho} 
\begin{equation}
\lim_{t_b \to \infty} \frac{dC }{dt_b} \sim \left(M-\Omega_+ J\right) - \left(M-\Omega_- J\right) \,. 
\end{equation} 
While for the BTZ case there was no obstruction to taking the extremal black hole limit, here the limit sits outside the allowed region of parameters, as depicted in fig.~\ref{fig:KerrParamters}.  Moving in the physical region one first gets to another interesting limiting value,  $\ell  \, \Omega_+ = 1$, which separates the region of solutions with sub-luminal rotation from the super-luminal ones.  As explained in the text, in terms of the parameters defining the geometry, this translates into either the critical  value $J/M \equiv a =\ell$, for black holes with $r_+ \geq \ell$, or into $r^2_+ = a \,\ell$, for black holes with $r_+ < \ell$.
We find that at fixed mass, the complexity growth rate only diverges in the limit $a \to \ell$, where the metric exhibits a parametric divergence. On the contrary, the late time complexification rate remains finite as $r^2_+ \to a \, \ell$ (see discussion around eq.~\ref{eq:lateCAK}). 
A similar behavior was observed for the complexity of formation that we were able to evaluate with the CV proposal. Again, at fixed mass one finds a divergent complexity of formation only for black holes with $r_+ \geq \ell$. That is, the result diverges only when the critical angular velocity limit $\Omega_+ \ell =1$ is reached via $a \to \ell$. This is shown in fig.~\ref{fig:DeltaCVKerrma} and fig.~\ref{fig:DeltaCVKerr}. In the grand canonical ensemble variables $T$ and $\Omega_+$, there are two branches of small and large black holes, and these exist for temperatures $T > T_{\rm min}$, with $\sqrt 3 /(2\pi) \geq \ell \, T_{\rm min} >1/(2\pi)$ for $0 \leq \ell\, \Omega_+ < 1$. The CV complexity of formation of large black holes has similarities with that of BTZ: it increases with the temperature and positively diverges in the critical limit $\ell\, \Omega_+ \to 1$. The main difference rests in the non linear dependence on the temperature. Small black holes have instead a distinct behavior: $\Delta C_V (t_b=0)$ goes to zero as $\ell \, T \to \infty$ and it is approximately constant in $\ell \, \Omega_+$. 

In  \cite{Balushi:2020wjt,Balushi:2020wkt}  a direct connection between CA, CV and thermodynamic volume
was highlighted for  odd-dimensional Myers-Perry AdS black holes with equal angular momenta in each orthogonal plane. We found that our Kerr-AdS$_4$ computation of the holographic complexity growth rate is indeed compatible with the claim of  \cite{Balushi:2020wjt,Balushi:2020wkt} that at   leading order for large black holes with $r_+ \gg \ell$
 \be
 \lim_{t_b\to \infty}   \frac{ d  C }{ dt_b } \propto   P\D V  \label{eq:manngrowth}
\ee
with $\D V  =  V_+ - V_-$  the difference between the inner and outer horizon thermodynamic volume. 
In the large black hole limit $r_+ \gg \ell$, the same authors also found that the complexity of formation is controlled by the thermodynamic volume rather than by the entropy, with a scaling that depends on the spacetime dimensionality $D$ (see \eqref{eq:CVMann}). Our result is compatible with this claim, but unfortunately we cannot test it independently. This is because in the large  black hole limit, within the region of parameters covered by the physical Kerr-AdS$_4$ solutions we considered,  the scaling of the entropy and thermodynamic volume is everywhere fixed, just as it happens for non-rotating and charged black holes.\\

In the last part of this work we employed Nielsen's approach to study QFT complexity in presence of rotation. We examined circuit complexity for thermofield double states of  2d free scalar fields on a circle with non-vanishing momentum along the compact spatial direction
\begin{equation} \label{eq:RTFDdiscuss}
\ket{rTFD}=\frac{1}{\sqrt{ Z\left(\beta,\Omega\right) }}\sum_n e^{-\beta\left(E_n+\Omega J_n\right)/2}e^{-i (E_n+\Omega J_n) t }\ket{E_n,J_n}_L \ket{E_n,J_n}_R \, .
\end{equation}
We adapted and extended the analysis of  the non-rotating TFD  state of   \cite{Chapman:2018hou}.  At the technical level, we  showed how factorization into momentum modes can be used to provide an effective description of \eqref{eq:RTFDdiscuss} in terms of  non-rotating TFD states. In particular one can draw a mode-by-mode correspondence between the rotating and non-rotating TFD by a mode dependent redefinition of temperature and time. This is similar to what happens for the charged TFD state analyzed in \cite{Chapman:2019clq}, with an important difference. In the charged case, for the effective description to make sense, one needs to impose on the mass parameter of the model   a lower bound, which is set by the chemical potential \cite{Chapman:2019clq}. An angular velocity $\Omega$ does not set such a bound instead, as long as $\Omega < 1$. 

Notice that given our analysis follows from the one in \cite{Chapman:2018hou}, the same caveats apply. In particular although we evaluated complexity for the $F_{\kappa =2}$ and the $F_{1}$ cost functions, the straight line circuit used in both cases was proven to be optimal only for the  $F_{\kappa =2}$. Thus  for the $F_{1}$  our result  only provides a upper bound on the value of circuit complexity. 

The presence of a potential affects differently positive and negative mode contributions to complexity. As compared to the $\Omega=0$ case, where positive and negative  $k$ pairs contribute equally, a non-vanishing $\Omega$  enhances the amplitude of negative $k$ modes and suppresses that of positive ones. The angular potential plays a similar role in the time evolution of the rotating TFD state \eqref{eq:RTFDdiscuss}.  It gives different periodicity to negative and positive modes, enlarging the period of the oscillations for the negatives modes as compared to positive ones (see eq.~\eqref{eq:periodicity}). Overall, the time dependence of the complexity variation $\d\mathcal{C}(t) \equiv \mathcal{C}(\ket{rTFD(t)})-\mathcal{C}(\ket{rTFD(0)})$ of both cost functions exhibits an oscillatory behavior  (see fig.~\ref{fig:Ftimeconformal} and \ref{fig:Ftimemass}). The zero-mode sets the overall shape, and the superposition of the other modes yields the oscillations. Their amplitude is amplified as we increase the temperature or the angular potential. While we cannot directly compare to the charged TFD state analyzed in the decompactified limit in \cite{Chapman:2019clq}, our results are fully compatible with the non-rotating TFD state on a circle studied in \cite{Chapman:2018hou} (see fig.~6 there). There the high temperature behavior, as shown in the rightmost panels of our fig.~\ref{fig:Ftimeconformal} at $\Omega \neq 0$, was interpreted as a saturation resulting from the presence of many modes non-trivially contributing to the circuit complexity sums at high temperatures. 

In studying the time dependence we also observed that from the point of view of the CFT  another natural way of evolving the TFD state is with the undeformed Hamiltonian, as in eq.~\eqref{eq:RTFDund}. This leads to a different oscillatory pattern of $\d\mathcal{C}(t)$, the main difference being that the potential $\Omega$ now only affects the amplitudes, but not the periodicity (see fig~\ref{fig:FtimeconformalHonly}). It would be interesting to explore how different time evolutions are  implemented in the dual out-of-equilibrium black hole description, and how such choices affect the holographic complexity evolution. 
 
Despite the clear differences between a free QFT and a strongly coupled chaotic CFT, the rotating TFD  model can be taken as a toy-model for a qualitative comparison with the holographic results obtained in the BTZ analysis. In fact, we can identify similarities between the holographic and QFT results in particular limits. 
The complexity of formation at high temperature increases linearly with the temperature both in the QFT model (eq.~\eqref{eq:largeTF1} and \eqref{eq:largeTK2})and in the BTZ holographic calculations for any of the complexity measures analyzed (\eqref{eq:CVBTZTO} and fig.~\ref{fig:GCcfCA},\ref{fig:GCcfCV}). The overall coefficient is proportional to $1/(1-\Omega^2)$ both for circuit complexity and CV. Also in the speed of light rotation limit, the complexity of formation diverges in all considered cases. However, the  leading divergence differs:  for the QFT case it goes like  $1/(1-\Omega)$, while in the holographic results it has an additional logarithmic factor (see  eq.~\eqref{eq:Omegadivholo} above). 

Finally, both CA and circuit complexity in the scalar model have arbitrary scales intrinsic to their definition. Although it is not clear yet how these scales are related on the two sides of the duality, we can make the following observation. In the QFT model the complexity of formation is independent from the arbitrary scale $\mu$ entering the definition of the reference state, exactly as in the neutral TFD case analyzed in \cite{Chapman:2018hou}. On the other hand here, in contrast with the non-rotating BTZ case, the holographic complexity of formation evaluated with CA  directly depends on the counterterm scale $L_{\rm ct}$. 
This contradicts the suggestion that $\mu$ and $L_{\rm ct}$ should be connected \cite{Chapman:2018lsv}, as was also argued based on different arguments in \cite{Bernamonti:2020bcf}.

\newpage 
\vskip 15pt \centerline{\bf Acknowledgments} \vskip 10pt \noindent 

We would like to thank Shira Chapman, Aldo Cotrone, Hugo Marrochio, Rob Myers and Domenico Seminara for useful comments and discussions. 
We thank the Galileo Galilei Institute (Florence) where part of this work has been carried out. This research was supported in part by Perimeter Institute for Theoretical Physics. Research at Perimeter Institute is supported by the Government of Canada through the Department of Innovation, Science and Economic Development and by the Province of Ontario through the Ministry of Research, Innovation and Science.  AB acknowledges support by the program ``Rita Levi Montalcini'' for young researchers and the INFN initiative GAST. 
FG has received funding from the European Union's Horizon2020 research and innovation programme under the Marie Sk\l{}odowska-Curie grant agreement No 754496.

\appendix

\section{BTZ:  complexity of formation in Boyer-Lindquist }\label{app:BTZBL}

In view of the study of 4d Kerr-AdS in sec.~\ref{sec:CVKerr}, we would now like to comment on a slightly different computation in which we work with BTZ in Boyer-Lindquist-like coordinates and evaluate the CV complexity formation. These coordinates are defined by \cite{Hawking:1998kw}  
\be
ds^2 = - \frac{\Delta_R}{R^2} \(dT - \frac{a}{\Xi} d\Phi\)^2 + \frac{R^2}{\Delta_R} \, dR^2 + \frac{1}{R^2} \( a  \,dT - \frac{R^2 +a^2}{\Xi} \, d\Phi\)^2\,, 
\ee
with
\bea
\Delta_R &\equiv& (R^2 + a^2) \(1+ \frac{R}{\ell^2}^2\) - 2 \tilde M R^2 \\
\Xi &\equiv& 1- \frac{a^2}{\ell^2} \,. 
\eea
This metric is related to \eqref{metrica} via the change of coordinates \cite{Hawking:1998kw}
\bea
T &=& t \\
R^2 &=& \Xi \,  r^2 - a^2 - \frac{2 \, a^2 \tilde M}{\Xi} \label{eq:KerrR}\\
\Phi &=& \phi - \frac{a}{\ell^2} \, t  
\eea
and parameters identifications
\bea
J &=& \frac{a\, \tilde M}{2 G_N \, \Xi^2} \\
M &=&  - \frac{1}{ 8 G_N} \(1 - \frac{2 \tilde M}{\Xi^2} \(1+\frac{a^2}{\ell^2}\)\)\,.
\eea
Notice that in these coordinates $\tilde M = 0$ parametrizes global AdS$_3$. 

\begin{figure}[h]
\begin{center}
\includegraphics[width=.5\linewidth]{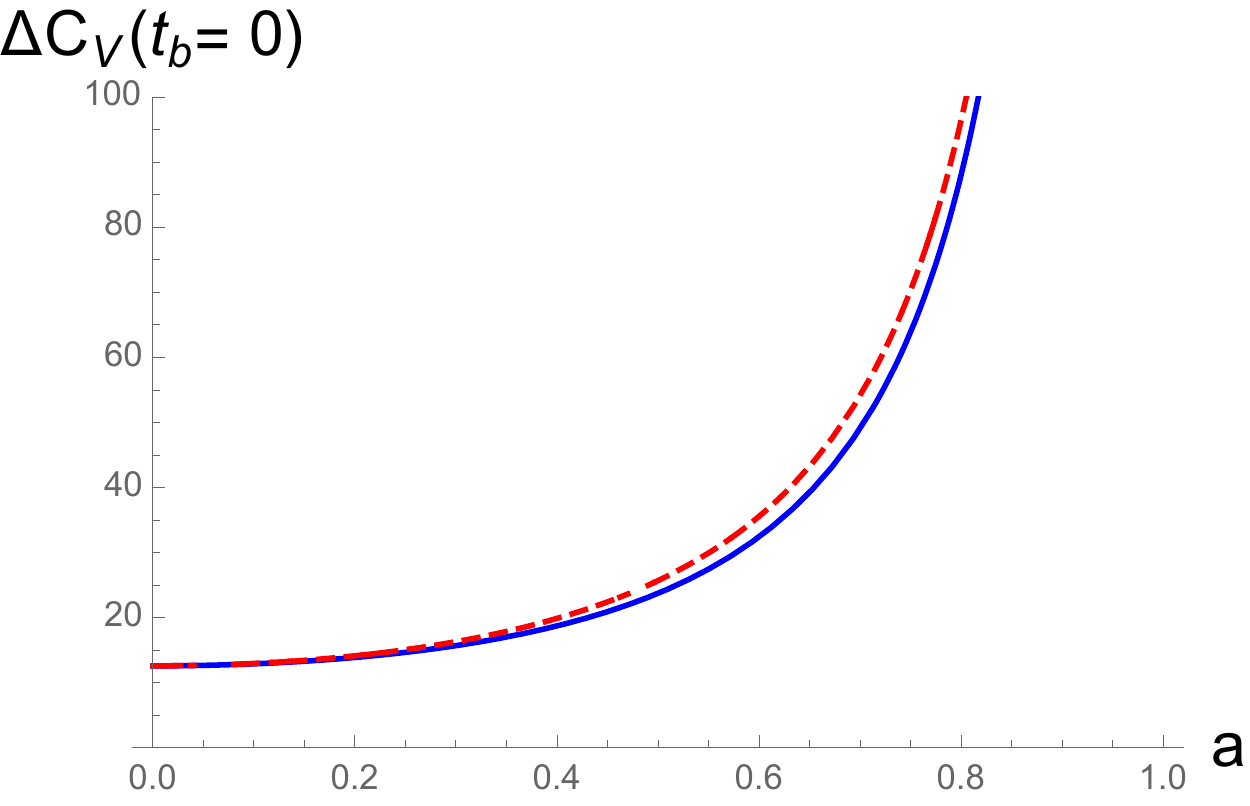}
\caption{In blue eq.~\eqref{eq:CVformationSt}  and in red dashed the BL result \eqref{eq:DeltaCVBTZBL}.}\label{fig:BTZBL}
\end{center}
\end{figure}

To evaluate CV in these coordinates we perform the integral 
\be
C_{\rm V}= \frac{4\pi}{G_N \ell \, \Xi}\int_{R_+}^{R_{\rm max}} dR \, \sqrt{\frac{(a^2 + R^2)^2}{\Delta_R} -a^2}\,,
\ee
where $R_+$ is the largest root of $\Delta_R = 0$ and maps to $r_+$ in \eqref{metrica}:
\be
R_\pm^2 \equiv  \frac{1}{2} \ell^2 \left[ \( 2\tilde M -1- \frac{a^2}{\ell^2} \) \pm \frac{1}{2} \sqrt{ \( 1+  \frac{a^2}{\ell^2} -2\tilde M \)^2 - \frac{4 a^2}{\ell^2}}  \right]\,.
\ee
Subtracting twice global AdS in these coordinates, that is the $\tilde M =0$ solution, we have
\bea
\Delta C_{\rm V} &=& \frac{4\pi}{G_N  \ell \, \Xi} \int_{R_+}^{R_{\rm max}} dR \, \sqrt{\frac{(a^2 + R^2)^2}{\Delta_R} -a^2} - \frac{4\pi}{G_N \ell } \int_{R_{\rm min}^{\rm AdS}}^{R_{\rm max}} dR \, \frac{R \ell}{\sqrt{\Xi (\ell^2 + R^2)}} 
\eea
in terms of a IR cutoff $R_{\rm max}$, which we can take to be same in both spacetimes. For $R_{\rm min}^{\rm AdS}$ in AdS, we would naturally set $R_{\rm min}^{\rm AdS} = 0$. Notice $\Delta C_{\rm V}$ evaluated in BL coordinates in this way differs by a finite term from  \eqref{eq:CVformationSt}, evaluated in standard coordinates. That is
\bea
\Delta C_{\rm V} &=& \frac{4\pi}{G_N \ell \, \Xi} \int_{R_+}^{R_{\rm max}} dR \, \sqrt{\frac{(a^2 + R^2)^2}{\Delta_R} -a^2} - \frac{4\pi}{G_N \ell  } \int_{0}^{R_{\rm max}} dR \, \frac{R \ell}{\sqrt{\Xi (\ell^2+ R^2)}} \\
&=& - \frac{4\pi  i \ell  }{G_N (1 - \frac{R_+^2 R_-^2}{\ell^4})} \Bigg\{\sqrt{-1+ \frac{R_+^2 R_-^2}{\ell^4}} - \frac{R_-}{\ell}(1+ \frac{R_+^2}{\ell^2}) E\(\sin^{-1} x, \frac{1}{x^2} \)\Bigg\} \label{eq:DeltaCVBTZBL}
\eea
with
\be
x= \frac{R_- (\ell^2 + R_+^2)}{\sqrt{-(R_+^2 - R_-^2)(\ell^4-R_+^2 R_-^2)}} \,. 
\ee
This expression can be written in explicitly real form using elliptic integrals identities and is plotted in figure~\ref{fig:BTZBL}.

To reproduce instead the result \eqref{eq:CVformationSt}, it is enough to formally extend the AdS integral to the complex value $R_{\rm min}^{\rm AdS} = \sqrt{-a^2}$, which we obtain setting $r=0$ and $\tilde M =0$ in \ref{eq:KerrR}. In this way we find
\bea
\Delta C_{\rm V} &=& \frac{4\pi}{G_N \, \Xi} \int_{R_+}^{R_{\rm max}} dR \, \sqrt{\frac{(a^2 + R^2)^2}{\Delta_R} -a^2} - \frac{4\pi}{G_N \ell } \int_{\sqrt{-a^2}}^{R_{\rm max}} dR \, \frac{R \ell }{\sqrt{\Xi (\ell^2 + R^2)}} \\
&=& \frac{4\pi \ell   }{G_N} \Bigg\{1 - i \ell \frac{R_-(\ell^2+ R_+^2) }{(\ell^4 - R_+^2 R_-^2 )}  E\(\sin^{-1} x, \frac{1}{x^2} \) \Bigg\} \,,
\eea
which after the appropriate parameters identifications can be verified to coincide with \eqref{eq:CVformationSt}.

\section{Kerr-AdS: UV cutoff and  complexity of formation}\label{app:div}

Here we study the boundary UV cutoff and CV divergences for the Kerr-AdS geometry. In the last paragraph, we also verify that the complexity of formation evaluated with respect to a rotating boundary metric by subtracting twice the complexity of the $m=0$ solution coincides with that evaluated with respect to a non-rotating frame by subtracting twice the complexity of the global AdS vacuum. 

\paragraph{UV cutoff.}  First we bring the Boyer-Lindquist metric \eqref{KerrAdSmetric} in Fefferman-Graham form \cite{fefferman1985elie,Fefferman:2007rka},
\be
ds^2 = \frac{\ell^2}{z^2} \(dz^2 + g_{ij} (x,z) dx^i dx^j\)\,,
\ee
with
\be \label{eq:gijFG}
g_{ij} (x,z) = g^{(0)}_{ij} (x) + z^2 g^{(1)}_{ij} (x) + z^3 g^{(3/2)}_{ij} (x) + \dots
\ee
through the asymptotic change of coordinates \cite{Papadimitriou:2005ii} 
\bea
r &=& \frac{\ell^2}{z} - \frac{1}{4} \(1 + \frac{a^2}{\ell^2} \sin^2 \bar \theta\) z + \frac{m}{3 \, \ell^2} z^2 + O(z^3) \label{eq:BLcutoff} \\
\theta &=& \bar \theta - \frac{a^2}{16 \, \ell^6} \,  \Delta_{\bar \theta} \sin (2\bar \theta) \, z^4 + O(z^6) \,. \label{eq:thetabar}
\eea
We also write explicitly the Taylor expansion \eqref{eq:gijFG} for $z \to 0$ of the non-vanishing components of $g_{ij} (x,z)$, which reads
\bea
g_{tt} &=& - 1 - \frac{ 1}{ 2\, \ell^2} \( \Delta_{\bar \theta} + \frac{a^2}{\ell^2}\) z^2 - \frac{ 4 m}{ 3 \, \ell^4} \, z^3  +  O(z^4) \\
g_{\bar \theta \bar \theta} &=& \frac{1}{\Delta_{\bar \theta}}  \left\{\ell^2 + \frac 1 2 \(2 -3 \Delta_{\bar \theta} - \frac{a^2}{\ell^2}\) z^2 + \frac{ 2 m}{ 3 \, \ell^2} \, z^3 \right\} +  O(z^4)\\
g_{t \varphi} &=&  \frac{a \sin^2 \bar \theta}{\Xi}   \left\{1 - \frac{ 1}{ 2 \, \ell^2} \( \Delta_{\bar \theta} - \frac{a^2}{\ell^2}\) z^2 - \frac{ 4 m}{ 3 \ell^2} \, z^3 \right\} +  O(z^4) \\
g_{\varphi \varphi} &=& \frac{\sin^2 \bar \theta}{\Xi}   \left\{ \ell^2 - \frac 1 2 \( \Delta_{\bar \theta} - \frac{a^2}{\ell^2}\) z^2 - \frac{ 2 m}{ 3 \, \ell^2 \, \Xi}  \( 2 - 3 \Delta_{\bar \theta} - 2\frac{a^2}{\ell^2}\)\, z^3 \right\} +  O(z^4) \,. 
\eea

Given these results, as per the standard holographic procedure, we set the UV cutoff $z = \delta$ in FG coordinates, which via \eqref{eq:BLcutoff}-\eqref{eq:thetabar} corresponds to the $\theta$-dependent BL cutoff:
\be
r_{\rm max} = \frac{\ell^2}{\delta} - \frac{\delta}{4} \(1 + \frac{a^2}{\ell^2} \sin^2 \theta\)  + \frac{m}{3 \, \ell^2} \, \delta^2 +\dots \,. \label{eq:UVcut}
\ee

\paragraph{Volume divergences.} 
To obtain the  divergences for the volume of the $t=0$ slice of  Kerr-AdS we follow the general analysis for AAdS spacetimes carried out in \cite{Carmi:2016wjl}.

We describe the codimension-1 submanifold  via the  embedding $X^\mu = X^\mu (\tau, \sigma^a)$, where $X^\mu = \{ z,x^i\}$ and $\{\tau, \sigma^a\}$ are intrinsic coordinates on the submanifold. In the $\tau =z$ gauge, the induced metric $h$ on the bulk $t=0$ surface has components
\bea
h_{zz} &\equiv & \frac{\ell^2}{z^2} \(1 + h^{(1)}_{zz} z^2 + \dots \) =  \frac{\ell^2}{z^2} \\
h_{\bar \theta \bar \theta} &\equiv&  \frac{\ell^2}{z^2} \(h^{(0)}_{\bar \theta \bar \theta} +h^{(1)}_{\bar \theta \bar \theta} z^2 + \dots \) =  \frac{\ell^2}{z^2} \,g_{\bar \theta \bar \theta}\\
h_{\varphi \varphi} &\equiv&  \frac{\ell^2}{z^2} \(h^{(0)}_{\varphi \varphi} +h^{(1)}_{\varphi \varphi} z^2 + \dots \) =  \frac{\ell^2}{z^2}\, g_{\varphi \varphi}
\eea
where $h^{(0)}_{ab}$ is the induced metric on the boundary time slice. We then introduce a cutoff at $z =\delta$ and evaluate explicitly the divergent terms worked out in \cite{Carmi:2016wjl}
\be
V = \frac{\ell^3}{2} \int_0^\pi d\bar \theta \int_0^{2\pi} d\varphi  \, \sqrt{h^{(0)}} \left[ \frac{1}{\delta^2} + \(\mathcal{R}_a^a - \frac{ \mathcal{R}}{ 2} \) \log\frac{\delta}{\ell}\right] + \dots \,.
\ee
Here $\mathcal{R}_a^a = h^{(0)ab} \mathcal{R}_{ab}$, and $\mathcal{R}_{ab}$ denotes the projection of the boundary Ricci tensor $\mathcal{R}_{ij} [g^{(0)}]$ into the time slice, while $\mathcal{R} [g^{(0)}]$ is the boundary Ricci scalar. These are related through \cite{Carmi:2016wjl}
\be
\mathcal{R}_a^a = \frac{\mathcal{R}}{2} - h^{(0)ab}h^{(1)}_{ab}
\ee
and hence 
\bea
V &=& \frac{\ell^3}{2} \int_0^\pi d\bar \theta \int_0^{2\pi} d\varphi  \, \sqrt{h^{(0)}} \left[ \frac{1}{\delta^2} - h^{(0)ab} h^{(1)}_{ab} \, \log\frac{\delta}{\ell}\right] + \dots \\
&=& \frac{\pi \ell^3 }{ \sqrt{\Xi}} \int_0^\pi d\bar \theta \, \frac{\sin \bar \theta}{\sqrt{ \Delta_{\bar \theta}}} \left[ \frac{\ell^2}{\delta^2} + \( 2\Delta_{\bar \theta} -1 \) \, \log\frac{\delta}{\ell}\right] +\dots \\
&=& \frac{2 \pi \ell^6}{a \sqrt{\Xi} \, \delta^2} \, \sin^{-1} \frac{a}{\ell} + 2\pi \ell^3 \, \log\frac{\delta}{\ell}+ \dots
\eea
The divergence structure of $V$ thus depends on the parameter $a$ but it is completely independent form the mass parameter $m$.

\paragraph{Schwarzschild-like coordinates.}

To make further contact with the BTZ case for which we mainly focused on the coordinates system \eqref{metrica}, we need to consider Schwarzschild-like coordinates. This is the situation in which the background boundary metric is not rotating and all rotation is in the states. We saw for BTZ in sec.~\ref{sec:CVBTZ} and  app.~\ref{app:BTZBL}  this accounts for an additional finite contribution in $\D C_{\rm V}(t_b=0)$. Instead of evaluating this contribution by performing a complicated change of coordinates to such coordinates system, here we can simply consider the implicit coordinate transformation \cite{Hawking:1998kw} 
\bea\label{newcoords} 
T&=&t  \\ 
\Phi &=& \varphi + \frac{a t}{\ell^2}  \\
y\cos\Theta&=& r\cos\theta \\
y^2 &=& \frac{1}{\Xi} \pq{r^2\D_{\theta}+a^2\sin^2\theta} 
\eea
that brings the $m=0$ Kerr-AdS$_4$ metric in Boyer-Lindquist coordinates to global AdS$_4$
\be 
ds^2 = - \(1+ \frac{y^2}{\ell^2}\) dT^2 + \frac{dy^2 }{1+ \frac{y^2}{\ell^2}} + y^2 \(d \Theta^2 + \sin^2 \Theta d\Phi^2\)\,. 
\ee
Notice this maps $y=0$ in global AdS to complex (purely imaginary) $r_{\rm min} \equiv \sqrt{ - a^2}, \theta = \frac \pi 2$ in BL coordinates. This same change of coordinates can be in principle applied to bring the Kerr-AdS metric to an asymptotically non rotating frame, and we can thus evaluate the complexity of formation of excited rotating thermofield double states on $\mathbb R \times S^2$ as
\bea\label{eq:CVKAdS-standard}
\D C_{\rm V}^{\rm Schw} (t_b=0) &=& 
\frac{4\pi}{G_N \,\ell} \int_0^{\pi}d\theta\sin\theta \int_{r_+}^{r_{\rm max}(y_{\rm max})} dr \, \frac{\r}{\Xi} \, \sqrt{\frac{(r^2+a^2)^2}{\D} - \frac{a^2 \sin^2\theta}{\D_{\theta}}} \nonumber \\
&& - \frac{4\pi}{G_N} \int_0^{\pi}d\Theta\sin\Theta \int_{0}^{y_{\rm max}} dy \, \frac{y}{\sqrt{\ell^2 + y^2}} \,, 
\eea
where 
\be
r_{\rm max}^2(y_{\rm max}) =\frac{ \Xi y_{\rm max}^2 - a^2 \sin^2 \theta}{\Delta_\theta}\,.
\ee
This coincides identically with $\D C_{\rm V} (t_b=0)$ \eqref{CVformationKAdS}, as follows from the independence of the divergences on $m$ analyzed above and the structure of the AdS term, which has vanishing contribution from the lower integration extremum $y = 0$.

\section{Circuit complexity: modes and $\mu$ dependence}\label{app:plots}

\begin{figure}[ht]
\centering
\includegraphics[width=.4\linewidth]{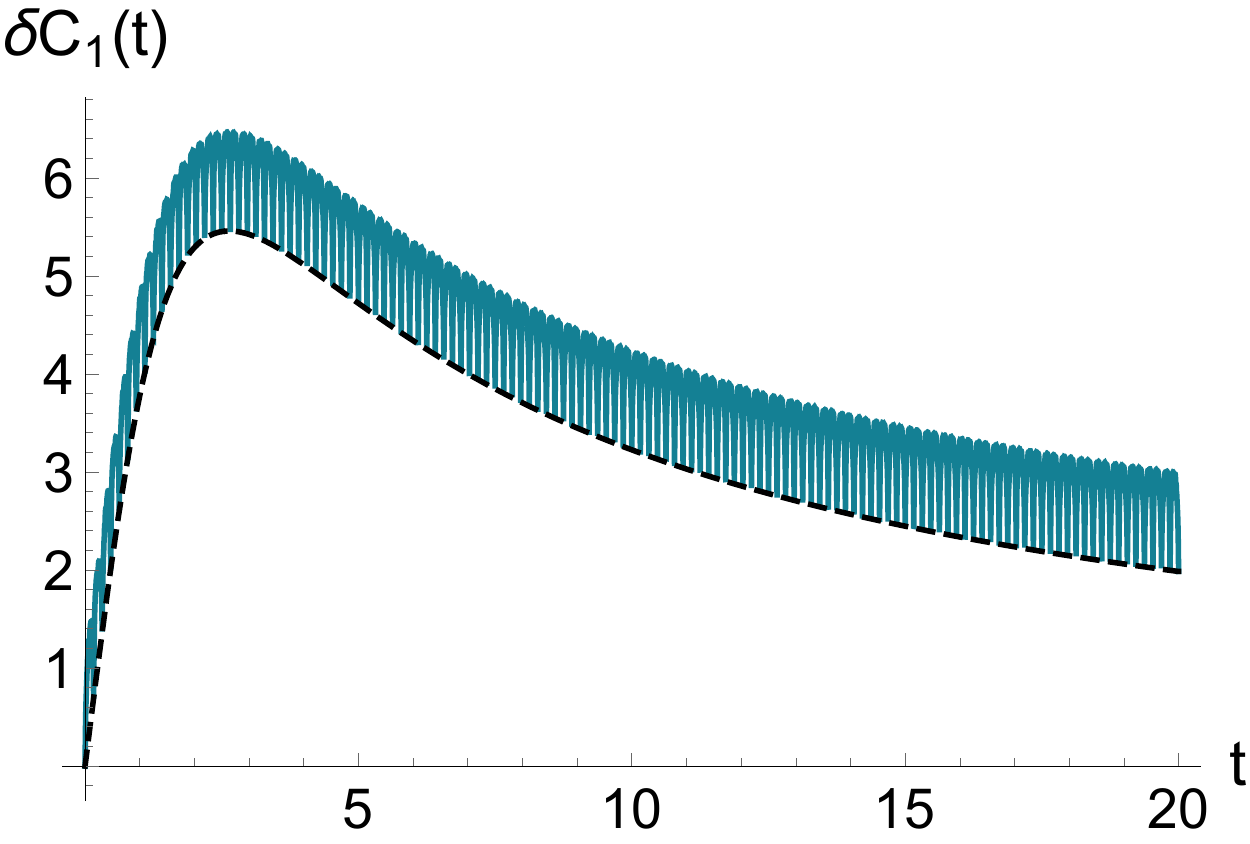} \hfill
\includegraphics[width=.4\linewidth]{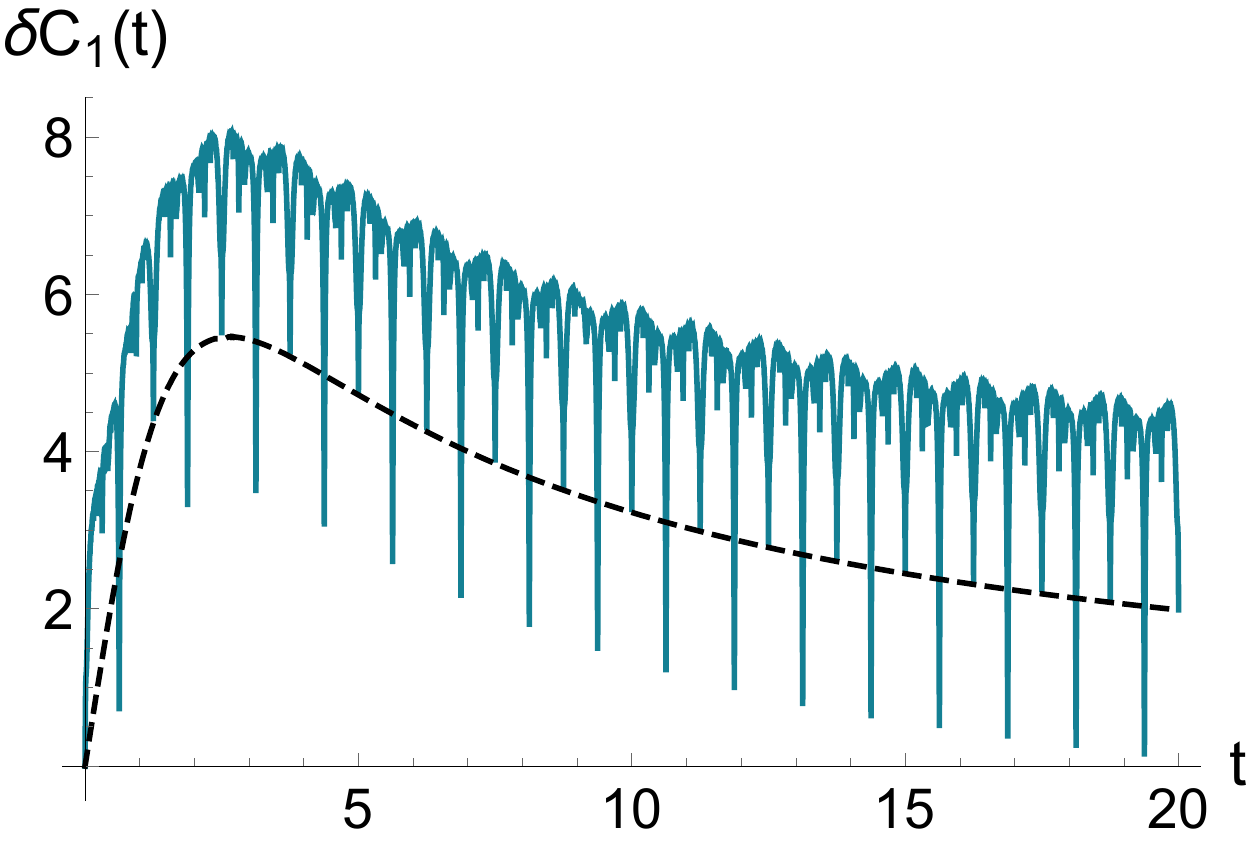} \\
\includegraphics[width=.4\linewidth]{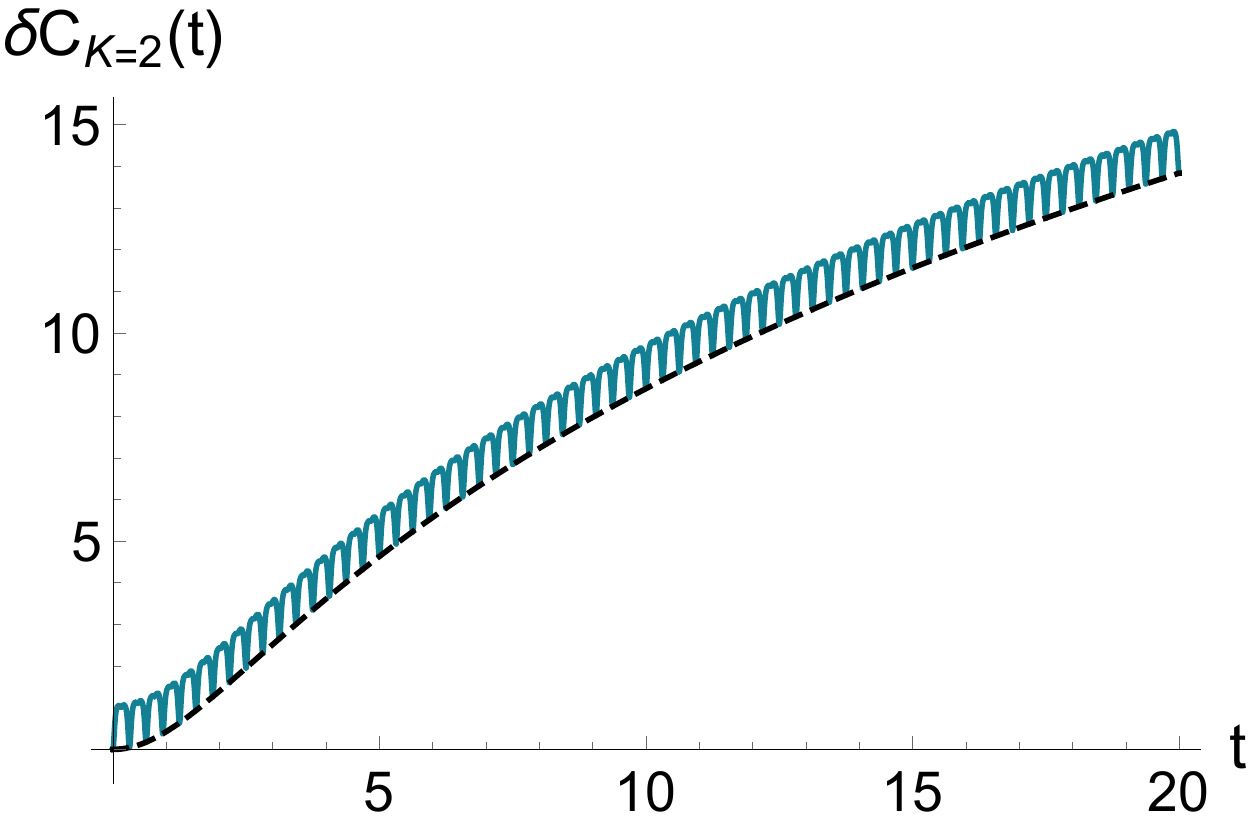} \hfill
\includegraphics[width=.4\linewidth]{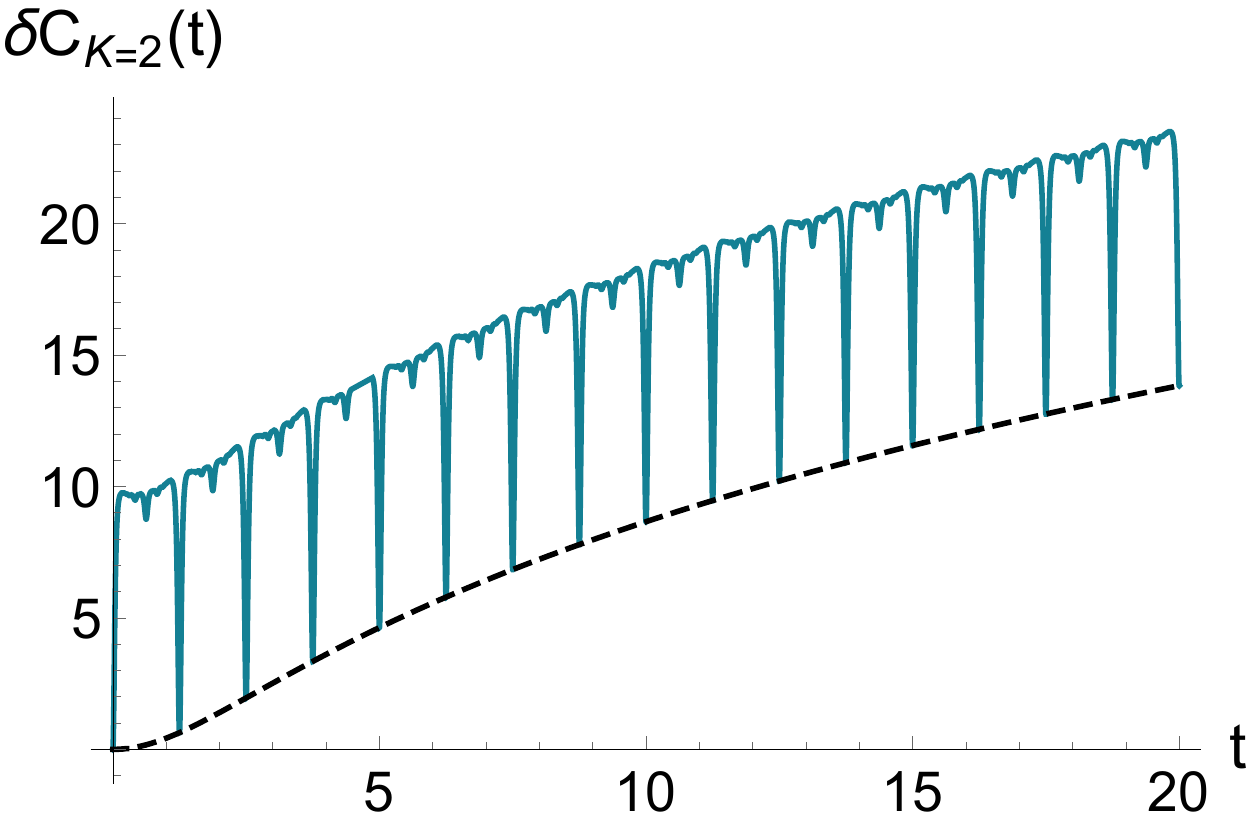} \\
\includegraphics[width=.4\linewidth]{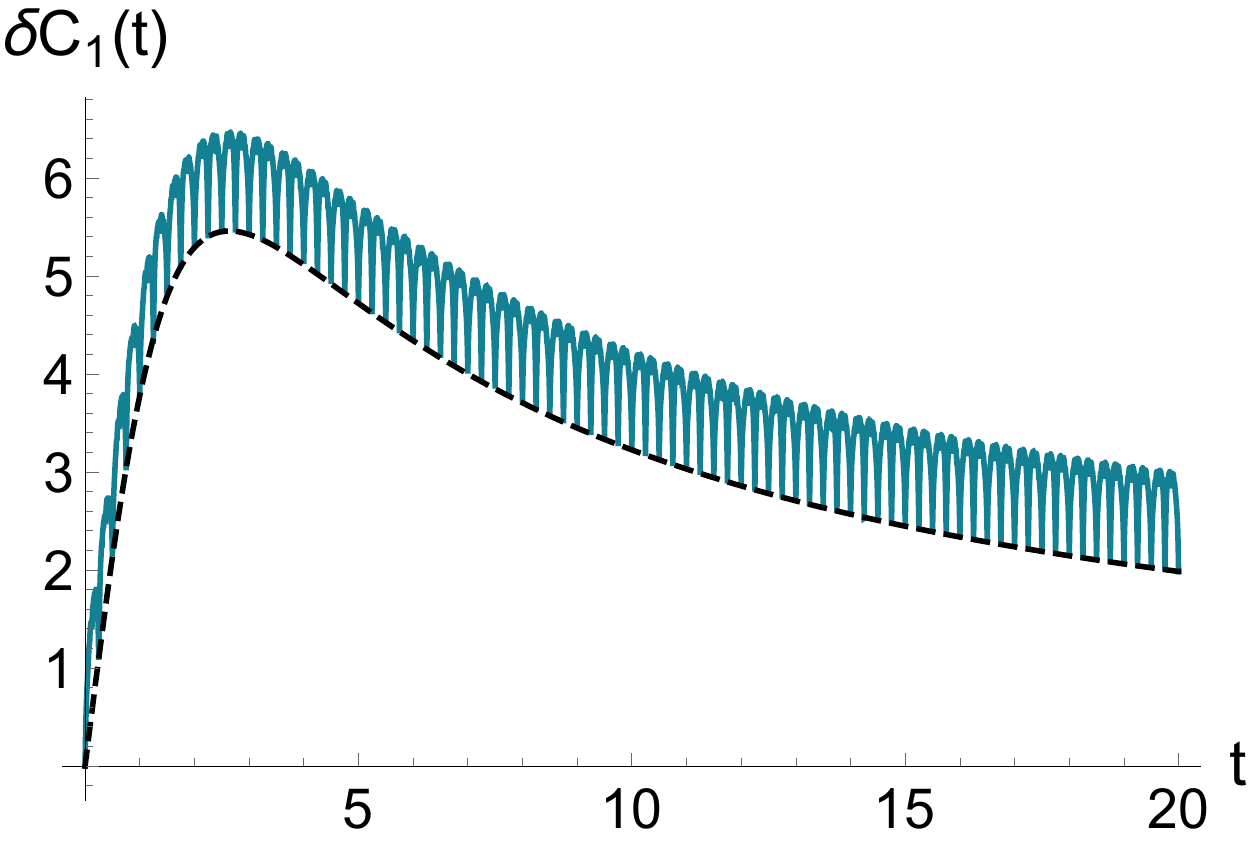} \hfill
\includegraphics[width=.4\linewidth]{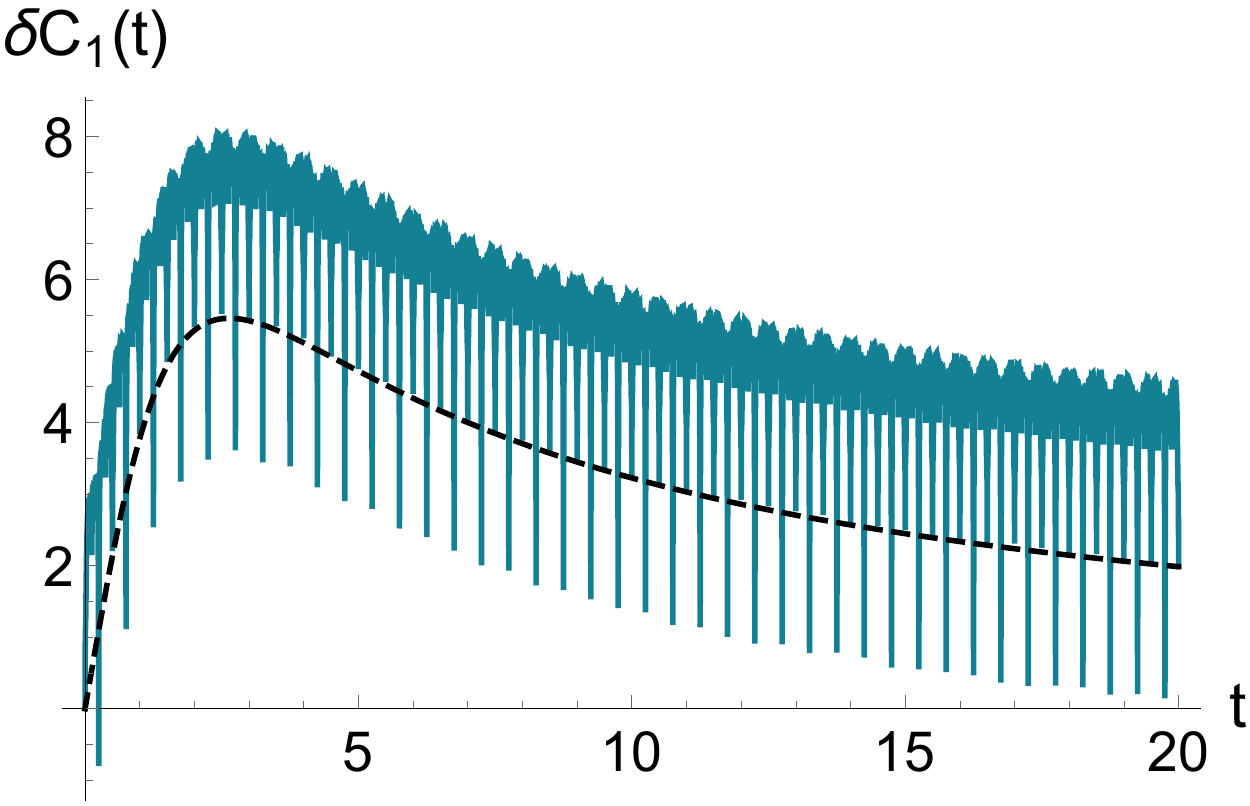} \\
\includegraphics[width=.4\linewidth]{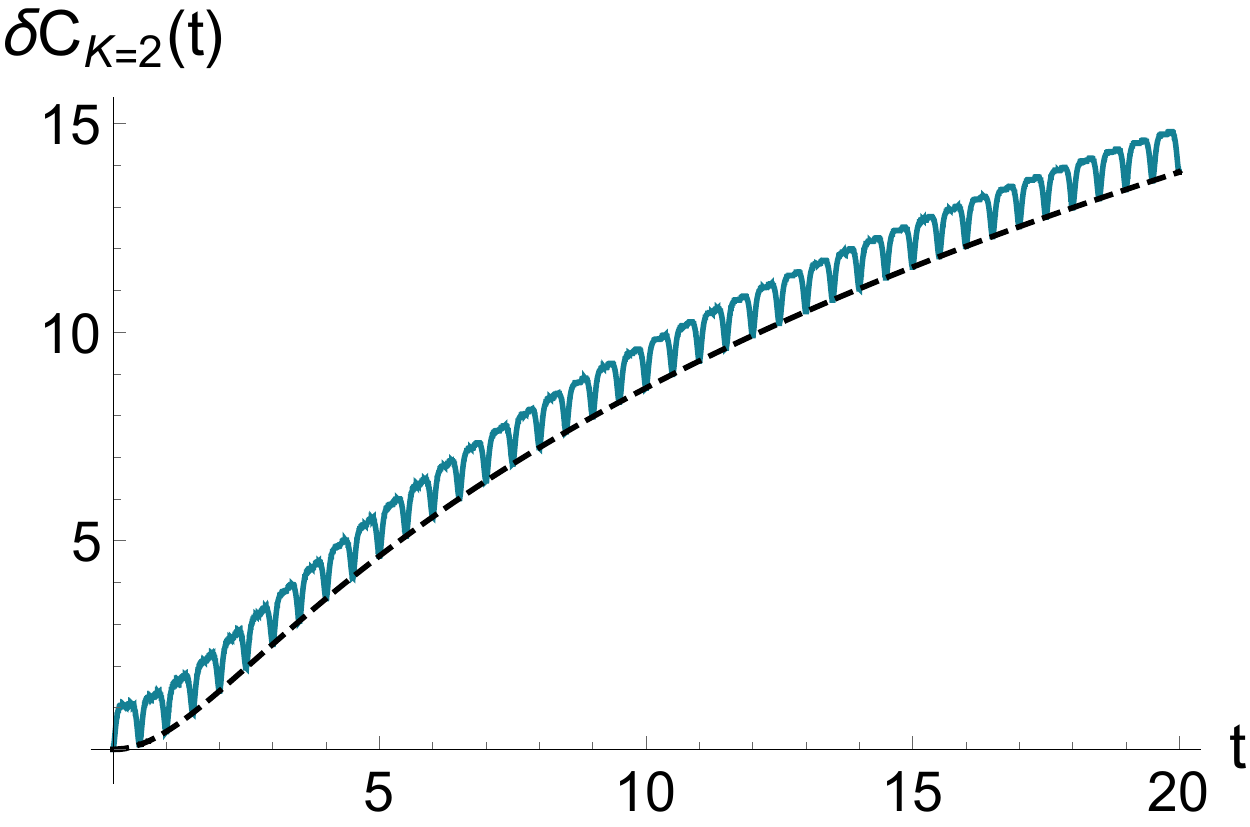} \hfill
\includegraphics[width=.4\linewidth]{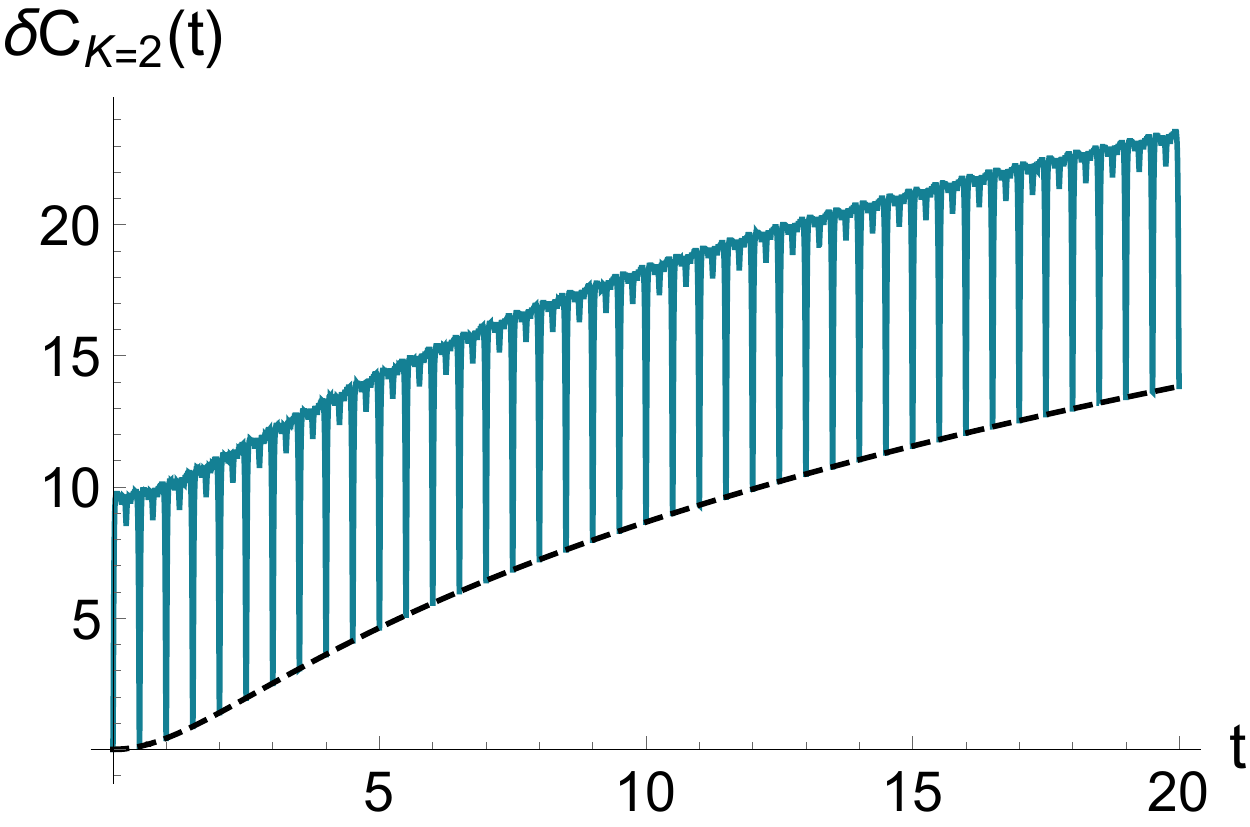}
\centering
\caption{Positive (left) and negative modes (right) contributions to the complexity variation $\d\mathcal{C}(t)$  for the Hamiltonian evolutions $H + \Omega J$ (first two rows) and $H$ (last two rows). Both positive and negative modes sum include the zero mode contribution (in dashed black). The plots have been produced choosing $L=1$, $N=200$, $\m=1$, $m=10^{-6}$, $\O=0.6$, $T=10$ as in the central panels of fig.~\ref{fig:Ftimeconformal} and in fig.~\ref{fig:FtimeconformalHonly}.}\label{fig:PosNegTime} 
\end{figure}

\paragraph{Mode analysis.} 
In fig.~\ref{fig:PosNegTime}  we show separately negative and positive modes contribution to the complexity variation $\d\mathcal{C}(t)$. The plots explicitly illustrate the different periodicity of negative and positive modes, as reported in  \eqref{eq:periodicity}, and the enhancement of the negative mode amplitudes over the positive ones, as discussed in the main text. For the time evolution with $H$ only (last two rows), there is no difference in the periodicity of negative and positive modes, but the amplitude enhancement persists.

\begin{figure}[h]
\centering
\includegraphics[width=.4\linewidth]{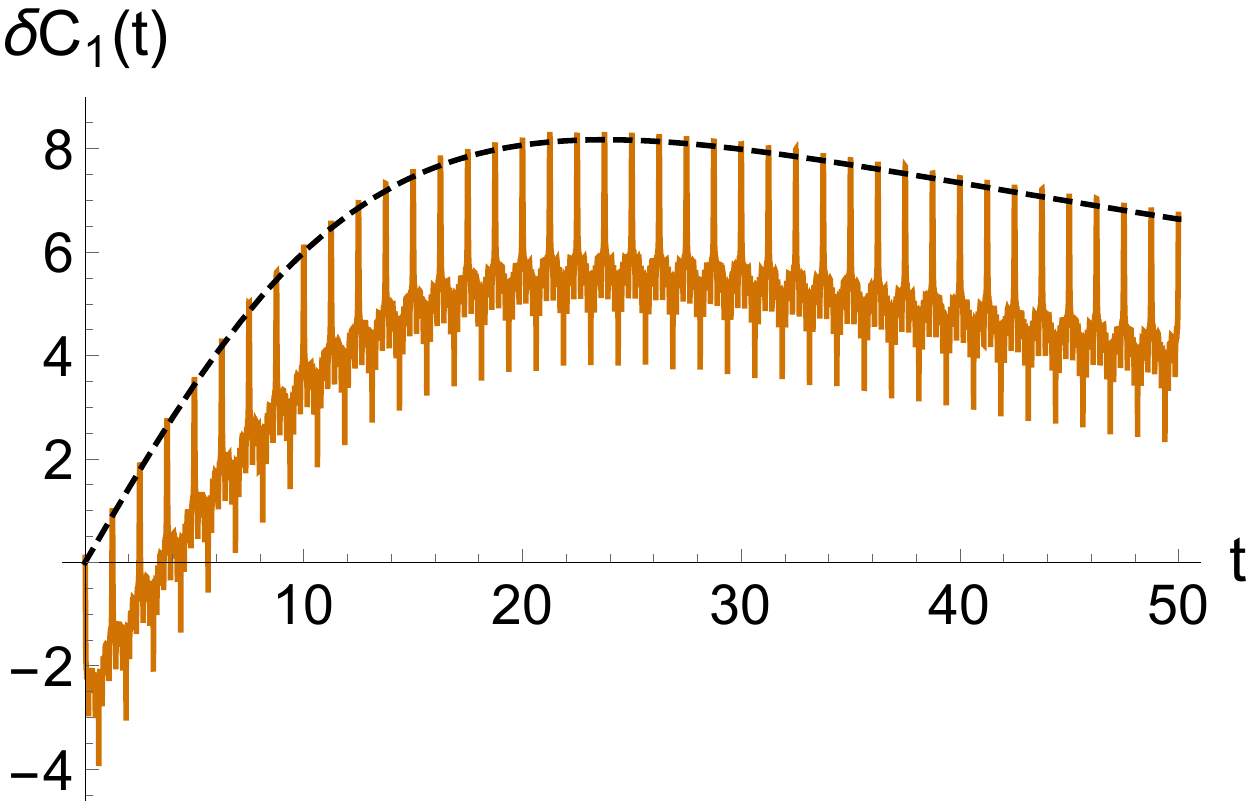} \hfill
\includegraphics[width=.4\linewidth]{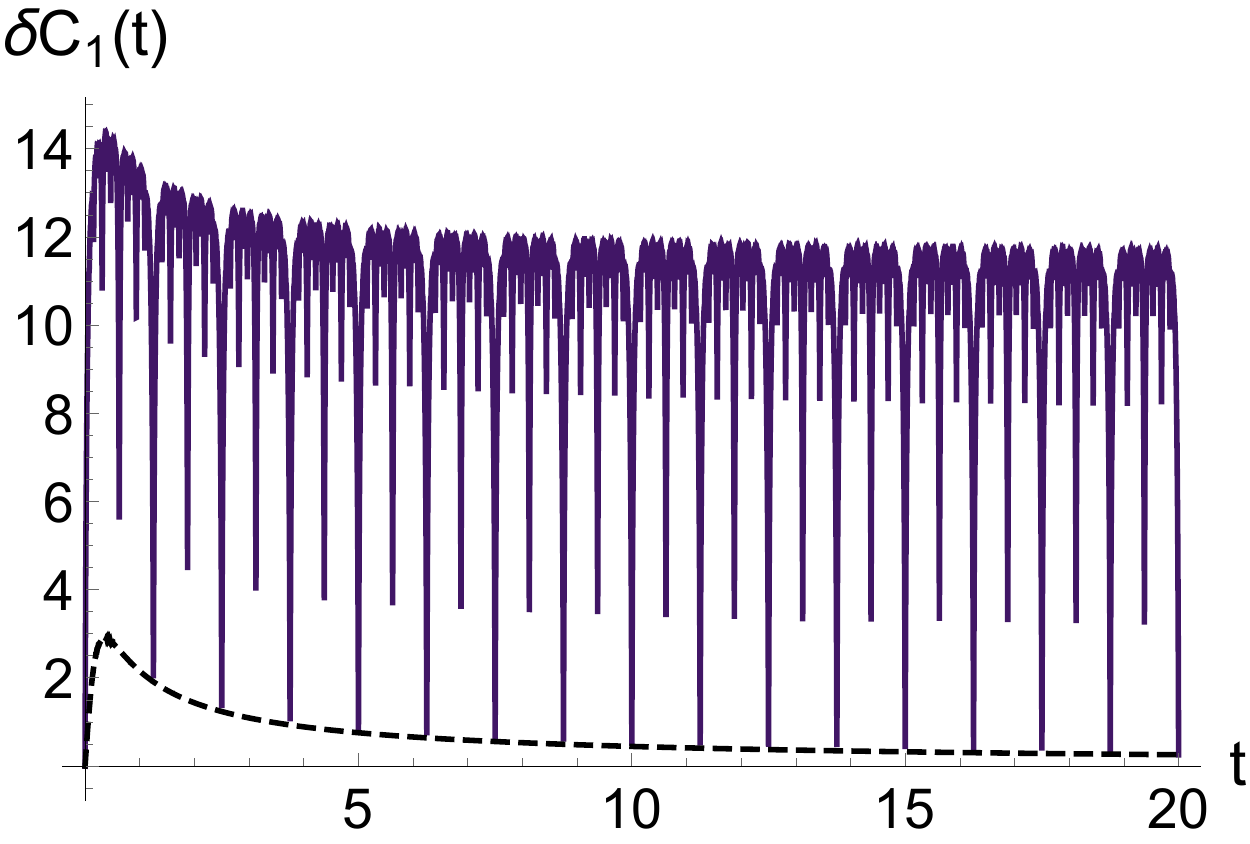} \\
\includegraphics[width=.4\linewidth]{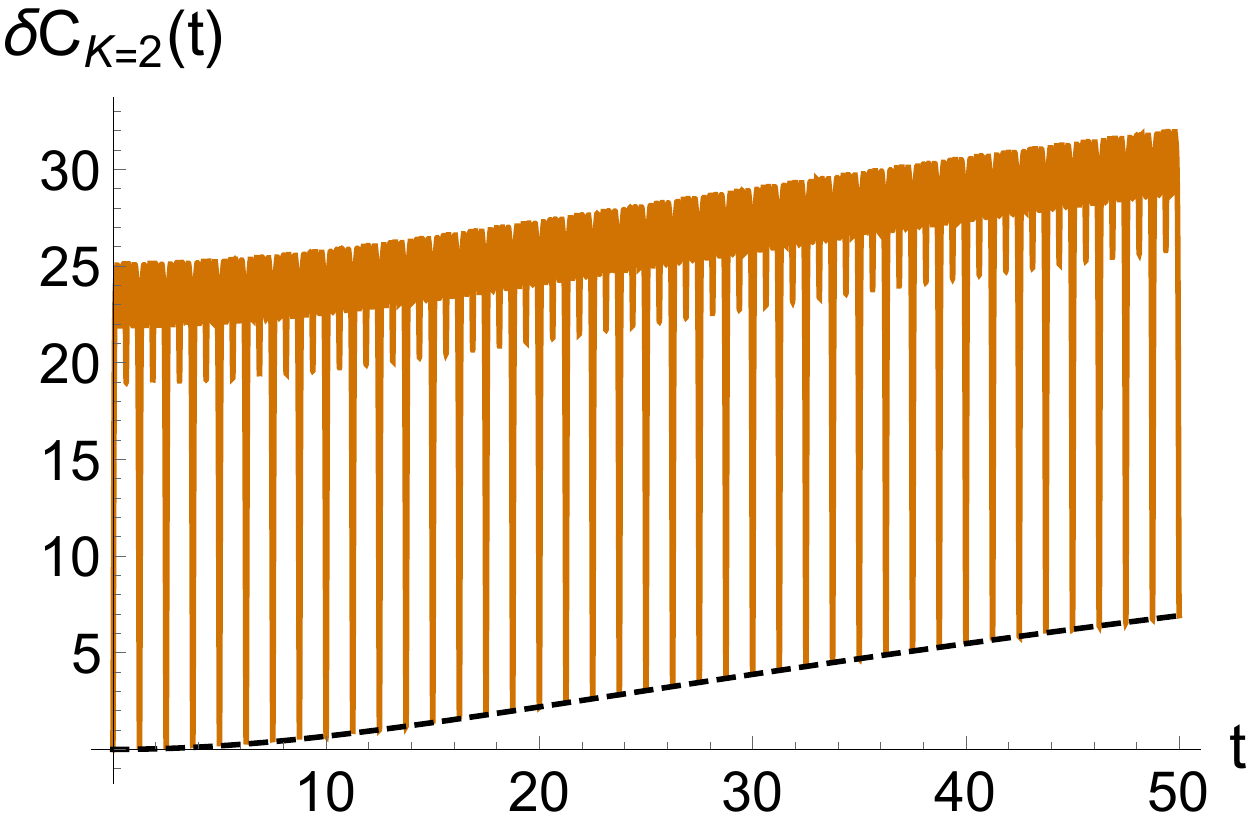} \hfill
\includegraphics[width=.4\linewidth]{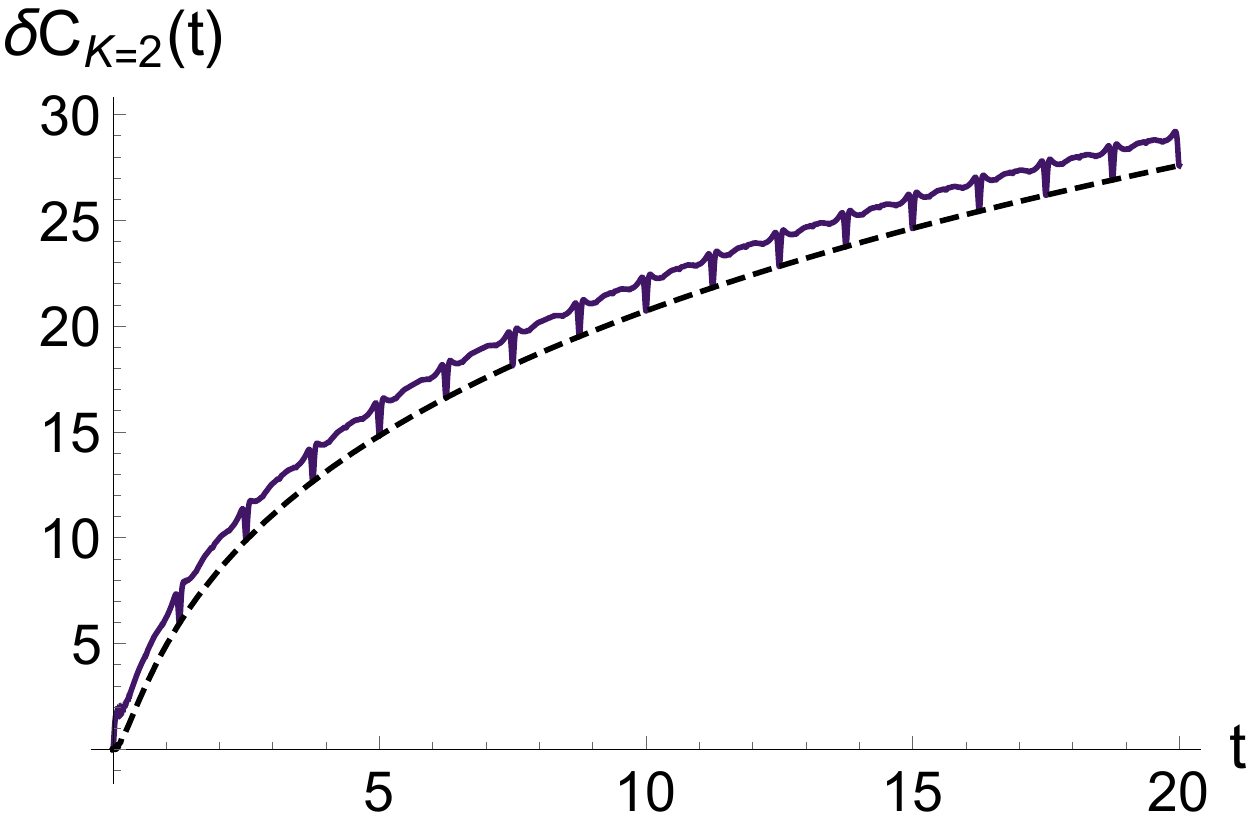} \\
\includegraphics[width=.4\linewidth]{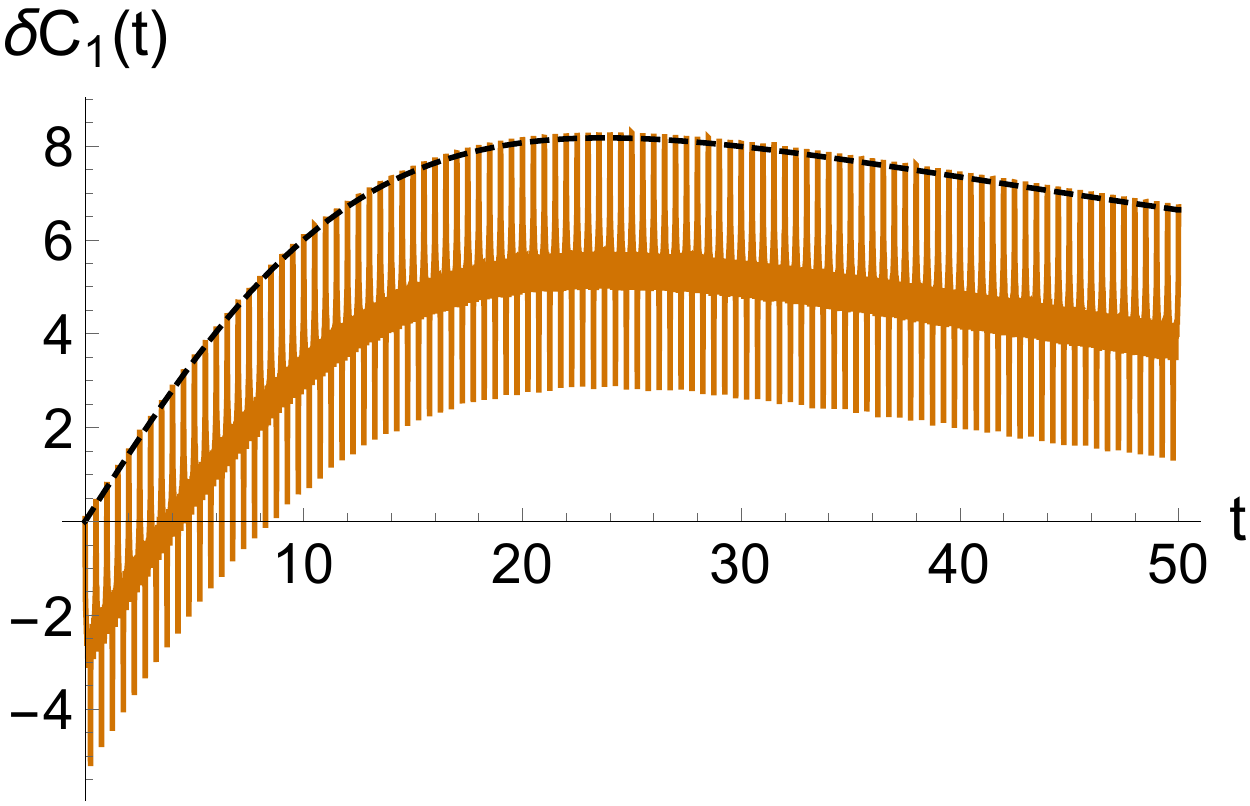} \hfill
\includegraphics[width=.4\linewidth]{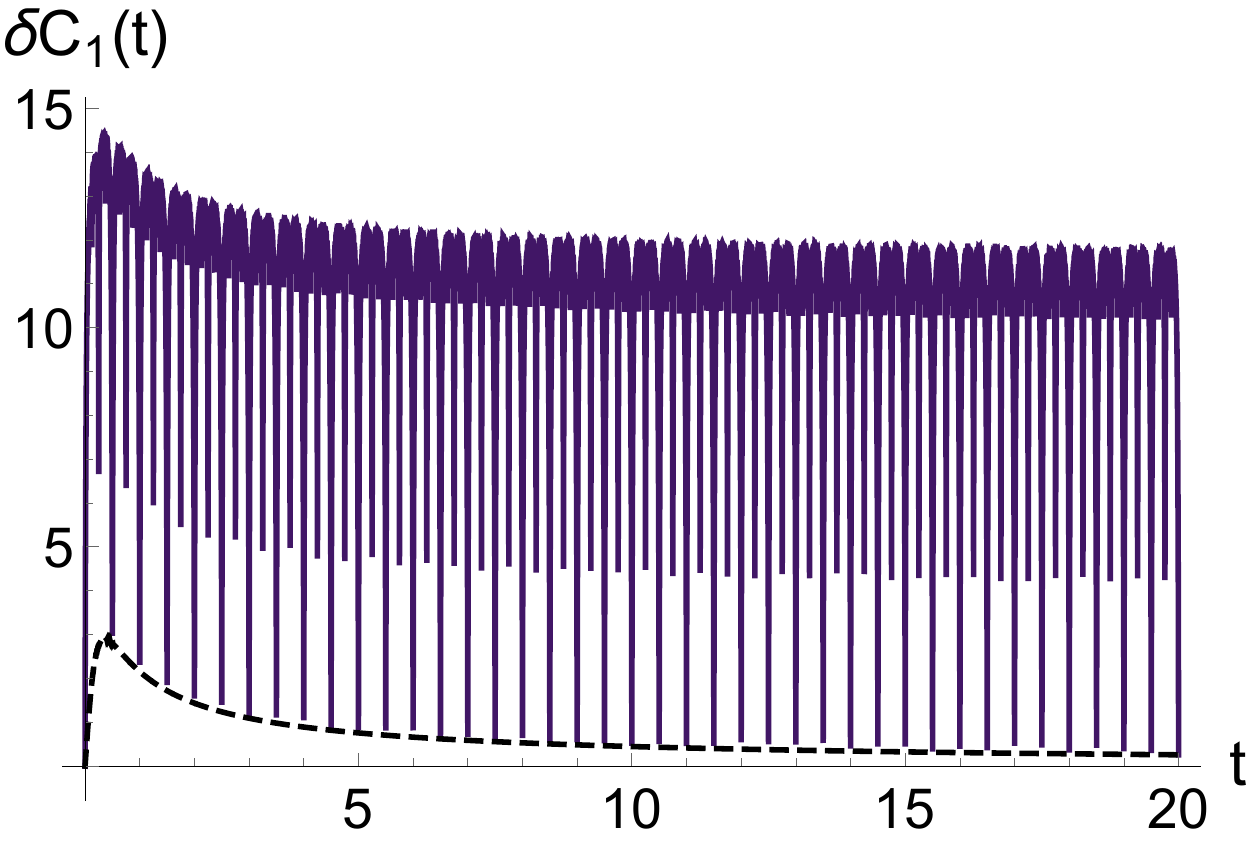} \\
\includegraphics[width=.4\linewidth]{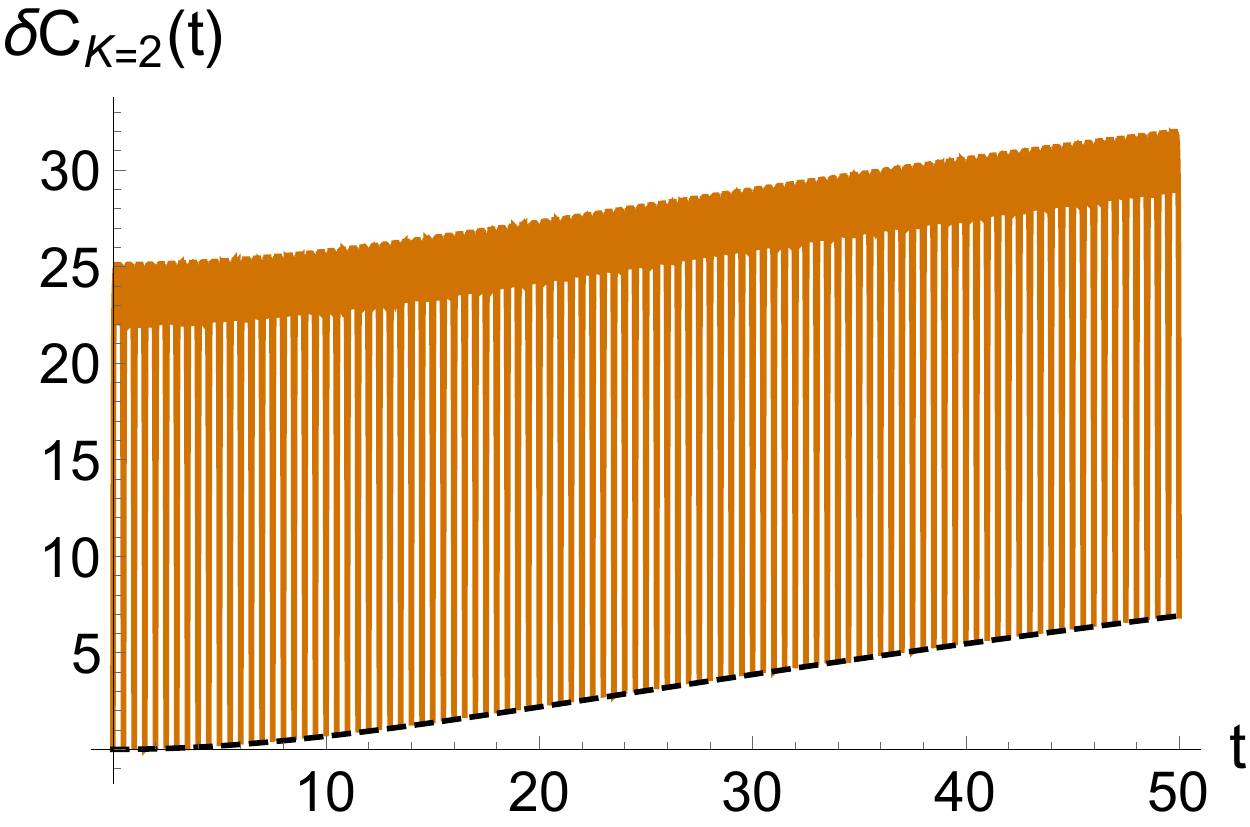} \hfill
\includegraphics[width=.4\linewidth]{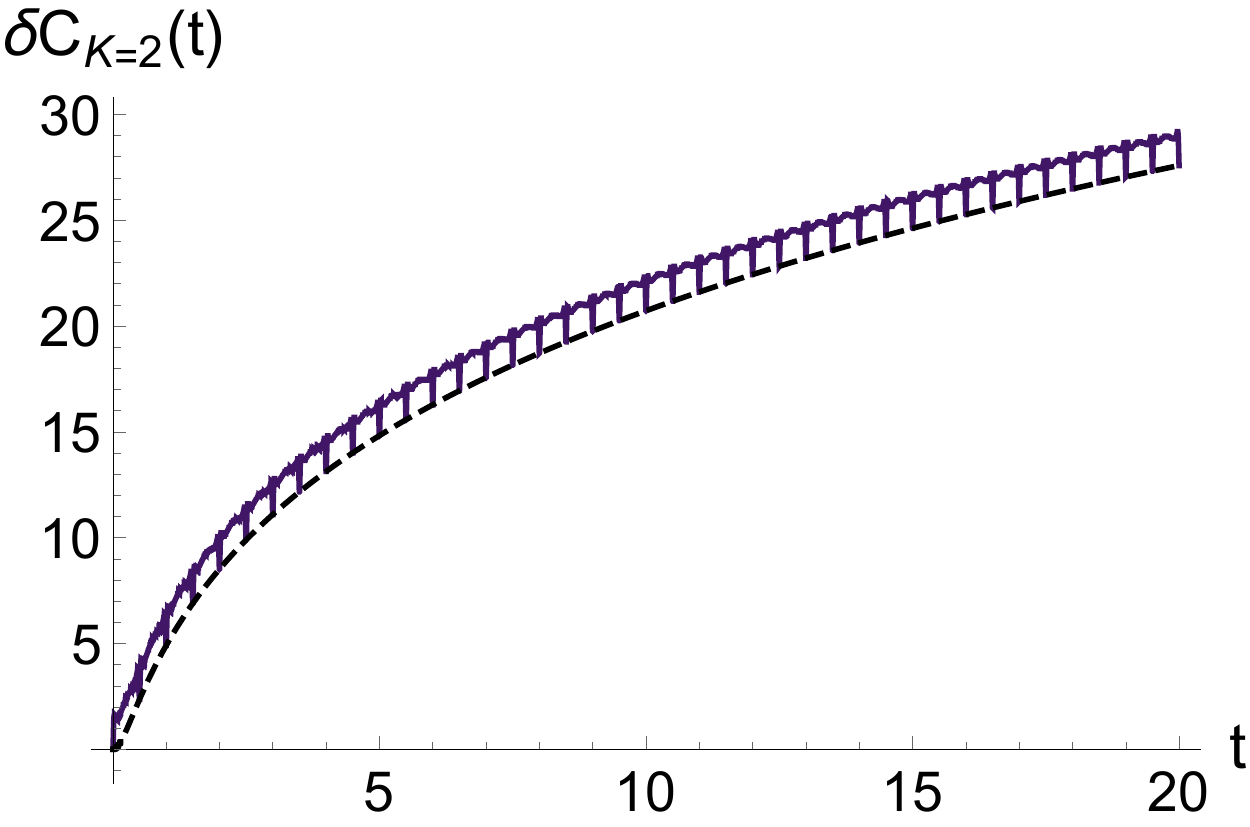}
\centering
\caption{Complexity variation $\d\mathcal{C}(t)$ obtained time-evolving with $H+\O J$ (first two rows) and $H$ (last two rows). The plots have been produced choosing $L=1$, $N=200$, $m=10^{-6}$, $\O=0.6$, $T=10$, as in the central panels of fig.~\ref{fig:Ftimeconformal}. The left panels show the result for $\mu=0.1$, and the right panels for  $\mu=10$. While the periodicity of the oscillations is independent from $\mu$,  the amplitude is clearly affected by it.}\label{fig:TdepHighMu} 
\end{figure}

\paragraph{Dependence on $\mu$.} 
In fig.~\ref{fig:TdepHighMu} 
we report sample plots illustrating how the value of the reference state scale $\mu$ influences the picture of the complexity time evolution.  
%

\clearpage

\bibliography{References}{}
\bibliographystyle{JHEP.bst}
\end{document}